\documentclass[a4paper,twoside,english]{report}
\usepackage{geometry}
\usepackage{graphicx}
\usepackage{epsfig}
\usepackage{amssymb}
\usepackage{stmaryrd}
\graphicspath{{eps/}}
\newskip\centreskip
\def\centre#1{\hbox to \hsize{\hskip\centreskip \hbox{#1}\hskip\centreskip}}

\def\color[#1]#2{}

\def\bfseries{\fontseries \bfdefault \selectfont \boldmath}

\def\del{\partial}

\def\frac#1#2{\mathinner{#1 \over #2}}

\def\Chi{\hbox{\raise 2pt \hbox{$\chi$}}}
\def\asctilde{\lower 3pt \hbox{\~{}}}

\def\mbf#1{\mathchoice{\hbox{\boldmath $\displaystyle #1$}}
	{\hbox{\boldmath $\textstyle #1$}}{\hbox{\boldmath $\scriptstyle #1$}}
	{\hbox{\boldmath $\scriptscriptstyle #1$}}}
\def\bDelta{\mbf\Delta}
\def\bdelta{\mbf\delta}

\def\pom{{\mathchoice{\hbox{$\displaystyle\mathbb P$}}
	{\hbox{$\textstyle\mathbb P$}}{\hbox{$\scriptstyle\mathbb P$}}
	{\hbox{$\scriptscriptstyle\mathbb P$}}}}
\def\odd{{\mathchoice{\hbox{$\displaystyle\mathbb O$}}
	{\hbox{$\textstyle\mathbb O$}}{\hbox{$\scriptstyle\mathbb O$}}
	{\hbox{$\scriptscriptstyle\mathbb O$}}}}
\def\reg{{\mathchoice{\hbox{$\displaystyle\mathbb R$}}
	{\hbox{$\textstyle\mathbb R$}}{\hbox{$\scriptstyle\mathbb R$}}
	{\hbox{$\scriptscriptstyle\mathbb R$}}}}

\def\Log{\mathop {\csname operator@font\endcsname Log}\nolimits}

\def\dblone{\hbox{$1\hskip -1.2pt\vrule depth 0pt height 1.6ex width 0.7pt
            \vrule depth 0pt height 0.3pt width 0.12em$}}

\def\shown#1{\expandafter\show\csname\string#1\endcsname}
\def\showthen#1{\expandafter\showthe\csname\string#1\endcsname}
\def\don#1{\csname\string#1\endcsname}
\def\defn#1{\expandafter\def\csname\string#1\endcsname}

\def\ct#1{{\em\rightskip=0pt plus 2em \spaceskip=.35em \xspaceskip=.35em
					\relax #1}}

\def\setsize{\csname @setfontsize\endcsname \setsize}

\def\putover#1#2{\mathrel{
\setbox0=\hbox{#1}\setbox1=\hbox{\scriptsize #2}
\dimen0=-0.5\wd0 \advance\dimen0 by -0.5\wd1
\dimen1=0.5\wd0 \advance\dimen1 by -0.5\wd1
\hbox{\box0\kern\dimen0%
\vbox to 0pt {\vss\hbox{\raise 0.7em \box1}}%
\kern\dimen1}
}}

\def\putunder#1#2{\mathrel{
\setbox0=\hbox{#1}\setbox1=\hbox{\scriptsize #2}
\dimen0=-0.5\wd0 \advance\dimen0 by -0.5\wd1
\dimen1=0.5\wd0 \advance\dimen1 by -0.5\wd1
\hbox{\box0\kern\dimen0%
\vbox to 0pt {\hbox{\lower 0.7em \box1}\vss}%
\kern\dimen1}
}}

\def\dmath#1#2{
$$\lineskiplimit=1000pt \advance\lineskip by #1\jot 
\mathsurround=0pt \tabskip=0pt plus 1000pt
\everycr{\noalign{\penalty\interdisplaylinepenalty}}
\halign to \displaywidth{
\hfil$\displaystyle{##}$\tabskip=0pt&%
\hfil $\displaystyle{{}##{}}$\hfil &%
$\displaystyle{##}$\hfil \tabskip=0pt plus 1000pt minus 1000pt&%
\refstepcounter{equation}\label{##}\llap{(\theequation)}\tabskip=0pt\cr
\noalign{\ifdim \prevdepth>-1000pt \vskip -#1\jot\fi}
#2\crcr}$$}

\def\crd{\crcr\noalign{\unpenalty
\multiply\interdisplaylinepenalty by 2 \penalty \interdisplaylinepenalty}}

\def\crl#1{\crcr\noalign{\unpenalty\penalty 10000
\nointerlineskip \vbox to 0pt {
\dimen0=\lineskip \vskip \dimen0 minus 1000pt \hbox to \displaywidth{%
\hfil \refstepcounter{equation}\label{#1}(\theequation)}
\vskip 0pt minus 1000pt} \penalty 10000}}

\def\mathtab#1#2{\vcenter{\openup #1\jot \mathsurround=0pt \tabskip=0pt
\halign{\hfil$\displaystyle{##}$&\hfil$\displaystyle{{}##{}}$\hfil&
                                      $\displaystyle{##}$\hfil\cr
#2\crcr}}}

\def\dmblock#1#2#3{
\begin{equation}\label{#2}\mathtab{#1}{#3}\end{equation}}

\begin{document}

\renewcommand{\textfraction}{0.01}
\renewcommand{\topfraction}{0.99}

\pagestyle{empty}

\vbox to \vsize
{
\begin{center}

\hfill HD-THEP-03-21 \break
\hbox to \hsize{\hfill hep-ph/0307326} \break

\vskip 0pt plus 1fill

{
\Huge\bf
High energy scattering
\vskip 3mm
in the Regge limit
}

\Large

\vskip 0pt plus 2fill

\vskip 0pt plus 2fill

Dissertation \\
of \\
Volker Schatz

\vskip 0pt plus 1fill

{
\sc
Institut f\"ur Theoretische Physik \\
Universit\"at Heidelberg
}

\vskip 0pt plus 1fill

\end{center}
}

\cleardoublepage

\vbox to \vsize
{
\fboxsep=4mm


\framebox{\vbox{
{\hsize=12.5cm \parskip=2pt
{\bf Hochenergiestreuung im Regge-Limes}

\parindent=0pt
Diese Arbeit umfa\ss t zwei Untersuchungen, die beide damit zu tun haben,
Regge-Theorie von QCD ausgehend zu verstehen.  Gegenstand der ersten ist, wie
das Proton an das Odderon koppelt, ein Reggeon, das keine Ladung tr\"agt und
unge\-rade unter Ladungskonjugation ist.  Die zweite betrifft
Unitarit\"atskorrekturen zur BFKL-Gleichung, die das Pomeron beschreibt, ein
Reggeon mit den Quantenzahlen des Vakuums.

Im ersten Teil dieser Arbeit wird der Odderon-Beitrag zur elastischen Streuung
von Protonen an Protonen und Antiprotonen berechnet.  Um den Einflu\ss\ der
Protonstruktur auf die Odderon-Proton-Kopplung zu untersuchen, wird ein
geometrisches Protonenmodell konstruiert.  Durch Vergleich mit experimentellen
Daten wird die durchschnittliche Gr\"o\ss e eines Diquark-Clusters im Proton
ermittelt.  Zwei weitere Odderon-Proton-Impaktfaktoren aus der Literatur werden
durch Vergleich mit dem Experiment getestet.

Der zweite Teil enth\"alt die Berechnung von vier-Pomeron-Vertizes, die im
Rahmen der Generalised Leading Logarithmic Approximation auftreten.  Diese
N\"aherung dient zur Unitarisierung von Streuamplituden, die den Austausch des
BFKL-Pomerons beschreiben.  Eine Anzahl grundlegender Funktionen, aus denen
solche Vertizes bestehen, werden systematisch im Impuls- und Ortsraum
behandelt, und ihre Transformationseigenschaften unter konformen
Transformationen im transversalen Ortsraum werden hergeleitet.  Die Vertizes
werden als eine Zahl von Integralen ausgedr\"uckt und auf die Form einer
konformen Vierpunktfunktion gebracht.
\par
}
}}

\vskip 0pt plus 1fill

\framebox{\vbox{
{\hsize=12.5cm \parskip=2pt
{\bf High energy scattering in the Regge limit}

\parindent=0pt
This thesis comprises two investigations, both connected with the attempt to
understand Regge theory in the framework of QCD.  The first is about how the
odderon, a Reggeon carrying no charge which is odd under charge conjugation,
couples to the proton.  The second concerns unitarity corrections to the BFKL
equation which describes the pomeron, a Reggeon with the quantum numbers of the
vacuum.

In the first part of this thesis, the odderon exchange contribution to elastic
proton-proton and proton-antiproton scattering is computed.  A geometrical
transverse-space model of the proton is constructed to investigate the
influence of a possible diquark cluster in the proton on the odderon-proton
coupling.  The average size of this cluster is determined by comparison with
experimental data.  Furthermore, the validity of two odderon-proton impact
factors from the literature is tested.

The second part consists in the derivation of four-pomeron vertices.  These
vertices occur in the Generalised Leading Logarithmic Approximation used to
unitarise scattering amplitudes describing the exchange of a perturbative
pomeron.  A number of basic functions of which such pomeron vertices are
composed is treated systematically in momentum and position space.  Their
conformal transformation properties in impact parameter space are derived.  The
vertices are expressed as a number of integrals and cast into the form of
conformally invariant four-point functions.
\par
}
}}


}

\cleardoublepage
\pagenumbering{roman}
\pagestyle{plain}
\tableofcontents

\cleardoublepage
\pagenumbering{arabic}
\pagestyle{headings}
\setcounter{secnumdepth}2

\addcontentsline{toc}{part}{Introduction}

\thispagestyle{empty}

\vbox to \vsize
{
\vfill

\begin{center}
\Huge\bfseries Introduction
\end{center}

\vskip 15mm

\leftskip=15mm
\rightskip=15mm

This introductory chapter gives an overview of the theories and physics issues
surrounding the two investigations contained in this thesis.  The basic notions
of high energy scattering and of descriptions of the strong interaction are
briefly portrayed.  This introduction is intended to give some background and
put the topics of this thesis into their context, which is why technical
details, notably of the BFKL equation, are left for later chapters.

\leftskip=0pt
\rightskip=0pt

\vfill
}
\pagebreak

\refstepcounter{chapter}
\chaptermark{Introduction}

\thispagestyle{plain}
{
\don{c@secnumdepth}=-1
\don{@makechapterhead}{1.1 \ Background and Motivation}
}

\refstepcounter{section}
\addcontentsline{toc}{section}{\numberline{\thesection}Background and Motivation}
\sectionmark{Background and Motivation}

Quantum Chromodynamics is now well established as the fundamental theory of the
strong interaction.  The quark model which it is based on provides the basis of
our understanding of matter.  The description of the quarks' interaction by the
gauge theory that is quantum chromodynamics has been successfully applied to a
very wide range of problems.  This interaction varies smoothly between the two
extremes of confinement, which makes it impossible to isolate quarks, and
asymptotic freedom, which makes the quarks inside scattering hadrons behave
like free particles at small distances.

Considering how successful and how satisfactory in view of the aim to
understand nature from first principles QCD is, it is not surprising that it
has eclipsed the attempts to describe the strong interaction which preceded it.
One of those is Regge theory~\cite{regge1,regge2,collins}.  It is based on
Lorentz invariance, unitarity and analyticity of the scattering matrix.
Developed before quarks and gluons were recognised as the fundamental degrees
of freedom of the strong interaction, it is a theory of hadrons.  It has been
quite successful in describing hadronic scattering processes in the so-called
Regge limit, for asymptotically large energies and momentum transfers of the
order of a hadronic mass scale.

In Regge theory, the exchange of particles in a scattering process is described
by the singularities of the scattering amplitude in the complex angular
momentum plane, the so-called Regge poles and cuts.  By way of crossing
symmetry, they can be related to the masses and spins of existing hadrons.  By
that token, every hadron is a Regge particle, or Reggeon.  However, some
Reggeons do not correspond to any known hadrons.  One such is the pomeron which
is the pole with the largest real part and therefore dominant at asymptotically
high energies.  It carries the quantum numbers of the vacuum, and the simplest
model for it consists of two gluons in a colour singlet state~\cite{low,nus}.
Another is its charge-conjugation-odd partner, the odderon~\cite{lunic}.  There
are some indications that both pomeron and odderon might be related to
glueballs.

To this day, concepts of Regge theory are used widely both in phenomenology,
eg~\cite{jarbkp,jar,dlfit,gln,dlnew} and theory,
eg~\cite{gribov,baker,bronsug,fadlip,lipglla,fadintegr,pesch}.  The existence
of the pomeron is universally accepted, and it is the topic of a wide range of
phenomenological work, for
instance~\cite{dlcond,dlnew,brandt,biapesch,nextbfkl,kharlev,munnav,kharkolev,stasto}.
The odderon, by contrast, has never been measured experimentally beyond doubt
and consequently is still a contentious topic.  A brief overview of odderon
physics will be given in the following; for an in-depth review, I refer the
reader to~\cite{oddreview}.  There are few processes in which the odderon is
the dominant or indeed only contribution.  One such process is the diffractive
photo- or electroproduction of $\eta_c$ or other heavy pseudoscalar mesons at
HERA~\cite{kwie_impac,bbcv,engeletac}.  But the cross-sections are estimated to
be too small to measurable in the foreseeable future.  Another
odderon-dominated process is the production of light pseudoscalar or tensor
mesons, the cross sections for which are expected to be
higher~\cite{knmeson,rysmeson,msvmeson,msvmeson2}.  But it has not been
observed experimentally~\cite{olsmeson,goldipl,berndt,berndtphd}.  This
discrepancy might be due to difficulties with the non-perturbative models which
are required to describe effects at such low scales.

However, the processes on which the search for the odderon has concentrated for
a long time are proton-proton and proton-antiproton scattering, a fact which
the large number of publications on the topic attests, for
instance~\cite{dlfit,gnldlcrit,gln,oddpoint,lev,des,bergnacht,leadtrue,desdips}.
These processes provide the hitherto only experimental evidence for the
odderon.  The elastic differential cross sections of these two processes differ
qualitatively: In proton-proton scattering, it shows a dip structure at a
squared momentum transfer $t$ of about $t=1$~GeV$^2$, while in
proton-antiproton scattering, it only flattens off at this point.  Odderon
exchange can account for this difference since it carries odd charge
conjugation parity and hence contributes with different signs to both
processes.  What is more, the odderon is the dominant $C=-1$ contribution
arising from Regge theory.  It is expected to have a Regge intercept of
slightly below~1, while $C$-odd meson trajectories typically have intercepts of
around 0.5.  Since the scattering amplitude of a Reggeon exchange with
intercept $\alpha_0$ is proportional to $s^{\alpha_0+\alpha't}$, the latter
contribute much less at high energies.  This difference in the differential
cross sections has been measured at the Intersecting Storage Rings at
CERN~\cite{bohm,nagy,ama,erhan,break}.  The experimental evidence is marred by
the fact that the statistics for $\bar pp$ scattering is low and that the
difference only shows in a few data points (see Section~\ref{sec:oddpp}).  A
good description of the data in the framework of Regge theory was given by
Donnachie and Landshoff~\cite{dlfit}.  It is characterised by fair accuracy and
considerable predictivity, having only a small number of free parameters.  A
few years later, it was extended~\cite{dlfitnew} to higher-energy
proton-antiproton data measured at the CERN SPS~\cite{bernard,bernard2}.
Proton-proton data at even higher energy from the Tevatron~\cite{amos} provide
no further check on the fit since they do not extend far enough in $-t$ to show
the dip.  In the Donnachie-Landshoff fit, the odderon-exchange contribution
produces the dip in proton-proton and its absence in proton-antiproton
differential cross sections.  Another description of $pp$ and $\bar pp$
scattering is due to Gauron, Nicolescu and Leader~\cite{gln}.  It uses the
so-called maximal odderon which corresponds to a different type of Regge
singularity from other odderon contributions.  As in the Donnachie-Landshoff
fit, the odderon contribution is instrumental in the description of the dip
region.  However, the maximal odderon itself already shows a rich structure
with a dip.  One finding of the work presented in this thesis is that each
odderon contribution is compatible only with its own fit, that exchanging
odderon contributions between fits is not possible (see
Section~\ref{sec:glndlodd}).  But no successful description of the data without
an odderon has been found so far.  A measurable quantity which is more
sensitive to odderon exchange than the cross sections is the ratio of the real
to imaginary part of the forward scattering amplitude.  But also here no
evidence for the odderon has been found~\cite{augier}.

If the thin experimental evidence for the odderon were all there is to it, one
might have discarded it as a misguided concept.  But other concepts of Regge
theory remain very valid today.  Furthermore, the odderon can in fact be
derived from perturbative QCD.  It is described by the
Bartels-Kwieci\'nski-Prasza\l owicz (BKP)
equation~\cite{bartbkp,kwiebkp,jarbkp}.  Odderon exchange amounts to a
simultaneous exchange of at least three gluons in a $C=-1$ state.  Since such
an exchange is clearly possible in QCD, a failure to find the odderon at all
would be a heavy blow to QCD.  Odderon research in QCD has received a boost
from the discovery that the perturbative odderon is equivalent to an integrable
model, the XXX Heisenberg model with spin zero~\cite{fadintegr,korintegr}.  A
prime interest in odderon physics has always been the determination of its
Regge intercept, see for
instance~\cite{gauinter2,gauinter,arminter1,arminter2,brauninter,brauninter2}.
Since the discovery of an additional conserved charge of the
odderon~\cite{lipq3}, work has concentrated on finding the spectrum of the
associated operator~\cite{korq3val,korq3val2,maasq3val,vegaq3val,derkq3val}.
Recently, this spectrum has been determined both by Korchemsky
et~al.~\cite{korground,korground2} and by de~Vega and
Lipatov~\cite{vegaground}.  To date, there are still unresolved differences
between the results of the two groups which centre on the question which
eigenvalues of the operator are physical.  Separately, explicit solutions of
the BKP equation have been found by two groups.  One, by Janik and
Wosiek~\cite{wojan,janwo}, has an intercept slightly below one~($\approx
0.96$).  It is of the general form derived by Lipatov and
collaborators~\cite{lipodd,glinodd,lipodd2}, in which an analytic auxiliary
function is left open.  The other solution, by Bartels, Lipatov and
Vacca~\cite{blvodd}, has an intercept of exactly one.  This means that its
contribution to forward scattering would not decrease with energy.  The
Bartels-Lipatov-Vacca (BLV) solution was contentious at first since it is not
of the general form found by Lipatov, at least not with an analytic auxiliary
function.  But now it looks like being accepted as legitimate by the community.
This is not just of academic theoretical interest, nor just a question about a
slightly different intercept.  Odderon solutions conforming to Lipatov's
general form cannot couple to some scattering particles, for instance a photon
fluctuating into an~$\eta_c$.  The BLV odderon, however, can.  Therefore, with
the advent of this new solution a number of new processes in which the odderon
may play a role becomes available.

A solution of the BKP equation, which describes the odderon's propagation, is
only one ingredient for calculating the odderon contribution to a given
process.  Besides, impact factors describing the odderon's coupling to
scattering particles are required.  {\em Pomeron} impact factors have always
aroused some interest~\cite{asaipomimp,klenpomimp,golpomimp} and have just been
computed in next-to-leading
order~\cite{ciapomimp,bartpomimp,giepomimp,fadpomimp}.  Less work has been done
on odderon impact factors~\cite{fk,lev,dloddimp,kwie_impac}.  While some impact
factors can be calculated perturbatively (for instance the odderon-$\eta_c$
impact factor~\cite{kwie_impac}), impact factors for proton scattering require
non-perturbative ans\"atze most of which have never been confronted with
experimental data.  As part of this work, the odderon-proton impact factors of
Levin and Ryskin~\cite{lev} and Kwieci\'nski et al.~\cite{fk,kwie_impac} were
used to calculate the elastic differential cross section for proton-proton
scattering.  The odderon was described simply as three gluons in a $C=-1$
colour singlet state.  The non-odderon contributions were taken from the
Donnachie-Landshoff fit~\cite{dlfit}.  This allows to fix the coupling constant
to be used with these impact factors.  My results imply that predictions of the
$\eta_c$ production cross section by Kwieci\'nski and two other group using his
impact factors~\cite{engeletac,bbcv} have to be revised downwards, see
Section~\ref{sec:momentimp}.

The main interest of the first part of this thesis (Chapters
{}\ref{chap:odderoncalc}, \ref{chap:oddresultspp}
and~\ref{chap:oddresultsppbar}) is however a different one.  It has been known
for some time that the structure of the proton has considerable effect on the
odderon-proton impact factor.  In the extreme case in which two of the proton's
valence quarks are clustered in a point-like diquark, the odderon-proton
coupling even vanishes~\cite{oddpoint}.  The odderon contribution for a diquark
cluster of finite size in the proton was computed non-perturbatively by R\"uter
and Dosch~\cite{rueter,rueterphd}.  They used it to determine the ratio of the
real to imaginary part of the forward scattering amplitude for proton-proton
and proton-antiproton scattering, respectively.  The {\em perturbative} odderon
contribution to these processes for a finite-sized diquark is presented in this
thesis, in Chapter~\ref{chap:odderoncalc}.  By computing the differential cross
section in the dip region for proton-proton scattering, the average size of the
diquark cluster is obtained from experimental data (Chapters
\ref{chap:oddresultspp} and~\ref{chap:oddresultsppbar}).  The geometrical
nature of the concept of a diquark cluster necessitates the use of a
position-space approach to high-energy scattering.  I choose the one developed
by Nachtmann~\cite{nacht}, which is otherwise used mainly with non-perturbative
models, notably the Model of the Stochastic Vacuum, see for
instance~\cite{doschmsv,dsmsv,simmsv,dfk,rueter,bergnacht,msvmeson,ssmsv,ssmsv2}.
My result of a small diquark size will benefit non-perturbative calculations:
Many of them treat the proton as a colour dipole~\cite{dfk,bergnacht,ssmsv3},
which is legitimate since soft gluons cannot resolve a small diquark.


Important though many results of Regge theory remain in present-day high-energy
physics, our understanding of the Regge limit from the first principles of QCD
is still far from complete.  This is because high-energy hadronic processes
tend to involve large parton densities and are dominated by soft scales.  Our
knowledge of non-perturbative QCD is not yet sufficient for a rigorous
treatment of the vast majority of high-energy processes.  However, some
processes are dominated by one hard scale and therefore allow the application
of perturbation theory.  One such is the scattering of heavy quarkonia.  The
process itself is not experimentally feasible.  But since photons tend to
fluctuate into $q\bar q$ states in high energy scattering, the scattering of
highly virtual photons is expected to provide a suitable substitute.  This is
predicted to be measurable~\cite{bartvirt1,bartvirt2,bhsvirt1,bhsvirt2} at a
future $e^+e^-$ collider like the planned Next Linear
Collider~\cite{nlc,nlczdr}.  A situation similar to the process just mentioned
occurs in the presence of hard forward jets in deep inelastic or in hadronic
scattering, called forward jets and Mueller-Navelet jets,
respectively~\cite{mueljet,muelnav}.  They have been investigated in depth by a
variety of groups, for
example~\cite{brljet,kmsjet,tangjet,bartjet2,pestruct,bartjet3,elvjet,munjet,andjet,andjet2,enbjet,bartjet4}.
Another process involving a hard scale is deep inelastic electron-proton
scattering at small values of the Bjorken scaling variable~$x$.  In this case
the hard scale is provided by the virtual photon which mediates the interaction
on the electron's side.

Even in the presence of a hard scale, perturbation theory in high-energy QCD is
not trivial.  Due to the large energies available, the phase space of
intermediate gluons becomes very large,~$\propto\log s$.  In QCD, this
logarithmic factor can compensate the smallness of the coupling constant, with
the effect that more complicated Feynman graphs are not always suppressed.
Therefore an infinite number of graphs have to be resummed to obtain a
consistent leading-order result.  This is called the Leading Logarithmic
Approximation.  The resummation was performed for the first time in QCD by
Balitsky and Lipatov~\cite{blbfkl}, using earlier work by Lipatov and
collaborators~\cite{fklbfkl}.  The result of the resummation is described by
the well-known BFKL equation.  The corresponding Reggeon exchange is called the
BFKL, perturbative or hard pomeron.  It represents a bound state of two
reggeised gluons and was found to correspond to a Regge cut in the complex
angular momentum plane.  A reggeised gluon is a colour octet state composed of
several QCD gluons~\cite{lipreggluon}.  It was discovered in 1985 that the BFKL
equation is conformally invariant in transverse position space~\cite{bfklconf}.
This was a finding of great importance since it allowed the solving of the
equation in the non-forward direction.  It also led to a novel interpretation
of BFKL evolution as a two-dimensional conformally invariant quantum mechanics
in which the complexified angular momentum is the energy-like and rapidity the
time-like variable.

Since the invention of the BFKL pomeron it has become clear that it violates a
basic principle of field theory: unitarity.  The Froissart-Martin theorem,
derived from this principle, states that hadronic cross sections can rise at
most with the squared logarithm of energy.  The growth of the
BFKL-pomeron-exchange cross section is power-like, with an exponent of
$\alpha_s 4 \ln 2 N_c/\pi\approx 0.5$.  But the BFKL pomeron represents only a
leading-order approximation.  Higher-order corrections tame this power-like
growth and restore unitarity.

A complete resummation of higher-order graphs relevant for pomeron exchange is
even today thought to be prohibitively difficult.  One has to be content with
adding specific classes of graphs which are chosen to restore compliance with
the Froissart-Martin Theorem and hence unitarity.  For restoring the unitarity
of the overall scattering matrix one has to add ladder graphs with more than
two reggeised gluons in the $t$ channel.  These additional terms are called
unitarity corrections.  The amplitudes (or propagators) of these larger bound
states of reggeised gluons are described by the Bartels-Kwieci\'nski-Prasza\l
owicz (BKP) equations~\cite{bartbkp,kwiebkp,jarbkp}.  They can be greatly
simplified by treating only the large $N_c$ limit.  Then a nearest-neighbour
interaction takes the place of the $n$-particle interaction between the
reggeised gluons.  This amounts to a one-dimensional solid, and was found to be
equivalent to the XXX Heisenberg model with spin zero, a completely integrable
model~\cite{lipq3,lipglla,fadintegr}.  This limited approach of demanding
unitarity only for the overall process is followed by Lipatov~\cite{lipglla}.

A more ambitious and more complete approach is due to
Bartels~\cite{bartglla,bartglla1,bartglla2}.  It demands unitarity not just in
the main process but also in all possible subprocesses.  This requires the
inclusion of graphs in which the number of reggeised gluons in the $t$ channel
is not constant.  Starting from two reggeised gluons coupling to the scattering
particles, amplitudes of more than two reggeised gluons emerge.  It was found
that the three- and five-gluon amplitudes are in fact just superpositions of
amplitudes with fewer gluons~\cite{bw,carlophd,be}, a phenomenon known as
reggeisation.  The four-gluon amplitude also contains a reggeising
part~\cite{bw}.  But besides that a vertex transforming two to four gluons
occurs, which gives rise to a new form of four-gluon amplitude.  The conformal
invariance of this new vertex was shown in~\cite{blw}.  The six-gluon
amplitude, investigated in~\cite{carlophd,be}, consists of a completely
reggeising part, a partially reggeising part, and a new $2\to6$ vertex.  The
completely reggeising part is a superposition of two-gluon amplitudes, the
partially reggeising part a superposition of four-gluon amplitudes.  The
$2\to6$ vertex again gives rise to a new amplitude.  If all transitions to a
larger number of reggeised gluons were conformally invariant, these unitarity
corrections would amount to a conformal field theory in $2+1$ dimensions.  In
analogy to the conformal quantum mechanics suggested by BFKL, rapidity would
serve as the time-like variable and the complex angular momentum as the
energy-like variable.

Recent work \cite{pesch,korconf,conf1_Nc,stringy} suggests an even more
intimate connection between string theory/conformal field theory and unitarity
corrections.  Some of this work involves pomerons as the fundamental degrees of
freedom and pomeron vertices.  Vertices of BFKL pomerons can be obtained by
projection of the reggeised-gluon transition vertices.  The simplest of these,
transforming one to two pomerons, has been investigated
in~\cite{lotterphd,pesch,korconf,braun3p,bart3p} and found to have the form of
a conformal three-point function.  There is some confusion about the existence
of higher vertices transforming one to $n>2$ pomerons.  Of course higher
vertices could be constructed by iterating the $1\to 2$ vertex.  But such
composed vertices would not be local in rapidity since the intermediate states
of pomerons would have to be allowed to evolve.  Peschanski~\cite{pesch} has
given an interpolated expression for local $1\to n$ pomeron vertices which,
like the $1\to2$ vertex, allows an interpretation as string-theoretic
Shapiro-Virasoro amplitudes.  Since then, Braun and Vacca~\cite{brvac} have
proven that such local vertices do not exist in the dipole approach of
Mueller~\cite{mueller,mueller2}.  This approach provides a derivation of the
BFKL equation which is much easier than the original one involving gluon
ladders.  However, it is not thought to be equivalent to the gluon-ladder
approach to all orders, not least because it does not allow the exchange of an
odderon.  While Peschanski also starts out from dipoles, the expression he
generalises is much closer to the gluon-ladder form of the result obtained by
Lotter~\cite{lotterphd}.

Some of the questions about higher pomeron vertices are addressed in the second
part of this work.  Chapter~\ref{chap:confbase} consists of preparing the
groundwork for treating reggeised-gluon and pomeron vertices in a systematic
way.  It deals with a function $G$ in terms of which the $2\to4$ and $2\to6$
reggeised gluon vertices can be expressed, and its components.  The properties
of $G$ under conformal transformations prove the conformal invariance of the
$2\to6$ vertex and also of higher vertices if they have an analogous form as
suggested in~\cite{ew2}.  This represents an important step towards proving the
existence of a conformal field theory of unitarity corrections.  In
Chapter~\ref{chap:p3p} the $1\to3$ pomeron vertex is derived in the
gluon-ladder approach and cast into the form of a conformal four-point
function.  Its existence does not contradict the result of Braun and Vacca but
rather reveals a discrepancy between the gluon-ladder and the dipole approach.
Their equivalence extends only to the $1\to2$ pomeron vertex but not to higher
orders.

\clearpage

\thispagestyle{plain}
{
\don{c@secnumdepth}=-1
\don{@makechapterhead}{Introduction}
}

\section{High-energy scattering}

When a number of particles get close to each other, they interact, possibly
break up and fly off in directions different from that of their initial
trajectories.  This is called a scattering process.  To simplify its
description, one assumes that the scattering particles approach each other from
a macroscopically large distance, interact in a microscopically small region,
and that the products of the interaction move away again to a macroscopically
large distance.  Before and after the interaction, the particles are thought of
as so far apart that they are effectively unaware of each other.  The
scattering operator $\mathcal S$ is defined as the operator which transforms
the initial multi-particle state in the infinite past to the final
multi-particle state in the infinitely remote future.
\dmath2{
\lim_{t\to+\infty}|t\rangle=|f\rangle&=&
\mathcal S\,|i\rangle=\mathcal S\,\lim_{t\to-\infty}|t\rangle
}

Since the time-evolution operator is linear, so is the scattering operator.  In
a suitable base, it can therefore be written as a matrix, the scattering
matrix, with elements denoted by~$\mathcal S_{fi}$.  More often than~$\mathcal
S$, one uses the $\mathcal T$ matrix.  It is obtained from $\mathcal S$ by
subtracting the unit matrix and removing some kinematic factors.  It therefore
can be thought of as describing the effect of the interaction, since a
scattering matrix equal to unity amounts to no interaction.
\dmath2{
\mathcal S_{fi}&=&\delta_{fi} + i\,(2\pi)^4\,\delta^4(P_f-P_i)\,\mathcal T_{fi}
}
Here $P_i$ and $P_f$ are the total four-momenta of all particles in the initial
and final states, respectively.

The $\mathcal T$ matrix element is also called the scattering amplitude and
denoted by~$T$.  It is not usually expressed as a function of the initial and
final states.  Rather, kinematic variables which characterise the states are
used as arguments of~$T$.  Easily the most important kinematic variable is~$s$.
It is the square of the centre-of-mass energy, or equivalently the square of
the invariant mass of the scattering system.
\dmath2{
s&=&P_i^2=P_f^2
}
In ``high-energy'' scattering, one treats the limit $s\to\infty$, or in
experimental reality the case where $s$ is significantly larger than the
squares of all the particle masses, $s\gg m_n^2\;\forall\,n$.

\begin{figure}
\centre{\Large\input{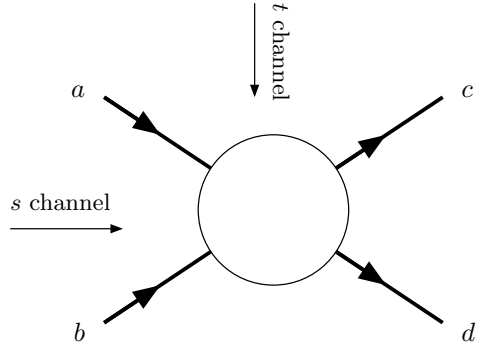}}
\caption{A two-particle to two-particle scattering process.}
\label{fig:scatter22}
\end{figure}

Let us now turn to the case of two particles scattering and two (possibly
different) particles emerging after the scattering process.  This situation is
displayed graphically in Figure~\ref{fig:scatter22}.  The diagram can be read
in two ways: from left to right or from top to bottom.  The left-to-right view
corresponds to the reaction
$$
a+b \longrightarrow c+d\,.\qquad\hbox{($s$ channel)}
$$
The top-to-bottom view corresponds to the reaction
$$
a+\bar c \longrightarrow \bar b+d\,.\qquad\hbox{($t$ channel)}
$$
$b$ and $c$ become antiparticles since they now propagate backwards in time.
This process has the centre-of-mass energy $(p_a-p_c)^2=:t$.  Looking at the
the $t$-channel process instead of the original $s$-channel reaction is called
``crossing''.  There is a third kinematic variable~$u$ and a corresponding
third process in which it is the squared centre-of-mass energy:
$$
a+\bar d \longrightarrow \bar b+c\qquad\hbox{($u$ channel)}.
$$
These three kinematic variables are very important in any of these processes.
They are called Mandelstam variables.  They are not independent but add up to
the sum of the mass squares of the four particles.  Here is their definition:
\dmath1{
s&=&(p_a+p_b)^2\cr
t&=&(p_a-p_c)^2&eq:mandelstam\cr
u&=&(p_a-p_d)^2\cr
}

In the $s$-channel process, $s$~is the squared centre-of-mass energy.  $t$ is
the squared four-momentum transfer: If one assumes the particle $a$ to emerge
as particle $c$, the four-momentum $p_a-p_c$ has been transferred to $b$ (which
becomes~$d$).  This is also the negative square of the invariant mass of the
(virtual) particle mediating the interaction.  One can show that in $s$-channel
processes, $t$ is always negative.  After crossing, the variables exchange
their roles.  In the $t$-channel process, $t$ is the squared centre-of-mass
energy and $s$ is the squared four-momentum transfer.  Now $s$ is negative and
$t$ is positive.  (It has to be taken into account that in the crossed process,
the momentum of $\bar c$ is $-p_c$.)

Interestingly, the same scattering amplitude applies in all crossed processes.
It just has to be continued analytically to the region in which the Mandelstam
variables have the right sign for the desired channel.  This is called crossing
symmetry.

Let us now turn from kinematics to calculating observables for $s$-channel
processes.  The observables of most universal interest are cross sections.  The
most widely used differential cross section is the derivative of the cross
section with respect to~$t$.  It is easily computed from the scattering
amplitude:
\dmath2{
\frac{d\sigma}{dt} &=& \frac{T(s,t)^2}{16\,\pi\, s^2}\,.
&eq:dsigdt}

The total cross section could of course be obtained by integrating over the
differential cross section.  But there is a simpler way which makes it possible
to obtain it directly from the scattering amplitude.  This is by using the
optical theorem.  It is a special case of the Cutkosky rules and relates the
forward elastic scattering amplitude to the total cross section.  In the
high-energy limit, the optical theorem has the following form:
\dmath1{
2\,\hbox{Im}\,T(s,0)&=&2\,s\,\sigma_{tot}(s)\,.
&eq:optheo}
This allows to compute the total cross section.

\section{The standard model, the strong interaction and QCD}

The standard model is a description of the three interactions relevant in
elementary particle physics --- the electromagnetic, weak and strong
interaction.  This description is based on quantum field theory.

The electromagnetic and weak interactions are jointly described by the
Weinberg-Salam model.  It is basically a gauge theory with four massless gauge
bosons and a scalar particle, the Higgs.  By fixing the vacuum expectation
value of the Higgs field, one induces spontaneous symmetry breaking.  As a
consequence, a gauge theory with three massive ($W^\pm$, $Z^0$) and one
massless ($\gamma$) gauge bosons emerges.  Even though the Higgs boson has not
yet been detected, this model is very successful.

The strong interaction is described by quantum chromodynamics (QCD).  Its
foundation is the quark model, the idea (supported by experiment) that hadrons
are composed of smaller point-like particles which cannot be isolated, the
quarks.  Quarks have two quantum numbers peculiar to them: flavour (which
accounts for different types of hadrons) and colour (which is the gauge charge
of QCD).  There are three different colours, and to date six flavours of quarks
have been found.  Observable particles are superpositions of systems of quarks
of all three colours, which renders them colour-neutral, ``white''.  The gauge
bosons of QCD are called gluons.  They are massless and come in the eight
existing colour-anticolour combinations, excluding white.

Making predictions on the basis of QCD is much harder than in QED, for two
reasons: One, gluons couple to each other.  There is a three-gluon and a
four-gluon vertex in QCD.  Two, QCD exhibits confinement.  The most obvious
effect of this property is that colour-carrying particles cannot be isolated.
Its deeper reason is that the strong coupling constant is large at small
momenta.  (Correspondingly, it is small at high momenta, which is called
asymptotic freedom.)  Taken together, these two points mean that gluons have a
tendency to split up into more gluons (or quark pairs) and that more
complicated Feynman graphs are not always significantly suppressed.

Problems concerning the strong interaction at low energies cannot be solved
with perturbative QCD.  Additional models or lattice calculations are required.
Perturbation theory does not work at low energies.  Perturbative calculations
of the evolution of the strong coupling constant lead to a divergence at the
scale $\Lambda_{\rm QCD}\approx 250$~MeV.  The expression for this evolution
probably starts to lose its meaning below two to four times that value.
Non-perturbative QCD is not the topic of this thesis, but the limitations of
perturbation theory have to be kept in mind in any QCD calculation.

\section{Before QCD: Regge theory}

\subsection{Regge theory and the scattering matrix}

Before the development of QCD as a quantum field theory of the strong
interaction, attempts were made to derive the properties and behaviour of
hadrons from reasonable assumptions about the scattering matrix.  The resulting
theory is called Regge theory after its inventor,
T.~Regge~\cite{regge1,regge2}.  (See
\cite{collins} for a thorough treatment.)

The assumptions about~$\mathcal S$ are the following:
\begin{enumerate}

\item $\mathcal S$ is Lorentz-invariant.  We have already made this assumption
in the first section when we wrote the scattering amplitude as a function of 
the Lorentz invariants $s$ and~$t$.

\item $\mathcal S$ is unitary, 
$\mathcal S \mathcal S^\dagger=\mathcal S^\dagger \mathcal S=\dblone$.  This is
the conservation of probability, ie the probability for a specific initial
state to end up in any final state at all must be one.

\item $\mathcal S$ depends analytically on Lorentz invariants.  Unitarity 
takes precedence over this, ie singularities which are necessary for
unitarity are allowed.

\end{enumerate}

The last point requires some explanation.  It may seem strange that unitarity
requires singularities in the scattering amplitude.  The reason lies in the
optical theorem~(\ref{eq:optheo}) which relates the forward elastic scattering
amplitude with the total cross section.  The Cutkosky rules from which it is
derived are a consequence of unitarity of the $\mathcal S$ matrix.  By way of
the optical theorem, contributions to the total cross section (from any
reaction) relate to contributions to the imaginary part of the forward elastic
amplitude.  Therefore one can learn something about the scattering amplitude by
looking at all the final states which might arise from two scattering
particles.

Two particles can merge into one only at discrete energies.  They can create
two different particles only at energies higher than the combined mass of the
lightest pair of particles allowed by quantum number conservation.  Therefore
there are intervals on the real $s$ axis where the total cross section, and
therefore the imaginary part of the forward scattering amplitude, is zero.
This allows to apply the Schwarz reflection principle from the theory of
complex functions: When a function is analytic in a given domain and real on a
straight line contained in that domain, then it takes values which are complex
conjugates of each other at points which are mirror reflections with respect to
that line.  In this context this means that
$$
T(s^*,t)=T(s,t)^*
$$
in its domain of analyticity.  (The straight line is one of the intervals on
the real $s$ axis where $T$ is real, so the mirror image of $s$ is~$s^*$.)

\begin{figure}[h]
\centre{\Large\input{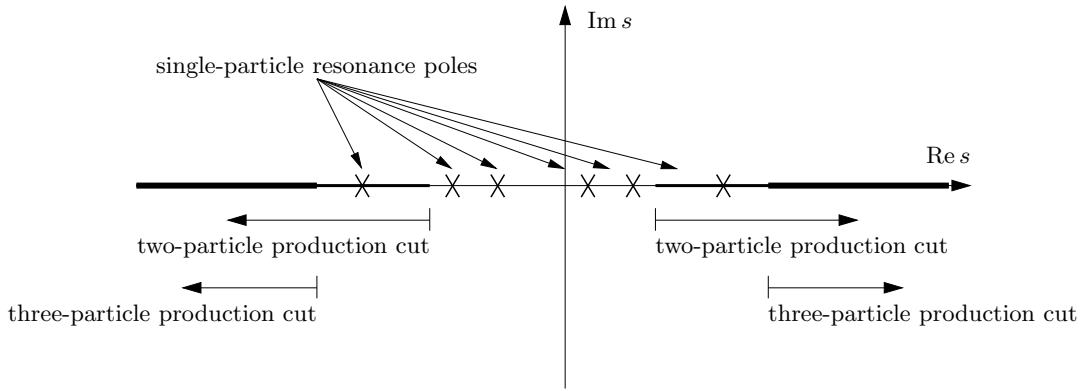}}
\caption{Singularity structure of the scattering amplitude in the $s$ plane.
		It is symmetric to a line parallel to the imaginary axis, but not 
		necessarily to the axis itself.}
\label{fig:singul}
\end{figure}

Since an analytic function has a unique power expansion in its domain of
analyticity, $T$ would have to be real on the whole real $s$ axis if it were to
be analytic there.  But that would mean $\sigma_{tot}=0$ for all energies,
which is clearly not usually true.  Therefore $T$ has to be singular at large
parts of the real axis.  The non-zero multiple-particle production amplitudes
are represented by branch cuts (since they extend all the way to infinity from
the $n$-particle-production threshold), while single particle production is
represented by isolated singularities, usually poles.  It can be shown from
crossing symmetry and the relation between the Mandelstam variables that $T$
has a structure symmetric with respect to a parallel to the imaginary axis, so
that the singularity structure shown in Figure~\ref{fig:singul} emerges.

\subsection{Complex angular momenta}
\label{sec:reggeon}

Let us consider two-particle to two-particle scattering.  It is well known from
quantum mechanics that one can express the scattering amplitude in terms of the
spherical harmonics with zero angular momentum, $Y_{l,0}$, which are Legendre
polynomials in the cosine of the scattering angle,~$\cos\theta$.  This is
called partial wave expansion.  In the $s$ channel, $\cos \theta=1+2s/t$ and
one can write:
\dmath2{
T(s,t)&=&\sum_{l=0}^\infty (2l+1)\,a_l(s)\,P_l(1+2s/t)\,,
&eq:partwav}
where $a_l$ are the coefficients of the expansion, the partial wave amplitudes.
This expression can be rewritten according to Cauchy's integral theorem as a
contour integral in the complex $l$ plane over a function with singularities at
$l\in\mathbb N_{\ge 0}$.  The value of the integral is the sum of the residues
at these poles, which are constructed to equal the terms in the sum
in~(\ref{eq:partwav}).  This is called Sommerfeld-Watson transform.  One
obtains:
\dmath2{
T(s,t)&=&\oint dl\, \frac1{\sin\pi l}\, (2l+1)\,a(l,s)\,P(l,1+2s/t)\,.
&eq:partwavint}
The integration contour extends to infinity and circles the positive real axis
clockwise.  The argument $l$ of $a$ and $P$ is no longer written as an index to
indicate that it is a now complex variable and that these functions have been
analytically continued over the whole complex $l$ plane.

The analytic continuation of the partial wave amplitude $a$ is not unique.
Adding a function with zeros at all non-zero integers to it does not change the
integral.  But such a function has to be a sine with a period of two divided by
an integer.  Such a sine function would rise exponentially on the imaginary
axis.  So there is a unique $a(l,s)$ which rises more slowly than
$\exp(\pi|l|)$ as $l$ approaches infinity.  However, some contributions to the
partial wave amplitude oscillate as $(-1)^l$.  This would be analytically
continued to a cosine, which also rises faster than our limit.  Since adding a
sine function cannot correct that on both the positive and negative imaginary
axis, a different route is chosen.  The analytic continuation is done for even
and odd angular momenta separately, and $a(l,s)$ is replaced by
\dmath2{
\tilde a(l,s)&=&\sum_{\eta=\pm1}\frac12\,(\eta+e^{-i\pi l})\,a^{(\eta)}(l,s)\,.
}
$\eta$ is called the signature of the partial wave.

To investigate the form of the scattering amplitude in the high energy (Regge)
limit, one transforms the integration contour so that the contour integral
vanishes in this limit.  What remains are the residues of the integrand's poles
the contour swept over during the deformation.  These are called Regge poles.
Using an asymptotic formula for the Legendre polynomials, one obtains for each
of the poles a contribution
\dmath2{
T(s,t)&\propto& s^{\alpha(t)}\,,
&eq:reggeampl}
where $\alpha(t)$ is the pole's position in the complex $l$ plane depending
on~$t$.  In view of this form, one can say that at high energies the scattering
amplitude is dominated by the Regge pole with the largest real part.  The
expression~(\ref{eq:reggeampl}) is usually interpreted as the exchange of a
``Reggeon'', an object with angular momentum equal to~$\alpha(t)$.
$\alpha(t)$~is called its Regge trajectory.

To gain some understanding of Regge trajectories, one continues them to the
region~$t>0$.  It was found that mesons plotted in a diagram of angular
momentum against square mass are arranged in straight lines.  Based on this
fact it is usually assumed in Regge theory that Regge trajectories are linear
functions also for~$t<0$, that is
$$
\alpha(t)=\alpha_0 + \alpha' t\,.
$$
$\alpha_0$ is called a Reggeon's Regge intercept, $\alpha'$ its Regge slope.
From~(\ref{eq:reggeampl}) one derives that the differential cross section for a
process in which a single Reggeon is exchanged is
$$
\frac{d\sigma}{dt}\propto s^{2\alpha_0 + 2\alpha't - 2}\,.
$$

\section{Pomeron and odderon}

Both pomeron and odderon are not particles but ``objects'' which can be
exchanged in scattering processes.  They are well described by Regge theory,
and hence are Reggeons.  This section will give a brief introduction to them
and to how to understand them in terms of QCD.  For a detailed treatment
see~\cite{fr} or~\cite{ddln}.  The state of the art of odderon research has
been reviewed very recently in~\cite{oddreview}.

The existence of the pomeron is universally accepted nowadays.  It is based on
the following path of reasoning: The Pomeranchuk theorem~\cite{pomtheo} states
that total cross sections of scattering processes in which charged particles
are exchanged have to vanish asymptotically for high energy.  Here, ``charge''
can mean any kind of charge, including electric charge, isospin, colour charge
etc.\,.  However, one observes experimentally that at the highest energies for
which measurements are available, total cross sections continue to rise slowly.
This made Pomeranchuk postulate an object, the pomeron, which carries no charge
whatsoever and therefore has the quantum numbers of the vacuum.  Pomeron
exchange can therefore account for the continuing rise of total cross sections.

The existence of the pomeron is well-founded now.  Pomeron exchange was found
to play a significant role also at lower, more accessible energies.  (The
original argument for the pomeron's existence is hampered by the fact that no
experimentally measured energy can reasonably be claimed to be ``asymptotically
high'', however high it may be compared to other experiments.)

The odderon is the partner of the pomeron with odd parity and charge
conjugation quantum number, $P=C=-1$.  All its other quantum numbers are those
of the vacuum.  There is much less experimental evidence for its existence than
for the pomeron.  The best evidence comes from elastic proton-proton and
proton-antiproton scattering and will be presented in Section~\ref{sec:oddpp}.
Nobody has been able to describe the difference between the differential cross
sections in these two processes without an odderon-exchange contribution, but
it is not reassuring that the odderon contributions used by different authors
are not compatible (see Section~\ref{sec:glndlodd}).

\begin{figure}[h]
\centre{\Large\input{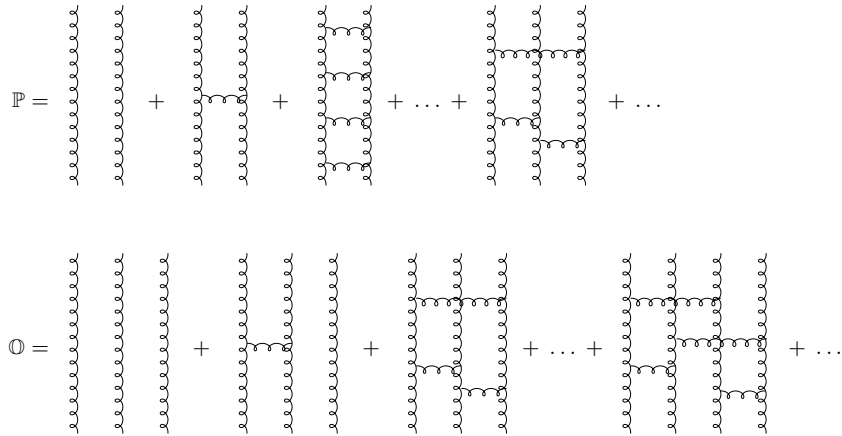}}
\caption{Pomeron and odderon in terms of QCD gluons.  Even though the same 
		Feynman graphs can contribute to both pomeron and odderon, the colour 
		structure is of course different.}
\label{fig:pomoddqcd}
\end{figure}

Even though pomeron and odderon were initially described by Regge theory, they
can of course be understood in the framework of QCD as well.  Both are
represented as the exchange of a bound state of several gluons.  In lowest
order, the pomeron consists of two $t$-channel gluons, and the odderon of
three.  Figure~\ref{fig:pomoddqcd} presents this diagrammatically.  Further
legitimacy is lent to this representation by the fact that some known glueballs
lie on the $t>0$ part of the Regge trajectories of pomeron and odderon.

The description of the odderon in the QCD framework provides the most important
rationale for investigating it.  There may be some doubt about whether the
odderon is observed experimentally.  But there is no doubt that the $C=-1$
multi-gluon exchanges which make up odderon exchange do exist in QCD.  If their
existence was to be disproved experimentally, it would have serious
implications for QCD, which is our best candidate to date for a theory of the
strong interaction.

\section{BFKL and the violation of unitarity}
\label{sec:bfklintro}

\begin{figure}
\centre{\Large\input{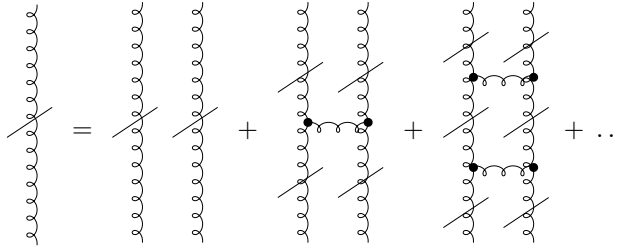}}
\caption{The reggeised gluon is a sum of ladders of which the vertical parts
		are themselves reggeised gluons.  The rungs are ordinary QCD gluons.
		The vertices are sums of QCD vertices which take into account rungs 
		crossing each other.}
\label{fig:reggluon}
\end{figure}

As described in the previous section, pomeron exchange amounts to the exchange
of a bound state of gluons.  The pomeron is more easily described in terms of
{\em reggeised} gluons~\cite{lipreggluon}.  These are gluon ladders not unlike
the pomeron itself, but in a colour octet state, ie with the quantum numbers of
a gluon.  They are graphically represented as gluon lines with slanted lines
through them, as shown in Figure~\ref{fig:reggluon}.

Sums of such ladder diagrams have an important property: At high energies, the
phase space for the additional rungs ($\propto\log s$) compensates the
additional couplings, so that all the ladder terms have the same order of
magnitude.  This means that in a consistent calculation, ladders with any
number of rungs have to be summed up.  In addition, there are genuinely
higher-order graphs in which there are more $t$-channel reggeised gluons or the
rung gluons split up into quark pairs.  Summing up all leading-order ladder
graphs is called the Leading Logarithmic Approximation and results in a sum of
the following form:
$$
\sum_n a_n\;\left(\alpha_s\,\log s\right)^n\,.
$$

In the case of the pomeron, the result of the resummation satisfies the BFKL
(Balitsky-Fadin-Kuraev-Lipatov) equation, an integral
equation~\cite{fklbfkl,blbfkl}.  It is presented in some detail in
Section~\ref{sec:BFKLeq}.  The BFKL equation has an important property: It is
invariant under global conformal transformations in transverse position space.
Only this made it possible eventually to solve it in the non-forward direction,
ie for $t\not =0$.  This also makes it possible to see the BFKL pomeron as a
state of two reggeised gluons evolving according to conformally invariant
two-dimensional quantum mechanics.  Rapidity serves as the time-like
variable because the rung gluons are strongly ordered in rapidity.\footnote{%
This is because everything else is suppressed by the propagators.  The pomeron
has to bridge the large difference in rapidity between the scattering
particles.  Because the gluon propagators punish large momenta (reggeised gluon
propagators are similar to QCD gluon propagators, with an additional factor),
the situation in which the momenta of the gluon rungs evolve monotonously
between the opposing scattering particles is preferred.}

The BFKL equation has one drawback, however: The scattering matrix for BFKL
pomeron exchange is not unitary.  The growth of the associated cross section
with the centre-of-mass energy is power-like rather than logarithmic, which
violates the Froissart-Martin unitarity bound.  This can be remedied by adding
a minimal set of (genuinely) higher-order terms.  The result is then called the
Generalised Leading Logarithmic Approximation (GLLA).  It involves including
some terms and rejecting others.  The selected terms are those which have the
strongest energy dependence and therefore dominate at high energies.

There are two approaches to restoring unitarity.  One restricts itself to
restoring unitarity to the overall scattering matrix
\cite{bartbkp,kwiebkp,lipglla,fadintegr,korintegr}.  This requires additional 
Feynman graphs in which an arbitrary number of reggeised gluons (but at least
two) are exchanged in the $t$ channel.

The more ambitious approach initiated by Bartels seeks to restore unitarity
also to the scattering matrices of subprocesses obtained by cutting up Feynman
graphs \cite{bartbkp,bartglla,bartglla1,bartglla2}.  This requires not only
more reggeised gluons in the $t$ channel, but also vertices which change the
number of $t$-channel gluons.  This approach is sometimes called the Extended
Generalised Leading Logarithmic Approximation (EGLLA).  Analogously to BFKL and
non-extended GLLA as two-dimensional quantum mechanics, this can be interpreted
as a two-dimensional quantum field theory, again with rapidity as the time-like
variable.  Its fundamental degrees of freedom are reggeised gluons which
interact through the number-changing vertices.  If one could prove that all
elements of this field theory are conformally invariant, it would be a
two-dimensional conformal field theory, about which much is known from the side
of mathematical physics.  By solving the theory, one could gain much better
understanding of high-energy scattering.

BFKL and its generalisations have traditionally concentrated on the scattering
of heavy quarkonia (ideally) resp.\ of highly virtual photons (in reality).
In onium-onium scattering, the pomeron couples to a quark loop.  It has been
shown~\cite{bw,carlophd,be} that the coupling of several reggeised gluons in a
colour singlet state to the quark loop can be described by a superposition of
two-reggeised-gluon amplitudes.  Therefore in this process the initial
condition for the evolution in rapidity can be assumed to be a two-gluon
amplitude.  In the EGLLA approach, this amplitude can give rise to amplitudes
containing a larger number of reggeised gluons.  The basis of this approach is
a tower of coupled integral equations describing these amplitudes.

The simplest way in which the number of $t$-channel gluons can change is by way
of a phenomenon known as reggeisation.  This refers to the fact that the
reggeised gluon is a superposition of graphs which themselves contain reggeised
gluons, see Figure~\ref{fig:reggluon}.  This allows a gluon to ``split up''
into several.  There is no vertex involved, it only constitutes a regrouping of
the QCD gluons contained in the reggeised gluon.  The resulting
$n$-reggeised-gluon amplitude can then be written as a superposition of
two-gluon amplitudes.  The EGLLA amplitudes with an odd number of gluons
investigated so far (three and five gluons) were found to reggeise
completely~\cite{bw,carlophd,be}.

The amplitudes with an even number of reggeised gluons are more complex.  So
far only the four- and six-gluon amplitude have been investigated.  The
four-gluon amplitude contains both a reggeising part and a new amplitude which
cannot be expressed as a superposition of amplitudes with fewer reggeised
gluons~\cite{bw}.  The new part arises from the initial two-gluon amplitude by
way of a $2\to4$ reggeised gluon vertex.  The structure of the four-gluon
amplitude can be represented graphically (diagrams from~\cite{be}):
$$
D_4 = \sum \lower 27pt \hbox{\begin{picture}(0,0)%
\includegraphics{solutiond41.pstex}%
\end{picture}%
\setlength{\unitlength}{3947sp}%
\begingroup\makeatletter\ifx\SetFigFont\undefined
\def\x#1#2#3#4#5#6#7\relax{\def\x{#1#2#3#4#5#6}}%
\expandafter\x\fmtname xxxxxx\relax \def\y{splain}%
\ifx\x\y   
\gdef\SetFigFont#1#2#3{%
  \ifnum #1<17\tiny\else \ifnum #1<20\small\else
  \ifnum #1<24\normalsize\else \ifnum #1<29\large\else
  \ifnum #1<34\Large\else \ifnum #1<41\LARGE\else
     \huge\fi\fi\fi\fi\fi\fi
  \csname #3\endcsname}%
\else
\gdef\SetFigFont#1#2#3{\begingroup
  \count@#1\relax \ifnum 25<\count@\count@25\fi
  \def\x{\endgroup\@setsize\SetFigFont{#2pt}}%
  \expandafter\x
    \csname \romannumeral\the\count@ pt\expandafter\endcsname
    \csname @\romannumeral\the\count@ pt\endcsname
  \csname #3\endcsname}%
\fi
\fi\endgroup
\begin{picture}(865,966)(2373,-2016)
\end{picture}
} 
+ \lower 36pt \hbox{\begin{picture}(0,0)%
\includegraphics{solutiond42.pstex}%
\end{picture}%
\setlength{\unitlength}{3947sp}%
\begingroup\makeatletter\ifx\SetFigFont\undefined
\def\x#1#2#3#4#5#6#7\relax{\def\x{#1#2#3#4#5#6}}%
\expandafter\x\fmtname xxxxxx\relax \def\y{splain}%
\ifx\x\y   
\gdef\SetFigFont#1#2#3{%
  \ifnum #1<17\tiny\else \ifnum #1<20\small\else
  \ifnum #1<24\normalsize\else \ifnum #1<29\large\else
  \ifnum #1<34\Large\else \ifnum #1<41\LARGE\else
     \huge\fi\fi\fi\fi\fi\fi
  \csname #3\endcsname}%
\else
\gdef\SetFigFont#1#2#3{\begingroup
  \count@#1\relax \ifnum 25<\count@\count@25\fi
  \def\x{\endgroup\@setsize\SetFigFont{#2pt}}%
  \expandafter\x
    \csname \romannumeral\the\count@ pt\expandafter\endcsname
    \csname @\romannumeral\the\count@ pt\endcsname
  \csname #3\endcsname}%
\fi
\fi\endgroup
\begin{picture}(865,1333)(3691,-2237)
\end{picture}
}\,.
$$
The ``splitting up'' of the reggeised gluons in the left diagram represents the
reggeisation.  The ellipse in the right diagram is the $2\to4$ vertex.  The
rectangle at the top of both diagrams represents the quark loop --- with
attached photon lines --- to which the pomeron couples.

The $2\to4$ reggeised gluon vertex has since been thoroughly investigated.  It
was found to be conformally invariant~\cite{blw}.  It was projected on pomeron
wave functions to obtain a $1\to2$ pomeron vertex~\cite{lotterphd,korconf}.
This pomeron vertex was found to allow an interpretation as a Shapiro-Virasoro
amplitude from string theory~\cite{pesch}.  The reggeising part of the
four-gluon amplitude can also be projected onto two pomerons.  However, it is
sub-leading in the large $N_c$ approximation~\cite{bart3p}.

The six-gluon amplitude also contains a reggeising part and a completely new
part.  In addition, there is a partly reggeising part which is a superposition
of irreducible four-gluon amplitudes (ie those resulting from the $2\to4$ gluon
vertex).  The six-gluon amplitude can like the four-gluon amplitude be
projected onto pomerons, and a $1\to3$ pomeron vertex can be obtained.  For the
irreducible and the partly reggeising part, this will be done in
Chapter~\ref{chap:p3p}.  This vertex is remarkable for being local in rapidity.
By contrast, if two $1\to2$ pomeron vertices occur after each other, the
intermediate four-gluon state evolves in rapidity, and as a result the
transition from one to three pomerons takes some interval on the rapidity axis.

Whether a local $1\to3$ pomeron vertex exists has been contentious.  Braun and
Vacca \cite{brvac} have shown that $1\to n$ pomeron vertices for $n>2$ do not
exist in Mueller's dipole approach.  This approach is a method of deriving the
BFKL equation which is much simpler than the original one involving gluon
ladders~\cite{mueller,mueller2}.  Mueller considers a scenario in which
scattering colour dipoles (onia) recursively split up into more dipoles which
finally interact by exchanging a single QCD gluon.  The dipole splitting leads
to the same expressions as the superposition of gluon ladders in the
traditional derivation of the BFKL equation.  However, the two approaches are
not equivalent to all orders.  For instance, there is no odderon in the dipole
approach: The squared amplitude for single-gluon exchange cannot describe the
exchange of a $C=-1$ object.  The existence of the $1\to3$ gluon vertices
derived in Chapter~\ref{chap:p3p} in the gluon ladder approach constitutes a
further discrepancy.

\cleardoublepage


\addcontentsline{toc}{part}
	{I\hskip 1em The perturbative odderon in $pp$ and $p\bar p$ scattering}

\thispagestyle{empty}

\vbox to \vsize
{
\vfill

\begin{center}
\Huge\bfseries The perturbative odderon in $pp$ and $p\bar p$ scattering
\end{center}

\vskip 15mm

\leftskip=15mm
\rightskip=15mm

This part is primarily concerned with the influence of the proton structure on
the odderon-proton coupling.  To that end, elastic proton-proton and
proton-antiproton scattering is investigated.  A geometric model of the
proton in transverse position space is constructed.  The odderon-proton
coupling depending on the size of a possible diquark cluster in the proton is
computed.  By adding different Reggeon-exchange contributions from a fit by
Donnachie and Landshoff to the perturbative odderon contribution based on the
proton model, it is compared to experimental data and the average size of a
diquark cluster in the proton is fixed within limits.

The same method --- substituting a perturbative odderon contribution within the
framework of the Donnachie-Landshoff fit --- is used to test the validity of
two other models for the odderon-proton coupling, one by Levin and Ryskin and
one by Fukugita and Kwieci\'nski.  The coupling constant to be used with these
models, which is the main parameter in these calculations, is obtained from
experiment.

Lastly, a fit for proton-proton and proton-antiproton elastic scattering by
Gauron et al.\ is compared to the Donnachie-Landshoff fit.  In particular, the
respective odderon contributions of the two fits are exchanged, which allows to
make a statement on the universality, or otherwise, of the odderon-exchange
contributions.

\leftskip=0pt
\rightskip=0pt

\vfill
}

\pagebreak

\chapter{The perturbative odderon contribution to elastic $pp$ scattering}
\label{chap:odderoncalc}

\section{The odderon in elastic $pp$ and $p\bar p$ scattering}
\label{sec:oddpp}

\begin{figure}[h]
\hbox{\input{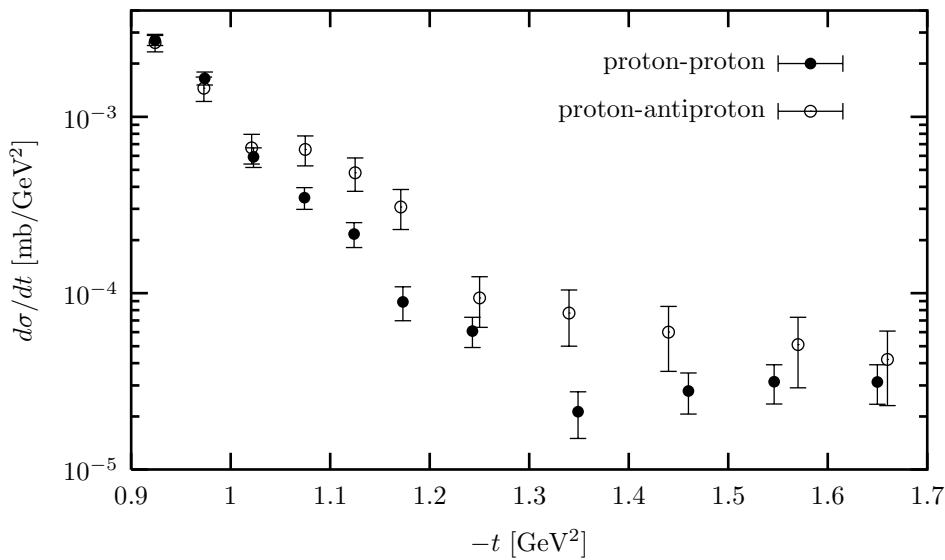}}
\caption{The qualitative difference between elastic $pp$ and $p\bar p$ 
		scattering. The plot shows the differential cross section for 
		$\sqrt s=53$~GeV. Data from~\cite{break}.}
\label{fig:compbar}
\end{figure}

The odderon plays an important part in elastic proton-proton and
proton-antiproton scattering.  As one can see from Figure~\ref{fig:compbar},
the differential cross section of the two processes differs qualitatively.  In
elastic proton-antiproton scattering, the differential cross section falls
continuously with rising squared momentum transfer~$-t$.  At one point, the
curve flattens, but it keeps on falling.  The differential cross section for
proton-proton scattering, on the other hand, shows a slight dip at the same
point in~$-t$.

This is obviously an interference effect.  It comes from different
contributions to the total scattering amplitude interfering constructively in
one case (proton-antiproton) and destructively in the other.  Since the
dominant contribution to elastic scattering at high energies is pomeron
exchange ($C=+1$), the contribution responsible for the difference must be odd
under charge conjugation (and hence couple to proton and antiproton with
different signs).

Odderon exchange is a $C=-1$ contribution.  It is the common consensus that it
is the odderon which is responsible for the difference between the
proton-proton and proton-antiproton elastic differential cross sections.  This
belief is based not so much on rigorous proof but on the fact that all authors
who have successfully described this state of affairs include a significant
odderon contribution.  Disturbingly, though, the different odderon
contributions seem to work only in the framework of one specific fit, see
Section~\ref{sec:glndlodd}.

Another point of caution concerns the experimental data, or more precisely the
lack of data.  53~GeV is the only energy at which both proton-proton and
proton-antiproton elastic differential cross sections have been measured.  The
difference between the two sets of data is between $3\,\sigma$ and $4\,\sigma$
at one point, and $2\,\sigma$ or less at a few others.  This is all the
evidence there is for this effect, and the best experimental evidence for the
existence of the odderon.

The relevant scale for investigating the dip region is the squared momentum
transfer at the dip, $-t\approx 1.4$~GeV$^2$.  We regard this as sufficiently
large for perturbation theory to be applicable.  All the same it would be
desirable to supplement out investigation with a non-perturbative calculation.
For the time being, the perturbative calculation to be presented is the first
step to completely understanding the phenomenon of the dip.

\section[Calculation with a geometric model]
		{Computing the odderon-proton coupling using a \hbox{geometric} model
		 for the \hbox{proton}}
\label{sec:starodd}

\subsection{The geometric model}
\label{sec:starmodel}

\begin{figure}[h]
\center
\includegraphics[height=5cm]{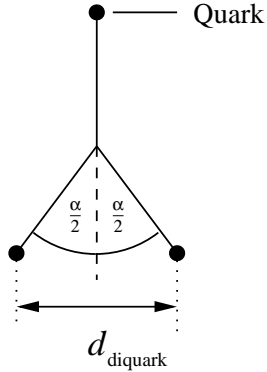}
\caption{The quark-diquark model for the proton}
\label{fig:starmodel}
\end{figure}

In our geometric model, the proton is assumed to be composed of three quarks
arranged in a symmetric star as depicted in Figure~\ref{fig:starmodel}.  The
angle $\alpha$ between the two lower quarks is a free parameter of the model.
For small $\alpha$ the two quarks at the bottom form a diquark cluster of the
size $d_{\rm diquark}$.  We will therefore also call this model the
``quark-diquark model''.  $\alpha=0$ corresponds to an exactly point-like
diquark.

In addition to the geometry just described, a wave function is ascribed to the 
proton:
\begin{equation}
\label{eq:wavefunc}
\psi(R)=\sqrt{\frac2\pi}\,\frac1S\,\exp(-R^2/S^2)\,,
\end{equation}
where $R$ is the distance from the centre of the proton. $S$ is a parameter
determining the size of the proton. 
In calculations of the electromagnetic form factor it was
determined to be 0.96~fm~\cite{paulusdip}. The size relevant for the
strong interaction ought not to be worlds away from that. 
Non-perturbative QCD calculations suggest the range 
$0.7\ldots0.9$~fm~\cite{dnpw,rueterphd}. We use $S=0.8$~fm.


The mean diquark size in transverse space, ie the typical distance between the 
two quarks of the diquark, can be easily computed from the geometry 
and the wave function. It is\footnote{%
This is the diquark size in transverse space. If one assumes the geometric 
star model to be the projection of a three-star rotated in three-dimensional 
space, one can calculate the mean three-dimensional diquark size by 
multiplying (\ref{eq:ddiq}) with $\sqrt2$, which is the average of the 
function~$|\sin|$ which occurs in the projection.
$$\langle d_{\hbox{\scriptsize diquark 3D}}\rangle=
		S \sqrt\pi\sin\left( \frac\alpha2\right)$$
While more accessible to the imagination, this three-dimensional diquark size
is of limited use since many calculations which depend on the smallness
of the diquark are done in transverse space.}
\begin{equation}
\label{eq:ddiq}
\langle d_{\rm diquark} \rangle = S\,\sqrt{\frac\pi2}\, 
									\sin\left(\frac\alpha2\right).
\end{equation}

An important reason for using this model is a common practice in
non-perturbative calculations: Often the proton is treated as a colour dipole.
The diquark plays the part corresponding to that of the antiquark in a meson,
which has the same colour structure.  This practice is justified if two of the
proton's quarks are clustered in a diquark so small that a soft gluon cannot
resolve them.  Using this model in a perturbative calculation allows us to
determine the diquark size and thus to test that hypothesis.


\subsection{High energy scattering in position space}
\label{sec:posscatt}

Performing calculations with the geometric model presented in the previous
section requires treating high energy scattering in position space.  This is
done using a framework developed by Nachtmann~\cite{nacht,nachtmsv}.  It was
created for evaluating non-perturbative models but is as suitable for a
perturbative calculation.  Due to its complexity, this section can give only a
brief sketch of Nachtmann's formalism.  For a more elaborate and didactic
introduction and applications in non-perturbative QCD, see~\cite{nachtmsv}.

\subsubsection{Quark-quark scattering}

One starts out by treating the elastic scattering of two quarks (even though
isolated quarks do not exist in QCD).  Later we will advance to colourless
quark clusters, ie hadrons.  The first step is to relate the $S$ matrix to a
four-point function in position space.  The $S$-matrix element can be
transformed into a Green function in position space with the help of the LSZ
reduction formalism:
\begin{eqnarray}
\label{ha1}
\langle\, p_3\,p_4^{\rm out}|p_1\,p_2^{\rm in}\,\rangle 
&=& Z_\psi^{-2} \int d^4x_1\cdots d^4x_4
\exp \left[i(p_3x_3+p_4x_4-p_1x_1-p_2x_2)\right] \times
\nonumber \\
&& \times \langle\, {\rm T} \bar u(p_3) f(x_3)\bar u(p_4) f(x_4)
\bar f(x_1) u(p_1) \bar f(x_2) u(p_2)\,\rangle\,.
\end{eqnarray}
Here the $p_i$ are the momenta of the incoming and outgoing quarks, 
$f(x) = (i\gamma^\mu\partial_\mu - m) \psi(x)$ and 
$Z_\psi$ is the wave function renormalisation. 

The four point function $\langle\, {\rm T} \psi(x_3)\psi(x_4)\bar\psi(x_1)\bar
\psi(x_2)\,\rangle$ contained in the expression on the right hand side of
Equation~\ref{ha1} can be expressed as a functional integral over the quark and
the gluon fields:
\begin{equation}
\langle\,{\rm T} \psi(x_3)\psi(x_4)\bar \psi(x_1)\bar \psi(x_2)\,\rangle = 
\int {\cal D}\psi\,
{\cal D} \bar\psi {\cal D} {\bf B} \,\psi(x_3)\psi(x_4)\bar \psi(x_1)
\bar \psi(x_2) \exp[-i S_{\rm full\;QCD}]\,, 
\label{ha2}
\end{equation}
where $S_{\rm full\;QCD}$ is the full QCD action.
Since the action is quadratic in the fermion fields, the fermion integration 
is Gaussian. Therefore the quark fields can be integrated out, which 
gives a determinant of the Dirac operator and several quark propagators:
\begin{eqnarray} \label{ha3}
&&\hspace{-3mm}
\langle\,{\rm T} \psi(x_3)\psi(x_4)\bar \psi(x_1)\bar \psi(x_2)\,\rangle =
\int {\cal D} {\bf B} \det[-i(i\gamma D - m)]\cdot  \\
&& \cdot \left[ S_F(x_3,x_1;{\bf B})\,S_F(x_4,x_2;{\bf B})+
S_F(x_3,x_2;{\bf B})\,S_F(x_4,x_1;{\bf B}) \right]
\exp[-iS_{\rm pure\;QCD}] \,.
\nonumber \end{eqnarray}
$S_F(x_i,x_j;{\bf B})$ are the quark propagators in the external colour 
potential ${\bf B}_\mu$. The remaining functional integration is to be 
performed only over the gluon fields with the pure QCD action in the
exponential, ie the one describing pure gluodynamics. 

In leading order the determinant of the Dirac operator can be set to one.
The two terms with propagators correspond to $t$-channel and $u$-channel 
exchange, respectively. Here we apply the eikonal approximation, ie
we assume that the four  momentum transfer is small compared to the 
total energy. Then the $u$-channel term can be neglected. 

After plugging (\ref{ha3}) into (\ref{ha1}), the propagators can be
wrapped up in scattering matrix elements for ``scattering'' one quark in an
external gluon field by following the above derivation in reverse.
\begin{equation}
\langle\, p_3\,p_4^{\rm out}|p_1\,p_2^{\rm in} \,\rangle
\approx Z_\psi^{-2}\, \int {\cal D} {\bf B}\,
{\cal S}(p_3,p_1;{\bf B}) {\cal S}(p_4,p_2;{\bf B}) 
\exp[-i S_{\rm pure\;QCD}] 
\label{ha4} 
\end{equation}
${\cal S}(p_i,p_j;{\bf B})$ is the $S$-matrix element for scattering a quark 
with momentum $p_j$ to one with momentum $p_i$ in an external colour field
${\bf B}_\mu$.

\subsubsection{A quark in a background field}

Our next goal is to find an expression for the $S$-matrix element 
${\cal S}(p_i,p_j;{\bf B})$. One can show \cite{nacht} that the quark 
scattering matrix elements in an external field can be expressed as a 
generalised WKB expression 
\begin{equation}
{\cal S}(p_i,p_j;{\bf B}) = \bar u(p_i) \gamma^\mu u(p_j) 
{\rm P}\exp\left[-ig\int_\Gamma {\bf B}_\rho\,dx^\rho\right]
\left(1+O\left(\frac{1}{p^0_i}\right)\right)\,,
\label{ha5}
\end{equation}
where {\bf B} is the gauge potential which takes values in $su(3)$. The
path-ordered integral is taken along the classical path $\Gamma$. 
For high-energy scattering the relevant paths are the following 
light-like paths:
\begin{equation}
\Gamma_1=(x^0,\mbf b/2, x^3=x^0) \quad \mbox{ and }\quad \Gamma_2=
(x^0, -\mbf b/2, x^3= -x^0)\,.
\label{3A2}
\end{equation}
They correspond to quarks moving in opposite directions along the 
$x^3$~axis at the speed of light. $\mbf b$ is the impact parameter.
Let ${\rm\bf V}_i$ be the phases picked up by the quarks along these paths:
\begin{equation}
{\rm\bf V}_i={\rm P}\exp\left[-ig \int_{\Gamma_{i}} 
{\rm\bf B}_\mu(z)\ dz^\mu\right] \,.
\label{3A3}
\end{equation}
It can be derived~\cite{nachtmsv} that the quark moving in positive 
$x^3$ direction has the wave function
\begin{equation}
\psi_{1}(x)= {\bf V}_1\;e^{-ip_1x}\,u(p_1)\,
				\left(1+O\left(\frac1{p_1^0}\right)\right)\,.
\end{equation}
An analogous expression is valid for the other quark. From them the 
S-matrix element~(\ref{ha5}) can be calculated. This can be plugged into
the quark-quark scattering matrix element~(\ref{ha4}). The ensuing 
calculations lead to rather lengthy formulas which will not be 
given here. They are presented very well in~\cite{nachtmsv}. 
The $S$-matrix element~(\ref{ha4}) is simplified by performing
the path-ordered integrations from~(\ref{3A3}). After again taking
the high energy limit, the result reads:
\begin{eqnarray}
\label{3A5}
\langle\, p_3\,p_4^{\rm out}|p_1\,p_2^{\rm in} \,\rangle
& \approx &  2is
\left\langle\int d^2 b\ e^{-i\mbf q \cdot \mbf b}\;
\left({\bf V}_1\right)_{\alpha_3\alpha_1}\left({\bf V}_2\right)_{\alpha_4\alpha_2}
\right\rangle\,,    \\
&&\hbox{where}\quad \langle \hbox{\it Expression}\rangle=
\int{\cal D}{\bf B}\;\hbox{\it Expression}
\,.
\nonumber
\end{eqnarray}
The non-Abelian phases ${\bf V}_i$ still depend on the impact parameter $\mbf
b$ and the gluon field ${\bf B}$.  The $\alpha_i$ are the colour indices of the
quarks and $\mbf q=\mbf p_1-\mbf p_3$ is the transverse momentum transfer.
(Small bold-face letters denote vectors in transverse space.)  Even though the
quark momenta are written explicitly here, the $S$-matrix element depends only
on $s=(p_1+p_2)^2$ and $t=(p_1-p_3)^2$.

\begin{figure}[h]
\begin{center}
\includegraphics[height=6cm]{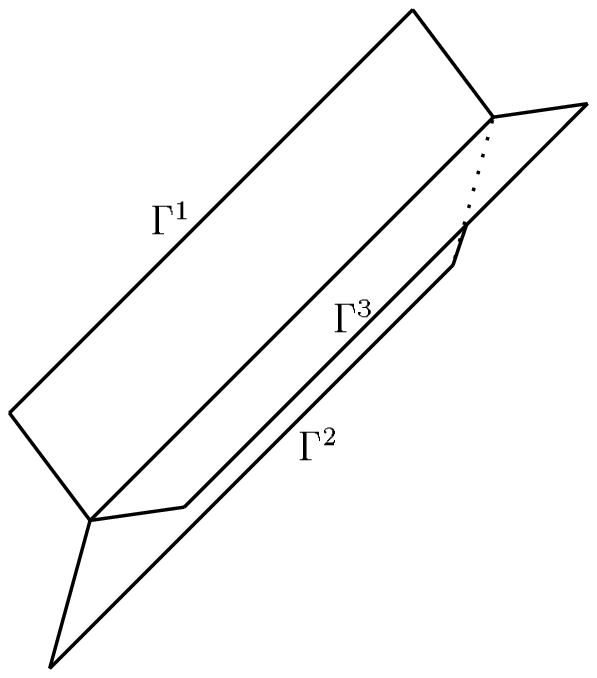}
\end{center}
\caption{The paths in a colour neutral three-quark cluster}
\label{fig:baryonpaths}
\end{figure}                                                                 

\subsubsection{Nucleon-nucleon scattering}
\label{sec:hadrhadr}

In order to come to the nucleon-nucleon scattering amplitude 
we first consider the scattering of two groups of three quarks 
moving on parallel light-like world lines of the form
\begin{equation}
\Gamma^a_1(x_0,\frac{\mbf b}2+\mbf x\,_1^a, x^3=x^0),~~
\Gamma^a_2(x_0,-\frac{\mbf b}2+\mbf x\,_2^a, x^3=-x^0),~~ a=1,2,3 \,.
\end{equation}
The $x\,_i^a$ are the positions of the quarks in cluster $i$ relative 
to its centre. In order to ensure that these quark clusters 
asymptotically form colour singlet states 
all colours are parallel-transported in the remote past and future to the
centre of the cluster and contracted there antisymmetrically.
This leads to the following $S$-matrix element
for the scattering of colour-neutral clusters~\cite{dfk}:
\begin{eqnarray}
\label{nuc1}
\lefteqn{S\left(\mbf b,\left\{\mbf{x}\,_i^a\right\}\right) = 
				\frac1{36}\frac1{Z_1 Z_2} \times} 
\\
&&
\times \left\langle\epsilon_{\alpha\beta\gamma}
\left({\bf V}^1_1\right)_{\alpha\alpha'}\left({\bf V}^2_1\right)_{\beta\beta'}
\left({\bf V}^3_1\right)_{\gamma\gamma'}\epsilon_{\alpha'\beta'\gamma'}
\epsilon_{\rho\mu\nu}\left({\bf V}^1_2\right)_{\rho\rho'}
\left({\bf V}^2_2\right)_{\mu\mu'}
\left({\bf V}^3_2\right)_{\nu\nu'}\epsilon_{\rho'\mu'\nu'}\right\rangle \,. 
\nonumber
\end{eqnarray}
The lower Greek indices are the colour indices of the quarks running from 1
to~3.  Because now the positions rather than the momenta of the quarks are
given, the Fourier integral from~(\ref{3A5}) is missing.  The non-Abelian phase
factors ${\bf V}_i^a$ are defined as in (\ref{3A3}) with the $\sqcup$-shaped
integration paths $\Gamma_i$ in Figure~\ref{fig:baryonpaths}.  The $Z_i$ denote
the wave function renormalisation for the clusters which in lowest order can be
set to one.

We will later use not the $S$-matrix element but the reduced 
scattering amplitude $J$. They are related in the following way:
\begin{equation}
\label{JisSminus1}
J(\mbf b,\{\mbf x\,_i^a\})= S(\mbf b,\{\mbf x\,_i^a\})-1\,.
\end{equation}

\subsection{Application to perturbative odderon exchange}
\label{sec:pathapply}

In our perturbative calculation we expand the ${\bf V}^a_i$ in $g$.
Up to order $g^3$ Equation~\ref{3A3} becomes
\begin{eqnarray}
\label{eq:Vexpa}
\left({\bf V}^a_i\right)_{\alpha\beta} &=&
\delta_{\alpha\beta} - i g  \hat B_{a,i}^c \tau^c_{\alpha\beta}
-\frac12 g^2 
\hat B_{a,i}^c\hat B_{a,i}^{c'}(\tau^c\tau^{c'})_{\alpha\beta} \nonumber \\
&&-\frac{i^3}{3!} g^3 
\hat B_{a,i}^c\hat B_{a,i}^{c'}\hat B_{a,i}^{c''}
(\tau^c\tau^{c'}\tau^{c''})_{\alpha\beta} + {\cal O}(g^4) \,,
\end{eqnarray}
where $\hat B_{a,i}^c$ are the coefficients of the expansion of the path 
integral
in the Gell-Mann matrices $\tau^c$:
\begin{equation}
\label{eq:integexpa}
\int_{\Gamma^a_i}dz^\mu~{\bf B}_\mu(z)= \hat B_{a,i}^c \tau^c \,.
\end{equation}

In general we would have to take into account path ordering
in~(\ref{eq:Vexpa}).  However, the odderon contribution which we want to
calculate has a symmetric colour structure.  This is because the $\gamma$
matrices which are the spinor space part of the vertices are antisymmetric.
This already gives a minus sign under charge conjugation.  Therefore the colour
structure has to be symmetric to give overall oddness under charge
conjugation.  Due to the symmetric colour structure permuting the $\hat
B_{a,i}^c$ changes nothing and we may discard the path ordering.

The next step is to expand the $S$-matrix element (\ref{nuc1}) in $g$ (using
the expansion (\ref{eq:Vexpa})) and to extract the terms which represent the
lowest order odderon contribution.  To this order the odderon is just composed
of three gluons in a $C=-1$ state.  Therefore the relevant terms are of
order~$g^6$.  Three of the six factors $g$ come from one hadron, three from the
other.  Otherwise some of the gluons would contribute to the self-energy of a
hadron rather than to the interaction between them.  As we have already argued,
the terms are symmetric in colour space.

The expansion of (\ref{nuc1}) is not given explicitly here because it is too
complex.  Even the odderon involves more than a hundred terms so that they
shall not be given in full.  It helps a lot to classify them according to the
ways the odderon couples to the proton at either end.  This will be done in the
next section on the basis of a Feynman graph description of the odderon
exchange.  This is the description which is easiest to implement in the
numerical calculations which will be necessary.  The derivation of an
equivalent formulation from Equation~\ref{nuc1} is described in
Appendix~\ref{sec:corrclass}.  It is rather involved and not of direct
relevance for the results of our work.

\subsection{Colour structure and combinatorics}
\label{sec:oddcolour}

\begin{figure}[tb]
\begin{center}
\includegraphics[width=\hsize]{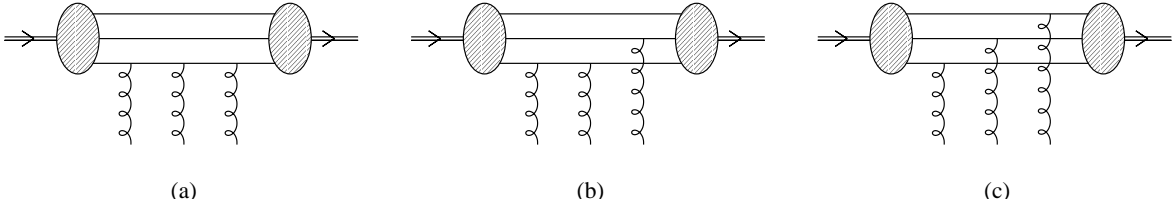}
\end{center}
\caption{The three ways in which the odderon can couple to the proton
			in the lowest order.}
\label{fig:oddcoup}
\end{figure}

Let us first consider one end of the odderon exchange, ie one half of
the expression in the brackets in Equation~\ref{nuc1}.
Three gluons can couple to the proton in the three ways displayed in
Figure~\ref{fig:oddcoup}. Type (a) corresponds to the expression which
contains the third-order term of one of the ${\bf V}_i^a$ and the 
$\delta_{\alpha\beta}$ of the two others (cf. Equation~\ref{eq:Vexpa}).
Type (b) represents one in which there is the second-order term, 
first-order term and $\delta$ symbol of one ${\bf V}_i^a$, respectively.
In type (c) the linear terms in $g$ of all three ${\bf V}_i^a$ appear.
Since three gluons couple to each hadron, the sum of the orders of
the contributing terms from the expansion of the three ${\bf V}_i^a$
has to be three.

The colour tensors differ according to the type of the coupling but are
independent of the permutations, that is which gluon couples to which quark.
They are calculated as follows:
\begin{eqnarray*}
\hbox{(a)}\quad C_a^{cc'c''} & = &
	\epsilon_{\alpha\beta\gamma}\;(\tau^c\tau^{c'}\tau^{c''})_{\alpha\alpha'}\;
	\delta_{\beta\beta'}\;\delta_{\gamma\gamma'}\;
	\epsilon_{\alpha'\beta'\gamma'}
	= \epsilon_{\alpha\beta\gamma}\;\epsilon_{\alpha'\beta\gamma}\;
	(\tau^c\tau^{c'}\tau^{c''})_{\alpha\alpha'}    \\
&&=	2\,\delta_{\alpha\alpha'}\,(\tau^c\tau^{c'}\tau^{c''})_{\alpha\alpha'}
	=2\,\hbox{Tr}(\tau^c\tau^{c'}\tau^{c''})
	=\hbox{Tr}([\tau^c,\tau^{c'}]\tau^{c''}+\{\tau^c,\tau^{c'}\}\tau^{c''})  \\
&&= \frac12if^{cc'c''}+\frac12d^{cc'c''}   \\
\hbox{(b)}\quad C_b^{cc'c''} & = & 
	\epsilon_{\alpha\beta\gamma}\;(\tau^c\tau^{c'})_{\alpha\alpha'}\;
	\tau^{c''}_{\beta\beta'}\;\delta_{\gamma\gamma'}\;
	\epsilon_{\alpha'\beta'\gamma'} 
	= \epsilon_{\alpha\beta\gamma}\;\epsilon_{\alpha'\beta'\gamma}\;
	(\tau^c\tau^{c'})_{\alpha\alpha'}\;\tau^{c''}_{\beta\beta'}  \\
&&= (\delta_{\alpha\alpha'}\;\delta_{\beta\beta'}-\delta_{\alpha\beta'}\;
	\delta_{\beta\alpha'}) \;(\tau^c\tau^{c'})_{\alpha\alpha'}\;
	\tau^c_{\beta\beta'} 
	=\hbox{Tr}(\tau^c\tau^{c'})\;\hbox{Tr}(\tau^{c''})-
	\hbox{Tr}(\tau^c\tau^{c'}\tau^{c''})		\\
&&= -\hbox{Tr}(\tau^c\tau^{c'}\tau^{c''})
	= -\frac14if^{cc'c''}-\frac14d^{cc'c''}  \\
\hbox{(c)}\quad C_c^{cc'c''} & = & 
	\epsilon_{\alpha\beta\gamma}\;\tau^c_{\alpha\alpha'}\;
	\tau^{c'}_{\beta\beta'}\;\tau^{c''}_{\gamma\gamma'}\;
	\epsilon_{\alpha\beta\gamma} 
	=\frac12d^{cc'c''}\qquad\hbox{(numerically)}
\end{eqnarray*}
As has been mentioned before, only the symmetric colour structure constants 
$d^{cc'c''}$ are relevant for the odderon contribution because the structure
in spinor space is antisymmetric. We obtain the following colour tensors for 
leading-order odderon exchange:
\begin{eqnarray}
\label{eq:coupcolour}
\hbox{(a)}\quad C_{a,\;{\rm odd}}^{cc'c''} & = & 
							\phantom{-}\frac12d^{cc'c''} \nonumber \\
\hbox{(b)}\quad C_{b,\;{\rm odd}}^{cc'c''} & = & -\frac14d^{cc'c''}  \\
\hbox{(c)}\quad C_{c,\;{\rm odd}}^{cc'c''} & = & 
							\phantom{-}\frac12d^{cc'c''}. \nonumber
\end{eqnarray}

Having completed the investigation of the coupling types, we can now classify
the Feynman graphs according to the types of the couplings at each hadron. 
\begin{figure}[h]
\begin{center}
\includegraphics[width=12cm]{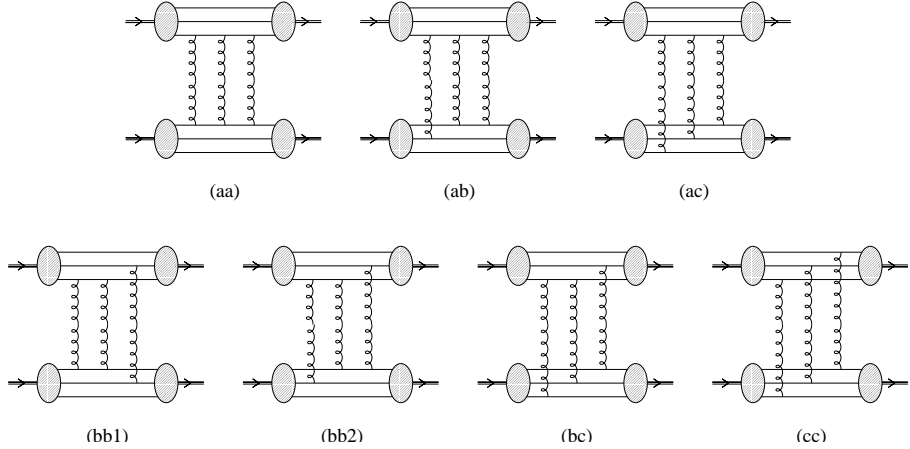}
\end{center}
\caption{Seven exemplary Feynman graphs representing the seven graph types.}
\label{fig:graphtypes}
\end{figure}
Combining three types of coupling at either end yields six possibilities,
ignoring which type occurs at which hadron. However, two (b)-type couplings 
can be combined in two ways: The pair of gluons coupling to the same quark 
may or may not be the same at both ends. Hence we obtain seven types of 
Feynman graphs. One representative of each type is displayed in 
Figure~\ref{fig:graphtypes}. This classification was already used by
R\"uter in non-perturbative calculations~\cite{rueterphd}.

The colour structure of the odderon gives an overall colour factor
of~$d^{cc'c''}\,d^{cc'c''}=\frac{40}3$.  In addition, the prefactors of the
symmetric structure constants in Equation~\ref{eq:coupcolour} contribute
different factors to different graph types.

Another type-dependent factor comes from combinatorics: To facilitate our
numerical calculations, we assume the gluons to be distinguishable.  To
compensate this overcounting, the result has to be divided by the number of
permutations of gluons which couple to the same quarks.  For instance, the
contribution of the graphs of type (bb1) will be overcounted by a factor of two
because for each graph an equivalent one will be calculated in which the two
first gluons are exchanged (see Figure~\ref{fig:graphtypes}).  By contrast, for
type (bb2) the gluons are distinguished by which quarks they couple to and no
overcounting occurs.

Table~\ref{tab:prefact} shows the colour and combinatorial factors for each 
graph type. The total type-specific prefactor $C_{\rm type}$ is arbitrarily 
defined as one for graph type (aa). That means that an overall 
factor of $\frac14\frac16=\frac1{24}$ has to be multiplied with
$C_{\rm type}$ to get the right factor for a specific graph type.
Also given in Table~\ref{tab:prefact} is the number of graphs belonging
to each type. They add up to the number of different Feynman graphs,~165.

A more rigorous derivation of the type-dependent prefactors is given in
Section~\ref{sec:corrclass} in Appendix~\ref{app:tofeyn}. They arise from
the coefficients in the expansion~(\ref{eq:Vexpa}) and from summing up 
equivalent terms in the expansion of the six-gluon correlator~(\ref{nuc1}).

\begin{table}[h]
\moveright 6mm \vbox{
\def\eol{\cr\noalign{\nointerlineskip}}
\def\m{\llap{$-$}}
\tabskip=0pt
\halign{
\vrule height2.75ex depth1.25ex # \tabskip=1em & #\hfil &#\vrule& 
	\hfil$#$\hfil &#\vrule& \hfil$#$\hfil 
	&#\vrule& \hfil $#$ & \hskip 60pt plus 1fil #\vrule & 
	\hfil # \qquad &#\vrule \tabskip=0pt\cr\noalign{\hrule}
& Type && \hbox{Colour factor} && \hbox{Comb. factor} &
	&\omit\hfil $C$\rlap{${}_{\rm type}$} &&\omit\hfil \# of graphs &\cr
\noalign{\hrule}
& (aa) && \frac14    && \frac16 && 1 \rlap{ (by definition)} 
                    					  &&  9  &\eol
& (ab) && \m\frac18  && \frac12 && -\frac32 &&  36 &\eol
& (ac) && \frac14    && 1       && 6        &&   6 &\eol
& (bb1)&& \frac1{16} && \frac12 && \frac34  &&  36 &\eol
& (bb2)&& \frac1{16} && 1       && \frac32  &&  36 &\eol
& (bc) && \m\frac18  && 1       && -3       &&  36 &\eol
& (cc) && \frac14    && 1       && 6        &&   6 &\cr
\noalign{\hrule}
}}
\caption{Prefactors of the different graph types and number of
		Feynman graphs belonging to each type. See text for explanation.}
\label{tab:prefact}
\end{table}

\subsection{The odderon-exchange scattering amplitude in the geometric model}

We can now put everything together and write down the reduced scattering
amplitude for odderon exchange between two protons described by our geometric
model. Since the quarks have a relative position fixed by the rigid star shape,
we can discard the six position variables ${\mbf x\,_i^a}$ and use just two
transverse-space vectors $\mbf R_i$, $i=1,2$, which describe the radius and
orientation of the star shape for each baryon.

The reduced scattering amplitude $J$ is defined as the scattering amplitude
of two protons with fixed size and star angle, minus a total factor $2is$.
It contains the overall factors we have computed so far: The colour 
factor $\frac{40}3$, a factor $\frac1{36}$ from Equation~\ref{nuc1} and a
conventional factor $\frac1{24}$ extracted from the type-dependent
prefactors. 

$J$ sums up all Feynman graphs with the appropriate prefactors. It depends 
on the impact parameter $\mbf b$ and the~$\mbf R_i$ which parametrise
the size and orientation of each proton.
\begin{equation}
\label{eq:j}
J(\mbf b, \mbf R_1, \mbf R_2) = 
\frac1{36} \frac{40}3 \frac1{24} g^6
\sum_{\rm types} C_{\rm type} \sum_{\{a_i\}\,\{b_i\}} 
\Chi_{a_1b_1}\,\Chi_{a_2b_2}\,\Chi_{a_3b_3}
\end{equation}
The second sum runs only over the sets of indices which are compatible with 
a specific graph type. The indices $a_i$ denote the quarks of the first 
proton the gluons couple to, the $b_i$ the quarks of the second proton.
$\Chi_{ab}$ is the gluon propagator in transverse space:
\begin{equation}
\Chi_{ab} = \frac i{2\pi}\,K_0(m\,|\mbf x_1^a-\mbf x_2^b|)\,.
\label{eq:chi}
\end{equation}
It is derived in Appendix~\ref{sec:intlight}. $m$ is a non-zero gluon
mass which we have to introduce intermediately to compute $\Chi$. In the
gauge invariant expressions which arise after summing up all graphs
it can be set to zero without causing problems.

The scattering amplitude is obtained by integrating over the reduced scattering
amplitude with a wave function~(\ref{eq:wavefunc}) for each proton.  Since we
want to obtain the odderon contribution as a function of the squared four
momentum transfer~$-t$, we have to Fourier transform it back into momentum
space.
\begin{equation}
T(s,\mbf q) = 2\,i\,s \int d^2 \mbf b \; e^{-i\,\mbf q\mbf b}
					 \int d^2 \mbf R_1 \int d^2 \mbf R_2 \;
                     |\psi(R_1, S)|^2 \; |\psi(R_2, S)|^2 \; 
					 J(\mbf b, \mbf R_1, \mbf R_2)
\end{equation}
The polar angle in one of the three planar integrations is redundant since the
forces between the hadrons depend only on their {\em relative} orientation.  We
can choose to measure the angles in the $\mbf R_1$ and $\mbf R_2$ planes
relative to the vector~$\mbf b$.  Then $J$ depends only on $b$~modulus.  Hence
we integrate out the angle in the $\mbf b$ plane and get a factor $2 \pi$ and a
Bessel function~$J_0$ instead of the exponential function from the Fourier
transform.  Using $|\mbf q|=\sqrt{-t}$, the result reads:
\begin{equation}
\label{eq:tstar}
T(s,t)  =  4 \pi\,i\,s \int db\,b \int d^2 \mbf R_1 \int d^2 \mbf R_2 \; 
                     J_0(\sqrt{-t} \, b) \; |\psi(R_1, S)|^2 \; 
                     |\psi(R_2, S)|^2 \;  J(b, \mbf R_1, \mbf R_2)\,.
\end{equation}

The remaining integrals can 
only be evaluated numerically. We do this using the {\sc Vegas} 
integration routine from Numerical Recipes, an adapting Monte-Carlo
method. In most cases, 6~iterations of the algorithm with 5\,000\,000
evaluations of the integrand each were sufficient. For some values of
the parameters, such as small angles in the star model, 8~iterations
\`a~7\,000\,000 evaluations were used.

To determine the errors of the integration, it was executed several 
times with a different seed value for the random number generator
on which the Monte-Carlo routine relies. The variation of the result proved 
a better indicator of the numerical errors than the built-in
error estimation. The typical error in the scattering amplitude is
2~\%, though it can be as much as 10~\% for very small star angles or 
large~$t$.

\section[Calculation from momentum-space impact factors]
		{Calculation of the odderon scattering amplitude from impact factors
		 in transverse momentum space}
\label{sec:impodd}

\subsection{Introduction}

Impact factors are expressions describing the coupling of a Reggeon being
exchanged to the scattering particles. By that token, the calculations
described in the previous section use impact factors in transverse position
space based on the geometric model. Impact factors in momentum space are
however far more common.

It can be shown that the scattering amplitude for a particular process
factorises into two impact factors and a propagator.  Hence, the odderon
scattering amplitude is calculated by convoluting two odderon-proton impact
factors with the propagators of the three gluons forming the odderon.
\begin{equation}
\label{eq:timpac}
T(s,t) = 2 s \frac{40}3 \frac1{3!} \frac1{(4\pi)^4} \int
        d^2\bdelta_{1t} \, d^2\bdelta_{2t} \,
        \Phi_p(\bdelta_{1t},\bdelta_{2t},\bDelta_t-\bdelta_{1t}-\bdelta_{2t})^2
        \frac1{\bdelta_{1t}^2\, \bdelta_{2t}^2\, 
			(\bDelta_t-\bdelta_{1t}-\bdelta_{2t})^2}
\end{equation}
$\bdelta_{it}, i=1,2,3$ are the transverse momenta of the gluons and
$\bDelta_t=\bdelta_{1t}+\bdelta_{2t}+\bdelta_{3t}$ is the transverse momentum 
of the odderon.
$40/3$~is the colour factor (equal to $d^{cc'c''}\,d^{cc'c''}$) 
and $1/3!$ the combinatorial factor reflecting the fact that the gluons
are indistinguishable.

In the high energy limit $\bDelta_t^2 = -t$. This defines only the modulus of
$\bDelta_t$. Since the scattering amplitude depends only on $t$ and not on the
vector $\bDelta_t$, the integral has to be independent of its direction.  This
is true because both the propagators and the impact factors depend only on the
relative orientation of the vectors $\bdelta_{it}$. Since $\bdelta_{1t}$ and
$\bdelta_{2t}$ are integrated over, the direction in which $\bDelta_t$, and
therefore $\bdelta_{3t}$, points is irrelevant.  Consequently in our
calculations we assume that $\bDelta_t$ points in one particular direction
(rather than averaging over the polar angle in the $\bDelta_t$ plane, for
instance).

The two two-dimensional integrals in Equation~\ref{eq:timpac} 
have to be evaluated numerically. We used the {\sc Vegas} routine from 
Numerical Recipes with 5~iterations \`a~5\,000\,000 evaluations of the 
integrand. The errors in the scattering amplitude were smaller than 0.5~\%.

\subsection{The general form of the odderon-proton impact factor}

It can be derived from the requirement of gauge invariance that the
odderon-proton impact factor has to be of the following form:
\begin{equation}
\label{eq:impact}
\Phi_p  =   g^3 \,[F(\bDelta_t,0,0) - 
                \sum_{i=1}^3 F(\bdelta_{it}, \bDelta_t-\bdelta_{it}, 0)
                +2F(\bdelta_{1t},\bdelta_{2t},\bdelta_{3t})]\,.
\end{equation}
The first expression in the brackets represents the diagram~(a) in 
Figure~\ref{fig:oddcoup}, the second diagram~(b), and the third corresponds to
diagram~(c). (Due to gauge invariance, all diagrams have to be there.)
$F(\bdelta_{1t},\bdelta_{2t},\bdelta_{3t})$ is the proton form factor which can
be chosen in different ways. 

The form factor has to be constructed so that the impact factor
vanishes when one of the $\bdelta_{it}$ is zero. This is required by gauge 
invariance and has the additional benefit that the integrand in 
Equation~\ref{eq:timpac} is analytic.

\subsection{The form factor of Levin and Ryskin}

A paper on proton-proton scattering by Levin and Ryskin~\cite{lev} uses a
Gaussian form factor for the proton.  It is constructed in such a way that it
gives a significant contribution only when all the gluon momenta are small.
\begin{equation}
\label{eq:formLev}
F(\bdelta_{1t},\bdelta_{2t},\bdelta_{3t}) = \exp\left(-R_p^2 \,
		\left(\bdelta_{1t}^2+\bdelta_{2t}^2+\bdelta_{3t}^2\right)\right)
\end{equation}
There is some confusion about the proton radius~$R_p$ since it is given in the
paper~\cite{lev} as \hbox{``$R_p^2=2.75$~GeV$^2$''}.  The more reasonable value
of~$R_p=0.32723$~fm is obtained when assuming the wrong sign to be in the 
exponent of the unit~GeV.

\subsection{The form factor of Kwieci\'nski et al.}

In 1997, Kwieci\'nski et al.\ published calculations of the odderon exchange in
the photo- and electroproduction of the $\eta_c$ meson~\cite{kwie_impac}, based
on earlier work by Fukugita and Kwieci\'nski~\cite{fk}.  To that end, they
compute odderon-photon and odderon-proton impact factors.  For us only the
odderon-proton impact factor is of interest.  It is constructed according to
Equation~\ref{eq:impact} from the following form factor:
\begin{equation}
\label{eq:formKwie}
F(\bdelta_{1t},\bdelta_{2t},\bdelta_{3t}) = \frac{A^2}{A^2 + \frac12
        [(\bdelta_{1t}-\bdelta_{2t})^2 + (\bdelta_{2t}-\bdelta_{3t})^2 + 
         (\bdelta_{3t}-\bdelta_{1t})^2]}.
\end{equation}
Its width is proportional to $A$ which was chosen by the authors to be half 
the rho mass, 384 MeV. Unlike the form factor of Levin and Ryskin, it can
still give a significant contribution when the gluon momenta are large.
This is the case when they are large but close to each other, ie when the
gluons move in the same direction.

\section[The Donnachie-Landshoff fit]
		{The Donnachie-Landshoff fit: A complete set of contributions to 
			elastic $pp$ and $p\bar p$ scattering}

In order to fit experimental data of proton-proton scattering, one has to
take into account other contributions besides the odderon. We decided to
obtain those from the phenomenological fit by A.~Donnachie and P.~Landshoff
published in~1984~\cite{dlfit}. It is well-known for reproducing
the dip in the differential cross section in proton-proton scattering.
Remarkably, the same parameter values fit the data for any centre-of-mass
energy, and the wandering of the dip with changing $s$ results 
automatically.

We use the Donnachie-Landshoff fit as a phenomenological input to our
calculations.  Since it depends only weakly on $\alpha_s$, in most cases we
leave $\alpha_s$ constant at 0.3, the value used in the original work.
Whenever it is changed, this will be noted in the results in chapters
\ref{chap:oddresultspp} and~\ref{chap:oddresultsppbar}.

Donnachie and Landshoff take into account a number of contributions
to the elastic proton-proton scattering amplitude. They will be 
described briefly in the following sections. For details see~\cite{dlfit}.

\subsection{All contributions of the DL fit}
\label{sec:dlfit}

\subsubsection{The pomeron}
\label{sec:dlpom}

The pomeron is described in the framework of Regge theory (see
section~\ref{sec:reggeon}).  Donnachie and Landshoff's pomeron has intercept
$\alpha_\pom(0)=1.08$ and slope $\alpha'_\pom=0.25\hbox{ GeV}^{-2}$.

To describe the pomeron-proton coupling, Donnachie and Landshoff use a proton
form factor of the following form:
$$
F_1(t)=\frac{4m^2-2.79t}{4m^2-t}\frac1{(1-t/0.71)^2}\,.
$$
The complete expression for the scattering amplitude is as follows.  (It
differs by a factor of two from \cite{dlfit} since our definition of the
scattering amplitude is different by that factor, {cf.}
Equation~\ref{eq:dsigdt}.)
$$
T_\pom(s,t)=
2i\,s\,\left(3\,\beta\,F_1(t)\right)^2\,
		\left(\frac{-i\,s}{m^2}\right)^{\alpha_\pom(t)-1}
$$

Since the pomeron is even under charge conjugation, it contributes equally
to proton-proton and proton-antiproton scattering.

\subsubsection{The Reggeon contribution}

The Reggeon contribution combines the effects of four different particle
exchanges: $\rho$ and $\omega$ (which are odd under charge conjugation), and
$f$ and $a_2$ (which are even).  All four particles lie on the Regge trajectory
with intercept 0.44 and slope 0.93~GeV$^{-2}$.  In addition to the form factor
and constant factors of the pomeron contribution there is a complex factor.  To
account for the fact that two of the described particles are odd under charge
conjugation, it has to be replaced by its complex conjugate when describing
proton-antiproton scattering.

\subsubsection{Three-particle contributions}
\label{sec:dl3}

Three different three-particle exchanges are taken into account:
triple-gluon exchange, triple-pomeron exchange and simultaneous exchange
of two gluons and one pomeron. For all of them, the scattering amplitude
has the following form:
$$
T_{3}(s,t)=
2s\frac K{\hbox{-}t+t_0}\int dx_1dx_2dx_3\,
		\delta(1\hbox{-}x_1\hbox{-}x_2\hbox{-}x_3)\,(x_1x_2x_3)^2
		V(x_1,x_2,x_3)^4\,D(s_1,t_1)D(s_2,t_2)D(s_3,t_3)\,.
$$
$x_i$ are the transverse momentum fractions of the three particles, 
and $s_i=x_i^2s$ and~$t_i=x_i^2t$.
$V(x_1,x_2,x_3)=Ax_1x_2x_3$ is the vertex function (wave function), 
and the $D(s,t)$ are the propagators of the particles. The constant~$A$ 
in the vertex function must be fitted so that the triple-gluon 
contribution approaches an asymptotic power law for the differential 
cross-section: $d\sigma/dt=0.09\hbox{ mb GeV}^{14}\,\cdot\,t^{-8}$.
(See \cite{dlt-8} for more on the power law.)

\begin{figure}
\begin{center}
\includegraphics[width=6cm]{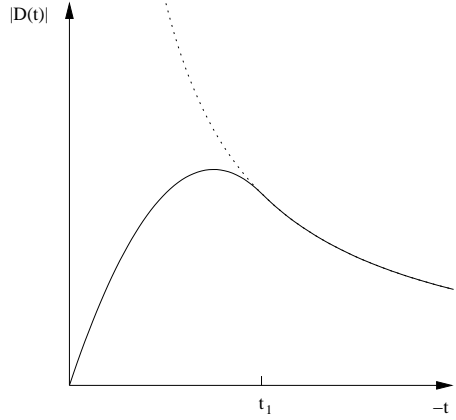}
\end{center}
\caption{The cut-off gluon propagator used by Donnachie and Landshoff.}
\label{fig:dlprop}
\end{figure}

The gluon propagator is $\alpha_s/t$ for large $|t|$. For small 
$|t|$ it is cut off with a parabola through the origin (see 
Figure~\ref{fig:dlprop}). It is independent of $s$. The following
ansatz is made for the parabola:
$$
D_g(t)=\lambda t (t+\tau)\qquad\hbox{for $-t<t_1$}\,.
$$
$\lambda$ and $\tau$ are fixed by requiring continuity of $D_g(t)$ and 
its derivative. This gives $\lambda=2/t_1^3$ and $\tau=-\frac32\;t_1$.

The pomeron propagator is derived from its Regge trajectory:
$$
D_\pom(s,t)=\frac{\beta^2}{4\pi}i
\left(\frac{-i\,s}{m^2}\right)^{\alpha_\pom(t)-1}\,.
$$

Depending on what particles are exchanged, three gluon propagators, three
pomeron propagators or one pomeron and two gluon propagators must be used.
The triple-gluon exchange contribution is odd under charge conjugation,
the other two contributions are even.

Even though the gluon propagator contains the coupling constant, the
Donnachie-Landshoff fit is nearly independent of it.  This is because the
constant $A$ which appears alongside the coupling constant is determined by
fitting the $ggg$ contribution to the asymptotic power law.  Therefore any
change in the coupling constant in the triple gluon contribution is compensated
by a change in~$A$.  The other three-particle exchanges depend on $A$ and hence
change when $A$ varies as a result of a change in~$\alpha_s$.  However, this
has little effect since they are rather less important.

\subsubsection{Reggeon-pomeron and double pomeron exchange}

To obtain an expression for simultaneous exchange of a Reggeons and a pomeron,
their trajectories are combined in the following way: 
$\alpha_{\reg\pom}(t)=\alpha_\reg(0)+\alpha_\pom(0)-1+
				\frac{\alpha_\reg'\alpha_\pom'}{\alpha_\reg'+\alpha_\pom'}t$.
In the case of  double pomeron exchange, the analogous formula is 
$\alpha_{\pom\pom}(t)= 2\alpha_\pom(0)-1+\frac12\alpha_\pom't$. The 
scattering amplitude is calculated from the trajectories of double pomeron and
Reggeon-pomeron  exchange in the following way:
\begin{eqnarray*}
T_{\pom\pom}(s,t)&=&\frac{-2iD\beta^4F_1(t)}{\log(s/m^2)}\,
				\left(\frac{-i\,s}{m^2}\right)^{\alpha_{\pom\pom}(t)-1}
\\
T_{\reg\pom}(s,t)&=&\frac{-4iD\beta^4F_1(t)}{\log(s/m^2)}\,
			(A+iB)\,\left(\frac{-i\,s}{m^2}\right)^{\alpha_{\reg\pom}(t)-1}\,,
\end{eqnarray*}
where $(A+iB)$ is the same complex prefactor which occurs in the Reggeon
contribution.

\subsubsection{How to produce the dip}

The dip in proton-proton scattering results from a three-way interference
between the following contributions: pomeron exchange, double pomeron exchange
and triple gluon (odderon) exchange.  The constant $D$ in the double pomeron
contribution is adjusted so that the imaginary part of pomeron and double
pomeron exchange cancel at the dip.  Since both are nearly imaginary in the
vicinity of the dip, only a small real part remains.  The odderon contribution
further diminishes the real part, which gives the dip.

The other contributions are much smaller in size.  They serve for fine-tuning
the depth and position of the dip.

\subsection{The odderon contribution in the Donnachie-Landshoff fit}
\label{sec:dlodd}

\begin{figure}[h]
\begin{center}
\includegraphics[height=2.5cm]{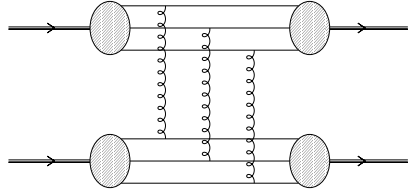}
\end{center}
\caption{The diagram corresponding to Donnachie and Landshoff's triple gluon
		 contribution.}
\label{fig:dlggg}
\end{figure}

This section will take a closer look at the triple gluon exchange contribution
of the Donnachie-Landshoff fit.  It evaluates a single Feynman diagram
displayed in Figure~\ref{fig:dlggg}.  This diagram corresponds to two type~(c)
couplings according to the classification in Section~\ref{sec:oddcolour}.

This has two implications.  First, since only a single diagram is included, the
expression is not gauge invariant.  This is the reason why the gluon propagator
has to be cut off to prevent infra-red divergences.

The second implication of choosing this specific diagram is its colour
structure: It is the purely symmetric $\frac12d^{cc'c''}$ (see
Equation~\ref{eq:coupcolour}).  Taking the antisymmetric structure in spinor
space into account, it follows that the contribution is odd under charge
conjugation.  Hence it is an odderon even though the authors do not call it
that.

A final remark concerning the triple gluon exchange contribution has to be made
and will be of great help to anybody reproducing the Donnachie-Landshoff fit.
There appears to be a misprint in the paper describing the fit~\cite{dlfit}.
The gluon propagator cutoff~$t_1$ and the scale~$t_0$ which corresponds to the
proton radius are given as ``$t_0=t_1=300\;\hbox{MeV}^2$''.  This value results
in a $ggg$ contribution which is far too large, dominating the differential
cross section for all~$t$.  In a later paper about a slightly modified
fit~\cite{dlfitnew} (described below), the authors mention $t_1$ as having
previously had the value 0.3~GeV$^2$.  Indeed choosing this value produces a
good fit.  Therefore, we use this parameter value for both $t_0$ and $t_1$.

\subsection{The improved fit of 1986}
\label{sec:dlnew}

While the DL fit is fairly good at ISR energies, it does not fit higher-energy
collider data.  In Section~\ref{sec:dlextend}, this will be shown in detail and
attempts to improve it will be presented.

This state of affairs was already remarked upon by Gauron, Leader and Nicolescu
in 1985~\cite{gnldlcrit}.  In response to their criticism, Donnachie and
Landshoff published an improved version of their fit which they claim both
extends to collider energies and shows some improvement at ISR
energies~\cite{dlfitnew}.  While it seems quite plausible that the
Regge-plus-dynamics description of Donnachie and Landshoff should be able to
accommodate high-energy data, I have not been able to reproduce the improved
fit.

In all probability the authors changed some aspects of the fit without
mentioning it in~\cite{dlfitnew}.  It is unfortunate but was to be expected
that today they cannot remember any details any more.  I followed the
prescription given in~\cite{dlfitnew} but start to disagree with the paper at
the point where according to the authors the double-pomeron contribution has to
be increased significantly.  In my reproduction of the fit, it in fact has to
be made smaller (though very slightly) to make the imaginary part of the total
amplitude vanish as prescribed in the paper.

\chapter{The results for $pp$ scattering}
\label{chap:oddresultspp}

\section{Experimental data}
\label{sec:ppdata}

Since we use the Donnachie-Landshoff fit as a framework for different odderon
contributions, we use data for proton-proton scattering in the range where we
know the fit is valid.  This is the case for the energy range investigated by
experiments at the Intersecting Storage Rings (ISR) at CERN.  Data from the ISR
were also used by Donnachie and Landshoff themselves.

The differential elastic proton-proton cross section has been measured by a
number of collaborations at the ISR.  Measurements have been made at the
centre-of-mass energies 23.5~GeV, 30.7~GeV, 44.7~GeV, 52.8~GeV and~62.5~GeV.
All the data we used were obtained from the Durham/RAL database~\cite{durdata}.
For all energies except 52.8~GeV, we used the data sets as collected by Amaldi
and Schubert~\cite{ama}.  Amaldi and Schubert did not do any measurements
themselves for this paper; rather, they collected all ISR data and applied a
normalisation to data which were not yet normalised. The data in the $t$ range
relevant for us were measured by Nagy and collaborators~\cite{nagy} and B\"ohm
and collaborators~\cite{bohm}.  The latter data were normalised by Amaldi and
Schubert.

For the centre-of-mass energy 52.8~GeV, we used the data published by Nagy et
al.~\cite{nagy}.  The data given by Amaldi and Schubert for that energy do not
extend to the $t$ region of interest to us.  We complement the data
from~\cite{nagy} with data measured by Breakstone and
collaborators~\cite{break} even though they were measured at 53~GeV
centre-of-mass energy.  It can be seen in the plots that they do not always
coincide, suggesting some experimental uncertainty.  The $s$ dependence of the
differential cross section is too small to explain the discrepancy.  We decided
against using the data from Erhan and collaborators~\cite{erhan} since they
were available only for relatively few values of~$t$ and tended to coincide
with the Breakstone data.

\section{The fit of Donnachie and Landshoff}

\subsection{Fixed coupling}
\label{sec:dlorig}

\begin{figure}[h]
\moveright 18mm \hbox{\input{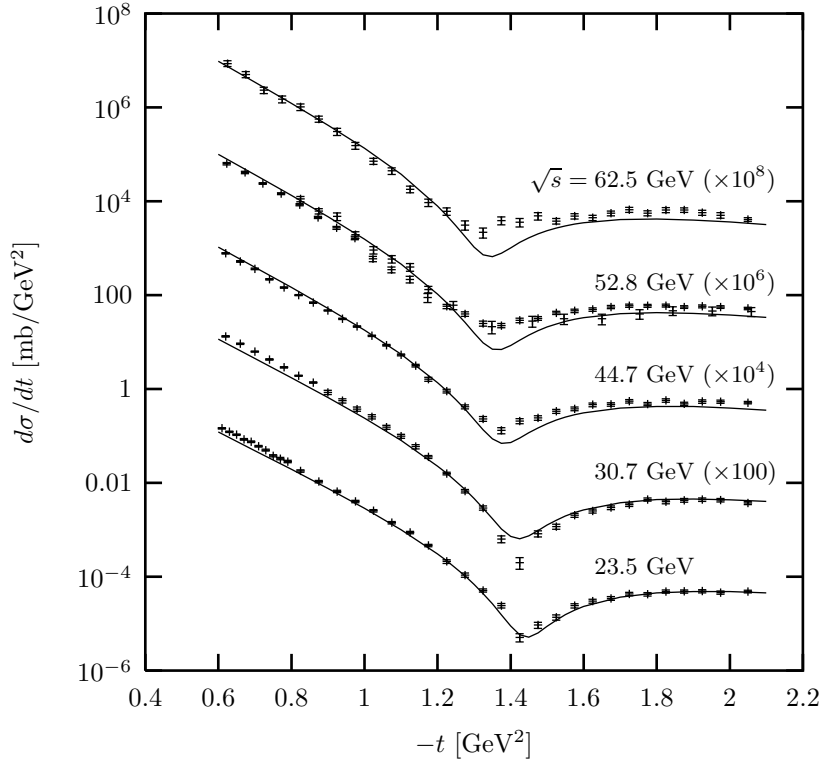}}
\caption{The original Donnachie-Landshoff fit with constant coupling.
		Data for five centre-of-mass energies are plotted simultaneously.
		The scale on the left applies to the lowest curve only.
		Successive curves are shifted upwards by a factor of 100.
		The points with error bars are experimental data. See 
		Section~\ref{sec:ppdata} for their sources.}
\label{fig:dlalls}
\end{figure}

Initially we had some problems reproducing the Donnachie-Landshoff fit.  This
was due to a misprint in~\cite{dlfit} (see Section~\ref{sec:dlodd}).  The
constant $t_0$ in the expression for the three-particle exchanges (which gives
the scale corresponding to the proton radius) and the the gluon propagator
cutoff $t_1$ is 0.3~GeV$^2$ rather than the value given in the paper.  The
original fit reproduced with these parameter values is displayed in
Figure~\ref{fig:dlalls}.

\subsection{Different cutoffs of the gluon propagator}
\label{sec:othercut}

\begin{figure}[h]
\begin{center}
\includegraphics[width=6cm]{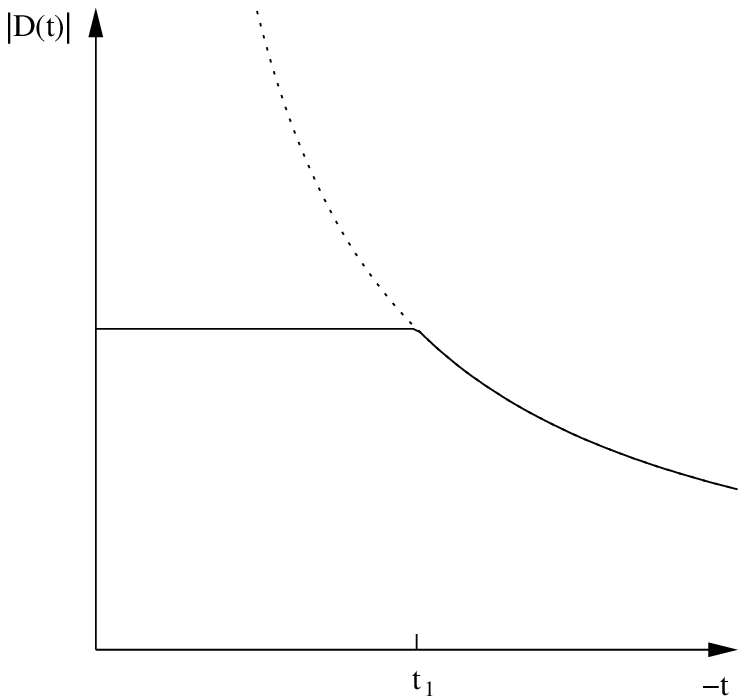}
\qquad
\includegraphics[width=6cm]{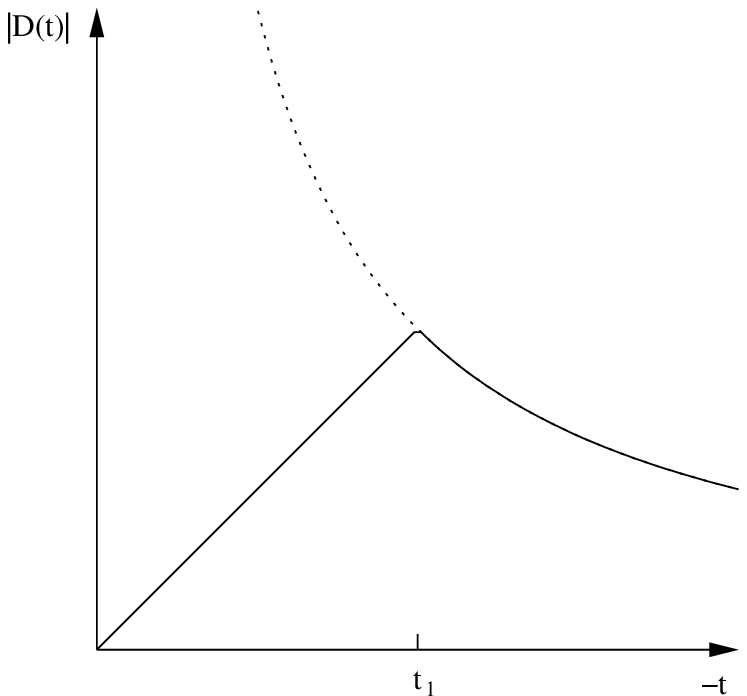}
\end{center}
\caption{The two alternative ways of cutting off the gluon propagator.}
\label{fig:othercut}
\end{figure}

To show that the Donnachie-Landshoff fit is independent of the precise
nature of the gluon propagator cutoff, alternative ways of cutting
off were tried. The two possibilities were a flat and a linearly 
rising propagator for small~$|t|$, see Figure~\ref{fig:othercut}.

Changes in the integral over the propagator can be compensated by changes in
the constants $t_1$ and~$A$.  The resulting curves are not shown here since
they are practically identical to the original fit shown in
Figure~\ref{fig:dlalls}.\footnote{%
At the time this calculation was done it was not yet clear how best to handle
the misprint in the paper of Donnachie and Landshoff (see
section~\ref{sec:dlodd}).  The scale~$t_0$ which corresponds to the proton
radius was chosen as 300~MeV$^2$.  Since this calculation was quite peripheral
to our main interest, investigating the effect of different models for the
odderon-proton coupling, it was not repeated when we decided on 0.3~GeV$^2$ as
the value most probably meant by the authors.  Since this change in $t_0$ did
not in the least modify the results of other calculations, it can be safely
assumed that this result is as valid.}

\subsection{Running coupling}
\label{sec:dlrun}

In a leading-order calculation, the coupling constant is just that, a constant.
However, it is known that in reality the strong coupling constant depends on
the scale.  It is possible with limited extra effort to take this into account
and see how it influences our results.  We replaced the strong coupling
constant with the following expression for the running coupling:
\begin{equation}
\label{eq:runcoup}
\alpha_s(t)=\frac{4\pi}{\left(\frac{11}3N_c-\frac23N_f\right)
						\log\left(\frac{|t|}{\Lambda_{\rm QCD}^2}+6\right)}\,.
\end{equation}
This is the one-loop expression for the running of $\alpha_s$ modified to
account for non-perturbative effects.  The constant~6 was added in the
logarithm to avoid the divergence of the perturbative running coupling, which
is called ``freezing'' of the coupling constant.  This specific way of freezing
$\alpha_s$ has been used before by Ewerz~\cite{ewmodalpha}.  The number of
colours~$N_c$ is~3, and we have set the number of flavours~$N_f$ also
to~3.  $\Lambda_{\rm QCD}$~was chosen to be 250~MeV\null.

\begin{figure}[h]
\moveright 18mm \hbox{\input{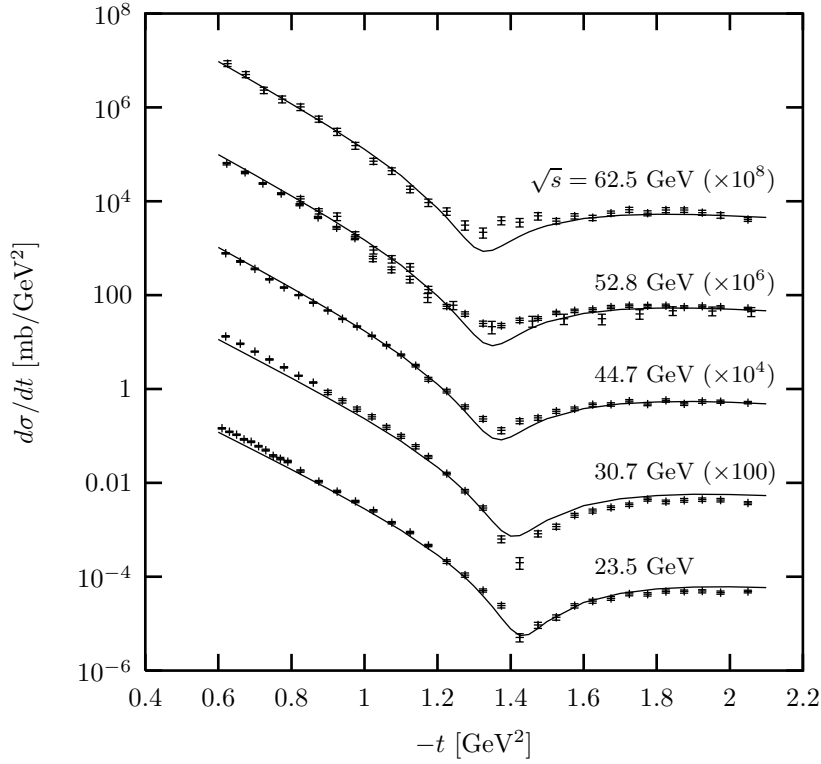}}
\caption{The Donnachie-Landshoff fit with running coupling.}
\label{fig:allsrun}
\end{figure}

The fit with the running coupling is as good as the original fit, see 
Figure~\ref{fig:allsrun}.

\begin{figure}[h]
\moveright 18mm \hbox{\input{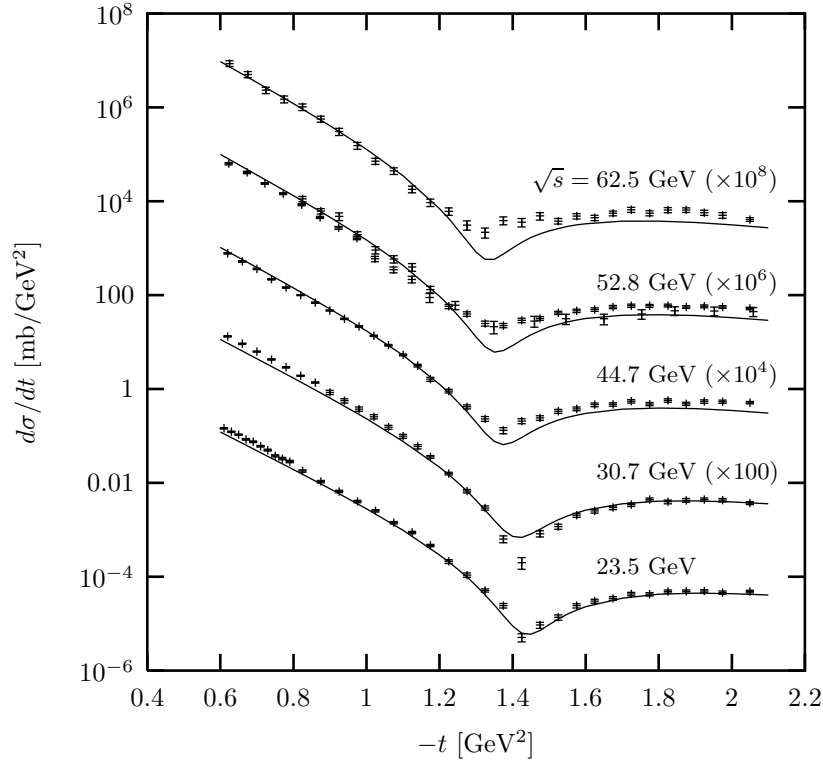}}
\caption{The fit with the odderon computed in the quark-diquark model, 
			with constant coupling.}
\label{fig:allsodd}
\end{figure}

\section{The fit with the geometric model odderon}

\subsection{Fixed coupling}

The main aim of our calculations in the quark-diquark model was to determine
the diquark size.  For that purpose we substituted the diquark model odderon
amplitude for the triple-gluon contribution in the Donnachie-Landshoff fit.
The coupling constant of the DL contributions was left at 0.3 since that is
what its authors chose.  (In fact only the $\pom gg$ and the $\pom\pom\pom$
contributions depend on $\alpha_s$.)  We use the DL fit as a framework in which
several odderon contributions are substituted for the original one.  For our
geometric odderon contribution, we chose the coupling constant 0.4 since it is
in the middle of the range $0.3\ldots0.5$ common in the literature at the scale
given by $-t$ at the dip.  See Section~\ref{sec:diqalpha} for a discussion of
other values for the coupling constant.

The diquark size (respectively the angle in Figure~\ref{fig:starmodel}) was
adjusted to give the best fit for a range of centre-of-mass energies.  The
result can be seen in Figure~\ref{fig:allsodd}.  The optimal angle is
0.14$\,\pi$, or approximately 25 degrees (in transverse space).  This
corresponds to a diquark size in transverse space of 0.22~fm according to
Formula~\ref{eq:ddiq}.  This diquark size justifies a practice common in
non-perturbative calculations.  Since soft gluons cannot resolve a diquark as
small as 0.3~fm, it is legitimate to treat the proton as a colour dipole, as in
\cite{dfk,bergnacht,ssmsv3}, for example.

\begin{figure}[h]
\moveright 10mm \hbox{\input{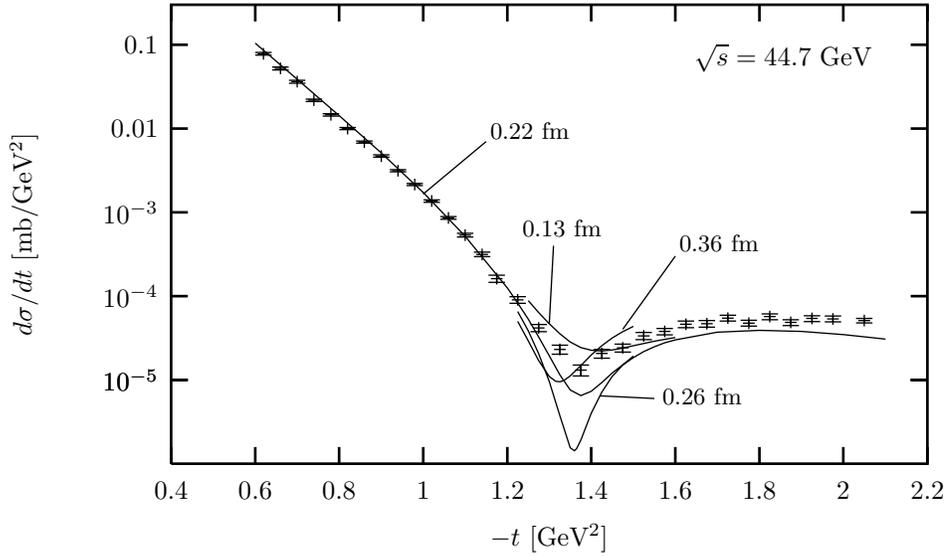}}
\caption{The fit with the quark-diquark model odderon for different
		star angles. The drawn-out curve is the best fit, 0.14$\,\pi$.
		The centre-of-mass energy is 44.7~GeV.}
\label{fig:fitangles}
\end{figure}

Choosing the correct angle is vital for obtaining a good fit.
Figure~\ref{fig:fitangles} shows that the fit gets considerably worse when the
angle is maladjusted.  The dependence of the odderon contribution on the
diquark size will be analysed in greater detail in the next section.

\subsection{The value of the coupling constant and the significance of the fit}
\label{sec:diqalpha}

The quality of the fit depends strongly on the size of the odderon contribution
in the dip region. That is why we were able to determine the diquark size (on
which the size of the odderon depends) from the fit of $d\sigma/dt$.  However,
the odderon contribution also depends strongly on the coupling constant: it is
proportional to $\alpha_s^6$. Therefore the optimal value for the diquark size
depends on the choice of the coupling constant (or the choice of
$\Lambda_{\hbox{\scriptsize QCD}}$ if $\alpha_s$ is running).

The difficulty here lies in the fact that in a leading-order calculation 
the value of the coupling can only be estimated. Determining the scale 
relevant for the coupling and the ``true'' value of the coupling constant 
requires higher-order calculations. We have therefore determined the 
optimal diquark size for several values of $\alpha_s$.

\begin{table}[h]
\moveright 1.2in \vbox{
\def\eol{\cr\noalign{\nointerlineskip}}
\tabskip=0pt
\halign{ 
\vrule height2.75ex depth1.25ex
#\tabskip=1em & \hfil #\hfil &\vrule # & \qquad$#\,\pi$\hfil &\vrule # &
		 \hfil # \hfil \tabskip=0pt&#\vrule \cr\noalign{\hrule}
& $\alpha_s$ &&\omit star angle && diquark size [fm] & \cr\noalign{\hrule}
& 0.3 && 0.22 && 0.34 &\eol
& 0.4 && 0.14 && 0.22 &\eol
& 0.5 && 0.095 && 0.15 &\cr\noalign{\hrule}
}}
\caption{Optimal diquark size for the odderon in the geometric model for 
		different coupling constants.}
\label{tab:couprdiq}
\end{table}

The values for $\alpha_s$ which are common in the literature at this scale 
in the momentum transfer lie in the range from 0.3 to~0.5. 
Table~\ref{tab:couprdiq} shows how the optimal diquark size for several
the coupling constants. Since the scattering amplitude rises both with
the diquark size and with $\alpha_s$, the diquark size has to be larger
for a smaller coupling constant. All resulting sizes for the diquark cluster 
in the proton are $\lesssim 0.35\;$fm, supporting the assumption of a 
reasonably small diquark unless the coupling constant is smaller than~0.3.

The other extreme, a rotationally symmetric ``Mercedes star'' proton,
can be excluded with some confidence. This would imply an unrealistically 
small coupling constant of~0.17.

\begin{figure}[h]
\moveright 10mm \hbox{\input{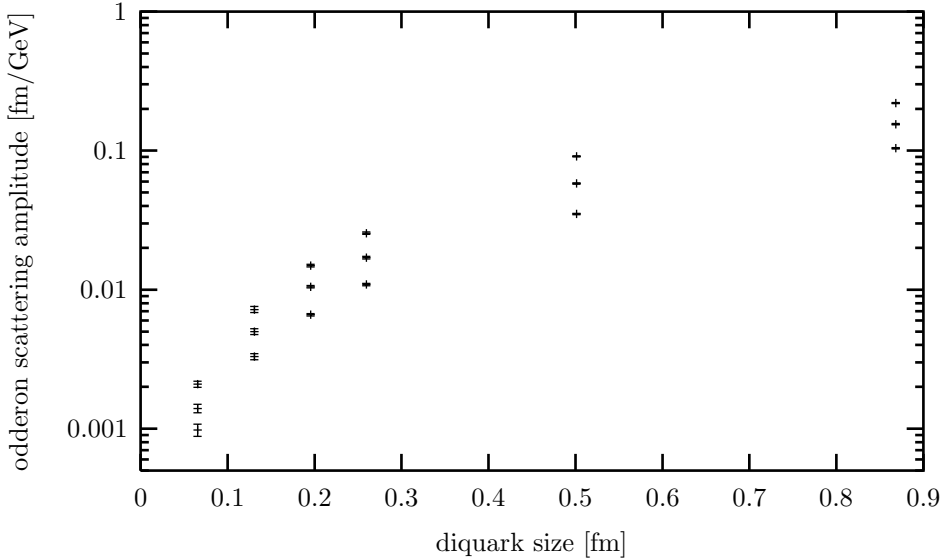}}
\caption{The dependence of the quark-diquark model odderon amplitude on the 
		diquark size. The error bars reflect the numerical errors of up 
		to 10\%. The three points for each value of the diquark size
		correspond to three different transverse momentum transfers. The 
		square of the momentum transfer $-t$ is 0.6~GeV$^2$, 1~GeV$^2$ and 
		1.6~GeV$^2$ from top to bottom.}
\label{fig:diqampl}
\end{figure}

\begin{figure}[p]
\moveright 10mm \hbox{\input{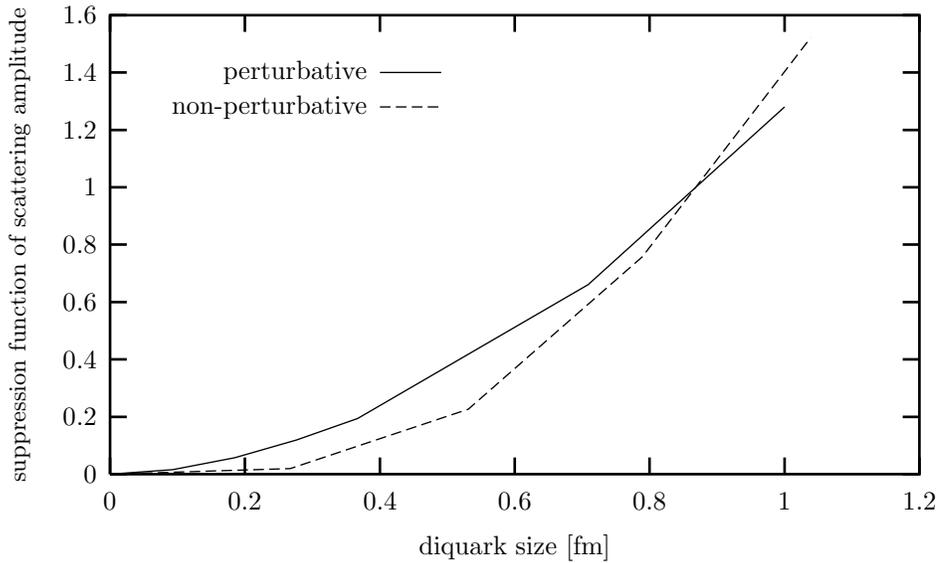}}
\caption{Comparison between the non-perturbative and the 
		perturbative odderon. The suppression function
		plotted here is defined as the scattering amplitude for a given
		diquark size divided by the scattering amplitude for the 
		symmetric three-star with angle~$2/3\,\pi$. The non-perturbative 
		odderon	amplitude vanishes significantly faster than the perturbative 
		one. The non-perturbative data were obtained from~\cite{rueterphd}.}
\label{fig:diqpnp}
\end{figure}

\begin{figure}[p]
\moveright 10mm \hbox{\input{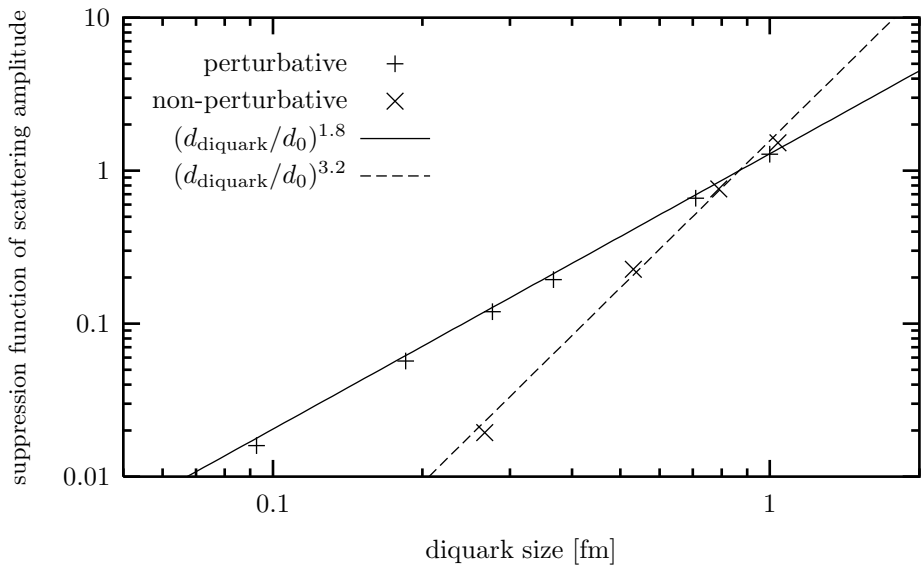}}
\caption{Double logarithmic plot of the data from Figure~\ref{fig:diqpnp}.
		The power laws which describe the vanishing of the amplitudes for 
		small diquark radii were obtained by fitting a straight line.
		The two lines intersect at a diquark size of $d_0=0.868$~fm which
		means a symmetric star and where the suppression function is 1.}
\label{fig:diqpnppow}
\end{figure}

\subsection{Properties of the geometric model odderon}

\subsubsection{Dependence on the diquark size}
\label{sec:diqdepend}

As can be seen in Figure~\ref{fig:fitangles}, the quality of the fit depends
strongly on the diquark size. The drawn-out curve in the figure is the 
overall best fit for all centre-of-mass energies. Though a better diquark 
radius could be found for one particular energy, it would give a worse fit
for other energies. 

The odderon amplitude's dependence on the diquark size is plotted in
Figure~\ref{fig:diqampl}. Data for three different values for $-t$ are in the
plot. As one can see the three sets of points differ by a constant factor.
This means that the suppression with vanishing diquark size is independent of
the transverse four-momentum transfer.  (The odderon is by construction
completely independent of the centre-of-mass energy.)

The odderon-proton coupling vanishes for zero diquark
size~\cite{oddpoint,rueter}, for the following reason: A proton with a
point-like diquark is a colour dipole.  Since the odderon is odd under charge
conjugation, it gives a contribution of opposite sign when the dipole is
rotated by 180 degrees.  After averaging over all orientations of the dipole
the coupling has to be zero because contributions from opposite orientations
cancel out.

Figure~\ref{fig:diqpnp} shows a comparison between the perturbative odderon in
the quark-diquark model and the non-perturbative odderon computed by M.~R\"uter
with the Model of the Stochastic Vacuum~\cite{rueterphd}.  The non-perturbative
odderon vanishes significantly faster than the perturbative one.  This can be
understood with physical arguments: The non-perturbative odderon which consists
of soft gluons cannot resolve small diquarks and takes them to be point-like.

To determine the power law according to which the odderon amplitudes vanish, 
a straight line was fitted to the computed amplitudes in a double logarithmic 
plot.  The result can be seen in Figure~\ref{fig:diqpnppow}.  The perturbative
odderon vanishes as $d_{\hbox{\scriptsize diquark}}^{\;1.8}$, the 
non-perturbative one as $d_{\hbox{\scriptsize diquark}}^{\;3.2}$.

\subsubsection{Relative importance of different graph types}

Since we classify the Feynman graphs in different types, it may seem easy to
investigate the contributions of graphs belonging to one particular
type. Unfortunately, the numerical integration over the reduced scattering
amplitude is much more difficult when only one graph type is computed.  Large
cancellations between the different graph types are to be expected since
contributions from single graph types are not gauge invariant.  However, the
reduced scattering amplitudes of single graph types seem to vary wildly, making
numerical integration difficult.

Nevertheless, calculations have been done to determine the relative size of
the contributions of different graph types. 10~times 7\,000\,000 
evaluations of the integrand were performed to evaluate the integral with 
the {\sc Vegas} algorithm. The results carry large errors
and are to be understood as indications of the order of magnitude rather
than numbers of any precision. Table~\ref{tab:typeampl} shows the results.

\begin{table}[h]
\moveright 1.2in \vbox{
\def\eol{\cr\noalign{\nointerlineskip}}
\tabskip=0pt
\halign{
\vrule height2.75ex depth1.25ex # \tabskip=1em & #\hfil &#\vrule& 
	\hfil$#$\phantom{tude [fm/GeV] } \tabskip=0pt &#\vrule\cr\noalign{\hrule}
& Graph type && $Scattering ampli\rlap{tude [fm/GeV]}$ &\cr
\noalign{\hrule}
& (aa)  &&  -70 &\eol
& (ab)  &&  375 &\eol
& (bb1) && -190 &\eol
& (bb2) && -430 &\eol
& (ac)  && -205 &\eol
& (bc)  &&  870 &\eol
& (cc)  && -215 &\eol
& total && $0\rlap{.0127}$ &\cr\noalign{\hrule}
}}
\caption{The very approximate scattering amplitudes corresponding to each 
		graph type. The transverse momentum transfer is 1~GeV$^2$; the star
		angle in transverse space is 0.14$\,\pi$, the value found to be 
		most realistic. The numerical errors are \bf 30 to 50~\%\rm, 
		except for the total.}
\label{tab:typeampl}
\end{table}

The graph type (bc) gives the largest contribution and (aa) the smallest.
The other graph types are clustered around 200 to 400~fm/GeV. The 
contributions add up to a number which is compatible with 0, considering
the large errors. This is reassuring since the total scattering amplitude
is much smaller than the individual contributions.

Donnachie and Landshoff \cite{dlfit} argue that the type (cc) contribution is
dominant at large~$t$.  They note that if the three gluons do not couple to
three different quarks, the quarks' trajectories are not parallel immediately
after the scattering.  Therefore a further gluon has to be exchanged before the
two triples of quarks can again form protons.  This and the fact that one of
the quarks has to be off-shell intermediately suppresses the couplings of types
other than (c).  Consequently, Donnachie and Landshoff's triple gluon
contribution contains only a graph corresponding to type (cc).

In our results, the (cc) contribution is not dominant.  This indicates that the
argument of Donnachie and Landshoff is not yet valid at $t=1$~GeV$^2$.  That
said, it has to be emphasised again that our results concerning the scattering
amplitude by graph type are very imprecise.

\begin{figure}[h]
\moveright 18mm \hbox{\input{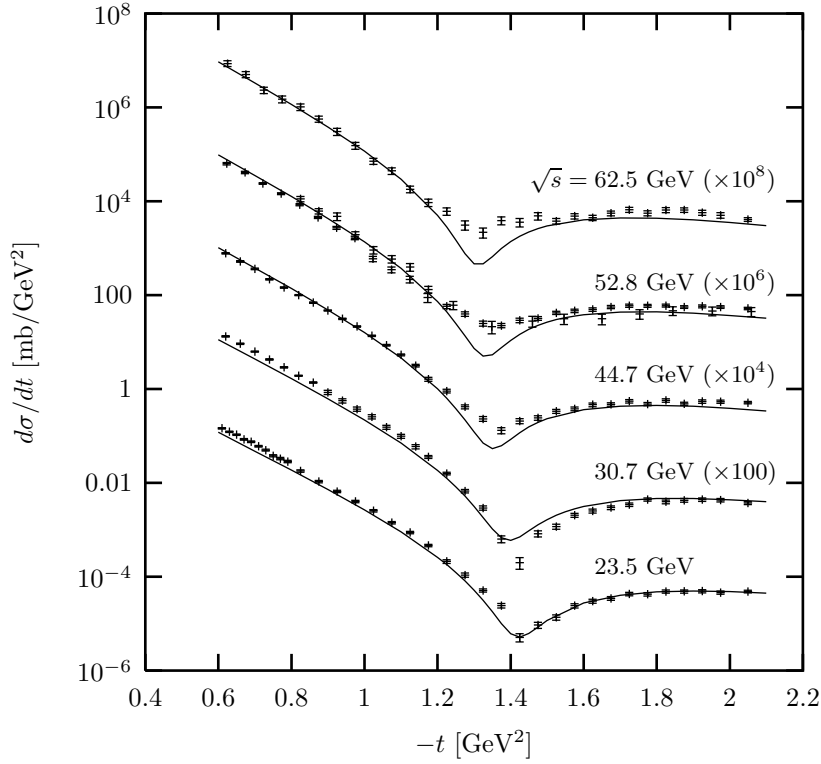}}
\caption{The fit with the odderon computed in the quark-diquark model, 
			with running coupling.}
\label{fig:allsorun}
\end{figure}

\subsection{Running coupling}

The diquark model odderon can equally be inserted into the Donnachie-Landshoff
fit with running coupling. The coupling constant in both the geometric
model odderon and the Donnachie-Landshoff contributions is substituted by 
the expression for the running coupling (\ref{eq:runcoup}).

The result is nearly as good as the previous fits,
though the dip tends to be too deep for the higher centre-of-mass energies.
The plot can be seen in Figure~\ref{fig:allsorun}.
The best fit is achieved with the same star angle
as for fixed coupling, $0.14\,\pi$, corresponding to a diquark size of 0.22~fm.

\section{The fit with momentum-space impact factors}
\label{sec:momentimp}

\subsection{The impact factor of Levin and Ryskin}

\begin{figure}[t]
\moveright 18mm \hbox{\input{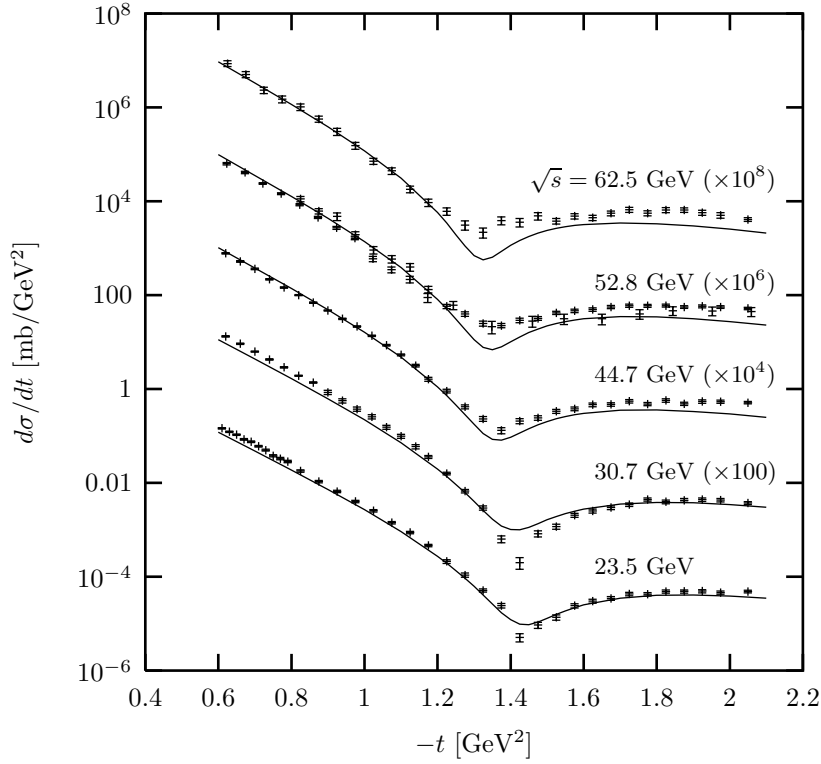}}
\caption{The fit with the odderon-proton impact factor by Levin and Ryskin.
		The coupling constant was 0.5.}
\label{fig:allslev}
\end{figure}

If one uses the impact factors with the Gaussian form factors as proposed
by Levin and Ryskin in~\cite{lev}, the coupling constant has to be
adapted slightly. To obtain the best fit, $\alpha_s$ has to be chosen
as~0.5. The curves for all centre-of-mass energies are displayed in 
Figure~\ref{fig:allslev}.

\begin{figure}[p]
\moveright 9mm \hbox{\input{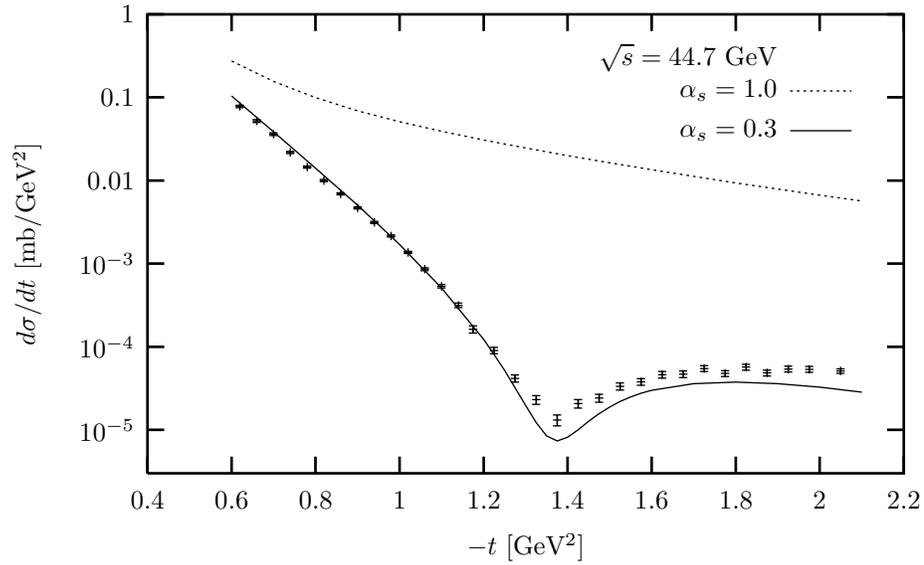}}
\caption{The fit with the odderon-proton impact factor by Kwieci\'nski et 
		al.\ with coupling constant $\alpha_s=1$ and 0.3, respectively.}
\label{fig:kwieorig}
\end{figure}

\begin{figure}[p]
\moveright 18mm \hbox{\input{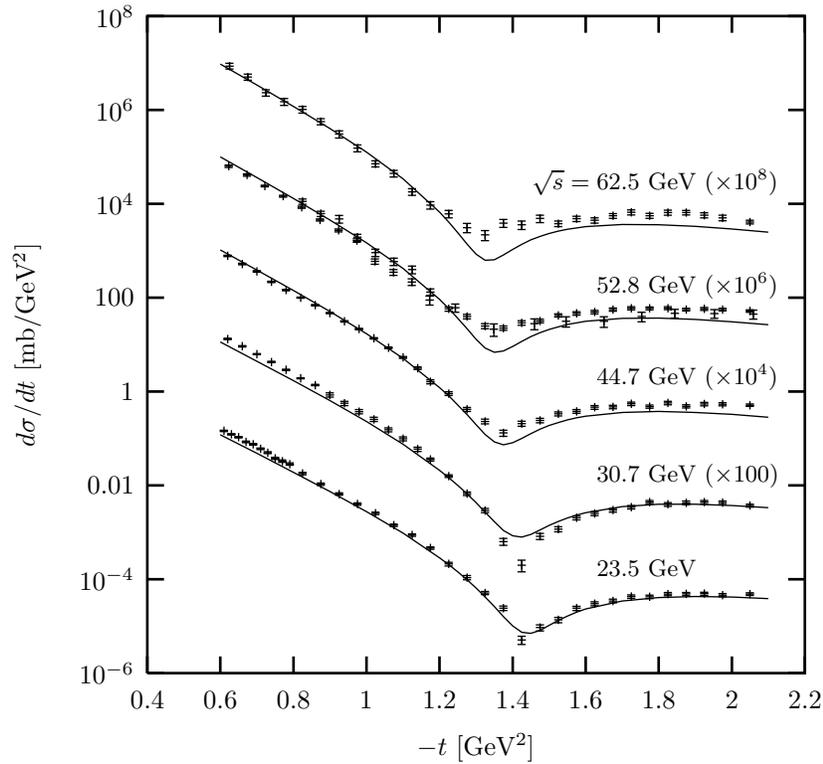}}
\caption{The fit with the odderon-proton impact factor by Kwieci\'nski et 
		al.\ with the modified coupling constant 0.3.}
\label{fig:allsimpa}
\end{figure}

\subsection{The impact factor of Kwieci\'nski et al.}

In~\cite{kwie_impac}, Kwieci\'nski et al.\ use the coupling constant
$\alpha_s=1$.  Using this value for proton-proton scattering makes the odderon
contribution far too large.  It dominates over all other contributions.  In
Figure~\ref{fig:kwieorig} on page~\pageref{fig:kwieorig} it is shown together
with data for $\sqrt s=44.7$~GeV.

A good fit can be obtained by changing the coupling constant.\footnote{It was
decided not to modify the constant $A$ which determines the width of the form
factor~(\ref{eq:formKwie}).  It is half the mass of the $\rho$~meson, the
lightest vector meson.  Since the structure of the proton is measured in
$\gamma p$ scattering among other processes and the photon has to fluctuate
into a vector meson to interact strongly with the proton, this mass is closely
connected to the structure of the proton.}
The best fit was obtained for $\alpha_s=0.3$ and is presented in 
Figure~\ref{fig:allsimpa}.  That such a small coupling constant is required
throws some doubt on results for diffractive $\eta_c$ production through
odderon exchange at HERA~\cite{kwie_impac,engeletac,bbcv}.  The enormous
dependence of the odderon on the coupling constant ($\propto\alpha_s^6$) means
that a factor of 0.3 in $\alpha_s$ changes the cross section by orders of
magnitude.  Furthermore, while the situation in the process of $\eta_c$
electroproduction could be said to be different from that in elastic $pp$
scattering, the coupling constant would be expected to be smaller still due to
the hard scale given by the $\eta_c$ mass.  Our results indicate that the
amplitude for $\eta_c$ production was overestimated by the three groups cited
above.

\section{Proton-proton summary}

\begin{figure}[t]
\hbox{\input{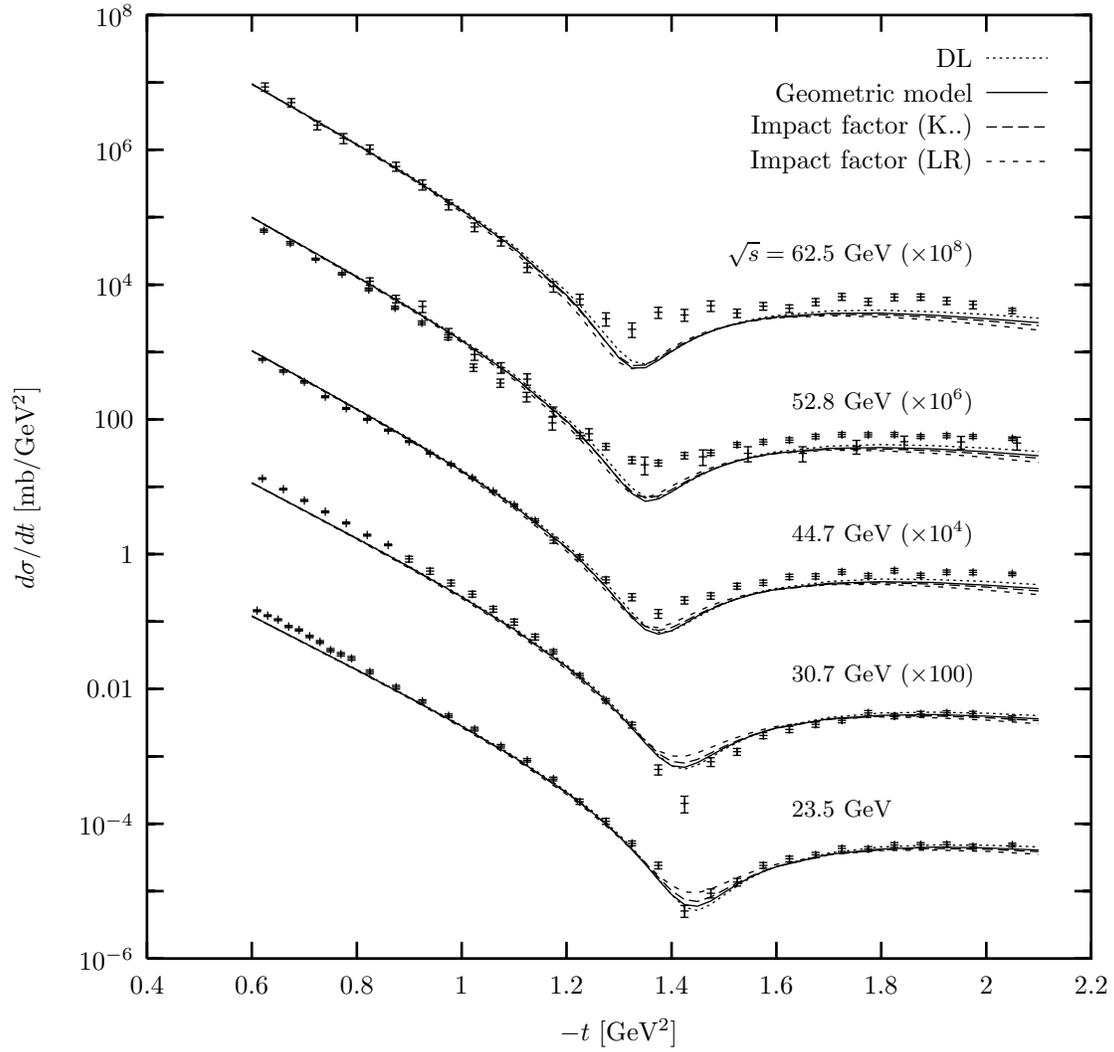}}
\caption{Summary of proton-proton scattering. All methods for computing the
		odderon contribution are compared for all centre-of-mass energies for 
		which experimental data are available.}
\label{fig:allsallmod}
\end{figure}

Figure~\ref{fig:allsallmod} shows a summary of the four methods used to compute
the odderon contribution: Donnachie and Landshoff's $ggg$ contribution, our
geometric model, and the impact factors with form factors by Kwieci\'nski et
al.\ and Levin and Ryskin, respectively.  The coupling is fixed in all
cases.  As can be seen, all methods yield good results after the coupling
constant is adapted.

Figure~\ref{fig:allsallmod} shows clearly that the quality of the data is not
sufficient to distinguish between different models for the odderon-proton
coupling.  However, the data can strongly restrict the range of parameter
values for each individual model.  The diquark size in the geometric model in
position space could be fixed to 0.22\,\,fm for a coupling constant of~0.4 and
was found to be $\lesssim 0.35\;$fm for all reasonable values of~$\alpha_s$.
It could be shown that the coupling constant used by several
groups~\cite{kwie_impac,engeletac,bbcv} would be far too large for the
situation of proton-proton scattering and that they probably overestimated the
cross section for $\eta_c$ production through odderon exchange.  For the impact
factors computed from the Gaussian form factor of Levin and Ryskin, the
coupling constant also had to be adapted, but less radically.

\chapter{The results for $p\overline p$ scattering}
\label{chap:oddresultsppbar}

\section{Experimental data}

For centre-of-mass energies in the range of our proton-proton data, the 
differential proton-antiproton cross section has been measured at only one 
energy, 53~GeV. The measurement was done at the CERN Intersecting Storage 
Rings (ISR) by Breakstone and collaborators~\cite{break}.

At higher energies measurements have been made at the $Sp\bar pS$ collider, 
also at CERN. The proton-antiproton differential elastic cross section
has been measured at the energy 546~GeV by Bozzo and 
collaborators~\cite{bozzo} and at 630~GeV by Bernard and 
collaborators~\cite{bernard}.

The measurement at Fermilab by Amos and collaborators at 1.8~TeV~\cite{amos} is
of no use to us, unfortunately.  It does not extend to the values of the
squared four momentum transfer $-t$ at which the data show a kink, where the
odderon begins to play a role.

\section{$p\bar p$ scattering at the ISR}

To treat proton-antiproton scattering, the sign of the odderon amplitude has to
be reversed relative to proton-proton scattering.  The contributions of the
Donnachie-Landshoff fit which describe $C$-odd particles have likewise to be
modified as described in Section~\ref{sec:dlfit} and in~\cite{dlfit}.

\begin{figure}[h]
\moveright 10mm \hbox{\input{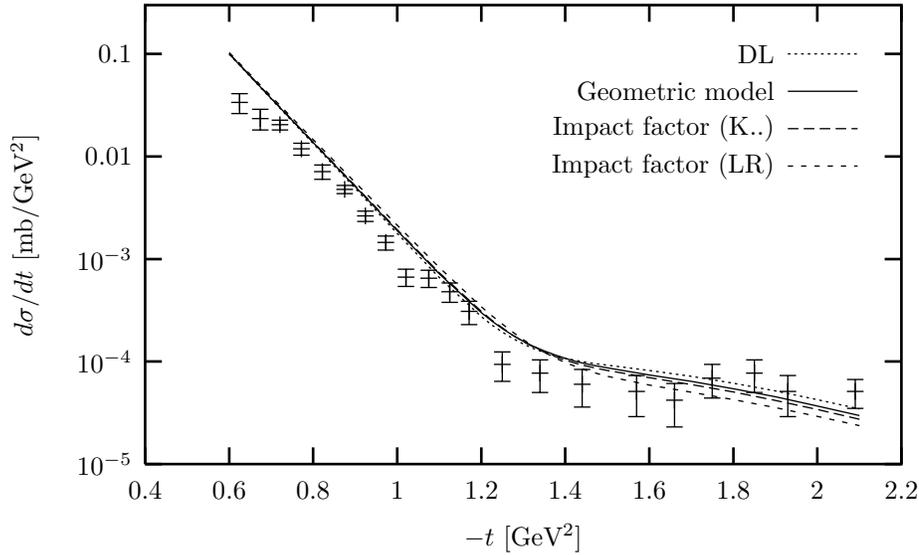}}
\caption{The computed proton-antiproton differential cross section compared to 
		experimental data from~\cite{break}. The centre-of-mass energy is
		53~GeV.}
\label{fig:ppbar3}
\end{figure}

Figure~\ref{fig:ppbar3} compares four calculated curves to the experimental
data at the energy 53~GeV. The dotted line is the original Donnachie-Landshoff
fit. The solid line shows the fit with the odderon computed in our geometric
model for the odderon-proton coupling. The diquark size was the same as for
proton-proton scattering, 0.22~fm. 

The two dashed lines show the fit with the odderon contribution computed from 
impact factors in momentum space, with form factors according to 
Kwieci\'nski et al.\ and Levin and Ryskin, respectively. The coupling constant
was the same as the one used for proton-proton scattering: 0.3 for 
Kwieci\'nski's form factor and 0.5 for Levin's and Ryskin's.

That the same parameter values as for proton-proton scattering describe
proton-antiproton data as well adds some weight to our results in the 
previous chapter. It is also a tribute to the Donnachie-Landshoff fit.

As was the case for proton-proton scattering, all models for the 
odderon-proton coupling fit the data equally well.

\section[Higher energies]
		{Describing higher-energy data with an energy dependent odderon}

\subsection{With the Donnachie-Landshoff fit as framework}
\label{sec:dlextend}

Since all the models for the odderon-proton coupling describe proton-antiproton
scattering at ISR energies so well, it is interesting to see whether our
calculations can be extrapolated to higher energies. Since all models led to
equivalent results, we restrict this investigation to the Donnachie-Landshoff
fit with the original triple-gluon contribution as an odderon.

\begin{figure}[h]
\moveright 10mm \hbox{\input{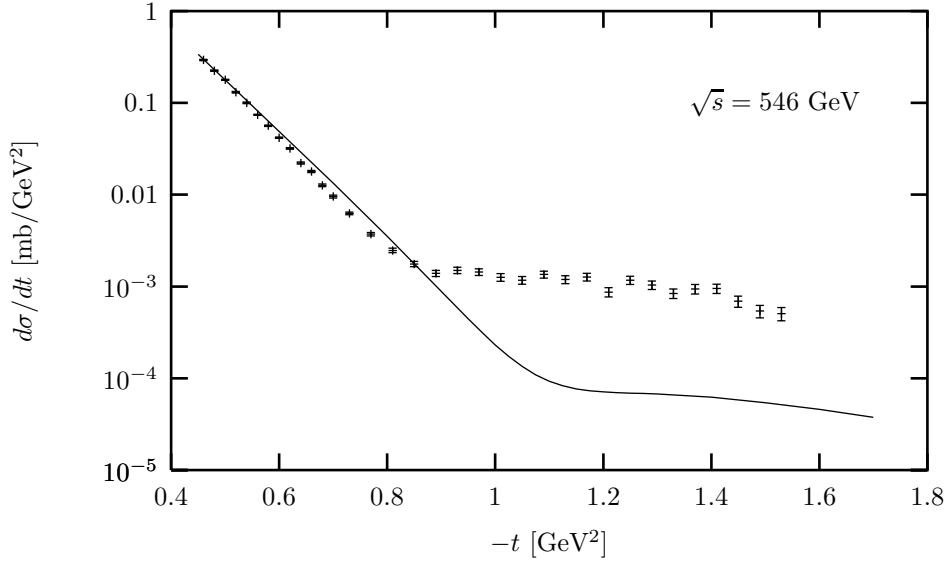}}
\caption{A pure extrapolation of the DL fit to higher energy $p\bar p$ 
			scattering matches the data poorly.  Data from~\cite{bernard2}.}
\label{fig:dl546}
\end{figure}

As Figure~\ref{fig:dl546} shows, a pure extrapolation is wide off the mark.
This may be explained by the fact that Donnachie and Landshoff's {\it ggg}
contribution (unlike other contributions) has no dependence on the
centre-of-mass energy $\sqrt s$.  In the relatively narrow range of energies
measured at the ISR this is not necessary, but when extrapolating to much
higher energies, an energy dependence for the odderon might improve the fit.

We attempted to improve the fit for the proton-antiproton data at higher 
energies by adding an energy dependence to the odderon. This could in
turn enable us to learn something about its $s$ dependence, specifically
its Regge intercept.

We gave the odderon contribution an $s$ dependence by multiplying its
scattering amplitude with the $s$ dependent factor known from Regge theory (see
Section~\ref{sec:reggeon}, Equation~\ref{eq:reggeampl}).  As a first step, only
the Regge intercept was used but not the slope:
\begin{equation}
\label{eq:oddsdep}
T_\odd(s,t)=\left(\frac s{s_0}\right)^{\alpha_0} T_\odd(t)\,.
\end{equation}
This amounts to taking into account the dependence of the odderon 
propagator on $s$ but not on $t$. All the $t$ dependence comes from 
the odderon-proton coupling contained in the original triple-gluon 
expression of Donnachie and Landshoff. Their expression corresponds
to an intercept $\alpha_0=1$.
Since the intercept is not expected to 
deviate much from 1, neglecting the $s$ dependence for the range of
ISR energies makes sense.

\begin{figure}[p]
\moveright 10mm \hbox{\input{ptex/dsigdt73_oddinter546.ptex}}
\caption{The fit with an $s$ dependent odderon for proton-antiproton 
		scattering at $\sqrt s=546$~GeV.
		The odderon intercept ranges from 0.8 to~1.5 in steps of 0.1.
		The reference energy was $\sqrt{s_0}=53$~GeV.
		Only the Regge intercept is taken into account, not the slope.}
\label{fig:inter546}
\end{figure}

\begin{figure}[p]
\moveright 10mm \hbox{\input{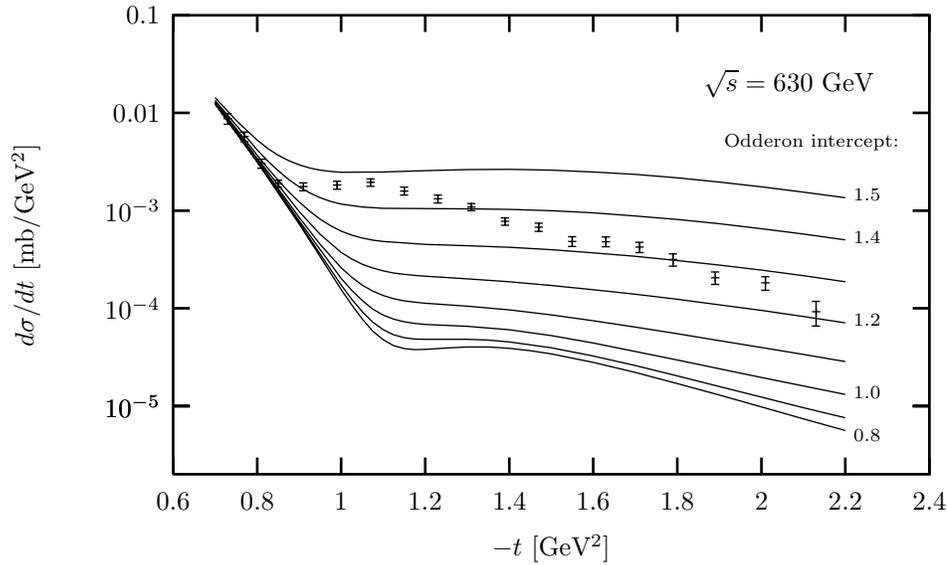}}
\caption{The fit with an $s$ dependent odderon for proton-antiproton 
		scattering at $\sqrt s=630$~GeV.
		The odderon intercept ranges from 0.8 to~1.5 in steps of 0.1.
		The reference energy was $\sqrt{s_0}=53$~GeV.
		Only the Regge intercept is taken into account, not the slope.}
\label{fig:inter630}
\end{figure}

Figures \ref{fig:inter546} and~\ref{fig:inter630} show the fit with the energy
dependent odderon for different intercepts.  As a reference energy we chose
$\sqrt{s_0}=53$~GeV since the fit works well at that energy.
Figure~\ref{fig:inter546} suggests a Regge intercept of~1.4 for the odderon.
This is the value that produces the best fit.  However,
Figure~\ref{fig:inter630} at $\sqrt s=630$~GeV which extends to larger $|t|$
shows clearly that none of the curves matches.  After the customary kink (which
looks a bit like a dip at this energy) the data points start to fall off more
rapidly again.  The computed curves which are of the right magnitude do not
reproduce this behaviour.  Furthermore, such a high intercept already has an
effect in the ISR range.  In proton-proton scattering it makes the dip deeper
for the higher ISR energies and flatter for the lower ones.  This is the
opposite of the behaviour shown by the data (see for instance
Figure~\ref{fig:dlalls}).

This disappointing result is not in contradiction with the little which is
known about the odderon intercept.  The odderon intercept calculated
perturbatively in the Leading Logarithmic Approximation is slightly below
one.\footnote{For the new solution of the BKP equation found be Bartels,
Lipatov and Vacca~\cite{blvodd}, it is exactly one.}  Given that the intercept
of the BFKL pomeron was made smaller rather than larger by
next-to-leading-logarithmic corrections, an intercept of 1.4 for the odderon
seems unlikely.  It is therefore no great surprise that calculations using this
intercept fail to match the data, even if the fit was satisfactory at one
energy.

One can go one step further and use the complete Regge trajectory of the
odderon, taking also the $t$ dependence of the odderon propagator into 
account. That has the effect that the $s$ dependence of the odderon
contribution is different for each value of $t$ (or vice versa). The 
expression for the scattering amplitude is the following:
\begin{equation}
\label{eq:oddintsl}
T_\odd(s,t)=\left(\frac s{s_0}\right)^{\alpha_0+\alpha't} T_\odd(t)\,.
\end{equation}

\begin{figure}[p]
\moveright 10mm \hbox{\input{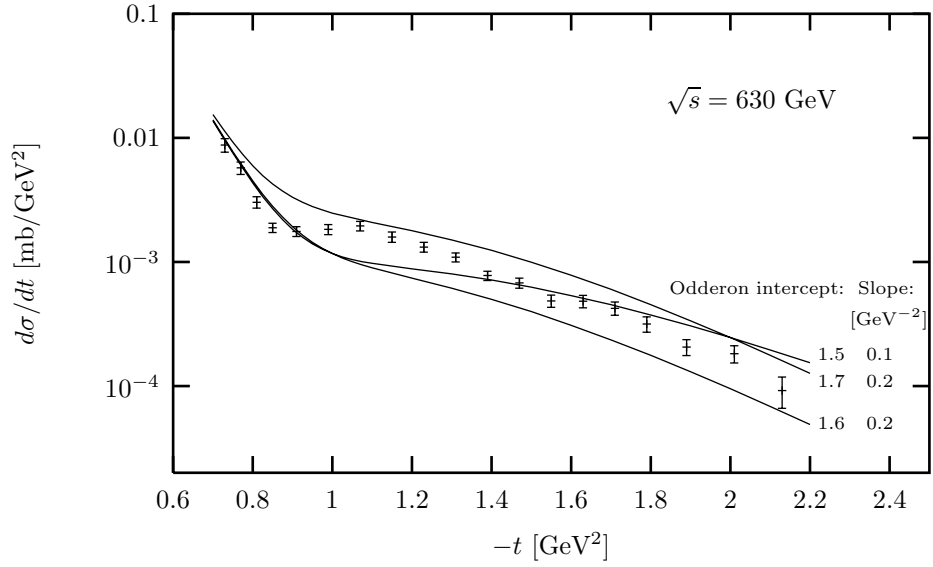}}
\caption{Taking the whole Regge trajectory of the odderon into account for 
		proton-antiproton scattering at $\sqrt s=630$~GeV.
		The plot displays the combinations of intercept and slope which came
		closest to matching the data.
		The reference energy was $\sqrt{s_0}=53$~GeV.}
\label{fig:slope630}
\end{figure}

\begin{figure}[p]
\moveright 10mm \hbox{\input{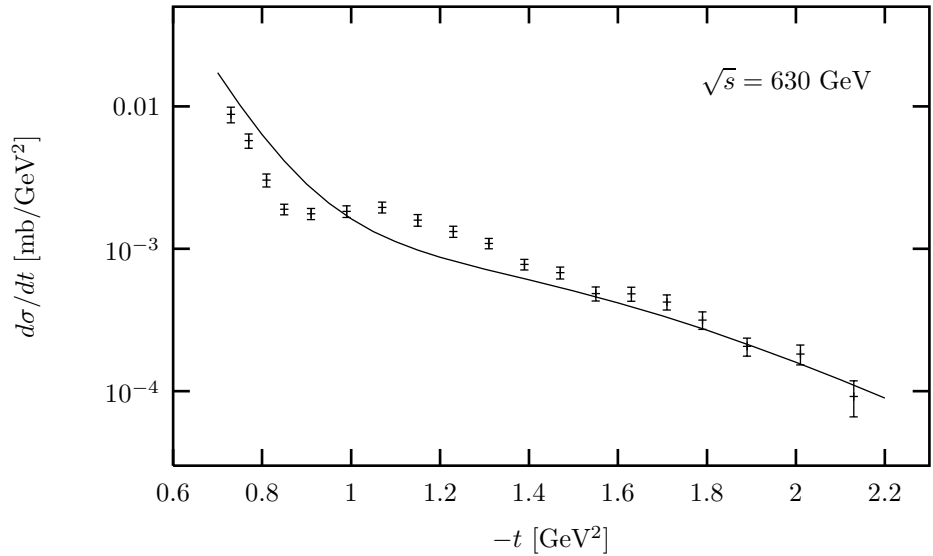}}
\caption{Fit with the energy dependent odderon, taking into account its
		Regge trajectory and the signature factor. The centre-of-mass
		energy is $\sqrt s=630$~GeV. The Regge intercept is 1.7, the slope 
		0.2~GeV$^{-2}$. The reference energy was again $\sqrt{s_0}=53$~GeV.}
\label{fig:sig630}
\end{figure}

Figure~\ref{fig:slope630} shows the result for some combinations of values for
the intercept and the slope. As was to be expected, the already very high
intercept had to be raised even more. The positive slope diminishes the odderon
contribution relative to $t=0$ since $t$ is negative.  To retain the same size
of the odderon contribution at $-t=1$~GeV, for instance, the intercept has to
be raised by 0.1 if the slope is set 0.1~GeV$^{-2}$ higher.

Figure~\ref{fig:slope630} suggests a Regge slope of 0.2~GeV$^{-2}$ and an
intercept between 1.6 and 1.7.  Such a large intercept seems extremely unlikely
for the reasons stated above.  Besides, the fit is far from perfect: The
flattening off in the data points between $-t=0.8$ and 1.2~GeV$^{-2}$ is not
reproduced well.

One final attempt was made to improve our description of higher-energy $p\bar
p$ scattering.  We have so far neglected the odderon's signature factor which
changes its phase according to its Regge trajectory (see
Section~\ref{sec:reggeon}).  This amounts to assuming that the trajectory stays
close to one, ie the intercept is close to one and the slope close to zero.
Since we found quite different values for both this is no longer consistent.
The expression of the odderon amplitude with the signature factor reads:
\begin{equation}
\label{eq:oddsig}
T_\odd(s,t)=\frac12\left(1-e^{-i\pi(\alpha_0+\alpha't)}\right)
			\left(\frac s{s_0}\right)^{\alpha_0+\alpha't} T_\odd(t)\,.
\end{equation}

The result is disappointing, as Figure~\ref{fig:sig630} shows. 
Far from improving the fit, the added signature factor makes it worse.
The curve falls off even more rapidly in the region between 0.8 and
1.2~GeV$^2$. 

It can be concluded that for extending the Donnachie-Landshoff fit to higher
energies, modifying the odderon contribution alone is not enough.  It seems
that some of the other contributions are no longer correct at energies above
the ISR range and also have to be modified.

\begin{figure}[p]
\moveright 10mm \hbox{\input{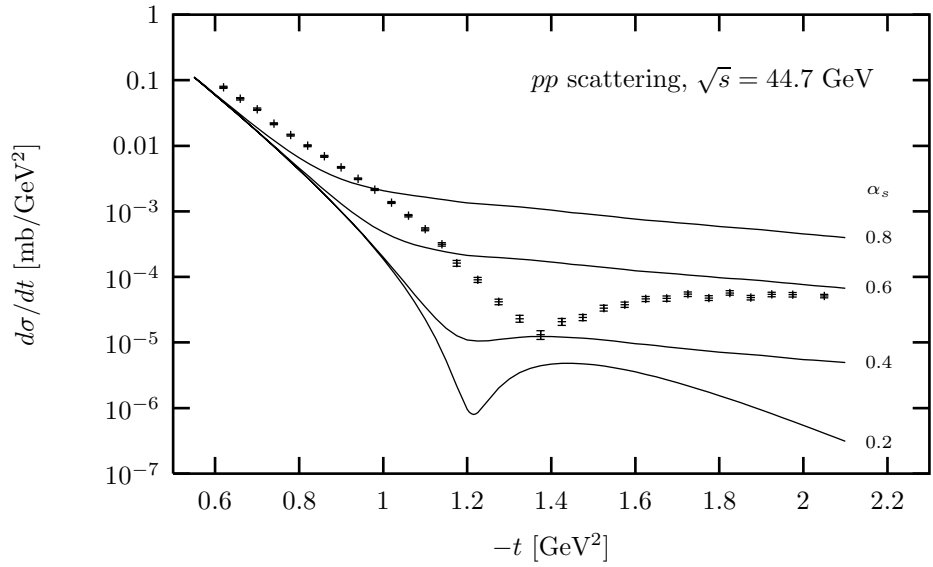}}
\caption{The GLN fit with our geometric odderon substituted for the three 
		odderon-related contributions. Proton-proton scattering at  
		$\sqrt s=44.7$~GeV.}
\label{fig:glngeo44.7}
\end{figure}

\begin{figure}[p]
\moveright 10mm \hbox{\input{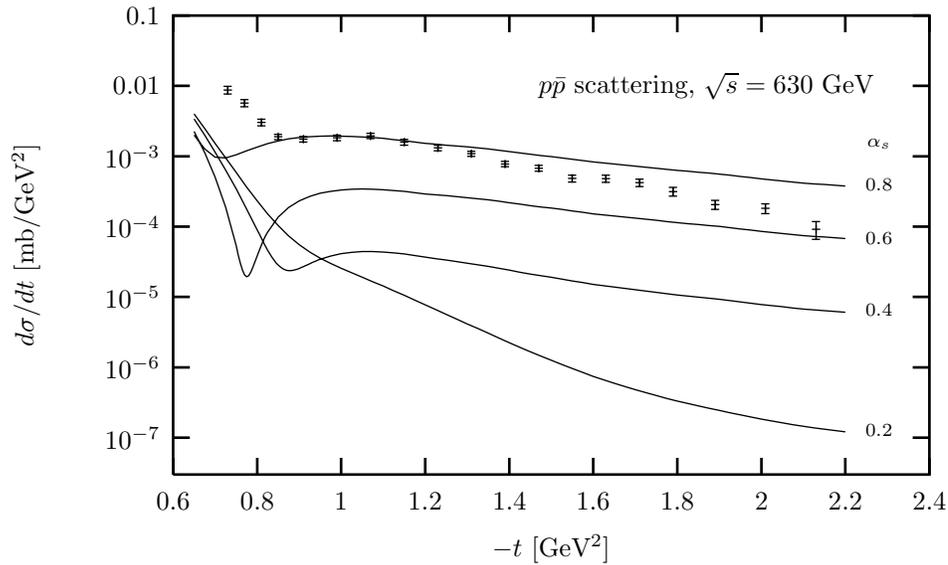}}
\caption{The GLN fit with our geometric odderon substituted for the three 
		odderon-related contributions. Proton-antiproton scattering at  
		$\sqrt s=630$~GeV.}
\label{fig:glngeo630}
\end{figure}

\subsection{With the fit by Gauron, Leader and Nicolescu}

Another attempt was made to determine the odderon intercept and slope.  A
different fit for $d\sigma/dt$ was used as a framework.  It was created by
Gauron, Leader and Nicolescu in 1989~\cite{gln}.  It provides a much better fit
than the one due to Donnachie and Landshoff~\cite{dlfit} for both
proton-antiproton and proton-antiproton at the cost of a large number of free
parameters.\footnote{Donnachie and Landshoff have improved their fit in
response to criticism by Nicolescu et al.~\cite{dlfitnew}.  However, I could
not reproduce the improved fit (see Section~\ref{sec:dlnew}).}  In particular,
it also fits higher-energy proton-antiproton data.

Gauron, Leader and Nicolescu take into account a rather large number of
contributions to the scattering amplitude. First, they include two
``asymptotic'' contributions which unlike the others remain significant in the
limit~$s\to\infty$. One of them is a $C=+1$ ``froissaron'', the other a $C=-1$
maximal odderon.%
\footnote{It is maximal in the sense that it is $\propto (\log s)^2$, the 
maximum growth allowed by asymptotic theorems.}
Besides there are contributions which play a role at lower (ISR) energies: a
pomeron pole, a double pomeron cut, an odderon pole, an odderon-pomeron cut,
three types of Reggeons and two Reggeon-pomeron cuts.

To substitute our odderon contribution, we leave out from the fit the three
contributions related to the odderon --- the maximal odderon, the odderon pole
and the odderon-pomeron cut --- and add our geometric odderon.  However, far
from being able to study the $s$~dependence, we failed even to produce a
satisfactory fit for a single energy.  Figures \ref{fig:glngeo44.7}
and~\ref{fig:glngeo630} show curves with a range of different coupling
constants for our odderon.

\begin{figure}[thb]
\moveright 10mm \hbox{\input{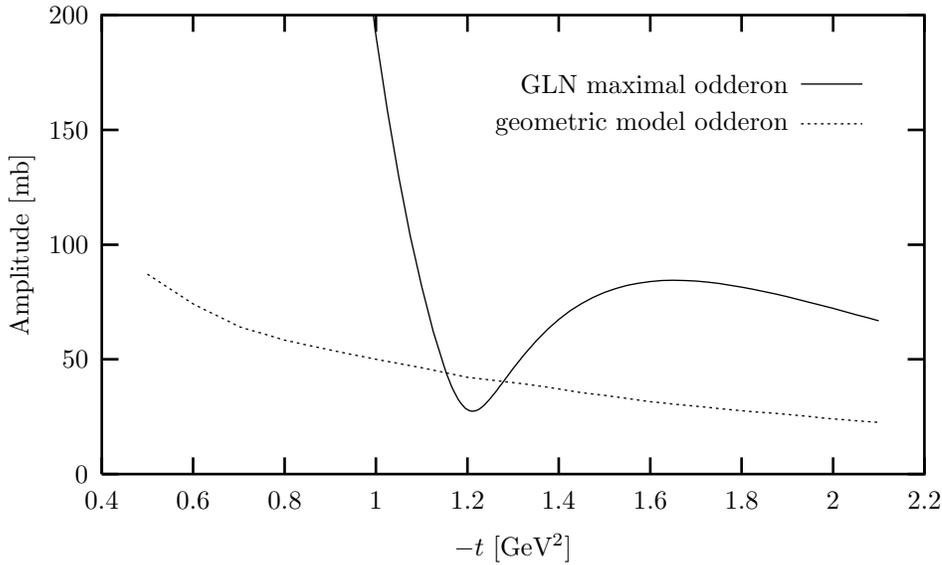}}
\caption{Modulus of the GLN fit's maximal odderon contribution compared to our
		geometric odderon for~$\alpha_s=0.4$.  The maximal odderon has a much 
		more complex structure.  The centre-of-mass energy is $\sqrt s=44.7$~GeV.}
\label{fig:glnoddcmp}
\end{figure}

A detailed analysis reveals that the maximal odderon is by far the most
important of the three odderon contributions in the GLN fit.  Due to
interference between the terms it contains, it already shows a dip structure
itself.  Figure~\ref{fig:glnoddcmp} shows a comparison between the modulus of
the maximal odderon amplitude and our odderon (which is
real-valued).  Apparently the odderon plays an even greater role in the GLN fit
than it does in the DL fit.  With an odderon with a much simpler structure such
as ours, we cannot reproduce their fit.

\subsection{Consequences for the odderon}
\label{sec:glndlodd}

As Figure~\ref{fig:glnoddcmp} shows, the maximal odderon used by Gauron et
{al.}  has a much richer structure than our geometric odderon and, by
implication, the odderon in the Donnachie-Landshoff fit.  It seems implausible
that the GLN odderon should work in the framework of the DL fit and vice versa.

\begin{figure}[h]
\moveright 10mm \hbox{\input{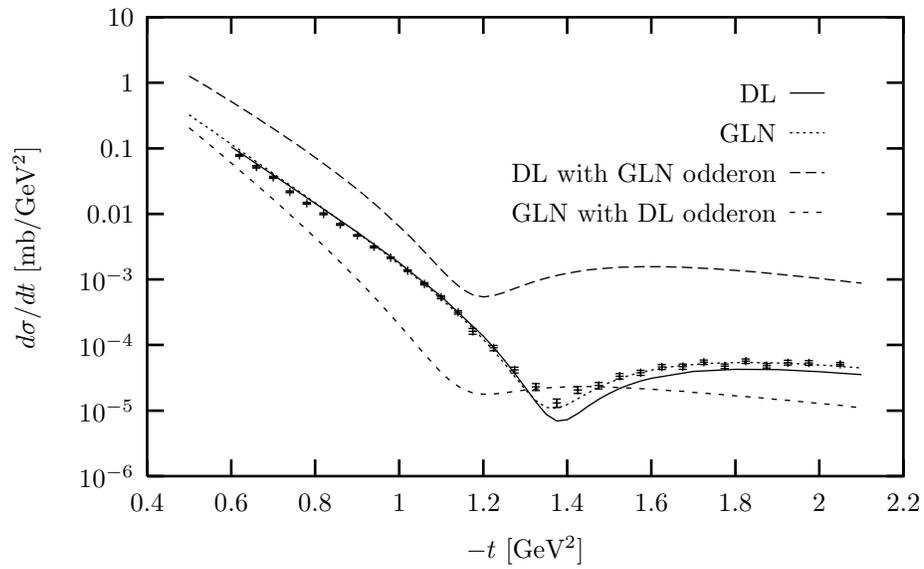}}
\caption{Comparison of the unchanged fits by Donnachie and Landshoff (DL) and 
		Gauron, Leader and Nicolescu (GLN) with fits in which the other group's
		odderon has been substituted.  This is proton-proton scattering at
		$\sqrt s=44.7\;$GeV.}
\label{fig:glndlodd}
\end{figure}

The fits indeed do not work with the wrong odderon contributions:
Figure~\ref{fig:glndlodd} shows the fits by Donnachie and Landshoff and Gauron,
Leader and Nicolescu, respectively, with the other group's odderon substituted,
together with the unchanged fits.  The fits with the wrong odderon match the
data quite badly.

This has disturbing implications for the odderon itself.  While many people
agree that there is an odderon, there is no consensus on what it should look
like.  This comparison shows that we do not know the universal odderon given by
the location of the odderon pole in the complex angular momentum plane.  Though
the odderon plays an important role in both fits, the form of the odderon
contribution is so different as to be incompatible.  While in principle
$C$-even and $C$-odd contributions can be determined independently of each
other, in practice the lack of more data and the freedom in the $C$-even terms
mean that it cannot.  The only explanation for the two very different odderons
in the two fits is that they make a different separation between the even and
odd contributions to the scattering amplitude.  (Their non-odderon odd
contributions are similar.)  Since both fit the data well, the data are not yet
able to decide between the two conflicting views.

\cleardoublepage

\def\rhotre#1#2#3{\rho_{#1#2}^{h_{#1}+h_{#2}-h_{#3}}}
\def\crhotre#1#2#3{\rho_{#1#2}^{*\;\bar h_{#1}+\bar h_{#2}-\bar h_{#3}}}
\def\Ez#1{{E^{(\nu_{#1},0)}}}
\def\eEz#1#2#3{{\left(\frac{|\rho_{#1#2}|}{|\rho_{#1#3}||\rho_{#2#3}|}\right)%
^{1+2i\nu_{#3}}}}
\def\cEz#1#2#3{{\left(\frac{|\rho_{#1#2}|}{|\rho_{#1#3}||\rho_{#2#3}|}\right)%
^{1-2i\nu_{#3}}}}
\def\Ph#1{\Phi_{h_{#1},\bar h_{#1}}(\rho_{#1},\rho_{#1}^*)}
\def\sumabc{\sum_{a\leftrightarrow b\leftrightarrow c}}
\def\crossrat#1#2#3#4#5#6#7#8{%
{\frac{\rho_{#1#2}\rho_{#3#4}}{\rho_{#5#6}\rho_{#7#8}}}}
\def\icrossrat#1#2#3#4#5#6#7#8{%
{(\rho_{#1#2}\rho_{#3#4})/(\rho_{#5#6}\rho_{#7#8})}}
\def\psithree#1#2#3#4#5#6#7#8{%
{\left(\frac{|\rho_{#1#2}||\rho_{#3#4}|}{|\rho_{#5#6}|}\right)^{1#72i\nu_{#8}}
}}
\def\onetworat#1#2#3#4#5#6#7#8{%
{\left(\frac{|\rho_{12}|}{|\rho_{1#8}||\rho_{2#8}|}
\frac{|\rho_{#1#2}||\rho_{#3#4}|}{|\rho_{#5#6}|}\right)^{1#72i\nu_{#8}}
}}
\def\onethreerat#1#2#3#4#5#6#7#8{%
{\left(\frac{|\rho_{13}|}{|\rho_{1#8}||\rho_{3#8}|}
\frac{|\rho_{#1#2}||\rho_{#3#4}|}{|\rho_{#5#6}|}\right)^{1#72i\nu_{#8}}
}}
\def\twothreerat#1#2#3#4#5#6#7#8{%
{\left(\frac{|\rho_{23}|}{|\rho_{2#8}||\rho_{3#8}|}
\frac{|\rho_{#1#2}||\rho_{#3#4}|}{|\rho_{#5#6}|}\right)^{1#72i\nu_{#8}}
}}

\def\Vallr{{\def\C{\alpha} \lower 18pt\hbox{\begin{picture}(0,0)%
\includegraphics{Vllr.pstex}%
\end{picture}%
\setlength{\unitlength}{1036sp}%
\begingroup\makeatletter\ifx\SetFigFont\undefined%
\gdef\SetFigFont#1#2#3#4#5{%
  \reset@font\fontsize{#1}{#2pt}%
  \fontfamily{#3}\fontseries{#4}\fontshape{#5}%
  \selectfont}%
\fi\endgroup%
\begin{picture}(1830,2537)(1111,-3152)
\put(2026,-1996){\makebox(0,0)[b]{\smash{\SetFigFont{8}{9.6}{\sfdefault}{\mddefault}{\updefault}{\color[rgb]{0,0,0}$\C$}%
}}}
\end{picture}
}}}
\def\Valrb{{\def\C{\alpha} \lower 23pt\hbox{\begin{picture}(0,0)%
\includegraphics{Vlrb.pstex}%
\end{picture}%
\setlength{\unitlength}{1036sp}%
\begingroup\makeatletter\ifx\SetFigFont\undefined%
\gdef\SetFigFont#1#2#3#4#5{%
  \reset@font\fontsize{#1}{#2pt}%
  \fontfamily{#3}\fontseries{#4}\fontshape{#5}%
  \selectfont}%
\fi\endgroup%
\begin{picture}(1830,2852)(1111,-3467)
\put(2026,-1996){\makebox(0,0)[b]{\smash{\SetFigFont{8}{9.6}{\sfdefault}{\mddefault}{\updefault}{\color[rgb]{0,0,0}$\C$}%
}}}
\end{picture}
}}}
\def\Vallb{{\def\C{\alpha} \lower 23pt\hbox{\begin{picture}(0,0)%
\includegraphics{Vllb.pstex}%
\end{picture}%
\setlength{\unitlength}{1036sp}%
\begingroup\makeatletter\ifx\SetFigFont\undefined%
\gdef\SetFigFont#1#2#3#4#5{%
  \reset@font\fontsize{#1}{#2pt}%
  \fontfamily{#3}\fontseries{#4}\fontshape{#5}%
  \selectfont}%
\fi\endgroup%
\begin{picture}(1830,2853)(1111,-3468)
\put(2026,-1996){\makebox(0,0)[b]{\smash{\SetFigFont{8}{9.6}{\sfdefault}{\mddefault}{\updefault}{\color[rgb]{0,0,0}$\C$}%
}}}
\end{picture}
}}}
\def\Valbb{{\def\C{\alpha} \lower 23pt\hbox{\begin{picture}(0,0)%
\includegraphics{Vlbb.pstex}%
\end{picture}%
\setlength{\unitlength}{1036sp}%
\begingroup\makeatletter\ifx\SetFigFont\undefined%
\gdef\SetFigFont#1#2#3#4#5{%
  \reset@font\fontsize{#1}{#2pt}%
  \fontfamily{#3}\fontseries{#4}\fontshape{#5}%
  \selectfont}%
\fi\endgroup%
\begin{picture}(1830,2852)(1111,-3467)
\put(2026,-1996){\makebox(0,0)[b]{\smash{\SetFigFont{8}{9.6}{\sfdefault}{\mddefault}{\updefault}{\color[rgb]{0,0,0}$\C$}%
}}}
\end{picture}
}}}
\def\Vbfkl{{\lower 23pt\hbox{\begin{picture}(0,0)%
\includegraphics{Vbfkl.pstex}%
\end{picture}%
\setlength{\unitlength}{1036sp}%
\begingroup\makeatletter\ifx\SetFigFont\undefined%
\gdef\SetFigFont#1#2#3#4#5{%
  \reset@font\fontsize{#1}{#2pt}%
  \fontfamily{#3}\fontseries{#4}\fontshape{#5}%
  \selectfont}%
\fi\endgroup%
\begin{picture}(1830,2853)(1111,-3468)
\put(2026,-1996){\makebox(0,0)[b]{\smash{\SetFigFont{7}{8.4}{\sfdefault}{\mddefault}{\updefault}{\color[rgb]{0,0,0}$K_{B\!F\!K\!L}$}%
}}}
\end{picture}
}}}
\def\Vallri#1{{\def\C{\alpha} \def\I{{#1}} \lower 23pt\hbox{\input{ptex/Vllri.pstex_t}}}}
\def\Valrbi#1{{\def\C{\alpha} \def\I{{#1}} \lower 30.5pt\hbox{\input{ptex/Vlrbi.pstex_t}}}}
\def\Vallbi#1{{\def\C{\alpha} \def\I{{#1}} \lower 30.5pt\hbox{\input{ptex/Vllbi.pstex_t}}}}
\def\Valbbi#1{{\def\C{\alpha} \def\I{{#1}} \lower 23pt\hbox{\input{ptex/Vlbbi.pstex_t}}}}
\def\VGllr{{\def\C{G} \lower 18pt\hbox{\begin{picture}(0,0)%
\includegraphics{Vllr.pstex}%
\end{picture}%
\setlength{\unitlength}{1036sp}%
\begingroup\makeatletter\ifx\SetFigFont\undefined%
\gdef\SetFigFont#1#2#3#4#5{%
  \reset@font\fontsize{#1}{#2pt}%
  \fontfamily{#3}\fontseries{#4}\fontshape{#5}%
  \selectfont}%
\fi\endgroup%
\begin{picture}(1830,2537)(1111,-3152)
\put(2026,-1996){\makebox(0,0)[b]{\smash{\SetFigFont{8}{9.6}{\sfdefault}{\mddefault}{\updefault}{\color[rgb]{0,0,0}$\C$}%
}}}
\end{picture}
}}}
\def\VGlrb{{\def\C{G} \lower 23pt\hbox{\begin{picture}(0,0)%
\includegraphics{Vlrb.pstex}%
\end{picture}%
\setlength{\unitlength}{1036sp}%
\begingroup\makeatletter\ifx\SetFigFont\undefined%
\gdef\SetFigFont#1#2#3#4#5{%
  \reset@font\fontsize{#1}{#2pt}%
  \fontfamily{#3}\fontseries{#4}\fontshape{#5}%
  \selectfont}%
\fi\endgroup%
\begin{picture}(1830,2852)(1111,-3467)
\put(2026,-1996){\makebox(0,0)[b]{\smash{\SetFigFont{8}{9.6}{\sfdefault}{\mddefault}{\updefault}{\color[rgb]{0,0,0}$\C$}%
}}}
\end{picture}
}}}
\def\VGllb{{\def\C{G} \lower 23pt\hbox{\begin{picture}(0,0)%
\includegraphics{Vllb.pstex}%
\end{picture}%
\setlength{\unitlength}{1036sp}%
\begingroup\makeatletter\ifx\SetFigFont\undefined%
\gdef\SetFigFont#1#2#3#4#5{%
  \reset@font\fontsize{#1}{#2pt}%
  \fontfamily{#3}\fontseries{#4}\fontshape{#5}%
  \selectfont}%
\fi\endgroup%
\begin{picture}(1830,2853)(1111,-3468)
\put(2026,-1996){\makebox(0,0)[b]{\smash{\SetFigFont{8}{9.6}{\sfdefault}{\mddefault}{\updefault}{\color[rgb]{0,0,0}$\C$}%
}}}
\end{picture}
}}}
\def\VGlbb{{\def\C{G} \lower 23pt\hbox{\begin{picture}(0,0)%
\includegraphics{Vlbb.pstex}%
\end{picture}%
\setlength{\unitlength}{1036sp}%
\begingroup\makeatletter\ifx\SetFigFont\undefined%
\gdef\SetFigFont#1#2#3#4#5{%
  \reset@font\fontsize{#1}{#2pt}%
  \fontfamily{#3}\fontseries{#4}\fontshape{#5}%
  \selectfont}%
\fi\endgroup%
\begin{picture}(1830,2852)(1111,-3467)
\put(2026,-1996){\makebox(0,0)[b]{\smash{\SetFigFont{8}{9.6}{\sfdefault}{\mddefault}{\updefault}{\color[rgb]{0,0,0}$\C$}%
}}}
\end{picture}
}}}
\def\Cddel{{\lower 5pt\hbox{\begin{picture}(0,0)%
\includegraphics{Cddel.pstex}%
\end{picture}%
\setlength{\unitlength}{1036sp}%
\begingroup\makeatletter\ifx\SetFigFont\undefined%
\gdef\SetFigFont#1#2#3#4#5{%
  \reset@font\fontsize{#1}{#2pt}%
  \fontfamily{#3}\fontseries{#4}\fontshape{#5}%
  \selectfont}%
\fi\endgroup%
\begin{picture}(2759,1093)(1111,-2580)
\put(2026,-1996){\makebox(0,0)[b]{\smash{\SetFigFont{8}{9.6}{\sfdefault}{\mddefault}{\updefault}{\color[rgb]{0,0,0}$d$}%
}}}
\end{picture}
}}}
\def\Cdpairs{{\lower 8pt\hbox{\begin{picture}(0,0)%
\includegraphics{Cdpairs.pstex}%
\end{picture}%
\setlength{\unitlength}{1036sp}%
\begingroup\makeatletter\ifx\SetFigFont\undefined%
\gdef\SetFigFont#1#2#3#4#5{%
  \reset@font\fontsize{#1}{#2pt}%
  \fontfamily{#3}\fontseries{#4}\fontshape{#5}%
  \selectfont}%
\fi\endgroup%
\begin{picture}(1830,1395)(1111,-2882)
\put(2026,-1996){\makebox(0,0)[b]{\smash{\SetFigFont{8}{9.6}{\sfdefault}{\mddefault}{\updefault}{\color[rgb]{0,0,0}$d$}%
}}}
\end{picture}
}}}
\def\Cdeldel{{\lower 8pt\hbox{\begin{picture}(0,0)%
\includegraphics{Cdeldel.pstex}%
\end{picture}%
\setlength{\unitlength}{1036sp}%
\begingroup\makeatletter\ifx\SetFigFont\undefined%
\gdef\SetFigFont#1#2#3#4#5{%
  \reset@font\fontsize{#1}{#2pt}%
  \fontfamily{#3}\fontseries{#4}\fontshape{#5}%
  \selectfont}%
\fi\endgroup%
\begin{picture}(538,1052)(3332,-2882)
\end{picture}
}}}
\def\Cddellink{{\lower 12pt\hbox{\begin{picture}(0,0)%
\includegraphics{Cddellink.pstex}%
\end{picture}%
\setlength{\unitlength}{1036sp}%
\begingroup\makeatletter\ifx\SetFigFont\undefined%
\gdef\SetFigFont#1#2#3#4#5{%
  \reset@font\fontsize{#1}{#2pt}%
  \fontfamily{#3}\fontseries{#4}\fontshape{#5}%
  \selectfont}%
\fi\endgroup%
\begin{picture}(2759,1627)(1111,-3114)
\put(2026,-1996){\makebox(0,0)[b]{\smash{\SetFigFont{8}{9.6}{\sfdefault}{\mddefault}{\updefault}{\color[rgb]{0,0,0}$d$}%
}}}
\end{picture}
}}}
\def\Clink{{\lower 12pt\hbox{\begin{picture}(0,0)%
\includegraphics{Clink.pstex}%
\end{picture}%
\setlength{\unitlength}{1036sp}%
\begingroup\makeatletter\ifx\SetFigFont\undefined%
\gdef\SetFigFont#1#2#3#4#5{%
  \reset@font\fontsize{#1}{#2pt}%
  \fontfamily{#3}\fontseries{#4}\fontshape{#5}%
  \selectfont}%
\fi\endgroup%
\begin{picture}(1649,1284)(2221,-3114)
\end{picture}
}}}
\def\Clowdel{{\lower 8pt\hbox{\begin{picture}(0,0)%
\includegraphics{Clowdel.pstex}%
\end{picture}%
\setlength{\unitlength}{1036sp}%
\begingroup\makeatletter\ifx\SetFigFont\undefined%
\gdef\SetFigFont#1#2#3#4#5{%
  \reset@font\fontsize{#1}{#2pt}%
  \fontfamily{#3}\fontseries{#4}\fontshape{#5}%
  \selectfont}%
\fi\endgroup%
\begin{picture}(512,286)(1320,-2882)
\end{picture}
}}}
\def\Cdcross{{\lower 8pt\hbox{\begin{picture}(0,0)%
\includegraphics{Cdcross.pstex}%
\end{picture}%
\setlength{\unitlength}{1036sp}%
\begingroup\makeatletter\ifx\SetFigFont\undefined%
\gdef\SetFigFont#1#2#3#4#5{%
  \reset@font\fontsize{#1}{#2pt}%
  \fontfamily{#3}\fontseries{#4}\fontshape{#5}%
  \selectfont}%
\fi\endgroup%
\begin{picture}(1830,1395)(1111,-2882)
\put(2026,-1996){\makebox(0,0)[b]{\smash{\SetFigFont{8}{9.6}{\sfdefault}{\mddefault}{\updefault}{\color[rgb]{0,0,0}$d$}%
}}}
\end{picture}
}}}
\def\Cddelcrlnk{{\lower 12pt\hbox{\begin{picture}(0,0)%
\includegraphics{Cddelcrlnk.pstex}%
\end{picture}%
\setlength{\unitlength}{1036sp}%
\begingroup\makeatletter\ifx\SetFigFont\undefined%
\gdef\SetFigFont#1#2#3#4#5{%
  \reset@font\fontsize{#1}{#2pt}%
  \fontfamily{#3}\fontseries{#4}\fontshape{#5}%
  \selectfont}%
\fi\endgroup%
\begin{picture}(2759,1624)(1111,-3111)
\put(2026,-1996){\makebox(0,0)[b]{\smash{\SetFigFont{8}{9.6}{\sfdefault}{\mddefault}{\updefault}{\color[rgb]{0,0,0}$d$}%
}}}
\end{picture}
}}}
\def\Cdbridge{{\lower 12pt\hbox{\begin{picture}(0,0)%
\includegraphics{Cdbridge.pstex}%
\end{picture}%
\setlength{\unitlength}{1036sp}%
\begingroup\makeatletter\ifx\SetFigFont\undefined%
\gdef\SetFigFont#1#2#3#4#5{%
  \reset@font\fontsize{#1}{#2pt}%
  \fontfamily{#3}\fontseries{#4}\fontshape{#5}%
  \selectfont}%
\fi\endgroup%
\begin{picture}(1830,1620)(1111,-3107)
\put(2026,-1996){\makebox(0,0)[b]{\smash{\SetFigFont{8}{9.6}{\sfdefault}{\mddefault}{\updefault}{\color[rgb]{0,0,0}$d$}%
}}}
\end{picture}
}}}
\def\VLa{{ \lower 19.5pt\hbox{\begin{picture}(0,0)%
\includegraphics{VLa.pstex}%
\end{picture}%
\setlength{\unitlength}{1036sp}%
\begingroup\makeatletter\ifx\SetFigFont\undefined%
\gdef\SetFigFont#1#2#3#4#5{%
  \reset@font\fontsize{#1}{#2pt}%
  \fontfamily{#3}\fontseries{#4}\fontshape{#5}%
  \selectfont}%
\fi\endgroup%
\begin{picture}(2730,2160)(661,-3557)
\put(2026,-1996){\makebox(0,0)[b]{\smash{\SetFigFont{8}{9.6}{\sfdefault}{\mddefault}{\updefault}{\color[rgb]{0,0,0}$L$}%
}}}
\end{picture}
}}}
\def\VLb{{ \lower 16pt\hbox{\begin{picture}(0,0)%
\includegraphics{VLb.pstex}%
\end{picture}%
\setlength{\unitlength}{1036sp}%
\begingroup\makeatletter\ifx\SetFigFont\undefined%
\gdef\SetFigFont#1#2#3#4#5{%
  \reset@font\fontsize{#1}{#2pt}%
  \fontfamily{#3}\fontseries{#4}\fontshape{#5}%
  \selectfont}%
\fi\endgroup%
\begin{picture}(2730,1935)(661,-3332)
\put(2026,-1996){\makebox(0,0)[b]{\smash{\SetFigFont{8}{9.6}{\sfdefault}{\mddefault}{\updefault}{\color[rgb]{0,0,0}$L$}%
}}}
\end{picture}
}}}
\def\VIa{{\def\C{I} \lower 16pt\hbox{\begin{picture}(0,0)%
\includegraphics{VIJa.pstex}%
\end{picture}%
\setlength{\unitlength}{1036sp}%
\begingroup\makeatletter\ifx\SetFigFont\undefined%
\gdef\SetFigFont#1#2#3#4#5{%
  \reset@font\fontsize{#1}{#2pt}%
  \fontfamily{#3}\fontseries{#4}\fontshape{#5}%
  \selectfont}%
\fi\endgroup%
\begin{picture}(2730,1935)(661,-3332)
\put(2026,-1996){\makebox(0,0)[b]{\smash{\SetFigFont{8}{9.6}{\sfdefault}{\mddefault}{\updefault}{\color[rgb]{0,0,0}$\C$}%
}}}
\end{picture}
}}}
\def\VIb{{\def\C{I} \lower 12.5pt\hbox{\begin{picture}(0,0)%
\includegraphics{VIJb.pstex}%
\end{picture}%
\setlength{\unitlength}{1036sp}%
\begingroup\makeatletter\ifx\SetFigFont\undefined%
\gdef\SetFigFont#1#2#3#4#5{%
  \reset@font\fontsize{#1}{#2pt}%
  \fontfamily{#3}\fontseries{#4}\fontshape{#5}%
  \selectfont}%
\fi\endgroup%
\begin{picture}(2730,1710)(661,-3107)
\put(2026,-1996){\makebox(0,0)[b]{\smash{\SetFigFont{8}{9.6}{\sfdefault}{\mddefault}{\updefault}{\color[rgb]{0,0,0}$\C$}%
}}}
\end{picture}
}}}
\def\VIbb{{\def\C{I} \lower 12.5pt\hbox{\begin{picture}(0,0)%
\includegraphics{VIJbb.pstex}%
\end{picture}%
\setlength{\unitlength}{1036sp}%
\begingroup\makeatletter\ifx\SetFigFont\undefined%
\gdef\SetFigFont#1#2#3#4#5{%
  \reset@font\fontsize{#1}{#2pt}%
  \fontfamily{#3}\fontseries{#4}\fontshape{#5}%
  \selectfont}%
\fi\endgroup%
\begin{picture}(2730,1710)(661,-3107)
\put(2026,-1996){\makebox(0,0)[b]{\smash{\SetFigFont{8}{9.6}{\sfdefault}{\mddefault}{\updefault}{\color[rgb]{0,0,0}$\C$}%
}}}
\end{picture}
}}}
\def\VIc{{\def\C{I} \lower 16pt\hbox{\begin{picture}(0,0)%
\includegraphics{VIJc.pstex}%
\end{picture}%
\setlength{\unitlength}{1036sp}%
\begingroup\makeatletter\ifx\SetFigFont\undefined%
\gdef\SetFigFont#1#2#3#4#5{%
  \reset@font\fontsize{#1}{#2pt}%
  \fontfamily{#3}\fontseries{#4}\fontshape{#5}%
  \selectfont}%
\fi\endgroup%
\begin{picture}(2730,1935)(661,-3332)
\put(2026,-1996){\makebox(0,0)[b]{\smash{\SetFigFont{8}{9.6}{\sfdefault}{\mddefault}{\updefault}{\color[rgb]{0,0,0}$\C$}%
}}}
\end{picture}
}}}
\def\VId{{\def\C{I} \lower 16pt\hbox{\begin{picture}(0,0)%
\includegraphics{VIJd.pstex}%
\end{picture}%
\setlength{\unitlength}{1036sp}%
\begingroup\makeatletter\ifx\SetFigFont\undefined%
\gdef\SetFigFont#1#2#3#4#5{%
  \reset@font\fontsize{#1}{#2pt}%
  \fontfamily{#3}\fontseries{#4}\fontshape{#5}%
  \selectfont}%
\fi\endgroup%
\begin{picture}(2730,1935)(661,-3332)
\put(2026,-1996){\makebox(0,0)[b]{\smash{\SetFigFont{8}{9.6}{\sfdefault}{\mddefault}{\updefault}{\color[rgb]{0,0,0}$\C$}%
}}}
\end{picture}
}}}
\def\VIe{{\def\C{I} \lower 16pt\hbox{\begin{picture}(0,0)%
\includegraphics{VIJe.pstex}%
\end{picture}%
\setlength{\unitlength}{1036sp}%
\begingroup\makeatletter\ifx\SetFigFont\undefined%
\gdef\SetFigFont#1#2#3#4#5{%
  \reset@font\fontsize{#1}{#2pt}%
  \fontfamily{#3}\fontseries{#4}\fontshape{#5}%
  \selectfont}%
\fi\endgroup%
\begin{picture}(2730,1935)(661,-3332)
\put(2026,-1996){\makebox(0,0)[b]{\smash{\SetFigFont{8}{9.6}{\sfdefault}{\mddefault}{\updefault}{\color[rgb]{0,0,0}$\C$}%
}}}
\end{picture}
}}}
\def\VIee{{\def\C{I} \lower 16pt\hbox{\begin{picture}(0,0)%
\includegraphics{VIJee.pstex}%
\end{picture}%
\setlength{\unitlength}{1036sp}%
\begingroup\makeatletter\ifx\SetFigFont\undefined%
\gdef\SetFigFont#1#2#3#4#5{%
  \reset@font\fontsize{#1}{#2pt}%
  \fontfamily{#3}\fontseries{#4}\fontshape{#5}%
  \selectfont}%
\fi\endgroup%
\begin{picture}(2730,1935)(661,-3332)
\put(2026,-1996){\makebox(0,0)[b]{\smash{\SetFigFont{8}{9.6}{\sfdefault}{\mddefault}{\updefault}{\color[rgb]{0,0,0}$\C$}%
}}}
\end{picture}
}}}
\def\VVmmA{{\def\C{V} \lower 22pt\hbox{\begin{picture}(0,0)%
\includegraphics{VVmmA.pstex}%
\end{picture}%
\setlength{\unitlength}{1036sp}%
\begingroup\makeatletter\ifx\SetFigFont\undefined%
\gdef\SetFigFont#1#2#3#4#5{%
  \reset@font\fontsize{#1}{#2pt}%
  \fontfamily{#3}\fontseries{#4}\fontshape{#5}%
  \selectfont}%
\fi\endgroup%
\begin{picture}(1830,2807)(1111,-3422)
\put(2026,-1996){\makebox(0,0)[b]{\smash{\SetFigFont{8}{9.6}{\sfdefault}{\mddefault}{\updefault}{\color[rgb]{0,0,0}$\C$}%
}}}
\end{picture}
}}}
\def\VVLmA{{\def\C{V} \lower 22pt\hbox{\begin{picture}(0,0)%
\includegraphics{VVLmA.pstex}%
\end{picture}%
\setlength{\unitlength}{1036sp}%
\begingroup\makeatletter\ifx\SetFigFont\undefined%
\gdef\SetFigFont#1#2#3#4#5{%
  \reset@font\fontsize{#1}{#2pt}%
  \fontfamily{#3}\fontseries{#4}\fontshape{#5}%
  \selectfont}%
\fi\endgroup%
\begin{picture}(1830,2807)(1111,-3422)
\put(2026,-1996){\makebox(0,0)[b]{\smash{\SetFigFont{8}{9.6}{\sfdefault}{\mddefault}{\updefault}{\color[rgb]{0,0,0}$\C$}%
}}}
\end{picture}
}}}
\def\VVllr{{\def\C{V} \lower 18pt\hbox{\begin{picture}(0,0)%
\includegraphics{VVllr.pstex}%
\end{picture}%
\setlength{\unitlength}{1036sp}%
\begingroup\makeatletter\ifx\SetFigFont\undefined%
\gdef\SetFigFont#1#2#3#4#5{%
  \reset@font\fontsize{#1}{#2pt}%
  \fontfamily{#3}\fontseries{#4}\fontshape{#5}%
  \selectfont}%
\fi\endgroup%
\begin{picture}(1830,2582)(1111,-3197)
\put(2026,-1996){\makebox(0,0)[b]{\smash{\SetFigFont{8}{9.6}{\sfdefault}{\mddefault}{\updefault}{\color[rgb]{0,0,0}$\C$}%
}}}
\end{picture}
}}}
\def\VVlmR{{\def\C{V} \lower 18pt\hbox{\begin{picture}(0,0)%
\includegraphics{VVlmR.pstex}%
\end{picture}%
\setlength{\unitlength}{1036sp}%
\begingroup\makeatletter\ifx\SetFigFont\undefined%
\gdef\SetFigFont#1#2#3#4#5{%
  \reset@font\fontsize{#1}{#2pt}%
  \fontfamily{#3}\fontseries{#4}\fontshape{#5}%
  \selectfont}%
\fi\endgroup%
\begin{picture}(1830,2582)(1111,-3197)
\put(2026,-1996){\makebox(0,0)[b]{\smash{\SetFigFont{8}{9.6}{\sfdefault}{\mddefault}{\updefault}{\color[rgb]{0,0,0}$\C$}%
}}}
\end{picture}
}}}
\def\VVlmA{{\def\C{V} \lower 18pt\hbox{\begin{picture}(0,0)%
\includegraphics{VVlmA.pstex}%
\end{picture}%
\setlength{\unitlength}{1036sp}%
\begingroup\makeatletter\ifx\SetFigFont\undefined%
\gdef\SetFigFont#1#2#3#4#5{%
  \reset@font\fontsize{#1}{#2pt}%
  \fontfamily{#3}\fontseries{#4}\fontshape{#5}%
  \selectfont}%
\fi\endgroup%
\begin{picture}(1830,2628)(1111,-3243)
\put(2026,-1996){\makebox(0,0)[b]{\smash{\SetFigFont{8}{9.6}{\sfdefault}{\mddefault}{\updefault}{\color[rgb]{0,0,0}$\C$}%
}}}
\end{picture}
}}}
\def\Cdxpairs{{\lower 8pt\hbox{\begin{picture}(0,0)%
\includegraphics{Cdxpairs.pstex}%
\end{picture}%
\setlength{\unitlength}{1036sp}%
\begingroup\makeatletter\ifx\SetFigFont\undefined%
\gdef\SetFigFont#1#2#3#4#5{%
  \reset@font\fontsize{#1}{#2pt}%
  \fontfamily{#3}\fontseries{#4}\fontshape{#5}%
  \selectfont}%
\fi\endgroup%
\begin{picture}(1830,1395)(1111,-2882)
\put(2026,-1996){\makebox(0,0)[b]{\smash{\SetFigFont{8}{9.6}{\sfdefault}{\mddefault}{\updefault}{\color[rgb]{0,0,0}$d$}%
}}}
\end{picture}
}}}
\def\Cdxcross{{\lower 8pt\hbox{\begin{picture}(0,0)%
\includegraphics{Cdxcross.pstex}%
\end{picture}%
\setlength{\unitlength}{1036sp}%
\begingroup\makeatletter\ifx\SetFigFont\undefined%
\gdef\SetFigFont#1#2#3#4#5{%
  \reset@font\fontsize{#1}{#2pt}%
  \fontfamily{#3}\fontseries{#4}\fontshape{#5}%
  \selectfont}%
\fi\endgroup%
\begin{picture}(1830,1395)(1111,-2882)
\put(2026,-1996){\makebox(0,0)[b]{\smash{\SetFigFont{8}{9.6}{\sfdefault}{\mddefault}{\updefault}{\color[rgb]{0,0,0}$d$}%
}}}
\end{picture}
}}}
\def\Cdxbridge{{\lower 12pt\hbox{\begin{picture}(0,0)%
\includegraphics{Cdxbridge.pstex}%
\end{picture}%
\setlength{\unitlength}{1036sp}%
\begingroup\makeatletter\ifx\SetFigFont\undefined%
\gdef\SetFigFont#1#2#3#4#5{%
  \reset@font\fontsize{#1}{#2pt}%
  \fontfamily{#3}\fontseries{#4}\fontshape{#5}%
  \selectfont}%
\fi\endgroup%
\begin{picture}(1830,1620)(1111,-3107)
\put(2026,-1996){\makebox(0,0)[b]{\smash{\SetFigFont{8}{9.6}{\sfdefault}{\mddefault}{\updefault}{\color[rgb]{0,0,0}$d$}%
}}}
\end{picture}
}}}
\def\VJa{{\def\C{J} \lower 16pt\hbox{\begin{picture}(0,0)%
\includegraphics{VIJa.pstex}%
\end{picture}%
\setlength{\unitlength}{1036sp}%
\begingroup\makeatletter\ifx\SetFigFont\undefined%
\gdef\SetFigFont#1#2#3#4#5{%
  \reset@font\fontsize{#1}{#2pt}%
  \fontfamily{#3}\fontseries{#4}\fontshape{#5}%
  \selectfont}%
\fi\endgroup%
\begin{picture}(2730,1935)(661,-3332)
\put(2026,-1996){\makebox(0,0)[b]{\smash{\SetFigFont{8}{9.6}{\sfdefault}{\mddefault}{\updefault}{\color[rgb]{0,0,0}$\C$}%
}}}
\end{picture}
}}}
\def\VJb{{\def\C{J} \lower 12.5pt\hbox{\begin{picture}(0,0)%
\includegraphics{VIJb.pstex}%
\end{picture}%
\setlength{\unitlength}{1036sp}%
\begingroup\makeatletter\ifx\SetFigFont\undefined%
\gdef\SetFigFont#1#2#3#4#5{%
  \reset@font\fontsize{#1}{#2pt}%
  \fontfamily{#3}\fontseries{#4}\fontshape{#5}%
  \selectfont}%
\fi\endgroup%
\begin{picture}(2730,1710)(661,-3107)
\put(2026,-1996){\makebox(0,0)[b]{\smash{\SetFigFont{8}{9.6}{\sfdefault}{\mddefault}{\updefault}{\color[rgb]{0,0,0}$\C$}%
}}}
\end{picture}
}}}
\def\VJbb{{\def\C{J} \lower 12.5pt\hbox{\begin{picture}(0,0)%
\includegraphics{VIJbb.pstex}%
\end{picture}%
\setlength{\unitlength}{1036sp}%
\begingroup\makeatletter\ifx\SetFigFont\undefined%
\gdef\SetFigFont#1#2#3#4#5{%
  \reset@font\fontsize{#1}{#2pt}%
  \fontfamily{#3}\fontseries{#4}\fontshape{#5}%
  \selectfont}%
\fi\endgroup%
\begin{picture}(2730,1710)(661,-3107)
\put(2026,-1996){\makebox(0,0)[b]{\smash{\SetFigFont{8}{9.6}{\sfdefault}{\mddefault}{\updefault}{\color[rgb]{0,0,0}$\C$}%
}}}
\end{picture}
}}}
\def\VJc{{\def\C{J} \lower 16pt\hbox{\begin{picture}(0,0)%
\includegraphics{VIJc.pstex}%
\end{picture}%
\setlength{\unitlength}{1036sp}%
\begingroup\makeatletter\ifx\SetFigFont\undefined%
\gdef\SetFigFont#1#2#3#4#5{%
  \reset@font\fontsize{#1}{#2pt}%
  \fontfamily{#3}\fontseries{#4}\fontshape{#5}%
  \selectfont}%
\fi\endgroup%
\begin{picture}(2730,1935)(661,-3332)
\put(2026,-1996){\makebox(0,0)[b]{\smash{\SetFigFont{8}{9.6}{\sfdefault}{\mddefault}{\updefault}{\color[rgb]{0,0,0}$\C$}%
}}}
\end{picture}
}}}
\def\VJd{{\def\C{J} \lower 16pt\hbox{\begin{picture}(0,0)%
\includegraphics{VIJd.pstex}%
\end{picture}%
\setlength{\unitlength}{1036sp}%
\begingroup\makeatletter\ifx\SetFigFont\undefined%
\gdef\SetFigFont#1#2#3#4#5{%
  \reset@font\fontsize{#1}{#2pt}%
  \fontfamily{#3}\fontseries{#4}\fontshape{#5}%
  \selectfont}%
\fi\endgroup%
\begin{picture}(2730,1935)(661,-3332)
\put(2026,-1996){\makebox(0,0)[b]{\smash{\SetFigFont{8}{9.6}{\sfdefault}{\mddefault}{\updefault}{\color[rgb]{0,0,0}$\C$}%
}}}
\end{picture}
}}}
\def\VJe{{\def\C{J} \lower 16pt\hbox{\begin{picture}(0,0)%
\includegraphics{VIJe.pstex}%
\end{picture}%
\setlength{\unitlength}{1036sp}%
\begingroup\makeatletter\ifx\SetFigFont\undefined%
\gdef\SetFigFont#1#2#3#4#5{%
  \reset@font\fontsize{#1}{#2pt}%
  \fontfamily{#3}\fontseries{#4}\fontshape{#5}%
  \selectfont}%
\fi\endgroup%
\begin{picture}(2730,1935)(661,-3332)
\put(2026,-1996){\makebox(0,0)[b]{\smash{\SetFigFont{8}{9.6}{\sfdefault}{\mddefault}{\updefault}{\color[rgb]{0,0,0}$\C$}%
}}}
\end{picture}
}}}
\def\VJee{{\def\C{J} \lower 16pt\hbox{\begin{picture}(0,0)%
\includegraphics{VIJee.pstex}%
\end{picture}%
\setlength{\unitlength}{1036sp}%
\begingroup\makeatletter\ifx\SetFigFont\undefined%
\gdef\SetFigFont#1#2#3#4#5{%
  \reset@font\fontsize{#1}{#2pt}%
  \fontfamily{#3}\fontseries{#4}\fontshape{#5}%
  \selectfont}%
\fi\endgroup%
\begin{picture}(2730,1935)(661,-3332)
\put(2026,-1996){\makebox(0,0)[b]{\smash{\SetFigFont{8}{9.6}{\sfdefault}{\mddefault}{\updefault}{\color[rgb]{0,0,0}$\C$}%
}}}
\end{picture}
}}}
\def\CJc{{\lower 12pt\hbox{\begin{picture}(0,0)%
\includegraphics{CJc.pstex}%
\end{picture}%
\setlength{\unitlength}{1036sp}%
\begingroup\makeatletter\ifx\SetFigFont\undefined%
\gdef\SetFigFont#1#2#3#4#5{%
  \reset@font\fontsize{#1}{#2pt}%
  \fontfamily{#3}\fontseries{#4}\fontshape{#5}%
  \selectfont}%
\fi\endgroup%
\begin{picture}(2759,1624)(1111,-3111)
\put(2026,-1996){\makebox(0,0)[b]{\smash{\SetFigFont{8}{9.6}{\sfdefault}{\mddefault}{\updefault}{\color[rgb]{0,0,0}$d$}%
}}}
\end{picture}
}}}
\def\CJd{{\lower 12pt\hbox{\begin{picture}(0,0)%
\includegraphics{CJd.pstex}%
\end{picture}%
\setlength{\unitlength}{1036sp}%
\begingroup\makeatletter\ifx\SetFigFont\undefined%
\gdef\SetFigFont#1#2#3#4#5{%
  \reset@font\fontsize{#1}{#2pt}%
  \fontfamily{#3}\fontseries{#4}\fontshape{#5}%
  \selectfont}%
\fi\endgroup%
\begin{picture}(2759,1651)(1111,-3138)
\put(2026,-1996){\makebox(0,0)[b]{\smash{\SetFigFont{8}{9.6}{\sfdefault}{\mddefault}{\updefault}{\color[rgb]{0,0,0}$d$}%
}}}
\end{picture}
}}}
\def\CJe{{\lower 12pt\hbox{\begin{picture}(0,0)%
\includegraphics{CJe.pstex}%
\end{picture}%
\setlength{\unitlength}{1036sp}%
\begingroup\makeatletter\ifx\SetFigFont\undefined%
\gdef\SetFigFont#1#2#3#4#5{%
  \reset@font\fontsize{#1}{#2pt}%
  \fontfamily{#3}\fontseries{#4}\fontshape{#5}%
  \selectfont}%
\fi\endgroup%
\begin{picture}(2759,1627)(1111,-3114)
\put(2026,-1996){\makebox(0,0)[b]{\smash{\SetFigFont{8}{9.6}{\sfdefault}{\mddefault}{\updefault}{\color[rgb]{0,0,0}$d$}%
}}}
\end{picture}
}}}
\def\CJee{{\lower 12pt\hbox{\begin{picture}(0,0)%
\includegraphics{CJee.pstex}%
\end{picture}%
\setlength{\unitlength}{1036sp}%
\begingroup\makeatletter\ifx\SetFigFont\undefined%
\gdef\SetFigFont#1#2#3#4#5{%
  \reset@font\fontsize{#1}{#2pt}%
  \fontfamily{#3}\fontseries{#4}\fontshape{#5}%
  \selectfont}%
\fi\endgroup%
\begin{picture}(2759,1624)(1111,-3111)
\put(2026,-1996){\makebox(0,0)[b]{\smash{\SetFigFont{8}{9.6}{\sfdefault}{\mddefault}{\updefault}{\color[rgb]{0,0,0}$d$}%
}}}
\end{picture}
}}}


\addcontentsline{toc}{part}
	{II\hskip 1em Four-pomeron vertices from unitarity corrections}

\thispagestyle{empty}

\vbox to \vsize
{
\vfill

\begin{center}
\Huge\bfseries Four-pomeron vertices from unitarity corrections
\end{center}

\vskip 15mm

\leftskip=15mm
\rightskip=15mm

This part concerns vertices occurring in unitarity corrections to the BFKL
equation.  A $1\to3$ pomeron vertex is computed from a $2\to6$ reggeised gluon
vertex occurring in the six-gluon amplitude in the Extended Generalised Leading
Logarithmic Approximation (EGLLA), part of the gluon-ladder approach to
higher-order BFKL.  This gluon vertex is part of the integral equation for the
six-reggeised-gluon amplitude in EGLLA.  It is an irreducible element in the
conjectured conformal field theory of unitarity corrections.

As a preparation, five basic functions of which the $1\to n$ pomeron vertices
are ultimately composed are presented in both momentum and configuration space.
A further function~$G$, which combines them and in terms of which the pomeron
vertices can be expressed, is shown to be conformally covariant.

The irreducible $1\to3$ pomeron vertex is computed in terms of integrals over
conformal eigenfunctions.  It is cast into the form of a conformal four-point
function.  The freedom left by this form, a function of an anharmonic ratio of
the four pomeron coordinates, is computed for all terms of the vertex and
expressed in terms of three integrals.  The partly reggeising part of the
six-gluon amplitude also gives rise to $1\to3$ pomeron vertices, which dominate
in the limit $N_c\to\infty$.  They are computed in the last sections of this
work.

These vertices are of great interest because $1\to n$ pomeron vertices for
$n>2$ were shown not to exist in the dipole approach to higher-order BFKL.  The
findings of this work, which were calculated in the gluon-ladder approach,
demonstrate a discrepancy between the two major approaches to higher-order
BFKL.

\leftskip=0pt
\rightskip=0pt

\vfill
}

\pagebreak

\chapter{Fundamental functions}
\label{chap:confbase}

\section{Introduction}

This chapter collects the preparatory work which has to be done before
we start on our derivation of pomeron vertices in the following chapter.
Section~\ref{sec:pre} contains an introduction to conformal transformations,
relevant elements of conformal field theory and the BFKL equation.  Sections
\ref{sec:abcst_m} and~\ref{sec:abcst_c} present five functions of which
reggeised gluon vertices are composed, and derive their form in configuration
space.  Section~\ref{sec:G} deals with the function $G$, a combination of these
five functions in terms of which reggeised gluon vertices are most easily
expressed, and determines its properties under conformal transformations.  None
of this is truly original work.  Notably the derivation of the transformation
properties of $G$ to some extent repeats the proof of the conformal invariance
of the $2\to4$ reggeised gluon vertex by Bartels, Lipatov and
W\"usthoff~\cite{blw}, which is a sum of nine terms of $G$ functions, and the
sometimes sketchy treatment of~$G$ by Braun and Vacca~\cite{brvac}.  Likewise,
the five functions of which $G$ is composed have been known in momentum space
for some time.  All the same, this is to my knowledge the first systematic
treatment of the five functions and $G$ in both momentum and position space and
will hopefully provide a clear and reproducible guide for future students of
the subject.

\section{Prerequisites}
\label{sec:pre}

\subsection{Conformal transformations}

The topic investigated in this and the following chapter is the behaviour of
expressions under conformal transformations.  Conformal transformations (or
M\"obius transformations) are those transformations in the complex plane which
preserve both angles and their orientation.  Rotations and dilatations
(rescalings) are conformal transformations.  So are translations.

The complex conjugation preserves angles but inverts their orientation and
therefore is no conformal transformation.  The inversion, $z\mapsto1/z$,
however, conserves orientations as well as angles.  It is not quite the
geometric inversion with respect to the unit circle.  The inversion with
respect to any circle changes the orientation of angles, just like the
reflection with respect to any straight line.  The geometric inversion with
respect to the unit circle can be written as $z\mapsto1/\bar z$ in complex
coordinates.  Hence the complex inversion is (graphically speaking) the
concatenation of the geometric inversion and the complex conjugation
(reflection with respect to the real axis), which is consistent with the fact
that it conserves orientations.

There is a general form for conformal transformations:
\dmath2{
z&\mapsto&\frac{a\,z+b}{c\,z+d}\qquad\qquad ad-bc\neq 0.
&eq:moebius}
However, it would not be easy to determine how an expression transforms under
such a relatively complicated transformation law with four parameters.

In fact, all M\"obius transformations can be generated from the four we have
already mentioned: rotations, dilatations, translations and the inversion.
Therefore it is sufficient to investigate their effect separately.  Their
formulaic expressions are as follows:
\dmath2{
z&\mapsto& e^{i\varphi}\,z\qquad \varphi\in\mathbb{R}\cr
z&\mapsto& c\,z\qquad\quad c\in\mathbb{R}_{>0}\crl{eq:fundaconf}
z&\mapsto& z+a\qquad a\in\mathbb{C}\cr
z&\mapsto& \frac1z\cr
}

There is one point which should be mentioned to avoid confusion with other work
concerning conformal transformations.  What we regard as conformal
transformations here differs from the definition of conformal transformations
in the theory of complex functions and in string theory.  There, conformal
transformations are defined as functions which are locally conformal, ie whose
Jacobi map preserves angles and their orientation.  This holds for any analytic
function with a non-vanishing derivative.  Here, however, we are interested
only in global conformal transformations.

\subsection{Conventions}

Before starting out in earnest, let me explain a few fundamental points of my
notation.  Variables with two indices are customarily defined to be the
difference between two coordinate vectors, ie the distance vectors between two
points.
$$
\rho_{12}:=\rho_1-\rho_2
$$

The derivative with respect to a coordinate is abbreviated by a differentiation
symbol with the same index, ie
$$
\del_1:=\frac{\del}{\del\rho_1}\,.
$$
The same convention will be used for Laplacians and other derivative operators.

I adopt the following convention for the constant factors in Fourier
transformations (for an arbitrary function $f$ depending on a two-dimensional
vector):
\begin{equation}
f(\mbf \rho)= \frac1{(2\pi)^2}\int d^2\mbf k\; f(\mbf k)\;e^{i\mbf k\mbf\rho}
\,;\qquad\qquad\qquad
f(\mbf k)= \int d^2\mbf\rho\; f(\mbf\rho)\; e^{-i\mbf k\mbf\rho}\,.
\end{equation}
The function and its Fourier transform will not be denoted by a different 
symbol; the distinction will be clear from the type of its argument.

\subsection[The vectorial and the complex formulation]
			{The vectorial and the complex formulation of expressions in 
			two-dimensional space}
\label{sec:vect_compl}

We will describe aspects of high-energy scattering in two-dimensional
transverse space.  There are two ways to describe a location in the plane: as a
two-vector or a complex number.  While the formulation in terms of vectors is
mostly the simpler and more intuitive, the complex notation is much more
convenient when applying conformal transformations.  In the following the two
notations will be distinguished by writing vectors as bold face characters and
complex numbers as plain ones.

Vectors can be transformed into complex numbers and vice versa.  The two
components of two-vectors are identified with the real and imaginary part of
the corresponding complex number.  In the context of conformal field theories,
however, the real and imaginary part are not customarily seen as the two
degrees of freedom of a complex number.  Instead, the complex number and its
complex conjugate are viewed as independent degrees of freedom.  While this
approach is hard to grasp intuitively, it is practical when dealing with
conformally invariant expressions.  After performing the calculations, one
imposes again the physical condition that $\rho^*$ be the complex conjugate
of~$\rho$.  The conversion formulas to and from the complex notation are:
\dmblock{2}{eq:veccomp}{
&\rho=\mbf e_1\cdot\mbf\rho+i\,\mbf e_2\cdot\mbf\rho\,,
\qquad
\rho^*=\mbf e_1\cdot\mbf\rho-i\,\mbf e_2\cdot\mbf\rho\,,&
\qquad
\hbox{where}\quad\mbf e_1={1\choose0},\quad\mbf e_2={0\choose1}\,;
\cr
&\mbf\rho=\frac12{\rho+\rho^* \choose -i(\rho-\rho^*)}\,.&
}

The scalar product of vectors can easily be transcribed into the complex
notation by using~(\ref{eq:veccomp}). To distinguish the scalar product from
the ordinary product of complex numbers, we write it with parentheses.
\begin{equation}
\label{eq:compsprod}
\mbf\rho_1\cdot\mbf\rho_2=
\textstyle\frac12(\rho_1\rho_2^*+\rho_1^*\rho_2)
=:(\rho_1,\rho_2)
\end{equation}

In addition to points in the plane, we need differential operators, notably
$\mbf\nabla$ and~$\Delta$. The derivation of the first-order differential
operators will not be performed explicitly here. It is straightforward; one can
obtain it for instance by requiring the first-order term in a Taylor expansion
of an arbitrary function to be equal in both notations. The Laplacian is
obtained as the scalar product of two $\mbf\nabla$s. The results are:
\begin{equation}
\label{eq:veccompdiff}
\vcenter{\openup 3mm
\halign{\hfil $\displaystyle #$ \hfil\cr
\del:=\frac\partial{\partial\rho}=
\frac12\mbf e_1\cdot\mbf\nabla-\frac i2\mbf e_2\cdot\mbf\nabla
\qquad
\del^*:=\frac\partial{\partial\rho^*}=
\frac12\mbf e_1\cdot\mbf\nabla+\frac i2\,\mbf e_2\cdot\mbf\nabla
\cr
\mbf\nabla={\del+\del^* \choose i(\del-\del^*)}
\cr
\Delta=4\,\del\,\del^*\cr}}
\end{equation}
Note that the conversion formula is different from the one for the coordinates.
In rare cases, I will use nabla operators in the complex notation to indicate
gradients.  They are converted from the vector nablas with the formula for
coordinates (\ref{eq:veccomp}), and consequently $\nabla=2\,\del^*$
and~$\nabla^*=2\,\del$.

Finally, there are integrals. The Jacobi determinant of the transformation
(\ref{eq:veccomp}) is $-2i$, hence the integral measure in both coordinate
systems differs by a factor of 2. Knowing that, we can relate vectorial
integrals and delta functions to integrals over and delta functions of the
complex coordinates:
\begin{equation}
\vcenter{\openup 2mm \tabskip 0pt
\halign{\hfil$\displaystyle#$ & = $\displaystyle#$\hfil\cr
\int d^2\mbf\rho\,f(\mbf\rho)\equiv\int d^2\rho\,f(\rho,\rho^*) &
\frac12\int d\rho\int d\rho^*\,f(\rho,\rho^*) \cr
\delta^2(\mbf\rho)\equiv\delta^2(\rho)&
2\,\delta(\rho)\,\delta(\rho^*) \cr
}}
\end{equation}
Two-dimensional integrals or delta functions of complex coordinates are
customarily defined to include the constant factors and are therefore
equivalent to the vectorial ones, as shown in the equations.

\subsection{Conformal eigenfunctions}

This and the following section will introduce the few elements of conformal
field theories we will need, conformal eigenfunctions and $n$-point functions.
For a thorough introduction to conformal field theory,
see~\cite{dotconf,ginsconf,schelconf}.

Eigenstates of operators which are invariant under conformal transformations
are severely restricted in their form.  They are called conformal
eigenfunctions.  Conformal eigenfunctions of integral operators with two
arguments (like the BFKL kernel, see below) have the following form (except for
constant factors independent of the~$\rho$s):
\dmath2{
E^{(\nu,n)}(\rho_1,\rho_2;\rho_a)&=&
\left(\frac{\rho_{12}}{\rho_{1a}\rho_{2a}}\right)^{\frac{1+n}2+i\nu}
\left(\frac{\rho_{12}^*}{\rho_{1a}^*\rho_{2a}^*}\right)^{\frac{1-n}2+i\nu}\,.
&eq:Enun}
They are parametrised with a real number~$\nu$ and an integer~$n$.  
$\nu$~is
called conformal dimension, $n$~conformal spin.  The exponents in
Equation~\ref{eq:Enun} are called conformal weights and denoted with the
symbols $h$ and~$\bar h$.  So we have:
\dmath1{
h = \frac{1+n}2+i\nu
&\qquad&
\bar h = \frac{1-n}2+i\nu = 1-h^*
\crl{eq:nunh}
n = h-\bar h
&\qquad&
\nu = \frac1{2i} (h+\bar h-1)
}

It is well known from the theory of complex functions that an arbitrary power
of a complex base to a complex exponent is not well defined.  The value of the
power $x^y:=\exp(y\,\log x)$ depends on which branch of the multi-valued
logarithm one chooses.  So there is a certain ambiguity in the
definition~(\ref{eq:Enun}) which is never remarked upon.  The most natural
choice is to choose the principal value logarithm denoted by $\Log$, the
imaginary part of which lies in the range $(-\pi,\pi]$.

Having defined the expression~(\ref{eq:Enun}), one can work out how the
$E^{(\nu,n)}$ transform under exchange of the first two arguments.  This
transformation amounts to changing the sign of the base of the power, ie the
argument of the logarithm.  The exact result is:
\dmath2{
E^{(\nu,n)}(\rho_2,\rho_1;\rho_a)&=&E^{(\nu,n)}(\rho_1,\rho_2;\rho_a)\cdot
\cases{
(-1)^n & for $r \in\mathbb C\setminus\mathbb R\,$,\cr
-\exp(-2\pi\nu) & for $r\in\mathbb R_{\ge0}\,$,\cr
-\exp(2\pi\nu) & for $r\in\mathbb R_{<0}\,$,\cr
}
&eq:Enunsymm}
where $r=\rho_{12}/(\rho_{1a}\rho_{2a})$.  For real $r$, $\Log r$ and $\Log
r^*$ transform in the same way ($\pi i$ is added to or subtracted from both).
This has the consequence that not the $\frac n2$ part but the rest of the
exponents is retained and determines the transformation property.

It can be safely assumed that the transformation law intended by the authors of
the conformal eigenfunctions was the first line of~(\ref{eq:Enunsymm}) only.%
\footnote{The original work~\cite{bfklconf,liprev} does not remark on the 
symmetry of the $E^{(\nu,n)}$, nor on the ambiguity of complex exponentials.}
Reluctantly in line with physicists' usual nonchalance towards mathematical
rigour, I will assume in the following that the eigenfunctions transform with a
factor $(-1)^n$, ie are symmetric for even $n$ and antisymmetric for odd~$n$.
A limited justification for this is that the deviation from that rule is
restricted to a set of measure zero, which can never be measured experimentally
and will furthermore be irrelevant under an integral.  However, this remains an
abuse of mathematics.  The desired transformation law could only be obtained be
choosing two different branches of the logarithm for the non-conjugated power
and the conjugated one, so that $\log^{(1)}(-r) -
\log^{(1)}r = -(\log^{(2)}(-r^*) - \log^{(2)}r^*)$ for all~$r$.  This is 
obviously mathematical nonsense.

In the later chapters, we will restrict ourselves to the ground state with zero
conformal spin.  Here is its explicit form, which depends only on the moduli of
vectors, not their direction:
\dmath0{
E^{(\nu,0)}(\rho_1,\rho_2;\rho_a)&=&
\left(\frac{|\rho_{12}|}{|\rho_{1a}||\rho_{2a}|}\right)^{1+2i\nu}\,.
&eq:Enu0}

The third coordinate argument of the eigenfunctions represents an external
coordinate of the corresponding fields.  We will often suppress it in the
calculations, but it is always implied.

\subsection{Conformal $n$-point functions}

In a conformal field theory there are special fields $\Phi_{h,\bar h}$ called
``primary fields''.  They are distinguished by satisfying the following
relation under conformal transformations $\rho\to\rho'$:%
\footnote{Most authors formulate (\ref{eq:primary}) as a transformation law
rather than an equation.  But since for physical reasons
$\Phi'(\rho',\rho^{\prime\,*})=\Phi(\rho,\rho^*)$ is required, the above
relation has to hold as an equation for primary fields.}
\dmath2{
\Phi_{h,\bar h}(\rho,\rho^*)&=&
\left(\frac{\del\rho'}{\del\rho}\right)^h
\left(\frac{\del\rho^{\prime\,*}}{\del\rho^*}\right)^{\bar h}
\Phi_{h,\bar h}(\rho',\rho^{\prime\,*})\,.
&eq:primary}
It is easily checked that the conformal eigenfunctions (\ref{eq:Enun}) fulfil
this requirement.  In this case, $h$ and $\bar h$ are related via
(\ref{eq:nunh}), even though in general they can be independent variables.

As a consequence of this transformation property, $n$-point functions of
primary fields are constrained in their form.  The two- and three-point
functions are fixed up to a constant:
\dmath2{
\big\langle \Ph1\,\Ph2\big\rangle&&{}=
\delta_{h_1,h_2}\,\delta_{\bar h_1,\bar h_2}
\frac{c_{h_1,\bar h_1}}{\rho_{12}^{2h_1}\,\rho_{12}^{*\;2\bar h_1}}
&eq:2point\cr
\big\langle \Ph1\,\Ph2\,\Ph3\big\rangle&&{}={}
&eq:3point\cr\crd
{}=c_{h_1,h_2,h_3,\bar h_1,\bar h_2,\bar h_3}\,
\big[\rhotre123\,\rhotre231\,&&\rhotre132\;
\crhotre123\,\crhotre231\,\crhotre132\big]^{-1}
}
The conformal eigenfunctions introduced in the previous section can themselves
be understood as conformal three-point functions.  In the context of the BFKL
equation, two of the participating fields can be interpreted as reggeised
gluons, one as a pomeron.  The conformal eigenfunctions, interpreted as
three-point functions, describe the pomeron in terms of reggeised gluons.  The
explicit formula is the following, where the correlation between $\nu$, $n$,
$h$ and $\bar h$ is as in (\ref{eq:nunh}):
\dmath2{
E^{(\nu,n)}(\rho_1,\rho_2;\rho_a)
&=&
\big\langle \Phi_{0,0}(\rho_1,\rho_1^*)\,\Phi_{0,0}(\rho_2,\rho_2^*)\,
\Phi_{h,\bar h}(\rho_a,\rho_a^*)\big\rangle\,.
&eq:E3pt}

The four-point function has some more freedom.  It is fixed only up to a
function which may depend on an anharmonic ratio of the four coordinates,
though not on the individual coordinates.
\dmath2{
\big\langle \Ph1\,\Ph2\,&&\Ph3\,\Ph4 \big\rangle={}\crd
&&{}=\Psi(x,x^*)\prod_{i<j}\left(\rho_{ij}^{-h_i-h_j+\frac13\sum_k h_k}
\rho_{ij}^{*\;-\bar h_i-\bar h_j+\frac13\sum_k \bar h_k}\right)\,,
&eq:4point\cr
&&\qquad x=\crossrat 12341324}

At this point some words about anharmonic ratios are helpful.  In principle,
one can construct six different anharmonic ratios from four coordinates: Both
numerator and denominator can contain three different pairs of differences of
coordinates.  If the fraction is not to be constant $=1$, numerator and
denominator must not be equal and six possibilities remain.  These are pairwise
inverses of each other.  They all can be expressed in terms of one anharmonic
ratio, eg~$x$.  Here are three of them (the other three being their inverses):
\dmath0{
x&=&\crossrat 12341324\,,
\qquad\qquad
1-x=\crossrat 14231324\,,
\qquad\qquad
\frac x{1-x}=\crossrat 12341423\,.
&eq:cross4}
For example, here is the proof for the last identity:
\dmath2{
\crossrat 12341423&=&\left(\crossrat 14231234\right)^{-1}=
\left(\frac{\rho_1\rho_2-\rho_4\rho_2-\rho_1\rho_3+\rho_4\rho_3}
{\rho_{12}\rho_{34}}\right)^{-1}={}\cr
&=&
\left(\frac1x+
\frac{-\rho_1\rho_2+\rho_3\rho_2+\rho_1\rho_4-\rho_3\rho_4
\;\;+\rho_1\rho_2-\rho_4\rho_2-\rho_1\rho_3+\rho_4\rho_3}
{\rho_{12}\rho_{34}}\right)^{-1}={}\cr
&=&
\left(\frac1x+
\frac{\rho_3\rho_2+\rho_1\rho_4-\rho_4\rho_2-\rho_1\rho_3}
{\rho_{12}\rho_{34}}\right)^{-1}=
\left(\frac1x-1\right)^{-1}=
\frac x{1-x}\,.
}


\subsection{The BFKL equation}
\label{sec:BFKLeq}

The BFKL equation describes the evolution in rapidity of the system of two
reggeised gluons which makes up pomeron exchange in QCD.  It is an integral
equation of Bethe-Salpeter type.  The unknown in the equation is a four-point
function of reggeised gluons, or equivalently the pomeron propagator.  The BFKL
equation in momentum space reads as follows:
\dmath2{
\omega\,\phi_\omega(\mbf k_1,\mbf k_2;\mbf k_1',\mbf k_2')&=&
\frac{\delta^2(\mbf k_1-\mbf k_1')}{\mbf k_1^2}
\frac{\delta^2(\mbf k_2-\mbf k_2')}{\mbf k_2^2}+{}
\crl{eq:BFKLeq}
&&{}+\int\frac{d^2\mbf k_1''\,d^2\mbf k_2''}{(2\pi)^3}
\frac1{\mbf k_1''^2}\frac1{\mbf k_2''^2}
K_{BFKL}\big(\mbf k_1'',\mbf k_2'';\mbf k_1',\mbf k_2'\big)
\,\phi_\omega\big(\mbf k_1,\mbf k_2;\mbf k_1'',\mbf k_2''\big)\,.
}
It takes the form of an eigenvalue equation for the amplitude $\phi_\omega$
with the eigenvalue $\omega$, the complex angular momentum.  The BFKL kernel
has the form:
\dmath2{
K_{BFKL}\big(\mbf k_1'',\mbf k_2'';\mbf k_1',\mbf k_2'\big)&=&
-N_c g^2\delta^2(\mbf k_1'+\mbf k_2'-\mbf k_1''+\mbf k_2'')
\left[(\mbf k_1'+\mbf k_2')^2
- \frac{\mbf k_1'^2\,\mbf k_2''^2}{(\mbf k_1'-\mbf k_1'')^2}
- \frac{\mbf k_1''^2\,\mbf k_2'^2}{(\mbf k_1'-\mbf k_1'')^2}\right]
\cr
&&{}+(2\pi)^3\,\mbf k_1''^2\,\mbf k_2''^2\,
\delta^2(\mbf k_1'-\mbf k_1'')\,\delta^2(\mbf k_2'-\mbf k_2'')\,
\big[\beta(\mbf k_1')+\beta(\mbf k_2')\big]\,,
&eq:BFKLkernel\cr
\beta(\mbf k)&=&-\frac{N_cg^2}2\int\frac{d^2\mbf l}{(2\pi)^3}
\frac{\mbf k^2}{\mbf l^2(\mbf l-\mbf k)^2}\,.
}
$\alpha(\mbf k^2)=1+\beta(\mbf k)$ is the Regge trajectory of the gluon.  For
that reason $\beta$ is also called, somewhat inaccurately, the trajectory
function.

The configuration-space representation of the BFKL equation is rather more
complicated.  The convolution of the BFKL kernel with a pomeron wave function
can be expressed in terms of three of the five fundamental functions we will
discuss later (see Section~\ref{sec:abcst_m} and Equation~\ref{eq:G_BFKL}).

\begin{figure}[h]
\begin{center}
\includegraphics[height=6cm]{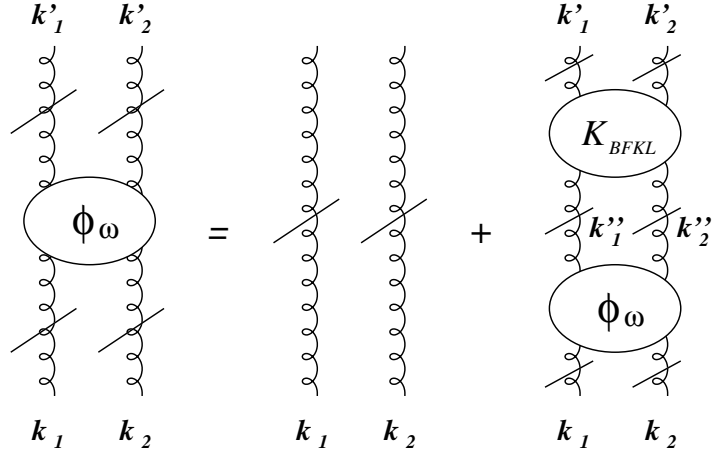}
\end{center}
\caption{Graphical representation of the BFKL equation.}
\label{fig:BFKLeq}
\end{figure}

The BFKL equation can also be presented in a graphical way, see
Figure~\ref{fig:BFKLeq}.  The vertical curly lines with slanted lines through
them represent reggeised gluons.  The BFKL kernel in the second term on the
right represents a superposition of exchanges of ordinary QCD gluons between
the two reggeised ones.  If one imagines solving the BFKL equation through
iteration, it becomes obvious that this adds up to a superposition of ladders
with an arbitrary number of rungs --- the pomeron.  While the $s$-channel gluon
which forms the rung is an ordinary QCD object, the vertices which couple it to
the reggeised gluons are not.  The BFKL kernel contains several terms for each
vertex, which has the effect that Feynman graphs in which rungs cross are also
included in the final superposition of ladders.

The solution of the BFKL equation was initially found only in the forward
direction, ie for~$q=0$.  Only when it was realised that the BFKL kernel is
invariant under global conformal transformations in position space, the
non-forward solution was found.  The consequence of this state of affairs is
that the conformal eigenfunctions (\ref{eq:Enun}) are eigenfunctions of the
kernel:
\dmath2{
\int d^2\rho_1'\,d^2\rho_2'\; K_{BFKL}(\rho_1,\rho_2;\rho_1',\rho_2')
\;E^{(\nu,n)}(\rho_1',\rho_2')&=&
\chi(\nu,n)\;E^{(\nu,n)}(\rho_1,\rho_2)\,,
&eq:BFKLeigeneq}
with the eigenvalue
\dmath2{
\chi(\nu,n)&=&\frac{N_c\alpha_s}{\pi}
\left[2\psi(1)-\psi\!\left(\frac{1+|n|}2+i\nu\right)
-\psi\!\left(\frac{1+|n|}2-i\nu\right)\right]\,,
&eq:bfkleigen}
where $\psi$ is the logarithmic derivative of the $\Gamma$ function.

The solution of the BFKL equation can be written in terms of the conformal
eigenfunctions:
\dmath2{
\phi_\omega(\rho_1,\rho_2;\rho_1',\rho_2')&=&
\sum\limits_{n=-\infty}^{\infty} \int \frac{d\nu}{2\pi}
\frac{16\nu^2+4n^2}{[4\nu^2+(n+1)^2]\,[4\nu^2+(n-1)^2]}\cdot{}
\crl{eq:BFKLsol}&&{}\cdot
\frac1{\omega-\chi(\nu,n)}
\int d^2\rho_0\;E^{(\nu,n)}(\rho_1,\rho_2;\rho_0)\,
E^{(\nu,n)\,*}(\rho_1',\rho_2';\rho_0)\,.
}
This is in effect the propagator of the pomeron, more precisely of two
reggeised gluons in a colour singlet state.

When pomeron exchange occurs in a scattering process mediated by photons, the
pomeron couples to a quark loop into which the photon fluctuates.  It can be
shown that the impact factor (which describes the coupling) of any number of
reggeised gluons in a colour singlet state coupling to a quark loop can be
written as a superposition of two-gluon state impact factors.\footnote{Strictly
speaking, one has to use a ``cut'' quark loop, in which quark lines are forced
to be on shell.  See for instance \cite{carlophd} or~\cite{be}.}  Therefore it
is sufficient to consider the coupling of two reggeised gluons to the quark
loop.

The amplitude of the pomeron can be obtained by convoluting an impact factor
with the pomeron propagator:
\dmath2{
D_2(\mbf k_1,\mbf k_2)&=&\int d^2\mbf k_1'\,d^2\mbf k_2'\;
\phi^0(\mbf k_1',\mbf k_2')\,
\phi_\omega(\mbf k_1,\mbf k_2;\mbf k_1',\mbf k_2')\,.
&eq:D2}
The impact factors $\phi^0$ are usually symmetric under exchange of their
arguments.  From this, Equation~\ref{eq:BFKLsol} and the properties of the
conformal eigenfunctions, it follows that $D_2$ is also symmetric.
Furthermore, it vanishes when its configuration-space arguments are set to be
equal, since the conformal eigenfunctions then vanish.  Besides, $D_2$ vanishes
when one momentum argument is set to zero or one configuration-space coordinate
is integrated out.  To sum up all relations in momentum and position space:
\dmath1{
D_2(\mbf k_1,\mbf k_2)=D_2(\mbf k_2,\mbf k_1)
&\qquad&
D_2(\mbf\rho_1,\mbf\rho_2)=D_2(\mbf\rho_2,\mbf\rho_1)
\cr
\int d^2\mbf l\,D_2(\mbf l,\mbf k-\mbf l)=0
&\qquad& D_2(\mbf\rho,\mbf\rho)=0
&eq:D2prop\cr
D_2(\mbf k,0)=0=D_2(0,\mbf k)
&\qquad& 
\int d^2\mbf\rho_2\,D_2(\mbf\rho_1,\mbf\rho_2)=0
=\int d^2\mbf\rho_1\,D_2(\mbf\rho_1,\mbf\rho_2)
}

Strictly speaking, $D_2$ is called the ``amputated'' amplitude because QCD
gluon propagators have to be added to attach it to anything.  The corresponding
amplitude with gluon propagators is denoted by~$\Phi_2$.  There is a third type
of amplitude called $C_2$ which differs from $D_2$ in that it does not contain
the {\em Reggeon} propagator $(\omega - \beta(\mbf k_1) - \beta(\mbf
k_2))^{-1}$, which is implicit in the BFKL equation.

\subsection{Useful formulas}
\label{sec:confformulas}

This section contains the derivation of some identities which will be used
frequently later on.  The first of these is the Fourier transformation of the
gluon propagator in (two-dimensional) transverse space.  The two-dimensional
gluon propagator has to be regularised with a gluon mass in order to perform
the Fourier integral:
\dmath1{
\frac1{(2\pi)^2}\int d^2\mbf k \frac1{\mbf k^2+m^2} e^{i\mbf k\mbf\rho}
	&=&\frac1{(2\pi)^2}\int\limits_0^\infty d k\;\frac k{k^2+m^2}
	  \int\limits_0^{2\pi}d\varphi\; e^{ik|\mbf\rho|\cos\varphi}
\cr
&=&\frac1{(2\pi)^2}\int\limits_0^\infty d k\;\frac k{k^2+m^2}\;
	2\pi\,J_0(k |\mbf\rho|) 
	= \frac1{2\pi}\,K_0(m|\mbf\rho|)\,.
&eq:prop2ft}
This integration is performed in somewhat more detail at the end of
Appendix~\ref{sec:intlight}. The integration formula by which the Bessel
function $K_0$ is obtained is cited there, in Equation~\ref{eq:bessJtoK}.%
\footnote{An observant reader will have noticed that the result is finite only
for $\mbf\rho\ne0$, ie it is still (logarithmically) ultraviolet divergent.  
This requires another regularisation if the propagator is integrated over.
We will perform it later.}

Since the argument of $K_0$ contains the gluon mass $m$ which will be set
to zero in physical expressions, we are interested in the expansion of
$K_0$ for small arguments:
\begin{equation}
K_0(z)=\ln 2 + \psi(1) - \ln z + \mathcal O(z^2)\,.
\label{eq:K0expand}
\end{equation}
Here $\psi$ is the logarithmic derivative of the Gamma function and
$\psi(1)=-\gamma\approx -0.5772$ is the negative of Euler's constant.  Since
all mass-regularised expressions imply the physical limit of $m\rightarrow0$,
$K_0(m|\mbf\rho|)$ will often be substituted by $[\ln2+\psi(1)-\ln
m-\ln|\mbf\rho|]$ and vice versa.

Many of the expression we will handle contain Laplace operators, some together
with the regularised configuration-space propagator we just calculated.
Therefore the Laplacian of $\ln|\mbf\rho|$ will be needed occasionally.  In
fact the logarithm is the Green's function of the Poisson equation in two
dimensions:
\begin{equation}
\Delta\,\ln|\mbf\rho|=2\pi\,\delta^2(\mbf\rho)\,.
\label{eq:greenpoiss2}
\end{equation}
To prove this, we look at the function $\ln(|\mbf\rho|+\epsilon)$.  We
calculate
\begin{equation}
\Delta\ln(|\mbf\rho|+\epsilon)=\nabla\cdot\frac{\mbf e_r}{|\mbf\rho|+\epsilon}
=\frac1{|\mbf\rho|\,(|\mbf\rho|+\epsilon)}-\frac1{(|\mbf\rho|+\epsilon)^2}
=\frac\epsilon{|\mbf\rho|\,(|\mbf\rho|+\epsilon)^2}\,,
\end{equation}
where $\mbf e_r$ is the unit vector in radial direction.  The third expression
was obtained by product differentiation and by using $\nabla\cdot\mbf
e_r=\frac1{|\mbf\rho|}$.  In the limit $\epsilon\rightarrow0$, the final
expression diverges if $\mbf\rho=0$ and approaches zero otherwise.  These are
the values the Dirac delta function takes.  To obtain the multiplicative
constant in Equation~\ref{eq:greenpoiss2}, we integrate over a disc with
radius~$R$ around the origin:
$$
\int\limits_{\rm disc}d^2\mbf\rho\,\Delta\ln(|\mbf\rho|+\epsilon)
=2\pi\int\limits_0^Rdr\,r\,\Delta\ln(r+\epsilon)
=2\pi\int\limits_0^Rdr\frac\epsilon{(r+\epsilon)^2}
=2\pi\left(-\frac\epsilon{R+\epsilon}+1\right)
\putunder{$-\!\!\!\longrightarrow$}{$\epsilon\!\to\!0$}2\pi\,.
$$
This accounts for the constant $2\pi$ in~(\ref{eq:greenpoiss2}).

Equation~\ref{eq:greenpoiss2} has an important implication for the derivatives
of complex coordinates.  To see it, we express the Laplacian in complex
derivatives (see~(\ref{eq:veccompdiff})).
$$
2\pi\,\delta^2(\mbf\rho)=\Delta\ln|\mbf\rho|=
4\,\del\del^*\ln|\rho|=4\,\del\del^*\frac12\ln(\rho\rho^*)=
2\,\del\frac1{\rho^*}
$$
By performing the derivation with respect to $\rho$ first, we obtain
$2\,\del^*{1\over\rho}$. Hence we have:
\begin{equation}
\label{eq:del1/rho}
\del\frac1{\rho^*}=\del^*\frac1{\rho}=\pi\,\delta^2(\rho)\,.
\end{equation}
Put into words, this means that $\rho$ and $\rho^*$ are not independent
variables at the origin.  Note that this relation holds only if the total power
of the complex conjugated variable in the expression to be differentiated
is~$-1$.  It is illegal to split off a factor $1/\rho^*$ from an expression and
use (\ref{eq:del1/rho}) in the product rule of differentiation.

\section{The five component functions in momentum space}
\label{sec:abcst_m}

Having presented a selection of basic formulas, we will now start to
investigate expressions closely related to the problems we intend to discuss.
The main aim of this and the following chapter is to investigate a vertex
describing the transition from two to six reggeised gluons.  It was shown
in~\cite{be} that such vertices can be expressed in terms of five fundamental
functions.  The starting point of these vertices is always a two-gluon
amplitude.  Therefore the amplitude $D_2$ is already included in the definition
of the five functions.  Their momentum-space representation is the following:
\dmath{2.5}{
a(\mbf k_1,\mbf k_2,\mbf k_3)&=&\int\frac{d^2\mbf l}{(2\pi)^3}
	\frac{\mbf k_1^2}{(\mbf l-\mbf k_2)^2(\mbf l-\mbf k_1-\mbf k_2)^2}
	D_2(\mbf l,\mbf k_1+\mbf k_2+\mbf k_3-\mbf l)
\cr
b(\mbf k_1,\mbf k_2)&=&\int\frac{d^2\mbf l}{(2\pi)^3}
	\frac{\mbf k_1^2}{\mbf l^2(\mbf l-\mbf k_1)^2}
	D_2(\mbf l,\mbf k_1+\mbf k_2-\mbf l)
\cr
c(\mbf k)&=&\int\frac{d^2\mbf l}{(2\pi)^3}
	\frac{\mbf k^2}{\mbf l^2(\mbf l-\mbf k)^2}D_2(\mbf l,\mbf k-\mbf l)
&eq:abcst_m\cr
s(\mbf k_1,\mbf k_2,\mbf k_3)&=&\int\frac{d^2\mbf l}{(2\pi)^3}
	\frac{\mbf k_1^2}{\mbf l^2(\mbf l-\mbf k_1)^2}
	D_2(\mbf k_1+\mbf k_2,\mbf k_3)
\cr
t(\mbf k_1,\mbf k_2)&=&\int\frac{d^2\mbf l}{(2\pi)^3}
	\frac{\mbf k_1^2}{\mbf l^2(\mbf l-\mbf k_1)^2}
	D_2(\mbf k_1,\mbf k_2)
}
The function~$a$ is symmetric in its last two arguments.  One can see this by
substituting $\mbf l\rightarrow\mbf k_1+\mbf k_2+\mbf k_3-\mbf l$ and using the
fact that $D_2$ is symmetric in its two arguments.  All the functions vanish
when their first argument vanishes.  $s$ and $t$ also vanish when their last
argument is zero, because the $D_2$ amplitude then vanishes according to the
third formula of~(\ref{eq:D2prop}).

There are some relations between the functions $a$, $b$ and~$c$ and between $s$
and $t$ and the trajectory function~$\beta$.
\dmblock2{eq:abcstrel_m}{
b(\mbf k_1,\mbf k_2)&=&a(\mbf k_1,0,\mbf k_2)=a(\mbf k_1,\mbf k_2,0)
\cr
c(\mbf k)&=&b(\mbf k,0)=a(\mbf k,0,0)
\cr
s(\mbf k_1,\mbf k_2,\mbf k_3)&=&
	-\frac2{N_cg^2}\beta(\mbf k_1)D_2(\mbf k_1+\mbf k_2,\mbf k_3)
\cr
t(\mbf k_1,\mbf k_2)&=&s(\mbf k_1,0,\mbf k_2)=
	-\frac2{N_cg^2}\beta(\mbf k_1)D_2(\mbf k_1,\mbf k_2)
}
The trajectory function~$\beta$, which was already introduced in
Section~\ref{sec:BFKLeq}, has the following form in momentum space:
\begin{equation}
\label{eq:beta_m}
\beta(\mbf k)=-\frac{N_cg^2}2\int\frac{d^2\mbf l}{(2\pi)^3}
\frac{\mbf k^2}{\mbf l^2(\mbf l-\mbf k)^2}\,.
\end{equation}
This integral can be performed after applying a regularisation scheme.  The two
regularisation schemes which have been used for this expression are mass
regularisation and dimensional regularisation using the dimension
$d=2+\epsilon$.  The derivations are lengthy and will not be performed here.
The results are as follows:
\dmath2{
\hbox{Mass regularisation:}\quad
\beta(\mbf k)&=&\lim_{m\to0}-\frac{N_c g^2}2\pi\ln\frac{\mbf k^2}{m^2}
&eq:beta_m_masreg\cr
\hbox{Dimensional regularisation:}\quad
\beta(\mbf k)&=&
\lim_{\epsilon\to0}-2\pi N_c g^2(\mbf k^2)^{\frac\epsilon2}\frac1\epsilon
&eq:beta_m_dimreg\cr
}

\begin{figure}
\moveright 10mm \hbox{\input{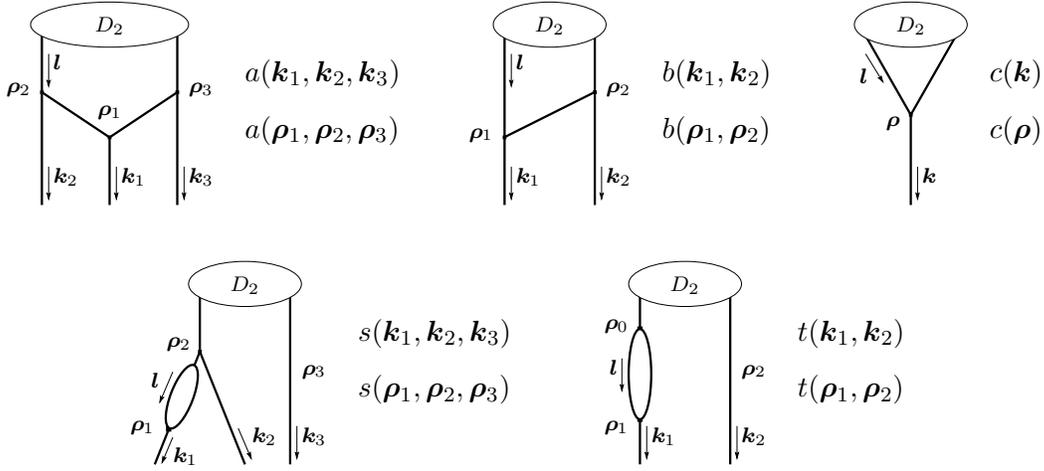}}
\caption{Diagrams representing the five functions $a$, $b$, $c$, $s$ and $t$
		with location of configuration space coordinates and momenta.
		Coordinates or momenta not appearing as arguments are integration 
		variables.}
\label{fig:abcst}
\end{figure}

The functions $a$, $b$ and $c$ represent real corrections, that is terms
composed of them describe the production of on-shell gluons.  $s$ and $t$
describe virtual corrections, self-energy corrections to the reggeised gluon
propagator.

Figure~\ref{fig:abcst} shows a diagrammatic representation of the five
functions.  With a view to the following sections, it shows both the momenta of
the gluon lines and the coordinates of the vertices.  The momenta of the
unlabelled lines are fixed by momentum conservation at the vertices.  The loop
momentum $\mbf l$ and the coordinate $\mbf\rho_0$ in the function~$t$ are
integration variables and therefore do not appear as arguments.  Note that the
middle momentum or spatial coordinate in the diagram for $a$ is conventionally
taken as the first argument of the function~$a$.

The formulas~\ref{eq:abcst_m} can be obtained from the diagrams with a simple
prescription~\cite{carlophd,be}:
\begin{itemize}
\item Find the lowest vertex in the diagram.
\item Write down propagators for the lines attached to it from above.
\item Put the square of the sum of the momenta attached from below in the 
		numerator.
\item Multiply with the amplitude $D_2$ with suitable arguments.
\item Integrate over the loop momentum $\mbf l$ and divide by $(2\pi)^3$.
\end{itemize}
Even though this is the simplest method, it is unfortunately somewhat
asymmetric: Not all vertices and lines are treated equally.  This has the
consequence that when attaching a wave function to the lower legs, the full
amplitude $\Phi_2$ has to be used even though the one attached to the upper
legs is an amputated amplitude,~$D_2$.

There is a more symmetric prescription for obtaining integrals from such
graphs~\cite{blw}.  Here all the lines get propagators and all vertices a
square of the total momentum attached from above or below.  As a consequence,
amputated amplitudes $D_2$ have to be attached both from above and below.
However, this does not lead to the same functions $a$, $b$, $c$, $s$ and $t$
which are defined in the asymmetric fashion described above.

\section{The five component functions in configuration space}
\label{sec:abcst_c}

\subsection{The real corrections $a$, $b$ and $c$}
\label{sec:abc_c}

We will be interested in the properties of gluon vertices under the conformal
transformations in transverse position space.  For that purpose we need to
express them in position space.  So we have to Fourier transform the five
functions $a$, $b$, $c$, $s$ and~$t$ of which they are composed.  Before we
start with the function~$a$, a remark concerning the wave functions is
required.  We will see that the configuration space functions are expressed
more conveniently in terms of the full amplitude $\Phi_2$ instead of the
amputated amplitude~$D_2$.  The two are related in the following way:
\dmath2{
\Phi_2(\mbf k_1,\mbf k_2)&=&
\frac1{\mbf k_1^2}\frac1{\mbf k_2^2}D_2(\mbf k_1,\mbf k_2)
&eq:Phi2m_D2m\cr
\Rightarrow\quad
D_2(\mbf k_1,\mbf k_2)&=&\mbf k_1^2\,\mbf k_2^2\;\Phi_2(\mbf k_1,\mbf k_2)
=\mbf k_1^2\,\mbf k_2^2\int d^2\mbf\rho_1\,d^2\mbf\rho_2\,
\Phi_2(\mbf\rho_1,\mbf\rho_2)\,
e^{-i(\mbf k_1\mbf\rho_1+\mbf k_2\mbf\rho_2)}
\cr
&=&\int d^2\mbf\rho_1\,d^2\mbf\rho_2\,
\Phi_2(\mbf\rho_1,\mbf\rho_2)\,
(-\Delta_1)(-\Delta_2)\,e^{-i(\mbf k_1\mbf\rho_1+\mbf k_2\mbf\rho_2)}
\cr
&=&\int d^2\mbf\rho_1\,d^2\mbf\rho_2\,
\big(\Delta_1\,\Delta_2\,\Phi_2(\mbf\rho_1,\mbf\rho_2)\big)\,
e^{-i(\mbf k_1\mbf\rho_1+\mbf k_2\mbf\rho_2)}
&eq:D2m_Phi2c}
The final result was arrived at by two-fold partial integration in each of the
integration variables.  Put into words, it means that the configuration space
$D_2$ amplitude is the double Laplacian of the configuration space $\Phi_2$.
We will make frequent use of this identity since the Laplacians will enable us
to perform the Fourier integrations easily and to derive the equivalents of the
relations (\ref{eq:abcstrel_m}) in configuration space.

Let us now turn to the Fourier transformation of the function~$a$. As a 
first step, we substitute the expression~(\ref{eq:D2m_Phi2c}) for $D_2$:
\dmath2{
a(\mbf\rho_1,\mbf\rho_2,\mbf\rho_3)
&=&\frac1{(2\pi)^6}\int d^2\mbf k_1\,d^2\mbf k_2\,d^2\mbf k_3\;
	a(\mbf k_1,\mbf k_2,\mbf k_3)\;
	e^{i(\mbf k_1\mbf\rho_1+\mbf k_2\mbf\rho_2+\mbf k_3\mbf\rho_3)}
\cr
&=&\frac1{(2\pi)^9}\int d^2\mbf k_1\,d^2\mbf k_2\,d^2\mbf k_3\,d^2\mbf l\;
	\int d^2\mbf\rho_1'\,d^2\mbf\rho_2'\;
	\frac{\mbf k_1^2}{(\mbf l-\mbf k_2)^2\,(\mbf l-\mbf k_1-\mbf k_2)^2}\cdot{}
\crd&&\qquad\qquad {}\cdot
	\big(\Delta_{1'}\,\Delta_{2'}\,\Phi_2(\mbf\rho_1',\mbf\rho_2')\big)\;
	e^{i(\mbf k_1\mbf\rho_1+\mbf k_2\mbf\rho_2+\mbf k_3\mbf\rho_3
	 -\mbf l\mbf\rho_1'-(\mbf k_1+\mbf k_2+\mbf k_3-\mbf l)\mbf\rho_2')}
\,.}
Now we express the numerator $\mbf k_1^2$ as a Laplacian which we can pull out
of the integral since the variable $\mbf\rho_1$ appears nowhere but in the
Fourier exponential.  Besides we can perform the integration over $\mbf k_3$,
which gives a delta function, and then integrate over~$\mbf\rho_2'$.
\dmath{3}{
\ldots&=&\frac1{(2\pi)^9}\,(-\Delta_1)
	\int d^2\mbf k_1\,d^2\mbf k_2\,d^2\mbf k_3\,d^2\mbf l\;
	\int d^2\mbf\rho_1'\,d^2\mbf\rho_2'\;
	\frac1{(\mbf l-\mbf k_2)^2\,(\mbf l-\mbf k_1-\mbf k_2)^2}\cdot{}
\crd&&\qquad\qquad {}\cdot
	\big(\Delta_{1'}\,\Delta_{2'}\,\Phi_2(\mbf\rho_1',\mbf\rho_2')\big)\;
	e^{i(\mbf k_1\mbf\rho_1+\mbf k_2\mbf\rho_2+\mbf k_3\mbf\rho_3
	 -\mbf l\mbf\rho_1'-(\mbf k_1+\mbf k_2+\mbf k_3-\mbf l)\mbf\rho_2')}
\cr&=&\frac{-\Delta_1}{(2\pi)^7}
	\int d^2\mbf k_1\,d^2\mbf k_2\,d^2\mbf l\;
	\int d^2\mbf\rho_1'\;
	\frac1{(\mbf l-\mbf k_2)^2\,(\mbf l-\mbf k_1-\mbf k_2)^2}\cdot{}
\crd&&\qquad\qquad {}\cdot
	\big(\Delta_{1'}\,\Delta_3\,\Phi_2(\mbf\rho_1',\mbf\rho_3)\big)\;
	e^{i(\mbf k_1\mbf\rho_1+\mbf k_2\mbf\rho_2
	 -\mbf l\mbf\rho_1'-(\mbf k_1+\mbf k_2-\mbf l)\mbf\rho_3)}
\cr&=&\frac{-\Delta_1}{(2\pi)^7}
	\int d^2\mbf k_1\,d^2\mbf k_2\,d^2\mbf l\;
	\int d^2\mbf\rho_1'\;
	\frac1{(\mbf l-\mbf k_2)^2\,(\mbf l-\mbf k_1-\mbf k_2)^2}\cdot{}
\crd&&\qquad\qquad {}\cdot
	\big(\Delta_{1'}\,\Delta_3\,\Phi_2(\mbf\rho_1',\mbf\rho_3)\big)\;
	e^{i(\mbf k_1\mbf\rho_{13}+\mbf k_2\mbf\rho_{23}
		+\mbf l\mbf\rho_{31'})}
}
After shifting the integration in $\mbf k_1$ and~$\mbf k_2$ ($\mbf{\hat
k}_1:=\mbf k_1+\mbf k_2-\mbf l,\ \mbf{\hat k}_2:=\mbf k_2-\mbf l$), the
momentum-space integral factorises into three separate integrals.  Two of them
are Fourier transformations of the two-dimensional gluon propagator which we
regularise with a gluon mass (\ref{eq:prop2ft}), the third gives a delta
function.
\dmath{3}{
\ldots&=&\frac{-\Delta_1}{(2\pi)^7}
\left(\int d^2\mbf{\hat k}_1\frac1{\mbf{\hat k}_1^2}
	e^{i\mbf{\hat k}_1\mbf\rho_{13}}\right)
\left(\int d^2\mbf{\hat k}_2\frac1{\mbf{\hat k}_2^2}
	e^{i\mbf{\hat k}_2\mbf\rho_{21}}\right)
\left(\int d^2\mbf l \int d^2\mbf\rho_1'\;
	\big(\Delta_{1'}\,\Delta_3\,\Phi_2(\mbf\rho_1',\mbf\rho_3)\big)\;
	e^{i\mbf l\mbf\rho_{21'}}\right)
\cr&=&
\lim_{m\to0}-\frac{\Delta_1}{(2\pi)^3}
K_0(m|\mbf\rho_{12}|)\,K_0(m|\mbf\rho_{13}|)\,
\Delta_2\,\Delta_3\,\Phi_2(\mbf\rho_2,\mbf\rho_3)
\cr&=&
\lim_{m\to0}-\frac{\Delta_1}{(2\pi)^3}
[\ln2+\psi(1)-\ln m-\ln|\mbf\rho_{12}|]\,
[\ln2+\psi(1)-\ln m-\ln|\mbf\rho_{13}|]\,
\Delta_2\,\Delta_3\,\Phi_2(\mbf\rho_2,\mbf\rho_3)
}
This is the configuration space representation of the function~$a$. Which of 
the last two expression to use is a matter of taste and convenience.

The transformation of the function~$b$ works quite similarly and will therefore
not be shown in such detail.  The steps are quite analogous up to the point
where we shifted the momentum integration variables.  In the case of the
$b$~function the integral does not factorise completely because there is one
Fourier integration less.
\dmath3{
b(\mbf\rho_1,\mbf\rho_2)&=&\ldots=
\frac{-\Delta_1}{(2\pi)^5}
\int d^2\mbf k_1\,d^2\mbf l \int d^2\mbf\rho_1'
\frac1{\mbf l^2\,(\mbf l-\mbf k_1)^2}
\big(\Delta_{1'}\,\Delta_2\,\Phi_2(\mbf\rho_1',\mbf\rho_2)\big)\;
e^{i(\mbf k_1\mbf\rho_{12}+\mbf l\mbf\rho_{21'})}
\cr&=&
\frac{-\Delta_1}{(2\pi)^5}\int d^2\mbf\rho_1'
\left(\int d^2\mbf{\hat k}_1\frac1{\mbf{\hat k}_1^2}
	e^{i\mbf{\hat k}_1\mbf\rho_{12}}\right)
\left(\int d^2\mbf l \frac1{\mbf l^2}
	e^{i\mbf l\mbf\rho_{11'}}\right)
\big(\Delta_{1'}\,\Delta_2\,\Phi_2(\mbf\rho_1',\mbf\rho_2)\big)
}
Here we substituted $\mbf{\hat k}_1:=\mbf k_1-\mbf l$.  We proceed by
performing the Fourier integrations over the propagators with mass
regularisation and perform a double partial integration in~$\mbf\rho_1'$,
shifting the Laplace operator to one of the $K_0$ Bessel functions.  Here it
proves useful that we use the amplitude $\Phi_2$ instead of~$D_2$.  The surface
terms vanish since the amplitude $\Phi_2$ vanishes at infinity.
\dmath3{
\ldots&=&
\lim_{m\to0}-\frac{\Delta_1}{(2\pi)^3}\int d^2\mbf\rho_1'
K_0(m|\mbf\rho_{12}|)\,K_0(m|\mbf\rho_{11'}|)\,
\Delta_{1'}\,\Delta_2\,\Phi_2(\mbf\rho_1',\mbf\rho_2)
\cr&=&
\lim_{m\to0}-\frac{\Delta_1}{(2\pi)^3}\int d^2\mbf\rho_1'
K_0(m|\mbf\rho_{12}|)\,
\big(\Delta_{1'}[\ln2+\psi(1)-\ln m-\ln|\mbf\rho_{11'}|]\big)\,
\Delta_2\,\Phi_2(\mbf\rho_1',\mbf\rho_2)
}
We have already expanded the $K_0$ Bessel function.  As we saw in
Section~\ref{sec:confformulas}, the Laplacian of the logarithm is proportional
to the two-dimensional delta function.  The constants vanish when performing
the differentiation.  Hence we can finally perform the integral
in~$\mbf\rho_1'$.
\dmath2{
b(\mbf\rho_1,\mbf\rho_2)&=&
\lim_{m\to0}\frac1{(2\pi)^2}\Delta_1
K_0(m|\mbf\rho_{12}|)\,\Delta_2\,\Phi_2(\mbf\rho_1,\mbf\rho_2)
\cr&=&
\lim_{m\to0}\frac1{(2\pi)^2}\Delta_1
[\ln2+\psi(1)-\ln m-\ln|\mbf\rho_{12}|]\,
\Delta_2\,\Phi_2(\mbf\rho_1,\mbf\rho_2)
}

In the case of the function~$c$, two configuration space integrations remain to
be done after transforming the propagators.  They are solved in the same way as
for the $b$ function, by shifting the Laplace operators from the amplitude to
the Bessel functions and integrating out the resulting delta functions.  So
Laplacians act on both Bessel functions and turn them into delta functions
which are integrated away.  The result is just a Laplacian of~$\Phi_2$
multiplied with a constant.  Here are the position space representations of all
three functions, with the Bessel functions expanded:
\dmath{2}{
a(\mbf\rho_1,\mbf\rho_2,\mbf\rho_3)&=&
-\frac1{(2\pi)^3}\Delta_1
[\ln2+\psi(1)-\ln m-\ln|\mbf\rho_{12}|]\,
[\ln2+\psi(1)-\ln m-\ln|\mbf\rho_{13}|]
\Delta_2 \Delta_3 \Phi_2(\mbf\rho_2,\mbf\rho_3)
\cr
b(\mbf\rho_1,\mbf\rho_2)&=&\frac1{(2\pi)^2}\Delta_1
[\ln2+\psi(1)-\ln m-\ln|\mbf\rho_{12}|]\,
\Delta_2\,\Phi_2(\mbf\rho_1,\mbf\rho_2)
&eq:abc_c\cr
c(\mbf\rho)&=&-\frac1{2\pi}\Delta\,\Phi_2(\mbf\rho,\mbf\rho)
}
The limit $m\to0$ is not written explicitly any more but is always implied.
The symmetry of the function $a$ in its latter two arguments is evident in
position space.

The relations between the three functions (\ref{eq:abcstrel_m}) have analogues
in configuration space.  While in momentum space $b$ could be obtained from
$a$, and $c$ from $b$, by setting one of the momenta to zero, in configuration
space the corresponding spatial coordinate has to be integrated out.
\dmblock{1.5}{eq:abcrel_c}{
b(\mbf\rho_1,\mbf\rho_2)&=&
\int d^2\mbf\rho'\,a(\mbf\rho_1,\mbf\rho',\mbf\rho_2)
=\int d^2\mbf\rho'\,a(\mbf\rho_1,\mbf\rho_2,\mbf\rho')
\cr
c(\mbf\rho)&=&\int d^2\mbf\rho'\,b(\mbf\rho,\mbf\rho')
}
This can be checked explicitly by performing the integrations analogously to
the last steps in the Fourier transformation of $b$ (or~$c$).  One Laplacian is
shifted from the amplitude $\Phi_2$ to one of the Bessel functions.  This
results in a delta function which allows performing the integration.  The
factor $2\pi$ in~(\ref{eq:greenpoiss2}) accounts for the different prefactors
in $a$, $b$ and $c$, and the minus sign in front of the logarithm in the
expansion of~$K_0$ accounts for the alternating sign.  Here is the explicit
calculation for deriving $c$ from $b$:
\dmath{1.5}{
\int d^2\mbf\rho'\,b(\mbf\rho,\mbf\rho')
&=&\frac1{(2\pi)^2}\int d^2\mbf\rho'\,\Delta
[\ln2+\psi(1)-\ln m-\ln|\mbf\rho-\mbf\rho'|]\,
\Delta'\,\Phi_2(\mbf\rho,\mbf\rho')
\cr
&=&\frac1{(2\pi)^2}\Delta\int d^2\mbf\rho'\,
\big(\Delta'\,[\ln2+\psi(1)-\ln m-\ln|\mbf\rho-\mbf\rho'|]\big)\,
\Phi_2(\mbf\rho,\mbf\rho')
\cr
&=&\frac1{(2\pi)^2}\Delta\int d^2\mbf\rho'\,
(-2\pi\,\delta^2(\mbf\rho-\mbf\rho'))\,\Phi_2(\mbf\rho,\mbf\rho')
\cr
&=&-\frac1{2\pi}\Delta\,\Phi_2(\mbf\rho,\mbf\rho)=c(\mbf\rho)\,.
}

\subsection{The virtual corrections $s$ and $t$ and the trajectory 
			function $\beta$}
\label{sec:st_c}

It is convenient to transform the expressions for $s$ and $t$ in terms of
$\beta$ first and to perform the Fourier transformation of $\beta$ separately.
To this end, we insert the Fourier transform of the configuration-space
trajectory function for the momentum-space $\beta$ contained in $s$ and~$t$.
Besides, we replace $D_2$ by the double Laplacian of $\Phi_2$ as we did in the
previous section.  We get the expression for $s$:
\dmath2{
s(\mbf\rho_1,\mbf\rho_2,\mbf\rho_3)&=&
\frac1{(2\pi)^6}\frac{-2}{N_cg^2}\int d^2\mbf k_1\,d^2\mbf k_2\,d^2\mbf k_3\;
\beta(\mbf k_1)\,D_2(\mbf k_1+\mbf k_2,\mbf k_3)\;
e^{i(\mbf k_1\mbf\rho_1+\mbf k_2\mbf\rho_2+\mbf k_3\mbf\rho_3)}
\cr
&=&\frac1{(2\pi)^6}\frac{-2}{N_cg^2}\int d^2\mbf k_1\,d^2\mbf k_2\,d^2\mbf k_3
\int d^2\mbf\rho'\,d^2\mbf\rho_1'\,d^2\mbf\rho_2'\;
\beta(\mbf\rho')\,
\big(\Delta_{1'}\,\Delta_{2'}\,\Phi_2(\mbf\rho_1',\mbf\rho_2')\big)\cdot{}
\cr
&&\qquad\qquad\qquad\qquad{}\cdot
e^{i(\mbf k_1\mbf\rho_1+\mbf k_2\mbf\rho_2+\mbf k_3\mbf\rho_3
	-\mbf k_1\mbf\rho'-(\mbf k_1+\mbf k_2)\mbf\rho_1'
	-\mbf k_3\mbf\rho_2')}
\,.}
Here we have already inserted the Fourier transform of the configuration-space
amplitude $\Phi_2$. All the momentum integrals give delta functions, which 
makes the spatial integrations trivial. The result is:
\begin{equation}
\label{eq:s_c}
s(\mbf\rho_1,\mbf\rho_2,\mbf\rho_3)=
-\frac2{N_cg^2}\beta(\mbf\rho_{12})\,\Delta_2\,\Delta_3\,
\Phi_2(\mbf\rho_2,\mbf\rho_3)\,.
\end{equation}

In the case of the $t$~function, there are only two Fourier integrals and
consequently one integral over a spatial coordinate remains in the final
expression.  Repeating the steps done for~$s$, we get the following expression:
\dmath2{
t(\mbf\rho_1,\mbf\rho_2)&=&\ldots=
\frac1{(2\pi)^2}\frac{-2}{N_cg^2}
\int d^2\mbf k_1 \int d^2\mbf\rho\,d^2\mbf\rho_0\;
\beta(\mbf\rho)\,\Delta_0\,\Delta_2\,\Phi_2(\mbf\rho_0,\mbf\rho_2)\;
e^{i\mbf k_1(\mbf\rho_1-\mbf\rho-\mbf\rho_0)}
\cr&=&
-\frac{2}{N_cg^2} \int d^2\mbf\rho_0\;
\beta(\mbf\rho_{10})\,\Delta_0\,\Delta_2\,
\Phi_2(\mbf\rho_0,\mbf\rho_2)
=\int d^2\mbf\rho_0\;s(\mbf\rho_1,\mbf\rho_0,\mbf\rho_2)\,.
&eq:t_c%
}
The relation between $s$ and $t$ (see (\ref{eq:abcstrel_m})) is already obvious
from comparison with Equation~\ref{eq:s_c}.

The main task remains to be performed, namely the Fourier transformation of the
trajectory function~$\beta$.  The techniques employed here are very similar to
those used for the real corrections $a$, $b$ and~$c$.  We express the $\mbf
k^2$ in the numerator as a Laplace operator and factorise the two-fold
momentum-space integral through a shift in the integration variable.  Both
factors give us a two-dimensional gluon propagator in position space, a $K_0$
Bessel function (in the mass regularisation scheme).
\dmath2{
\beta(\mbf\rho)&=&
-\frac{N_cg^2}2\frac1{(2\pi)^5}\int d^2\mbf k\,d^2\mbf l\,
\frac{\mbf k^2}{\mbf l^2\,(\mbf l-\mbf k)^2}\,e^{i\mbf k\mbf\rho}
=\frac{N_cg^2}2\frac\Delta{(2\pi)^5}
\left(\int d^2\mbf{\hat k}\frac1{\mbf{\hat k}^2}\,e^{i\mbf{\hat k}\mbf\rho}
\right)
\left(\int d^2\mbf l\frac1{\mbf l^2}\,e^{i\mbf l\mbf\rho}\right)
\cr&=&
\lim_{m\to0}\frac{N_cg^2}2\frac1{(2\pi)^3}\Delta\big(K_0(m|\mbf\rho|)\big)^2
=\lim_{m\to0}N_cg^2 \frac1{(2\pi)^3}
\big[K_0(m|\mbf\rho|)\Delta K_0(m|\mbf\rho|)+|\nabla K_0(m|\mbf\rho|)|^2\big]
}
After expanding the Bessel functions we can perform the differentiations.
Using~(\ref{eq:greenpoiss2}) we obtain:
\dmath0{
\beta(\mbf\rho)&=&\lim_{m\to0}N_cg^2 \frac1{(2\pi)^3}
\left(-2\pi\,\delta^2(\mbf\rho)\,[\ln2+\psi(1)-\ln m-\ln |\mbf\rho|]
+\frac1{|\mbf\rho|^2}\right)
&eq:beta_c}

This formula contains two terms which are ultraviolet divergent under an
integral and will later have to be regularised: $\delta(\mbf\rho)\ln\mbf\rho$
and $1/|\mbf\rho|^2$.  However, we will ignore this issue in most of this
chapter.  We will write down integrals over $\beta$ as formal expressions.  In
section~\ref{sec:Greg}, the problems relating to regularisation will be
discussed.

\subsection{Summary of Results}

This section contains a summary of our results so far: the configuration space
representations of the five component functions.  Figure~\ref{fig:abcst} shows
how the coordinates correspond to the location of the vertices in the diagrams.
\dmath{2}{
a(\mbf\rho_1,\mbf\rho_2,\mbf\rho_3)&=&
-\frac1{(2\pi)^3}\Delta_1
[\ln2+\psi(1)-\ln m-\ln|\mbf\rho_{12}|]\,
[\ln2+\psi(1)-\ln m-\ln|\mbf\rho_{13}|]
\Delta_2 \Delta_3 \Phi_2(\mbf\rho_2,\mbf\rho_3)
\cr
b(\mbf\rho_1,\mbf\rho_2)&=&\frac1{(2\pi)^2}\Delta_1
[\ln2+\psi(1)-\ln m-\ln|\mbf\rho_{12}|]\,
\Delta_2\,\Phi_2(\mbf\rho_1,\mbf\rho_2)
\cr
c(\mbf\rho)&=&-\frac1{2\pi}\Delta\,\Phi_2(\mbf\rho,\mbf\rho)
&eq:abcst_c%
\cr
s(\mbf\rho_1,\mbf\rho_2,\mbf\rho_3)&=&
\frac2{(2\pi)^3}
\left(2\pi\,\delta^2(\mbf\rho_{12})\,[\ln2+\psi(1)-\ln m-\ln |\mbf\rho_{12}|]
-\frac1{|\mbf\rho_{12}|^2}\right)
\,\Delta_2\,\Delta_3\,\Phi_2(\mbf\rho_2,\mbf\rho_3)
\cr
t(\mbf\rho_1,\mbf\rho_2)&=&
\frac2{(2\pi)^3}\int d^2\mbf\rho_0
\left(2\pi\,\delta^2(\mbf\rho_{10})\,[\ln2+\psi(1)-\ln m-\ln |\mbf\rho_{10}|]
-\frac1{|\mbf\rho_{10}|^2}\right)
\Delta_0\,\Delta_2\,\Phi_2(\mbf\rho_0,\mbf\rho_2)
}
The limit $m\to0$ is implied in all expressions even though it is not written
there.

The relations between the five functions and the trajectory function have the
following form in position space:
\dmath{1.5}{
b(\mbf\rho_1,\mbf\rho_2)&=&
\int d^2\mbf\rho'\,a(\mbf\rho_1,\mbf\rho',\mbf\rho_2)
=\int d^2\mbf\rho'\,a(\mbf\rho_1,\mbf\rho_2,\mbf\rho')
\cr
c(\mbf\rho)&=&\int d^2\mbf\rho'\,b(\mbf\rho,\mbf\rho')
=\int d^2\mbf\rho'\,d^2\mbf\rho''\,a(\mbf\rho,\mbf\rho',\mbf\rho'')
\crl{eq:abcstrel_c}
s(\mbf\rho_1,\mbf\rho_2,\mbf\rho_3)&=&
-\frac2{N_cg^2}\beta(\mbf\rho_{12})\,\Delta_2\,\Delta_3\,
\Phi_2(\mbf\rho_2,\mbf\rho_3)
\cr
t(\mbf\rho_1,\mbf\rho_2)&=&
\int d^2\mbf\rho_0\;s(\mbf\rho_1,\mbf\rho_0,\mbf\rho_2)=
-\frac{2}{N_cg^2} \int d^2\mbf\rho_0\;
\beta(\mbf\rho_{10})\,\Delta_0\,\Delta_2\,\Phi_2(\mbf\rho_0,\mbf\rho_2)
}

\section{The building block of vertices: the function $G$}
\label{sec:G}

\subsection{The function $G$ in momentum and configuration space}
\label{sec:thisisG}

It has already been mentioned that $2\to n$ reggeised gluon vertices, one of
which we will investigate in the following chapter, can be expressed in terms
of the five functions which were the topic of the previous section.  However,
the vertices do not seem to be arbitrary combinations of them.  The two known
vertices ($V_{2\to4}$ and $V_{2\to6}$) are sums of terms of one function which
can in turn be expressed through the five fundamental ones
{}\cite{bw,vaccaphd,brvac,ew2}.  It is conjectured~\cite{ew2} that higher
vertices also can be expressed in terms of this function.  The function is
denoted by $G$ and has the following form:
\dmath{0.5}{
G(\mbf k_1,\mbf k_2,\mbf k_3)&=&
\frac{g^2}2\big[2\,c(\mbf k_1+\mbf k_2+\mbf k_3)
-2\,b(\mbf k_1+\mbf k_2,\mbf k_3)-2\,b(\mbf k_2+\mbf k_3,\mbf k_1)
+2\,a(\mbf k_2,\mbf k_1,\mbf k_3)
\cr&&\qquad
{}+t(\mbf k_1+\mbf k_2,\mbf k_3)+t(\mbf k_2+\mbf k_3,\mbf k_1)
-s(\mbf k_2,\mbf k_1,\mbf k_3)-s(\mbf k_2,\mbf k_3,\mbf k_1)\big]\,.
&eq:G_m}
To avoid writing so many sums of momenta, one can express this in a
short-hand notation:
\dmath{0.5}{
G(1,2,3)&=&
\frac{g^2}2\big[2\,c(123)-2\,b(12,3)-2\,b(23,1)+2\,a(2,1,3)
\crl{eq:G_symb}
&&\qquad
{}+t(12,3)+t(23,1)-s(2,1,3)-s(2,3,1)\big]\,.
}
Each number stands for the momentum of the corresponding reggeised gluon, that
is 1 for $\mbf k_1$, 2 for $\mbf k_2$ and so on.  A set of several numbers, for
instance 123, stands for the sum of the momenta, $\mbf k_1+\mbf k_2+\mbf k_3$.
An empty argument, which will be denoted by ``$-$'', stands for zero momentum.
This notation has the advantage that it can be easily transformed into a
configuration-space expression as well.  In configuration space, a number
stands for the corresponding coordinate.  A set of numbers means that the
coordinates are identified with a delta function with which the expression has
to be multiplied.  This follows from the identity
$$
\frac1{(2\pi)^4}\int d^2\mbf k_1\,d^2\mbf k_2\,f(\mbf k_1+\mbf k_2)
e^{i(\mbf k_1\mbf\rho_1+\mbf k_2\mbf\rho_2)}
=\frac1{(2\pi)^4}\int d^2\mbf k_1'\,d^2\mbf k_2\,f(\mbf k_1')
e^{i(\mbf k_1'\mbf\rho_1+\mbf k_2(\mbf\rho_2-\mbf\rho_1))}
=\delta^2(\mbf\rho_{12})f(\mbf\rho_1)\,,
$$
which holds for an arbitrary function~$f$.  An empty argument ``$-$''
translates as an integral over the corresponding coordinate argument in
position space.

It can be easily seen from~(\ref{eq:G_symb}) that $G$ is symmetric under
exchange of its first and third argument.  The symmetry does not extend to the
second argument.  However, it has a different interesting property: When the
second momentum argument is set to 0, the $a$ and $s$ functions vanish and $G$
reduces to a derivative of the BFKL kernel multiplied with a constant.
\dmath2{
\crl{eq:G_BFKL}
G(1,-,3)&=&
\frac{g^2}2\big[2\,c(13)-2\,b(1,3)-2\,b(3,1)+t(1,3)+t(3,1)\big]=
\frac1{N_c}\Delta_1\,\Delta_3\, (K_{BFKL}\otimes \Phi_2)(1,3)
}
Besides it can be proven that $G$ vanishes when its first or last momentum
argument is zero.  This is because then the component functions either vanish
or turn into a different component function with fewer arguments (according
to~(\ref{eq:abcstrel_m})).  They then cancel among each other.  Here is the
explicit calculation if the first argument vanishes:
\dmath{0.5}{
G(-,2,3)&=&
\frac{g^2}2[2\,c(23)-2\,b(2,3)-2\,b(23,-)+2\,a(2,-,3)\crd
&&\qquad
{}+t(2,3)+t(23,-)-s(2,-,3)-s(2,3,-)]\crl{eq:G-23}
&=&
\frac{g^2}2[2\,c(23)-2\,b(2,3)-2\,c(23)+2\,b(2,3)\crd
&&\qquad
{}+t(2,3)+0-t(2,3)-0]\cr
&=&0\,.}

For our investigation of $2\to n$ gluon vertices, we want to know how $G$
transforms under conformal transformations in transverse position space.
Therefore we need the configuration space representation of~$G$.  It can easily
be obtained from the coordinate-free symbolic notation as described above.  The
configuration-space expression for $G$ is:
\dmath1{
G(\mbf\rho_1,\mbf\rho_2,\mbf\rho_3)&=&
\frac{g^2}2\big[2\,\delta^2(\mbf\rho_{12})\delta^2(\mbf\rho_{23})c(\mbf\rho_2)
-2\,\delta^2(\mbf\rho_{12})b(\mbf\rho_1,\mbf\rho_3)
-2\,\delta^2(\mbf\rho_{23})b(\mbf\rho_3,\mbf\rho_1)
+2\,a(\mbf\rho_2,\mbf\rho_1,\mbf\rho_3)
\cr&&\qquad
{}+\delta^2(\mbf\rho_{12})t(\mbf\rho_1,\mbf\rho_3)
+\delta^2(\mbf\rho_{23})t(\mbf\rho_3,\mbf\rho_1)
-s(\mbf\rho_2,\mbf\rho_1,\mbf\rho_3)
-s(\mbf\rho_2,\mbf\rho_3,\mbf\rho_1)\big]\,.
&eq:G_c%
}

\subsection{The conformal transformation properties of the function $G$}

\subsubsection{Initial simplifications and properties under translations, 
				rotations and dilatations}

The function $G$ is a component of expressions for $2\to n$ reggeised gluon
vertices which occur in unitarity corrections to BFKL.  Since the BFKL kernel
was found to be conformally invariant in transverse position space, it is a
legitimate question to ask whether these vertices, too, are conformally
invariant.  This has been proven for the $2\to4$ gluon vertex~\cite{blw}.  As
the higher $2\to n$ gluon vertices contain a significantly larger number of
terms of the five functions $a$, $b$, $c$, $s$ and $t$, to approach this
question directly for them would be all but impossible.  But since the largest
known vertex, $V_{2\to6}$, and quite possibly higher vertices, are composed of
$G$ functions, their conformal transformation properties can be inferred from
the properties of~$G$.  Therefore investigating $G$'s conformal transformation
properties is of great importance.

Conformal $n$-point functions are defined as expectation values of a product of
$n$ conformal fields.  By contrast, $G$ and the gluon vertices are defined to
contain only one conformal field $\Phi_2$ (which is a superposition of
$E^{(\nu,n)}$s).  The expression we will identify with a conformal $n$-point
function is a convolution of the $2\to n$ gluon vertices with several conformal
fields.  Since the convolution integrals themselves transform under dilatations
and the inversion, we will not expect the gluon vertices, and by implication
$G$, to be strictly invariant.  Rather, $G$ should transform in such a way that
it is invariant together with convolution integrals over its arguments.  In the
following, we will call this property conformal covariance.

The amplitude $\Phi_2$ defined into $G$ has to be exempted from our
investigation.  The transformation properties of the conformal eigenfunctions
are given exclusively by their external coordinates.  This (or the
corresponding transformation properties of general conformal fields) is what
leads to the specific form for conformal $n$-point functions (\ref{eq:2point}),
(\ref{eq:3point}) and~(\ref{eq:4point}).  Hence we need not take the
transformation properties of $\Phi_2$ into account when investigating~$G$.

To investigate how $G$ transforms under conformal transformations, we start out
from Equation~\ref{eq:G_c}.  However, expressing $G$ in terms of the five
functions $a$, $b$, $c$, $s$ and $t$ is of limited use for determining its
conformal transformation properties.  Of these functions, only $c$ is
conformally covariant on its own.  Therefore we will rewrite $G$ in a way which
may seem far more complicated at first sight, but which will allow us to find
conformally covariant groups of terms.

Let us first group the terms in (\ref{eq:G_c}) somewhat differently:
\dmath{1.5}{
\label{}
G(\mbf\rho_1,\mbf\rho_2,\mbf\rho_3)&=&
\frac{g^2}2[2\delta^2(\mbf\rho_{12})\delta^2(\mbf\rho_{23})\,c(\mbf\rho_2)
\cr&&\qquad
{}+\delta^2(\mbf\rho_{12})\,(-2b+t)(\mbf\rho_1,\mbf\rho_3)
+\delta^2(\mbf\rho_{23})\,(-2b+t)(\mbf\rho_3,\mbf\rho_1)
&eq:Gregroup\cr&&\qquad
{}+(2a(\mbf\rho_2,\mbf\rho_1,\mbf\rho_3)
-s(\mbf\rho_2,\mbf\rho_1,\mbf\rho_3)
-s(\mbf\rho_2,\mbf\rho_3,\mbf\rho_1))]\,.
}

It is well known that the combination $-2b+t$ and the expression containing $a$
and $s$ in the last line are infrared finite.  We will now simplify these two
expressions.  After integrating out the delta function in the first part of
$t$, we obtain for $-2b+t$:
\dmath2{
(-2b+t)(\mbf\rho_1,\mbf\rho_3)&=&
\frac1{(2\pi)^3}\bigg(
-2\,(2\pi)\Delta_1[\ln2+\psi(1)-\ln m-\ln|\mbf\rho_{13}|]\,
\Delta_3\,\Phi_2(\mbf\rho_1,\mbf\rho_3)
\crd&&\qquad\quad\quad
{}+2\,(2\pi)\,[\ln2+\psi(1)-\ln m]\,
\Delta_1\,\Delta_3\,\Phi_2(\mbf\rho_1,\mbf\rho_3)
\crd&&\qquad\quad\quad
{}-2\int d^2\mbf\rho_0\left(2\pi\,\delta^2(\mbf\rho_{10})\ln|\mbf\rho_{10}|+
\frac1{|\mbf\rho_{10}|^2}\right)
\Delta_0\,\Delta_3\,\Phi_2(\mbf\rho_0,\mbf\rho_3)\bigg)
\cr
&=&\frac2{(2\pi)^3}\bigg(
2\pi\,\Delta_1\ln|\mbf\rho_{13}|\,\Delta_3\,\Phi_2(\mbf\rho_1,\mbf\rho_3)
\crd&&\qquad\quad\quad
{}-\int d^2\mbf\rho_0\left(2\pi\,\delta^2(\mbf\rho_{10})\ln|\mbf\rho_{10}|+
\frac1{|\mbf\rho_{10}|^2}\right)
\Delta_0\,\Delta_3\,\Phi_2(\mbf\rho_0,\mbf\rho_3)\bigg)
\cr
&=&\frac2{(2\pi)^3}\bigg(
\left[(2\pi)^2\delta^2(\mbf\rho_{13})\,\Delta_3
+4\pi\frac{\mbf\rho_{13}\cdot\mbf\nabla_1}{|\mbf\rho_{13}|^2}\,\Delta_3
+2\pi\ln|\mbf\rho_{13}|\,\Delta_1\Delta_3
\right]\Phi_2(\mbf\rho_1,\mbf\rho_3)
\crd&&\qquad\quad\quad
{}-\int d^2\mbf\rho_0\left(2\pi\,\delta^2(\mbf\rho_{10})\ln|\mbf\rho_{10}|+
\frac1{|\mbf\rho_{10}|^2}\right)
\Delta_0\,\Delta_3\,\Phi_2(\mbf\rho_0,\mbf\rho_3)\bigg)\,.
&eq:-2b+t}

The last line in~(\ref{eq:Gregroup}) can also be simplified.  Since the first
Laplacian in the $a$~function affects only the $K_0$ Bessel functions, we can
expand them and perform the differentiations with the product rule.  The
resulting terms can be combined with the terms from the $s$~functions to form a
concise result.
\dmath2{
2a(\mbf\rho_2,\mbf\rho_1,\mbf\rho_3)
&-&s(\mbf\rho_2,\mbf\rho_1,\mbf\rho_3)
-s(\mbf\rho_2,\mbf\rho_3,\mbf\rho_1)={}
\cr
&=&\frac2{(2\pi)^3}\bigg[
-\Delta_2\,K_0(m|\mbf\rho_{12}|)\,K_0(m|\mbf\rho_{23}|)\,
-\left(2\pi\,\delta^2(\mbf\rho_{12})\,K_0(m|\mbf\rho_{12}|)
	-\frac1{|\mbf\rho_{12}|^2}\right)
\crd&&\qquad\qquad
-\left(2\pi\,\delta^2(\mbf\rho_{23})\,K_0(m|\mbf\rho_{23}|)
	-\frac1{|\mbf\rho_{23}|^2}\right)
\bigg]
\Delta_1\,\Delta_3\,\Phi_2(\mbf\rho_1,\mbf\rho_3)
\cr
&=&\frac2{(2\pi)^3}\bigg[
-\left(-2\pi\,\delta^2(\mbf\rho_{12})\,K_0(m|\mbf\rho_{23}|)
+2\frac{\mbf\rho_{21}\cdot\mbf\rho_{23}}{|\mbf\rho_{12}|^2|\mbf\rho_{23}|^2}
-2\pi\,\delta^2(\mbf\rho_{23})\,K_0(m|\mbf\rho_{12}|)\right)
\crd&&
{}-2\pi\,\delta^2(\mbf\rho_{12})\,K_0(m|\mbf\rho_{12}|)
-2\pi\,\delta^2(\mbf\rho_{23})\,K_0(m|\mbf\rho_{23}|)
+\frac1{|\mbf\rho_{12}|^2}+\frac1{|\mbf\rho_{23}|^2}
\bigg]
\Delta_1\,\Delta_3\,\Phi_2(\mbf\rho_1,\mbf\rho_3)
\cr
&=&\frac2{(2\pi)^3}\bigg[
-2\pi\big(\delta^2(\mbf\rho_{12})-\delta^2(\mbf\rho_{23})\big)
\big(K_0(m|\mbf\rho_{12}|)-K_0(m|\mbf\rho_{23}|)\big)
\crd&&\qquad\qquad
+\left|\frac{\mbf\rho_{12}}{|\mbf\rho_{12}|^2}
		+\frac{\mbf\rho_{23}}{|\mbf\rho_{23}|^2}\right|^2
\bigg]
\Delta_1\,\Delta_3\,\Phi_2(\mbf\rho_1,\mbf\rho_3)
\crl{eq:2a-s-s}
&=&\frac2{(2\pi)^3}\bigg[
2\pi\big(\delta^2(\mbf\rho_{12})-\delta^2(\mbf\rho_{23})\big)
\ln\frac{|\mbf\rho_{12}|}{|\mbf\rho_{23}|}
+\left|\frac{\mbf\rho_{12}}{|\mbf\rho_{12}|^2}
		+\frac{\mbf\rho_{23}}{|\mbf\rho_{23}|^2}\right|^2
\bigg]
\Delta_1\,\Delta_3\,\Phi_2(\mbf\rho_1,\mbf\rho_3)
}
In the last step we have again used the expansion~(\ref{eq:K0expand}) of $K_0$ for small arguments, assuming $m$ to be small.

We can now put everything together and write down the whole $G$~function:
\dmath{2.5}{
G(\mbf\rho_1,\mbf\rho_2,\mbf\rho_3)&=&
\frac{g^2}{(2\pi)^3}\bigg[
(2\pi)^2\delta^2(\mbf\rho_{12})\delta^2(\mbf\rho_{23})
\big({-\Delta_2}\,\Phi_2(\mbf\rho_2,\mbf\rho_2)
+(\Delta_1+\Delta_3)\,\Phi_2(\mbf\rho_1,\mbf\rho_3)\big)
\cr&&
{}+4\pi\,\delta^2(\mbf\rho_{12})
\frac{\mbf\rho_{23}\cdot\mbf\nabla_1}{|\mbf\rho_{23}|^2}\Delta_3\,
\Phi_2(\mbf\rho_1,\mbf\rho_3)
-4\pi\,\delta^2(\mbf\rho_{23})
\frac{\mbf\rho_{12}\cdot\mbf\nabla_3}{|\mbf\rho_{12}|^2}\Delta_1\,
\Phi_2(\mbf\rho_1,\mbf\rho_3)
\cr&&
{}-\delta^2(\mbf\rho_{12})\int d^2\mbf\rho_0
\left(2\pi\,\delta^2(\mbf\rho_{10})\ln|\mbf\rho_{10}|
		+\frac1{|\mbf\rho_{10}|^2}\right)
\,\Delta_0\,\Delta_3\,\Phi_2(\mbf\rho_0,\mbf\rho_3)
\cr&&
{}-\delta^2(\mbf\rho_{23})\int d^2\mbf\rho_0
\left(2\pi\,\delta^2(\mbf\rho_{30})\ln|\mbf\rho_{30}|
		+\frac1{|\mbf\rho_{30}|^2}\right)
\,\Delta_0\,\Delta_1\,\Phi_2(\mbf\rho_0,\mbf\rho_1)
\cr&&
{}+2\pi\big(\delta^2(\mbf\rho_{12})\,\ln|\mbf\rho_{12}|+
\delta^2(\mbf\rho_{23})\ln|\mbf\rho_{23}|\big)
\,\Delta_1\,\Delta_3\,\Phi_2(\mbf\rho_1,\mbf\rho_3)
\cr&&
{}+\left|\frac{\mbf\rho_{12}}{|\mbf\rho_{12}|^2}
		+\frac{\mbf\rho_{23}}{|\mbf\rho_{23}|^2}\right|^2
\Delta_1\,\Delta_3\,\Phi_2(\mbf\rho_1,\mbf\rho_3)
\bigg]\,.
&eq:Gexpvec}
The second term in the first line comes from the first term
of~(\ref{eq:-2b+t}).  (Note that there are two terms containing $-2b+t$
in~(\ref{eq:Gregroup}).)  The following three lines are from~(\ref{eq:-2b+t}),
too; the minus sign in the second line arises because we have replaced
$\mbf\rho_{21}$ with~$\mbf\rho_{12}$.  The third term in~(\ref{eq:-2b+t})
cancels part of the first term in~(\ref{eq:2a-s-s}), leaving the last two
lines.

At this point we can already say that $G$ is invariant under translations and
rotations (ignoring the amplitude~$\Phi_2$).  This is obvious because $G$
contains only relative vectors and differential operators.  Both are invariant
under translations.  Furthermore, all expressions are rotationally invariant
since the vectors occur only as moduli or scalar products.

We can also already determine $G$'s transformation properties under
dilatations.  Let us first look at how elementary expressions of which $G$ is
composed change under dilatations:
\dmath2{
\mbf\rho&\longrightarrow&\lambda\,\mbf\rho\cr
\mbf\nabla&\longrightarrow&\lambda^{-1}\,\mbf\nabla\cr
\Delta&\longrightarrow&\lambda^{-2}\,\Delta\cr
\int d^2\mbf\rho&\longrightarrow&\lambda^2\int d^2\mbf\rho  &eq:elemdil\cr
\delta^2(\mbf\rho)&\longrightarrow&\lambda^{-2}\,\delta^2(\mbf\rho)\cr
\ln|\mbf\rho|&\longrightarrow&\ln|\mbf\rho|+\ln\lambda
}

Using these rules, it is easy to see from (\ref{eq:Gexpvec}) that all terms of
$G$ get a factor $\lambda^{-6}$ under dilatations (ignoring the transformation
of~$\Phi_2$).  In addition, the terms containing a logarithm of a coordinate
give an additional term~$\ln\lambda$.  Taken together, these extra terms
cancel:
\dmath2{
-\delta^2(\mbf\rho_{12})\int d^2\mbf\rho_0
\;2\pi\,\delta^2(\mbf\rho_{10})&&\ln\lambda\, 
\Delta_0\,\Delta_3\,\Phi_2(\mbf\rho_0,\mbf\rho_3)
-\delta^2(\mbf\rho_{23})\int d^2\mbf\rho_0
\;2\pi\,\delta^2(\mbf\rho_{30})\ln\lambda\,
\Delta_0\,\Delta_1\,\Phi_2(\mbf\rho_0,\mbf\rho_1)
\crd
{}+2\pi\big(\delta^2(\mbf\rho_{12})\,\ln\lambda+{}
\delta^2(\mbf\rho_{23})&&\ln\lambda\big)
\,\Delta_1\,\Delta_3\,\Phi_2(\mbf\rho_1,\mbf\rho_3)
\cr&&=2\pi\,\ln\lambda\,
\big(
-\delta^2(\mbf\rho_{12})\,\Delta_1\,\Delta_3\,\Phi_2(\mbf\rho_1,\mbf\rho_3)
-\delta^2(\mbf\rho_{23})\,\Delta_3\,\Delta_1\,\Phi_2(\mbf\rho_3,\mbf\rho_1)
\crd&&\qquad\qquad
{}+\delta^2(\mbf\rho_{12})\,\Delta_1\,\Delta_3\,\Phi_2(\mbf\rho_1,\mbf\rho_3)
+\delta^2(\mbf\rho_{23})\,\Delta_1\,\Delta_3\,\Phi_2(\mbf\rho_1,\mbf\rho_3)
\big)
\crd&&=0\,.
&eq:Gdillogcancel}

This result is what we expected.  It means that when we convolute $G$ with
conformal fields, the transformation properties of the resulting expression are
given exclusively by the attached fields.  The three convolution integrals
provide a factor $\lambda^6$ under dilatations which cancels the $\lambda^{-6}$
we obtained from~$G$.  Therefore the transformation properties of the whole
expression under dilatations are given by moduli of the external coordinates of
the fields (including the $\Phi_2$ contained in~$G$), raised to a power given
by the conformal weights.  This is a property of conformal $n$-point functions.
One can conclude that vertices constructed from $G$ in the way presented in the
following section are conformal four-point functions.  It only remains to be
shown that the same applies under inversion, which will be done in the
following section.

\subsubsection{The complex representation and properties under inversion}

Unlike the other basic conformal transformations, the inversion cannot easily
be expressed in the vectorial notation. Therefore we will rewrite $G$ in terms
of complex variables. The result looks very similar to~(\ref{eq:Gexpvec}).
\dmath{2.5}{
G(\rho_1,\rho_2,\rho_3)&=&
\frac{g^2}{(2\pi)^3}\bigg[
(2\pi)^2\delta^2(\rho_{12})\delta^2(\rho_{23})
\big({-\Delta_2}\,\Phi_2(\rho_2,\rho_2)
+(\Delta_1+\Delta_3)\,\Phi_2(\rho_1,\rho_3)\big)
\cr&&
{}+8\pi\,\delta^2(\rho_{12})
\frac{(\rho_{23},\del_1^*)}{|\rho_{23}|^2}\Delta_3\,
\Phi_2(\rho_1,\rho_3)
-8\pi\,\delta^2(\rho_{23})
\frac{(\rho_{12},\del_3^*)}{|\rho_{12}|^2}\Delta_1\,
\Phi_2(\rho_1,\rho_3)
\cr&&
{}-\delta^2(\rho_{12})\int d^2\rho_0
\left(2\pi\,\delta^2(\rho_{10})\ln|\rho_{10}|
		+\frac1{|\rho_{10}|^2}\right)
\,\Delta_0\,\Delta_3\,\Phi_2(\rho_0,\rho_3)
\cr&&
{}-\delta^2(\rho_{23})\int d^2\rho_0
\left(2\pi\,\delta^2(\rho_{30})\ln|\rho_{30}|
		+\frac1{|\rho_{30}|^2}\right)
\,\Delta_0\,\Delta_1\,\Phi_2(\rho_0,\rho_1)
\cr&&
{}+2\pi\big(\delta^2(\rho_{12})\,\ln|\rho_{12}|+
\delta^2(\rho_{23})\ln|\rho_{23}|\big)
\,\Delta_1\,\Delta_3\,\Phi_2(\rho_1,\rho_3)
\cr&&
{}+\left|\frac1{\rho_{12}}
		+\frac1{\rho_{23}}\right|^2
\Delta_1\,\Delta_3\,\Phi_2(\rho_1,\rho_3)
\bigg]
&eq:Gexpcom} 

Now we can investigate how $G$ changes under the inversion transformation.  As
we did above for dilatations, we first write down how simple expressions
occurring in $G$ transform:
\dmath2{
\rho&\longrightarrow&\rho^{-1}\cr
\rho_{12}&\longrightarrow&-\frac{\rho_{12}}{\rho_1\rho_2}\cr
\del&\longrightarrow&-\rho^2\,\del\cr
\Delta&\longrightarrow&|\rho|^4\,\Delta          &eq:eleminv\cr
\int d^2\rho&\longrightarrow&\int d^2\rho\;|\rho|^{-4}\cr
\delta^2(\rho_{12})&\longrightarrow&|\rho_1|^4\,\delta^2(\rho_{12})\cr
\ln|\rho_{12}|&\longrightarrow&\ln|\rho_{12}|-\ln|\rho_1\rho_2|
}

Let us now look at the terms in (\ref{eq:Gexpcom}) line by line.  The terms in
the first line all get a factor of a coordinate modulus to the power of twelve.
Eight powers of coordinates come from the delta functions, the remaining four
from the Laplacians.  Since the delta functions identify all three coordinates,
we can express this factor in terms of any of them.  For reasons which will
become clear shortly, we will choose to write it as
$|\rho_1|^4|\rho_2|^4|\rho_3|^4$.

For the second line, we will make use of the explicit form of the derivative of
the conformal eigenfunction~$E^{(\nu,n)}$.  Since $\Phi_2$ is a linear
combination of these eigenfunctions, it is sufficient to treat inversion of the
expression for $\Phi_2=E^{(\nu,n)}$.  The derivatives of $E^{(\nu,n)}$ are
calculated in Appendix~\ref{app:dE}.  For the first term in the second line
of~(\ref{eq:Gexpcom}), we have:
$$
\del_1^*\,\Delta_3\,E^{(\nu,n)}(\rho_1,\rho_3)=
\frac12 \nabla_1\Delta_3\,E^{(\nu,n)}(\rho_1,\rho_3)=
(-1-n+2\,i\nu)\frac{\rho^*_{3}}{\rho^*_{1}\rho^*_{13}}
\Delta_3\,E^{(\nu,n)}(\rho_{1},\rho_{3})\,.
$$
Leaving out the $E^{(\nu,n)}$ and constant factors from the first term of the
second line, we obtain the expression we have to investigate. Taking into
account the definition of the scalar product (\ref{eq:compsprod}) and the delta
function, it can be simplified as follows:
\dmath0{
\delta^2(\rho_{12})
\left(\rho_{23},\frac{\rho^*_{3}}{\rho^*_{1}\rho^*_{13}}\right)
\frac1{|\rho_{23}|^2}\Delta_3
&=&
\delta^2(\rho_{12})
\frac12\left(\frac{\rho_3}{\rho_1}+\frac{\rho^*_3}{\rho^*_1}\right)
\frac1{|\rho_{23}|^2}\Delta_3\,.
&eq:Gline2.1}
Under inversion, this becomes:
\dmath2{
|\rho_2|^4\,\delta^2(\rho_{12})
\frac12\left(\frac{\rho_1}{\rho_3}+\frac{\rho^*_1}{\rho^*_3}\right)
\frac{|\rho_2|^2|\rho_3|^2}{|\rho_{23}|^2}\,|\rho_3|^4\,\Delta_3
&=&
|\rho_1|^2|\rho_2|^4|\rho_3|^6\,\delta^2(\rho_{12})
\frac12\left(\frac{\rho_1}{\rho_3}+\frac{\rho^*_1}{\rho^*_3}\right)
\frac1{|\rho_{23}|^2}\,\Delta_3
\cr
&=&
|\rho_1|^4|\rho_2|^4|\rho_3|^4\,\delta^2(\rho_{12})
\frac12\left(\frac{\rho_3^*}{\rho_1^*}+\frac{\rho_3}{\rho_1}\right)
\frac1{|\rho_{23}|^2}\,\Delta_3\,.
}
The last expression is the same as~(\ref{eq:Gline2.1}) with the additional
factor $|\rho_1|^4|\rho_2|^4|\rho_3|^4$.  It has been obtained by pulling a
factor $|\rho_1|^2/|\rho_3|^2$ out of the scalar product.  The second term in
the second line of~(\ref{eq:Gexpcom}) can be treated analogously.

Now come the most difficult terms, the integral terms in the third and fourth
line.  We will start with the second half of the term in the third line, the
term $1/|\rho_{10}|^2$ in the parentheses.  Under inversion, the delta function
yields a factor $|\rho_1|^4$, the integral $|\rho_0|^{-4}$, the fraction
$|\rho_1|^2|\rho_0|^2$, and the Laplacians $|\rho_0|^4|\rho_3|^4$.  From our
experience with the previous terms of the $G$ function, we expect to obtain
$|\rho_1|^4|\rho_2|^4|\rho_3|^4$ multiplied with the original term.  Therefore,
we rewrite the factors with the help of the delta function as:
$|\rho_1|^2|\rho_2|^4|\rho_3|^4\int\cdots|\rho_0|^2\cdots$.  What remains to be
done is to convert the $|\rho_0|^2$ in the integral to a $|\rho_1|^2$.  To that
end, we have to prove that
$$
\int d^2\rho_0\,(|\rho_1|^2-|\rho_0|^2)\,
\frac1{|\rho_{10}|^2}
\Delta_0\,\Delta_3\,\Phi_2(\rho_0,\rho_3)=0\,.
$$
We write the difference down in a different way:
$$
|\rho_1|^2-|\rho_0|^2=\rho_1\rho_1^*-\rho_0\rho_0^*
=\rho_1\rho_{10}^*+\rho_{10}\rho_0^*\,.
$$
For the first term, we obtain by writing out the first Laplacian and with 
partial integration:
$$
\int d^2\rho_0\frac{\rho_1}{\rho_{10}}
\,4\del_0\del_0^*\,\Delta_3\,\Phi_2(\rho_0,\rho_3)
=
-\int d^2\rho_0\underbrace{\left(\del_0^*\frac{\rho_1}{\rho_{10}}\right)}_{=0}
\,4\del_0\,\Delta_3\,\Phi_2(\rho_0,\rho_3)\,.
$$
Note that the derivative in parentheses is zero even though there is a $\rho_0$
contained in the denominator of the fraction.  Since partial integration is
based on the differentiation of a product, relation~(\ref{eq:del1/rho}) does
not apply.  The second term can be shown analogously to vanish under the
integral.  This tells us that we can replace the $|\rho_0|^2$ under the
integral by a $|\rho_1|^2$.  Hence we get the same factor as for the previous
lines, $|\rho_1|^4|\rho_2|^4|\rho_3|^4$.

The fraction term in the fourth line of~(\ref{eq:Gexpcom}) can be treated
analogously.  The logarithm terms in the third, fourth and fifth line also get
a factor $|\rho_1|^4|\rho_2|^4|\rho_3|^4$ (taking into account the delta
functions).  In addition, the logarithms give rise to additional terms
according to~(\ref{eq:eleminv}).  Because of the delta functions, they can all
be written as logarithms of $|\rho_2|^2$.  As was the case under dilatations,
they cancel among each other (cf.~(\ref{eq:Gdillogcancel})).

That leaves the last term in $G$, the sixth line in~(\ref{eq:Gexpcom}).  The
square modulus takes the following form under inversion:
\dmath2{
\left|\frac1{\rho_{12}}+\frac1{\rho_{23}}\right|^2
&\longrightarrow&
\left|\frac{\rho_1\rho_2}{\rho_{12}}+\frac{\rho_2\rho_3}{\rho_{23}}\right|^2
=
\left|\frac{\rho_1\rho_2\rho_{23}+\rho_2\rho_3\rho_{12}}{\rho_{12}\rho_{23}}
\right|^2
=
\left|\frac{\rho_1\rho_2^2-\rho_1\rho_2\rho_3+
			\rho_1\rho_2\rho_3-\rho_2^2\rho_3}{\rho_{12}\rho_{23}}
\right|^2
\cr&&
=
|\rho_2|^4
\left|\frac{\rho_1-\rho_3}{\rho_{12}\rho_{23}}\right|^2
=
|\rho_2|^4
\left|\frac{\rho_{12}+\rho_{23}}{\rho_{12}\rho_{23}}\right|^2
=
|\rho_2|^4
\left|\frac1{\rho_{12}}+\frac1{\rho_{23}}\right|^2\,.
}
Together with the factors from the two Laplacians, this again gives a factor
$|\rho_1|^4|\rho_2|^4|\rho_3|^4$.

When one convolutes $G$ with conformal fields, the factor
$|\rho_1|^4|\rho_2|^4|\rho_3|^4$ is exactly cancelled by a factor
$|\rho_i|^{-4}$ from each convolution integral.  It follows that $G$ is
conformally covariant in the sense defined above, and gluon vertices composed
of $G$-functions are conformal $n$-point functions.

\subsection{Regularising $G$}
\label{sec:Greg}

In the previous section, we have proven that $G$ is conformally covariant, that
is its conformal transformation properties are precisely the opposite of the
properties of convolution integrals over its arguments.  However, we derived
this only for the unregularised, formal expression of~$G$.  This is sufficient
for proving that the vertices constructed from it are conformal $n$-point
functions.  Since the vertices are physical quantities, any regularisation
terms must cancel.  Therefore the transformation properties of the vertices can
be derived from the formal non-regularised expression for~$G$ that we have
investigated.  For instance, the proof of the conformal invariance of the
(regularised) $2\to4$ reggeised gluon vertex in~\cite{blw} shows explicitly
that the extra terms from regularisation cancel.

As a matter of interest, we will now look into whether and how regularisation
changes $G$'s conformal transformation properties.  In this section, we will
always imply convolution integrals over $G$'s arguments.  We will see that this
is necessary to treat all terms equally.

As we have remarked in Section~\ref{sec:st_c}, the reggeised gluon trajectory
function $\beta$ contains two terms which are ultraviolet divergent under the
integral of the function~$t$.  We regularise the fraction $1/|\rho|^2$ with a
theta function $\theta(|\rho|-\epsilon)$, ie a UV cutoff.  The term
$\delta^2(\rho)\ln|\rho|$ also constitutes a logarithmic divergence which we
regularise by replacing $\ln|\rho|$ with $\ln\epsilon$.  This causes these two
divergences to cancel on Fourier transformation back into momentum space.
$\beta$ then takes the form:
\dmath2{
\beta(\rho)&=& \frac{N_cg^2}{(2\pi)^3}
\left(-2\pi\,\delta^2(\rho)\,[\ln2+\psi(1)-\ln m-\ln\epsilon]
+\frac{\theta(|\rho|-\epsilon)}{|\rho|^2}\right)\,.
&eq:beta_reg}

In the $G$ function, the terms $\delta^2(\rho)\ln\epsilon$ cancel, that is the
first terms in the third and fourth line of~(\ref{eq:Gexpcom}) cancel with the
fifth.  The result is:
\dmath{2.5}{
G(\rho_1,\rho_2,\rho_3)&=&
\frac{g^2}{(2\pi)^3}\bigg[
(2\pi)^2\delta^2(\rho_{12})\delta^2(\rho_{23})
\big({-\Delta_2}\,\Phi_2(\rho_2,\rho_2)
+(\Delta_1+\Delta_3)\,\Phi_2(\rho_1,\rho_3)\big)
\cr&&
{}+8\pi\,\delta^2(\rho_{12})
\frac{(\rho_{23},\del_1^*)}{|\rho_{23}|^2}\Delta_3\,
\Phi_2(\rho_1,\rho_3)
-8\pi\,\delta^2(\rho_{23})
\frac{(\rho_{12},\del_3^*)}{|\rho_{12}|^2}\Delta_1\,
\Phi_2(\rho_1,\rho_3)
\cr&&
{}-\delta^2(\rho_{12})\int d^2\rho_0
\frac{\theta(|\rho_{10}|-\epsilon)}{|\rho_{10}|^2}
\,\Delta_0\,\Delta_3\,\Phi_2(\rho_0,\rho_3)
\cr&&
{}-\delta^2(\rho_{23})\int d^2\rho_0
\frac{\theta(|\rho_{30}|-\epsilon)}{|\rho_{30}|^2}
\,\Delta_0\,\Delta_1\,\Phi_2(\rho_0,\rho_1)
\cr&&
{}+\left(\frac{\theta(|\rho_{12}|-\epsilon)}{|\rho_{12}|^2}
+\frac2{|\rho_{12}\rho_{23}|}
+\frac{\theta(|\rho_{23}|-\epsilon)}{|\rho_{23}|^2}\right)
\Delta_1\,\Delta_3\,\Phi_2(\rho_1,\rho_3)
\bigg]\,.
&eq:Gexpreg}
The limit $\epsilon\to0$ is implied in this expression, even though we do not
write it down explicitly.  We split up the expression in the last line since
not all of it requires regularisation.

Let us now look at what changes under conformal transformations.  The
invariance under translations and rotations is unchanged.  Under dilatations,
however, the arguments of the theta functions change.  The theta function in
the third line becomes
$\theta(|\lambda\rho_{10}|-\epsilon)=\theta(|\rho_{10}|-\epsilon/\lambda)$.
This gives rise to an additional logarithmic term according to the following
identity:
\dmath2{
&&\lim_{\epsilon\to0}\int d^2\rho_0
\left(
\frac{\theta(|\rho_{10}|-\epsilon/\lambda)}{|\rho_{10}|^2}f(\rho_0)
-\frac{\theta(|\rho_{10}|-\epsilon)}{|\rho_{10}|^2}f(\rho_0)
\right)=
\cr&&\qquad\qquad=
\lim_{\epsilon\to0}\int d^2\rho_0
\frac{\theta(|\rho_{10}|-\epsilon/\lambda)-\theta(|\rho_{10}|-\epsilon)}
{|\rho_{10}|^2}
\big(f(\rho_1)+\mathcal O(\epsilon)\big)
\cr&&\qquad\qquad=
2\pi\,f(\rho_1)\,\lim_{\epsilon\to0}\int\limits_{\epsilon/\lambda}^{\epsilon}
d|\rho_{10}|\frac1{|\rho_{10}|} 
=2\pi\,\ln\lambda\,f(\rho_1)
=2\pi\,\ln\lambda\int d^2\rho_0\,\delta^2(\rho_{10})\,f(\rho_0)\,.\qquad
&eq:tregchange}
We have used the abbreviation
$f(\rho_0)=\Delta_0\,\Delta_3\,\Phi_2(\rho_0,\rho_3)$, which is an analytic
function in the vicinity of the point $\rho_0=\rho_1$.  The domain of
integration is limited by the theta functions to a ring around $\rho_1$ between
the radii $\epsilon/\lambda$ and $\epsilon$.  Since both are small quantities,
we can approximate $f$ by its value at $\rho_1$.

So we get two logarithmic terms from the third and fourth line
in~(\ref{eq:Gexpreg}):
$-2\pi\,\ln\lambda\,(\delta^2(\rho_{12})+\delta^2(\rho_{23}))\Delta_1\,\Delta_3\,\Phi_2(\rho_1,\rho_3)$.
If we imply an integral over the arguments of $G$ (which we have whenever we
use $G$), the derivation (\ref{eq:tregchange}) applies also to the two
regularised terms in the last line.  Using the last expression
in~(\ref{eq:tregchange}), we obtain the terms
$2\pi\,\ln\lambda\,(\delta^2(\rho_{12})+\delta^2(\rho_{23}))\Delta_1\,\Delta_3\,\Phi_2(\rho_1,\rho_3)$.
That is exactly the same with the opposite sign.

The situation under inversion is more complicated.  The $|\rho_{10}|$ in the
theta function (third line in~(\ref{eq:Gexpreg})) transforms as
$|\rho_{10}|/(|\rho_1||\rho_0|)$.  The $|\rho_1|$ is a constant with respect to
the integration over~$\rho_0$, so we can use relation~(\ref{eq:tregchange}) for
$\lambda=1/|\rho_1|$.  That we can do the same thing with the $|\rho_0|$ is not
so clear.  But we can argue that it can be replaced by a $|\rho_1|$.

The theta function $\theta(|\rho_{10}|/|\rho_0|-\epsilon)$ restricts the domain
of integration to the outside of the curve on which
$|\rho_{10}|=|\rho_0|\epsilon$.  Before the inversion, the corresponding curve
was given by $|\rho_1-\rho_0|=\epsilon$.  The $\rho_0$ on this curve differs
only infinitesimally (of order~$\epsilon$) from~$\rho_1$.  Since the inversion
is a continuous mapping, this still has to be the case after the inversion.
Therefore
$$
\theta(|\rho_{10}|/|\rho_0|-\epsilon)=\theta(|\rho_{10}|-\epsilon\,|\rho_0|)
=\theta(|\rho_{10}|-\epsilon\,|\rho_1|+\mathcal O(\epsilon^2))\,.
$$
This means that replacing $|\rho_0|$ by $|\rho_1|$ in the theta function only
introduces an error of order~$\epsilon^2$, which we can ignore.  We can then
pull the $\rho_1$ out of the theta function as above, using
formula~(\ref{eq:tregchange}).%
\footnote{The authors of~\cite{blw} use an analogous procedure, without 
remarking on why it is legitimate.}

Hence we get an additional term $\delta^2(\rho_{12})\,4\pi\,\ln|\rho_1|$ from
the third line in~(\ref{eq:Gexpreg}), which came from a $t$ function.  The
fourth line (also from a $t$ function) gives an analogous term.  Implying
integrations over the arguments of~$G$, the fifth line (which came from an $s$
function) also gives two logarithms, but with a minus sign.  As was the case
for dilatations, the extra terms from $s$ and $t$ functions cancel.  So we
obtain the result that the regularised $G$ function is conformally covariant.

This is somewhat surprising given the fact that the regularised $G$ contains a
fixed scale, the UV cutoff~$\epsilon$.  One would have expected this to destroy
covariance under dilatations and the inversion, unless the terms containing
$\epsilon$ themselves cancel.  We argued at the beginning of this section that
the terms containing $\ln\epsilon$ do indeed cancel.  However, the fraction
terms do not cancel.  One can see this by writing the terms from the third and
fourth line of~(\ref{eq:Gexpreg}) in a form similar to the fifth line.  This
requires convoluting the three arguments of $G$ with conformal fields, which we
will collectively denote by~$\Phi_3$.  We can then integrate out the delta
function and rename the integration variable $\rho_0\to\rho_1$.
\dmath2{
&&-\int d^2\rho_1\,d^2\rho_2\,d^2\rho_3\,
\Phi_3(\rho_1,\rho_2,\rho_3)\,
\delta^2(\rho_{12})\int d^2\rho_0
\frac{\theta(|\rho_{10}|-\epsilon)}{|\rho_{10}|^2}
\,\Delta_0\,\Delta_3\,\Phi_2(\rho_0,\rho_3)=
\cr&&\qquad\qquad\qquad\qquad
=-\int d^2\rho_2\,d^2\rho_3\,
\Phi_3(\rho_2,\rho_2,\rho_3)\,
\int d^2\rho_0
\frac{\theta(|\rho_{20}|-\epsilon)}{|\rho_{20}|^2}
\,\Delta_0\,\Delta_3\,\Phi_2(\rho_0,\rho_3)
\cr&&\qquad\qquad\qquad\qquad
=-\int d^2\rho_1\,d^2\rho_2\,d^2\rho_3\,
\Phi_3(\rho_2,\rho_2,\rho_3)\,
\frac{\theta(|\rho_{12}|-\epsilon)}{|\rho_{12}|^2}
\,\Delta_1\,\Delta_3\,\Phi_2(\rho_1,\rho_3)
}
This is almost but not quite the negative of the first term from the fifth
line, convoluted with~$\Phi_3$.  The fifth-line term would lead to the same
integral, but with the first argument of $\Phi_3$ being $\rho_1$, not~$\rho_2$.

So why does the presence of the fixed scale $\epsilon$ not destroy scale
invariance?  The answer is that, despite all appearances to the contrary,
$\epsilon$ does not constitute a fixed scale.  Changing $\epsilon$ gives rise
to the same type of logarithmic terms as a dilatation transformation, according
to~(\ref{eq:tregchange}).  These cancel out among the different terms of~$G$,
as we have seen.  So the regularised expression for $G$ is in fact independent
of the choice of~$\epsilon$.  It just cannot be written down without it.

\chapter{The $1\to3$ pomeron vertices}
\label{chap:p3p}

\section{The irreducible $2\to6$ reggeised gluon vertex}
\label{sec:V2to6}

As explained in Section~\ref{sec:bfklintro}, the Extended Generalised Leading
Logarithmic Approximation to BFKL involves amplitudes of more than two
reggeised gluons in a colour singlet state.  The six-gluon amplitude is a sum
of a reggeising part, a partly reggeising part and a new irreducible part.  The
sum of the latter two parts fulfils an integral equation which was first
derived in~\cite{be}:
\dmath2{
\left(\omega-\sum_{i=1}^6\beta(i)\right)&&
\big(D_6^{I,R}+D_6^{I,I}\big)^{a_1a_2a_3a_4a_5a_6}(1,2,3,4,5,6)={} \cr
{}={}&&(V_{2\to6}^{a_1a_2a_3a_4a_5a_6}D_2)(1,2,3;4,5,6) \cr
&&{}+\sum f_{a_1a_2a_3}\,f_{a_4a_5a_6}\;L(1,2,3;4,5,6) \cr
&&{}+\sum d^{a_1a_2a_3a_4}\,\delta_{a_5a_6}\;I(1,2,3,4;5,6) &eq:D6eq\cr
&&{}+\sum d^{a_2a_1a_3a_4}\,\delta_{a_5a_6}\;J(1,2,3,4;5,6) \cr
&&{}+\sum K_{2\to4}^{\{b\}\to\{a\}}\otimes D_4^{I\,b_1b_2b_3b_4}
+\sum K_{2\to3}^{\{b\}\to\{a\}}\otimes D_5^{I\,b_1b_2b_3b_4b_5} \cr
&&{}+\sum K_{2\to2}^{\{b\}\to\{a\}}\otimes 
\big(D_6^{I,R}+D_6^{I,I}\big)^{a_1a_2a_3a_4a_5a_6}\,.
}
$D_6^{I,R}$ denotes the partly reggeising six-gluon amplitude, $D_6^{I,I}$ the
completely irreducible part.  The arguments of the spatial functions are
written as the indices of the reggeised gluons instead of the momenta.  So 1
stands for $\mbf k_1$ and so on.  A set of several indices (eg 134) for one
argument would mean a sum of momenta.  This notation has the advantage that it
can easily be transcribed into configuration space.  There an index stands for
the corresponding coordinate, and a set of indices means that the coordinates
are identified with a delta function.  This correspondence can be derived by
performing the Fourier transform of a function taking a sum of momenta as its
argument (see Section~\ref{sec:thisisG}).  An argument without any index,
denoted by a hyphen ``$-$'', would mean a zero argument in momentum space and
an integral over the respective coordinate in position space.

The first term on the right hand side of~(\ref{eq:D6eq}) is a $2\to6$
(reggeised) gluon vertex which is local in rapidity.  Since it cannot be
expressed in terms of smaller vertices, it constitutes a new element of the
conjectured conformal field theory of unitarity corrections.  It has the
following form:
\dmath2{
(V_{2\to6}^{a_1a_2a_3a_4a_5a_6}D_2)(1,2,3;4,5,6)=&&
	d_{a_1a_2a_3}d_{a_4a_5a_6}(WD_2)(1,2,3;4,5,6) \crl{eq:V2to6}
	&+& d_{a_1a_2a_4}d_{a_3a_5a_6}(WD_2)(1,2,4;3,5,6)+\ldots
}
The colour structure is given by the symmetric structure constants $d_{abc}$ of
$su(3)$.  The sum runs over all ten possibilities of dividing six momenta and
corresponding colour indices into two groups of three.  The function $(WD_2)$
is a convolution of a function $W$ and a pomeron amplitude $D_2$.  The
amplitude $D_2$ represents the pomeron attached to the vertex from above.
Since this vertex occurs in unitarity corrections to BFKL, there is always a
pomeron coming from above, and we will always consider the combination $(WD_2)$
rather than $W$ alone.\footnote{So, strictly speaking, this is not a $2\to6$
gluon vertex but a 1~pomeron${}\to{}$6 gluon vertex.  However, we will follow
the nomenclature in~\cite{carlophd,be}.}  $(WD_2)$ can be expressed in terms of
the function $G$ introduced in Section~\ref{sec:G} or in (quite many) terms of
the functions $a$, $b$, $c$, $s$ and~$t$.  Here is its representation in terms
of $G$:
\dmath2{
(WD_2)(1,2,3;4,5,6)&=&
\frac{g^4}8 \sum_{M\in\mathcal P(\{1,\ldots,6\})}
(-1)^{\#M}\;G(123\backslash M,M,456\backslash M)\,.
&eq:WD2}
This sum runs over all subsets $M$ of the set of indices. The notation
``123$\backslash M$'' means the indices 1, 2 and 3 except those contained
in~$M$.  $\#M$~is the number of indices contained in~$M$.  This sum and the
symmetry of the function~$G$ under exchange of its first and third argument
cause the function $(WD_2)$ to be symmetric under all permutations of its first
three and its last three arguments, and under exchange of the two groups of
arguments.  Together with the sum in~(\ref{eq:V2to6}), this makes the $2\to6$
vertex symmetric under arbitrary permutations of the gluons.

The representation~(\ref{eq:WD2}) is easily generalised to a different (even)
number of reggeised gluons.  In fact the $2\to4$ reggeised gluon vertex can be
written as a sum over functions $(VD_2)$ which are defined analogously
to~$(WD_2)$~\cite{bw}.  The sum runs over all (three) possibilities of
dividing four indices into two groups of two.  It is conjectured that there
exist higher $2\to2n$ reggeised gluon vertices which are also new elements of
the field theory of unitarity corrections and which are of analogous
form~\cite{ew2}.

\section{The $1\to3$ pomeron vertex from the irreducible $2\to6$ gluon vertex}
\label{sec:Wp3p}

\subsection{Projecting the $2\to6$ reggeised gluon vertex}

Since the pomeron is a bound state of two reggeised gluons, a $1\to3$ pomeron
vertex may be obtainable by projection from the $2\to6$ reggeised gluon vertex.
This is analogous to the projection of the $2\to4$ reggeised gluon vertex to
obtain a $1\to2$ pomeron vertex~\cite{lotterphd,korconf}.  It is not clear a
priori that the projection is not zero, in which case no pomeron vertex can be
derived from the $2\to6$ gluon vertex.  The projection is performed by
convoluting the spatial part of the gluon vertex with three pomeron amplitudes
and by contracting the colour tensors with three Kronecker deltas with
corresponding indices.  Because of the symmetry of the vertex, it does not
matter which pomeron has which arguments resp.\ colour indices.  We will choose
one pomeron to have the arguments 1 and~2, the second 3 and~4 and the last 5
and~6.

These three pomerons shall have the external coordinates $\rho_a$, $\rho_b$ and
$\rho_c$, respectively.  The fourth pomeron, whose amplitude has been
defined into the function $G$ and its component functions, shall have external
coordinate $\rho_d$.  

We will restrict ourselves to the ground state of the pomeron.  This amounts to
setting the full amplitude $\Phi_2=E^{(\nu,0)}$, both for the three pomerons
attached to the vertex from below and the one contained in~$G$.  We will not
write down the external coordinates as arguments to save space; which amplitude
is which will be clear from the conformal dimensions $\nu_a$, $\nu_b$, $\nu_c$
and~$\nu_d$.

The pomeron with index $d$ is special in that a corresponding amplitude is
already defined into the vertex $V_{2\to6}$ and its component functions.  This
amplitude is a $D_2$ which contains a pomeron impact factor (see
Equation~\ref{eq:D2}).  However, the principle of Regge factorisation states
that impact factors are independent of the evolution of the amplitude rapidity.
This allows us to discard the impact factor and attach an amplitude $\Ez{}$,
thereby obtaining a pomeron vertex which is independent of the impact factor.
This $\Ez{}$ corresponds not to $D_2$ but to the non-amputated
amplitude~$\Phi_2$.

Hence our projection of the vertex can be written in the following way:
\dmath2{
V_{\pom\to3\pom}&=&
\delta_{a_1a_2}\,\delta_{a_3a_4}\,\delta_{a_5a_6}\,
\Big[(V_{2\to6}^{a_1a_2a_3a_4a_5a_6}D_2)(1,2,3;4,5,6)\Big]_{\Phi_2\sslash\Ez d}
\otimes{}\cr&&{}\qquad\qquad\qquad
\otimes\Ez a^*(1,2)\,\Ez b^*(3,4)\,\Ez c^*(5,6)
}

\subsection{The colour structure}

The colour structure of the pomeron is $\delta_{ab}$.  Because the symmetric
structure constants $d_{abc}$ vanish when two of their indices are contracted,
this has the consequence that less than half of the terms in the sum
in~(\ref{eq:V2to6}) contribute to the $1\to3$ pomeron vertex.  The only
remaining terms are those where the two colour indices of each pomeron are
contracted with different $d$ tensors.  Since we chose as the pomerons'
arguments (12), (34) and (56), this leaves the following four permutations of
(colour, momentum or coordinate) indices:
\dmath2{
(1,3,5;2,4,6),\quad(1,3,6;2,4,5),\quad(1,4,5;2,3,6),\hbox{ and }(1,4,6;2,3,5)
\,.&&&eq:projind}

Each of these permutations gives rise to the same colour factor because of the
symmetry of $d_{abc}$.  Since each $d_{abc}$ tensor is contracted with one
index of each of the $\delta$s, the result is the contraction of two $d$
tensors.  Explicitly for the first permutation:
$$
d_{a_1a_3a_5}d_{a_2a_4a_6}\delta_{a_1a_2}\delta_{a_3a_4}\delta_{a_5a_6}=
d_{a_1a_3a_5}d_{a_1a_3a_5}=\frac{40}3\,.
$$

\subsection{The spatial part}
\label{sec:p3pspatial}

\subsubsection{Introduction}

The lion's share of the work to be done for the projection is sorting out all
the terms of the spatial part of $V_{2\to6}$.  This work is complicated by the
fact that the $1\to3$ pomeron vertex cannot be expressed in terms of~$G$.
Because $\Ez{}(\mbf\rho,\mbf\rho)=0$, some of the component functions of $G$
vanish on projection, others don't.

The brute-force way of extracting those that remain would be to write out
$(WD_2)$ in terms of the functions $a$, $b$, $c$, $s$ and~$t$ and cancel those
that vanish.  This is indeed how it was first done, but there is a more
systematic way.  First, we observe that the four permutations
(\ref{eq:projind}) remaining after the contraction of colour tensors all lead
to the same term.  The reason for this is that they differ only by the exchange
of two coordinates of the same pomeron wave function, which is symmetric.
Hence it is sufficient to consider only one of the permutations and multiply
the result by four.

The second important observation is that the conformal eigenfunctions~$\Ez{}$
vanish if their two coordinate arguments are equal.  Therefore terms in which
both argument indices of the same pomeron occur in the same argument of~$G$
vanish under projection.  This is the key to finding out which terms remain
after projection.  We now have a look at the first permutation
in~(\ref{eq:projind}).  The spatial part of the vertex for this permutation has
the following form:
\dmath2{
(WD_2)(1,3,5;2,4,6)&=&
\frac{g^4}8 \sum_{M\in\mathcal P(\{1,\ldots,6\})}
(-1)^{\#M}\;G(135\backslash M,M,246\backslash M)\,.
&eq:WD2_135}
We can see immediately that a term from this sum vanishes if two coordinates of
the same pomeron are contained in the middle argument of~$G$.  The first and
last argument of~$G$ contain only coordinates or momenta belonging to different
pomerons and hence cannot cause the term to vanish under projection.  Therefore
all the terms for which $M$ does not contain both indices of a pomeron remain:
\dmath2{
(WD_2)_{\pom\to3\pom}(1,3,5;2,4,6)&=&
\frac{g^4}8 \sum_{{M\in\mathcal P(\{1,\ldots,6\}) \atop 
M\not\supset\{1,2\}\wedge M\not\supset\{3,4\}\wedge M\not\supset\{5,6\}}} 
(-1)^{\#M}\;G(135\backslash M,M,246\backslash M)\,.
&eq:WD2proj}
To say it in words: $M$ may contain either 1 or 2 or none of these two indices,
and 3 or 4 or neither of them, and 5 or 6 or neither.  This yields
$3\cdot3\cdot3=27$ terms.  However, two of these vanish because of a property
of the $G$ function: It vanishes when its first or third argument is empty, see
Equation~\ref{eq:G-23}.  This kills the terms for $M=\{1,3,5\}$ and
$M=\{2,4,6\}$.  Hence 25 terms remain.

The function $G$ is composed of functions with one, two and with three
arguments.  The function with one argument ($c$) does not contribute at all to
the pomeron vertex.  Since its argument always contains all indices, $c$ does
not survive the projection.  The functions with two arguments, $b$ and $t$, are
also components of the BFKL kernel.  We will discuss them in the next part of
this section.  The functions with three arguments, $a$ and $s$, will be dealt
with after that.

\subsubsection{The BFKL term}

The functions $b$ and $t$ survive only if they are convoluted with pomerons in
one specific way, namely if each pomeron's arguments are convoluted with
different arguments of $b$ or~$t$.  This means that both arguments of $b$
resp.\ $t$ contain three indices, one from each pomeron.  First of all, this is
the case for the term in (\ref{eq:WD2}) where $M=\varnothing$.  

Besides, there are terms in the sum~(\ref{eq:WD2proj}) where $M$ contains only
indices from the first or only from the second group of arguments of~$(WD_2)$.
Since one argument of $b$ and $t$ contains two arguments of $G$ taken together
(see~(\ref{eq:G_symb})), one of the two $b$ and $t$ functions would have
momenta of different pomerons in each argument and hence would contribute.
However, by replacing $M$ by its complement with respect to the group of
arguments of $(WD_2)$ of which it is a subset, an identical term with opposite
sign can be obtained (unless $M$ is identical to the whole group of arguments,
but these terms vanish, see above).  Therefore all such terms cancel out.

To illustrate this, let us look at an example.  We again consider the first of
the permutations in~(\ref{eq:projind}) with the spatial
function~(\ref{eq:WD2_135}).  For example, $M=\{1,3\}$ contains only indices
from the first three arguments of~$(WD_2)$.  The corresponding term in the sum
contains the following $G$ function:
\dmath{0.5}{
G(5,13,246)&=&
\frac{g^2}2\big[2\,c(123456)-2\,b(135,246)-2\,b(12346,5)+2\,a(13,5,246)
\cr&&\qquad
{}+t(135,246)+t(12346,5)-s(13,5,246)-s(13,246,5)\big]\,.
}
The terms with more than three indices in one argument, such as $c$ and the
second $b$ and~$t$ functions, vanish because this identifies two coordinates of
a pomeron.  For now, we ignore the functions $a$ and~$s$; we will deal with
them in the next section.  The remaining terms, $-2\,b(135,246)$ and
$t(135,246)$ do not vanish.  However, the same terms occur in the term for the
complement of $M$, $M'=\{1,3,5\}\backslash M=\{5\}$.  The corresponding $G$
function is:
\dmath{0.5}{
G(13,5,246)&=&
\frac{g^2}2\big[2\,c(123456)-2\,b(135,246)-2\,b(2456,13)+2\,a(5,13,246)
\cr&&\qquad
{}+t(135,246)+t(2456,13)-s(5,13,246)-s(5,246,13)\big]\,.
}
Again, $-2\,b(135,246)$ and $t(135,246)$ remain.  But this term has the
opposite sign because of the prefactor~$(-1)^{\#M}$ in~(\ref{eq:WD2proj}).  So
the $t$ and $b$ functions for $M\not=\varnothing$ cancel out.\footnote{The
deeper reason for this is that $6/2=3$ is odd.  It can be conjectured that the
same mechanism of cancellation occurs in the projection of higher $1\to n$
pomeron vertices from $2\to2n$ reggeised gluon vertices for odd~$n$.}

Only for $M=\varnothing$ the functions $b$ and~$t$ contribute to the $1\to3$
pomeron vertex.  Knowing that $G$ becomes a derivative of the BFKL kernel when
its middle momentum argument vanishes (Equation~\ref{eq:G_BFKL}), we can
immediately write out this term.  Since there is one such term for each of the
four permutations whose colour structure does not vanish, we have to multiply
the result by~4.  We obtain the following BFKL term:
\dmath2{
V_{\pom\to3\pom}^{BFKL}&=&
\frac{40}3 \frac{g^4}8 4
\int d^2\rho_1\,d^2\rho_2\,
\Ez	a^*(\rho_1,\rho_2)\,
\Ez	b^*(\rho_1,\rho_2)\,
\Ez	c^*(\rho_1,\rho_2)\cdot{}
\crd&&\qquad\qquad\qquad
{}\cdot\frac1{N_c}
\Delta_{1}\,\Delta_{2}\,\big(K_{BFKL}\otimes \Ez d\big)(\rho_1,\rho_2)
\cr&=&
\frac{40}3 \frac{g^4}8 4
\frac{\chi(\nu_d,0)}{N_c}
\int \frac{d^2\rho_1\,d^2\rho_2}{|\rho_{12}|^4}
\Ez	a^*(\rho_1,\rho_2)\,
\Ez	b^*(\rho_1,\rho_2)\,
\Ez	c^*(\rho_1,\rho_2)\,
\Ez d(\rho_1,\rho_2)
\cr&=&
\frac{40}3 \frac{g^6}{16} 4 \frac2{(2\pi)^3}
\xi(\nu_d)\,(4\,\nu_d^2+1)^2
\cdot{}
&eq:p3pbfkl\cr\crd
&&{}\cdot
\int \frac{d^2\rho_1\,d^2\rho_2}{|\rho_{12}|^4}
\cEz12a\cEz12b\cEz12c\eEz12d\,.
}
We have evaluated the convolution with the BFKL kernel using the BFKL
eigenvalue~(\ref{eq:bfkleigen}) and put in the derivative of the wave
function~(\ref{eq:LapLapE}).

In the last step, we have rewritten the BFKL eigenvalue $\chi$ as a function
$\xi$ which was already used in~\cite{lotterphd}.  It differs from $\chi$ only
in that it does not contain factors of $N_c$ and the coupling constant:
\dmath2{
\xi(\nu)&=&2\pi
\left[2\,\psi(1)-\psi\!\left(\frac12+i\nu\right)
-\psi\!\left(\frac12-i\nu\right)\right]
=\frac{8\pi^3}{N_c g^2}\chi(\nu,0)\,.
&eq:xi}

\begin{figure}
\centre{\input{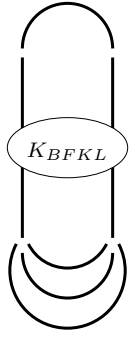}}
\caption{Graphical representation of the part of the $1\to3$ pomeron vertex 
		which contains the BFKL kernel. The half-circular lines represent 
		pomeron wave functions.}
\label{fig:p3pbfkl}
\end{figure}

Figure~\ref{fig:p3pbfkl} shows a graphical representation of this part of the
vertex.

\subsubsection{The $\alpha$ terms}

\begin{figure}
\centre{\Large\input{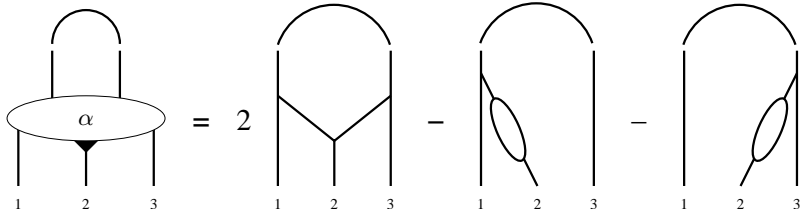}}
\caption{Graphical representation of Equation~\ref{eq:alpha}.  The arcs
		represent pomeron wave functions defined into the functions $a$, $s$
		and $\alpha$.  They were drawn as ellipses labelled ``$D_2$'' in
		Figure~\ref{fig:abcst} but we will drop that here since we will have
		many more of them.}
\label{fig:alpha}
\end{figure}

What remains now are the functions with three arguments, $a$ and~$s$.  Since
they have the same properties (both vanish only when their first argument, ie
the second argument of $G$, is empty), they always occur together, and the
result of the projection can be expressed in terms of the function
\dmath0{
\alpha(2,1,3)&:=&2a(2,1,3)-s(2,1,3)-s(2,3,1)\,.
&eq:alpha}
$\alpha$ is symmetric in its last two arguments.  Figure~\ref{fig:alpha} shows
a graphical representation of this equation.  The middle leg of $\alpha$, which
corresponds to its first argument, is marked to remind you of the fact that it
is symmetric in the other two arguments, but not this one.

We will now divide the remaining terms from (\ref{eq:WD2proj}) into groups that
lead to the same type of integral after convolution with the wave functions.
There are 24 terms since the case $M=\varnothing$ has already been dealt with
above.  The most obvious classification criterion is the number of elements $M$
contains, ie the number of momenta or coordinates in the first argument of
$\alpha$.  There are six terms for which $M$ has one element, twelve for which
it has two and six with three elements.  (Normally there would be eight in the
last class, but for two of them $M$ contains a whole group of arguments and $G$
vanishes, see above.)  It makes sense that the numbers of terms in all classes
are divisible by six.  Six is the number of permutations of three objects.
Because of the high symmetry of the $2\to6$ reggeised gluon vertex, there are
terms with every pomeron coupling to every pair of arguments of the
function~$G$.  Terms which differ only by permutations of the labels
($={}$external coordinates) of the pomerons belong to the same class since they
lead to the same type of integral.  The terms from different classes contribute
with different signs due to the factor $(-1)^{\#M}$ in~(\ref{eq:WD2_135}).  The
class in which $M$ contains two elements has positive sign, the others negative
sign.

\begin{figure}
\centre{\Large\input{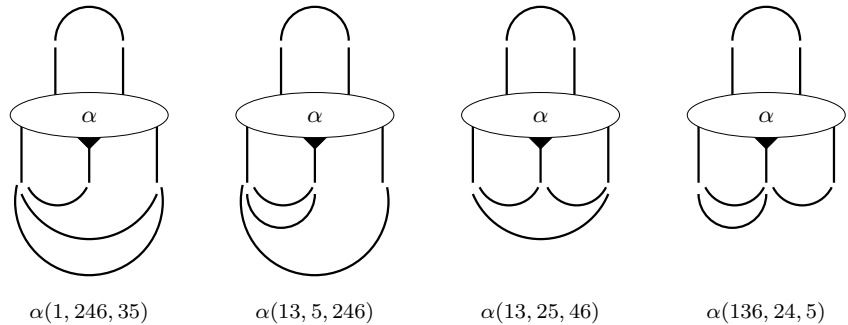}}
\caption{Graphical representation of the classes of terms with the $\alpha$ 
		function.  One representative term is printed underneath each graph
		(assuming the pomerons (12), (34) and (56)).  The half-circular lines 
		represent pomeron wave functions.}
\label{fig:p3palpha}
\end{figure}

The class in which $M$ contains two elements can be subdivided further.  It
contains those terms for which the first argument of $\alpha$ contains two
indices.  It makes a difference whether the second and third arguments also
contain two or whether one has three and the other only one.  With this
subdivision, we have four classes in total, with six terms each.  There is an
intuitive interpretation of these classes which is presented graphically in
Figure~\ref{fig:p3palpha}.  The classes differ in how many pomerons are
attached to each leg of the $\alpha$ function (ie how many are convoluted with
the corresponding coordinate in position space; or how many there are whose
momentum occurs in the corresponding sum of momenta in momentum space).

Bearing in mind that $\alpha$ is symmetric in its last two arguments but not
its first, it is easy to see that the four classes are indeed disjoint.  What
is more, they are also complete.  It is impossible to construct a term which
does not belong to one of the classes (taking the properties of $\alpha$ and
the pomeron amplitude into account). 

\subsubsection{A closer look at the $\alpha$ terms}

Before writing down the integrals containing $\alpha$ which make up the largest
part of the $1\to3$ pomeron vertex, let us recall the configuration space form
of $\alpha$ which was already calculated in the previous chapter
(Equation~\ref{eq:2a-s-s}).  In complex notation, it can be written as:
\dmath2{
\alpha(\rho_2,\rho_1,\rho_3)
&=&\frac2{(2\pi)^3}\bigg[
2\pi\big(\delta^2(\rho_{12})-\delta^2(\rho_{23})\big)
\ln\frac{|\rho_{12}|}{|\rho_{23}|}
+\left|\frac{\rho_{13}}{\rho_{12}\rho_{23}}\right|^2
\bigg]
\Delta_1\,\Delta_3\,\Phi_2(\rho_1,\rho_3)\,.
&eq:alpha_c}

The two delta functions identify the second or third coordinate argument of $G$
with its first.  When this function is convoluted with further pomeron wave
functions as illustrated in Figure~\ref{fig:p3palpha}, one or both of the delta
function terms vanish after integration because the delta functions identify
two coordinates of a pomeron.  For the latter two groups both delta function
terms vanish (since there are pomerons between the middle and both outer legs
of $\alpha$), for the others only one.

Several terms of the function $\alpha$ are potentially divergent under an
integral.  The terms $\delta^2(\rho_{12})\,\ln|\rho_{12}|$ are obviously
dangerous, but also the fraction may cause a logarithmic divergence with
respect to the integration over $\rho_{12}$ or~$\rho_{23}$.  The pomeron wave
function for zero conformal spin $\Ez{}(\rho_1,\rho_2)$ contains one power of
$|\rho_{12}|$ and can hence regularise these divergences.  Eliminating the
delta functions requires that a pomeron is attached between the middle and one
of the outer legs of~$\alpha$.  Therefore the terms from the last two groups in
Figure~\ref{fig:p3palpha} are finite while in the others one divergence remains
and has to be regularised.

We will now write down the integrals resulting from each of the four groups of
terms.  We will denote them (and the groups) with an upper index equal
to~$\#M$, the number of elements in~$M$.  The two groups with $\#M=2$ will be
distinguished by the minimal number of coordinates in the other arguments of
the $G$ function and denoted by ``(2,1)'' and ``(2,2)''.  The divergent terms
are regularised in the same way as the function $G$ in Section~\ref{sec:Greg}.
The logarithm next to the delta function is replaced by $\ln\epsilon$, and the
fraction is regularised with a theta function.  For the first group ($\#M=1$),
we get:
\dmath2{
V_{\pom\to3\pom}^{(1)}&=&
-\frac{40}3 \frac{g^6}{16} 4 \sumabc
\int d^2\rho_1\,d^2\rho_2\,d^2\rho_3\,d^2\rho_4\,d^2\rho_5\,d^2\rho_6\cdot{}
\crd&&\qquad\qquad
{}\cdot
\Ez a^*(\rho_1,\rho_2)\,\Ez b^*(\rho_3,\rho_4)\,\Ez c^*(\rho_5,\rho_6)
\;\delta^2(\rho_{24})\,\delta^2(\rho_{26})\,\delta^2(\rho_{35})\cdot{}
\crd&&\qquad\qquad
{}\cdot
\frac2{(2\pi)^3}
\left[2\pi\,\delta^2(\rho_{13})\,\ln\frac\epsilon{|\rho_{12}|}
+\frac{|\rho_{23}|^2}{|\rho_{12}|^2|\rho_{13}|^2}\theta(|\rho_{13}|-\epsilon)
\right]\,\Delta_2\,\Delta_3\,\Ez d(\rho_2,\rho_3)
\cr&=&
-\frac{40}3 \frac{g^6}{16} 4
 \frac2{(2\pi)^3}\sumabc\int d^2\rho_1\,d^2\rho_2\,d^2\rho_3\,
\Ez a^*(\rho_1,\rho_2)\,\Ez b^*(\rho_2,\rho_3)\,\Ez c^*(\rho_2,\rho_3)\cdot{}
\crd&&\qquad\qquad\qquad
{}\cdot
\left[2\pi\,\delta^2(\rho_{13})\,\ln\frac\epsilon{|\rho_{12}|}
+\frac{|\rho_{23}|^2}{|\rho_{12}|^2|\rho_{13}|^2}\theta(|\rho_{13}|-\epsilon)
\right]
\,\Delta_2\,\Delta_3\,\Ez d(\rho_2,\rho_3)\,.
}
The fraction $40/3$ is the colour factor, the factor 4 the combinatorial
factor from the four permutations~(\ref{eq:projind}).  The factor $g^6/16$ is
the product of the prefactors of $(WD_2)$ and~$G$.  We have rewritten the sum
over the six terms of this group as a sum over the permutations of the pomeron
indices which is equivalent to it by way of renaming the integration variables.
It is denoted by $\sumabc$.  We can now insert the explicit form of the pomeron
wave functions.  This allows us to evaluate the double Laplacian which was
calculated in Appendix~\ref{app:dE} (Equation~\ref{eq:LapLapE}).
\dmath3{
V_{\pom\to3\pom}^{(1)}&&=\ldots =
-\frac{40}3 \frac{g^6}{16} 4 \frac2{(2\pi)^3}(4\,\nu_d^2+1)^2
\sumabc\int d^2\rho_1\,d^2\rho_2\,d^2\rho_3\,\cdot
\crd&&\qquad\qquad\qquad{}\cdot
\cEz12a\,\cEz23b\,\cEz23c\cdot{}
\crd&&\qquad\qquad\qquad{}\cdot
\left[2\pi\,\delta^2(\rho_{13})\,\ln\frac\epsilon{|\rho_{12}|}
+\frac{|\rho_{23}|^2}{|\rho_{12}|^2|\rho_{13}|^2}\theta(|\rho_{13}|-\epsilon)
\right]
\frac1{|\rho_{23}|^4}\eEz23d
\cr&&\hskip -5mm
=-\frac{40}3 \frac{g^6}{16} 4 \frac2{(2\pi)^3}(4\,\nu_d^2+1)^2
\sumabc\int \frac{d^2\rho_2\,d^2\rho_3}{|\rho_{23}|^4}
\cEz23b\cEz23c\hskip -7mm \cdot{}
&eq:p3p1\cr
&&\hskip -5mm
{}\cdot\eEz23d\int d^2\rho_1\,
\left[2\pi\,\delta^2(\rho_{13})\,\ln\epsilon
+\frac{|\rho_{23}|^2}{|\rho_{12}|^2|\rho_{13}|^2}
\theta\!\left(\frac{|\rho_{13}|}{|\rho_{12}|}-\epsilon\right)\right]
\,\cEz12a
}
We have converted a logarithm into a factor inside the theta function.  We had
done the same thing in Section~\ref{sec:Greg} when deriving the conformal
transformation properties of the regularised $G$ function.  Here we have
applied relation~(\ref{eq:tregchange}) with respect to the integration
over~$\rho_1$, for $\lambda=1/|\rho_{12}|$ and $f(\rho_1)=\Ez
a^*(\rho_1,\rho_2)$.  As we argued in Section~\ref{sec:Greg},
(\ref{eq:tregchange}) can be applied even when the factor $\lambda$ contains
the integration variable.

It is easily shown that $V_{\pom\to3\pom}^{(1)}$ alone has the
transformation properties of a conformal four-point function.  Invariance under
translations and rotations is trivial.  Invariance of the integral operator
under dilatation with a factor $\lambda$ is not hard to prove: The three
integrals give a factor $\lambda^6$, the denominator $|\rho_{23}|^4$ a
factor~$\lambda^{-4}$.  Both the delta function and the fraction in the
rectangular brackets give a factor~$\lambda^{-2}$.%
\footnote{
The theta function is invariant under dilatations.  This was the main reason
for absorbing the logarithm into the theta function.  In our derivation of the
transformation properties of the regularised $G$ function, we did not do that
because the logarithm terms had cancelled with others from the component
functions $b$ and~$t$.  We aimed at the simplest possible expression for $G$.}

There remains inversion.  The first two integrals together with the denominator
are invariant.  So is the $\rho_1$ integral together with the delta function
and the fraction.  However, the theta function gives rise to an extra
logarithmic term inside the brackets.  Using relation~(\ref{eq:tregchange})
with respect to the $\rho_1$ integration for $\lambda=|\rho_2/\rho_3|$, it is
computed to $|\rho_{23}|^2/|\rho_{12}|^2 \cdot
2\pi\,\delta(\rho_{13})\ln|\rho_2/\rho_3|$.  After performing the $\rho_1$
integration, everything except this logarithm is symmetric under the exchange
$\rho_2\leftrightarrow\rho_3$.  Hence the additional term is overall
antisymmetric and vanishes under the integration over $\rho_2$ and $\rho_3$.

The integrals resulting from the other groups of terms are derived analogously.
In all cases, the remaining integration variables are $\rho_1$, $\rho_2$ and
$\rho_4$, which we will rename~$\rho_3$.  The second group, for which $\#M=2$
and one of the arguments of $G$ contains only one coordinate, also requires
regularisation.  We deal with that as shown for $V_{\pom\to3\pom}^{(1)}$,
including absorbing the logarithm into the theta function.  Like
$V_{\pom\to3\pom}^{(1)}$, the vertices belonging to the other groups are also
conformal four-point functions on their own.  Here they are:
\dmath3{
V_{\pom\to3\pom}^{(2,1)}&=&
\frac{40}3 \frac{g^6}{16} 4 \frac2{(2\pi)^3}(4\,\nu_d^2+1)^2
\sumabc \int d^2\rho_1\,d^2\rho_2\,d^2\rho_3\,
\frac1{|\rho_{12}|^4}
\cEz12a\cEz13b\cdot{}
\crd&&
\hskip -5mm {}\cdot\cEz13c\left[2\pi\,\delta^2(\rho_{23})\ln\epsilon+
\frac{|\rho_{12}|^2}{|\rho_{13}|^2|\rho_{23}|^2}
\theta\!\left(\frac{|\rho_{23}|}{|\rho_{13}|}-\epsilon\right)\right]
\eEz12d
&eq:p3p21\cr
V_{\pom\to3\pom}^{(2,2)}&=&
\frac{40}3 \frac{g^6}{16} 4 \frac2{(2\pi)^3} (4\,\nu_d^2+1)^2
\sumabc\int d^2\rho_1\,d^2\rho_2\,d^2\rho_3\,
\frac1{|\rho_{12}|^2|\rho_{13}|^2|\rho_{23}|^2}
\cEz12a\cdot{}
\crd
&&\qquad\qquad\qquad\qquad {}\cdot\cEz13b\cEz23c\eEz23d
&eq:p3p22\cr
V_{\pom\to3\pom}^{(3)}&=&
-\frac{40}3 \frac{g^6}{16} 4 \frac2{(2\pi)^3} (4\,\nu_d^2+1)^2
\sumabc\int d^2\rho_1\,d^2\rho_2\,d^2\rho_3\,
\frac1{|\rho_{12}|^2|\rho_{13}|^2|\rho_{23}|^2}
\cEz12a\cdot{}
\crd
&&\qquad\qquad\qquad\qquad {}\cdot\cEz13b\cEz13c\eEz23d 
&eq:p3p3}

\subsection{Simplifying the spatial part and the function $\Psi$}

\subsubsection{The $1\to3$ pomeron vertex as a conformal four-point function}

In the previous section we have learnt that each of the five terms of the
$1\to3$ pomeron vertex is a conformal four-point function.  This means that
they are of the form~(\ref{eq:4point}).  One small difference arises from the
fact that we used complex conjugated wave functions for the pomerons attached
to the vertex from below.  This is equivalent to replacing $\nu\to-\nu$ and
$n\to-n$.  Besides, the formula~(\ref{eq:4point}) can be simplified somewhat
because we use only ground state wave functions with $n=0$, and therefore
$h=\bar h=i\nu$.  The resulting formula for our four-point function reads:
\dmath2{
\left\langle \Ez a^*(\rho_a)\,\Ez b^*(\rho_b)\,\Ez c^*(\rho_c)\,\Ez d(\rho_d) 
\right\rangle
&=&\Psi(x,x^*;\{\nu_i\})\prod_{i<j}
|\rho_{ij}|^{2i\left(-\tilde\nu_i-\tilde\nu_j+\frac13\sum_k \tilde\nu_k\right)}
\,,\qquad\quad\quad
\crl{eq:our4point}
x&=&\crossrat abcdacbd\,,} where $\tilde\nu_i$ is $-\nu_i$ for $i=a,b,c$ and
$\nu_d$ for $i=d$.  The single argument of the conformal eigenfunctions is the
external coordinate of the pomeron state.  We have omitted the coordinates of
the reggeised gluons contained in the pomerons since they are integrated over
to obtain the expectation value on the left-hand side.  By writing down all the
factors, one can obtain the following explicit form of the four-point function:
{
\def\fourrat#1#2#3#4#5#6#7#8#9{
\left(\frac{|\rho_{a#1}|^2|\rho_{#2#3}|^2|\rho_{#4d}|^2}
{|\rho_{#5#6}||\rho_{#5#7}||\rho_{#6#7}|}\right)^{-\frac13#8\frac23i\nu_{#9}}}
\dmath2{
\Big\langle \Ez a^*(\rho_a)&&\,\Ez b^*(\rho_b)\,\Ez c^*(\rho_c)\,\Ez d(\rho_d) 
\Big\rangle ={}&eq:our4exp\cr
{}=\Psi&&(x,x^*;\{\nu_i\})
\fourrat bacabcd+a  \fourrat bbcbacd+b  \cdot{}
\cr&&\qquad\qquad{}\cdot
\fourrat cbccabd+c \fourrat dbdcabc-d\,.
}
}
Now we rewrite the ratios in brackets as follows (shown here for the first
expression):
\dmath2{
\frac{|\rho_{ab}|^2|\rho_{ac}|^2|\rho_{ad}|^2}
{|\rho_{bc}||\rho_{bd}||\rho_{cd}|}&=&
\left|\crossrat abcdacbd\right|
\frac{|\rho_{ab}||\rho_{ac}|^3|\rho_{ad}|^2}
{|\rho_{bc}||\rho_{cd}|^2}=
\left|\crossrat abcdacbd\right|
\left|\crossrat abcdadbc\right|
\frac{|\rho_{ac}|^3|\rho_{ad}|^3}{|\rho_{cd}|^3}\,.
&eq:rewrite4frac}
The first of the anharmonic ratios appearing here is what we defined as $x$ in
Equation~\ref{eq:cross4}, the second is~$x/(1-x)$.  For rewriting the next two
factors in (\ref{eq:our4exp}) we use (\ref{eq:rewrite4frac}) with the indices
$a$, $b$ and $c$ permutated in order: $a\to b\to c$.  This has the consequence
that the constant exponents $-\frac13$ of the first three anharmonic ratios
cancel.  From the last factor, we pull out the two anharmonic ratios which have
$\rho_{ad}\rho_{bc}$ in the numerator, so that the variable $\rho_a$ does not
occur any more in the remaining part.
{\def\fourthree#1#2#3#4#5#6#7#8{%
{\left(\frac{|\rho_{#1#2}||\rho_{#3#4}|}{|\rho_{#5#6}|}\right)^{-1#72i\nu_{#8}}
}}
\dmath2{
\Big\langle\ldots\Big\rangle&=&
\Psi(x,x^*;\{\nu_i\})\;
\left|\crossrat abcdacbd\right|^{\frac23i(\nu_a-\nu_c)}
\left|\crossrat abcdadbc\right|^{\frac13+\frac23i(\nu_a-\nu_b+\nu_d)}
\left|\crossrat adbcacbd\right|^{-\frac13+\frac23i(\nu_b-\nu_c-\nu_d)}
\cdot{}\cr&&{}\cdot
\fourthree acadcd+a \fourthree abbdad+b \fourthree bccdbd+c \fourthree bdcdbc-d
\qquad\cr&=&
\Psi(x,x^*;\{\nu_i\})\;\;
|x|^{\frac13+\frac23i(2\nu_a-\nu_b-\nu_c+\nu_d)}\;
|1-x|^{-\frac23+\frac23i(-\nu_a+2\nu_b-\nu_c-2\nu_d)}
\cdot{}\cr&&{}\cdot
\fourthree acadcd+a \fourthree abbdad+b \fourthree bccdbd+c \fourthree bdcdbc-d
}}
To obtain the $\Psi$ function corresponding to our four-point function, we
solve this equation for~$\Psi$.  We get the formula:
%
\dmath2{
\Psi&=&
|x|^{-\frac13+\frac23i(-2\nu_a+\nu_b+\nu_c-\nu_d)}\;
|1-x|^{\frac23+\frac23i(\nu_a-2\nu_b+\nu_c+2\nu_d)}
\cdot{}\cr&&{}\cdot
\psithree acadcd-a \psithree abbdad-b \psithree bccdbd-c \psithree bdcdbc+d
\cdot{}
&eq:getpsi\cr
&&{}\cdot
\Big\langle \Ez a^*(\rho_a)\,\Ez b^*(\rho_b)\,\Ez c^*(\rho_c)\,\Ez d(\rho_d) 
\Big\rangle\,.
}
From now on we will not write the arguments of $\Psi$ explicitly any more.  It
depends on the $\nu_i$ as well as on $x$ and~$x^*$.

\subsubsection{The BFKL term}

Since each of the integral expressions derived in Section~\ref{sec:p3pspatial}
is a conformal four-point function, the $\Psi$ function of the vertex can be
written as a sum over five $\Psi$ functions associated with the BFKL term and
the four $\alpha$ terms.  They can be computed by inserting the integrals into
Equation~\ref{eq:getpsi}.  We will start by calculating the $\Psi$ function of
the BFKL term,~$\Psi_{BFKL}$.  Inserting (\ref{eq:p3pbfkl}) into the formula
for $\Psi$, we get:
\dmath2{
\Psi_{BFKL}&=&
\frac{40}3 \frac{g^6}{16} 4 \frac2{(2\pi)^3} \xi(\nu_d)\,(4\nu_d^2+1)^2\;
|x|^{-\frac13+\frac23i(-2\nu_a+\nu_b+\nu_c-\nu_d)}\;
|1-x|^{\frac23+\frac23i(\nu_a-2\nu_b+\nu_c+2\nu_d)}
\cdot{}\crd&&{}\cdot
\psithree acadcd-a \psithree abbdad-b \psithree bccdbd-c \psithree bdcdbc+d
\cdot{}\crd&&{}\cdot
\int \frac{d^2\rho_1\,d^2\rho_2}{|\rho_{12}|^4}
\cEz 12a\,\cEz 12b\,\cEz 12c\,\eEz 12d
\cr&=&
\frac{40}3 \frac{g^6}{16} 4 \frac2{(2\pi)^3} \xi(\nu_d)\,(4\nu_d^2+1)^2\;
|x|^{-\frac13+\frac23i(-2\nu_a+\nu_b+\nu_c-\nu_d)}\;
|1-x|^{\frac23+\frac23i(\nu_a-2\nu_b+\nu_c+2\nu_d)}
\cdot{}\crd&&{}\cdot
\int \frac{d^2\rho_1\,d^2\rho_2}{|\rho_{12}|^4}
\onetworat acadcd-a \onetworat abbdad-b
\cdot{}
\cr\crd
&&\qquad\qquad\qquad{}\cdot
\onetworat bccdbd-c \onetworat bdcdbc+d
\cr
&=:&\frac{40}3 \frac{g^6}{16} 4 \frac2{(2\pi)^3} \xi(\nu_d)\,(4\nu_d^2+1)^2\;
\Xi(-\nu_a,-\nu_b,-\nu_c,\nu_d;x,x^*)\,.
&eq:psip3pbfkl}
Note that all the factors in~(\ref{eq:psip3pbfkl}) are conformally invariant by
themselves.  The function $\Xi$ is defined as the integral multiplied with the
powers of the anharmonic ratios.  It was defined with the arguments $-\nu_a$,
$-\nu_b$ and~$-\nu_c$ since in our case the first three wave functions are
complex conjugated.  This is not the case in a general four-point function, so
this notation facilitates comparison and generalisation of our result.

The function~$\Xi$ is symmetric under simultaneous exchange of the pomeron
coordinates and the $\tilde\nu\,$s ($\tilde\nu_{a/b/c}=-\nu_{a/b/c},
{}\tilde\nu_d=\nu_d$).  This follows from the fact that both the part of the
vertex without the $\Psi$ function (see (\ref{eq:our4point})) and the integral
of the BFKL term (\ref{eq:p3pbfkl}) have this symmetry.  Note that a
permutation of the coordinates changes the anharmonic ratio which is the last
argument of~$\Xi$.  Which anharmonic ratio results can be seen with the help of
Equation~\ref{eq:cross4}.

\subsubsection{The $\alpha$ terms}

In this section we will calculate the $\Psi$ functions corresponding to the
terms containing the function~$\alpha$.  We will see that they can be written
in a form quite similar to~$\Psi_{BFKL}$.

We will start with the term which arose from the set $M$ having one element,
$V^{(1)}$.  It contained a sub-integral with an expression which had to be
regularised.  The same integral has already occurred in a term of the $1\to2$
pomeron vertex.  The associated integral operator is conformally invariant and
was found to have the eigenvalue~$1/2\,\xi(\nu)$~\cite{lotterphd}.
\dmath2{
\int d^2\rho_1\,
\left[2\pi\,\delta^2(\rho_{13})\,\ln\epsilon
+\frac{|\rho_{23}|^2}{|\rho_{12}|^2|\rho_{13}|^2}
\theta\!\left(\frac{|\rho_{13}|}{|\rho_{12}|}-\epsilon\right)\right]
\,\cEz12a
&=&
\frac12\xi(\nu_a)\,\cEz23a
\cr\noalign{\vskip5pt}
\crl{eq:bfkllike}}
Inserting this relation, we get for this part of the vertex:
\dmath2{
V_{\pom\to3\pom}^{(1)}&=&
-\frac{40}3 \frac{g^6}{16} 4 \frac2{(2\pi)^3}(4\,\nu_d^2+1)^2
\sumabc\int \frac{d^2\rho_2\,d^2\rho_3}{|\rho_{23}|^4}
\cEz23b\cEz23c \cdot{}
\crd
&&\qquad\qquad\qquad\qquad
{}\cdot\eEz23d\,\frac12\xi(\nu_a)\,\cEz23a
\cr&=&
-\frac{40}3 \frac{g^6}{16} 4 \frac2{(2\pi)^3}
(\xi(\nu_a)+\xi(\nu_b)+\xi(\nu_c))\,(4\,\nu_d^2+1)^2
\cdot{}&eq:p3p1xi\cr\crd&&\hskip -5mm{}\cdot
\int \frac{d^2\rho_2\,d^2\rho_3}{|\rho_{23}|^4}
\cEz23a\cEz23b\cEz23c\eEz23d\,.
}
Since the integrand except for the $\xi$ function is symmetric in the pomeron
indices $a$, $b$ and $c$, we could remove the sum over the permutations and
insert a sum over $\xi$s instead.  The result closely resembles the BFKL
term~(\ref{eq:p3pbfkl}), except for the sign and the argument of the $\xi$
function.  Therefore we can immediately write down the $\Psi$ function of this
part of the vertex.
\dmath2{
\Psi_{(1)}&=&
-\frac{40}3 \frac{g^6}{16} 4 \frac2{(2\pi)^3}
(\xi(\nu_a)+\xi(\nu_b)+\xi(\nu_c)) \,(4\nu_d^2+1)^2\;
\Xi(-\nu_a,-\nu_b,-\nu_c,\nu_d;x,x^*)\qquad
&eq:psip3p1}

The next group of $\alpha$ terms, $V^{(2,1)}$ contains a new integral.  We will
see later that this integral occurs in none of the other terms.  Therefore we
will just write down $\Psi_{(2,1)}$ without defining a function corresponding
to~$\Xi$.  Plugging (\ref{eq:p3p21}) into~(\ref{eq:getpsi}), we get:
\dmath2{
\Psi_{(2,1)}&=&
\frac{40}3 \frac{g^6}{16} 4 \frac2{(2\pi)^3}(4\,\nu_d^2+1)^2\;
|x|^{-\frac13+\frac23i(-2\nu_a+\nu_b+\nu_c-\nu_d)}\;
|1-x|^{\frac23+\frac23i(\nu_a-2\nu_b+\nu_c+2\nu_d)}
\cdot{}\cr&&{}\cdot
\psithree acadcd-a \psithree abbdad-b \psithree bccdbd-c \psithree bdcdbc+d
\cdot{}\cr&&{}\cdot
\sumabc \int d^2\rho_1\,d^2\rho_2
\frac1{|\rho_{12}|^4}
\cEz12a\eEz12d\cdot{}
&eq:psip3p21\cr
&&{}\cdot
\int d^2\rho_3\,\left[2\pi\,\delta^2(\rho_{23})\ln\epsilon+
	\frac{|\rho_{12}|^2}{|\rho_{13}|^2|\rho_{23}|^2}
	\theta\!\left(\frac{|\rho_{23}|}{|\rho_{13}|}-\epsilon\right)\right]
\cEz13b\cEz13c\,.
}

There remain the integrals $V^{(2,2)}$ and $V^{(3)}$.  Looking at the formulas
(\ref{eq:p3p22}) and~(\ref{eq:p3p3}), one can see that they are very similar.
One has to be aware that the only distinction between different terms is the
number of pomerons attached to a given pair of coordinates, (12), (13) or~(23).
Which of the three pomerons $a$, $b$ and $c$ is attached to which does not
matter since all permutations are summed up.  Also, the indices of the
integration variables are immaterial because renaming them changes nothing.
Only one pair of indices is distinguished from the others by the fact that the
upper pomeron with index $d$ is attached to it.

Taking all this into account, the only difference between $V^{(2,2)}$ and
$V^{(3)}$ is the following: In $V^{(2,2)}$, each of the three pomerons attached
to the vertex from below has a different pair of coordinates as their
arguments.  In $V^{(3)}$, none of the lower pomerons has the same pair of
coordinate arguments as the upper ($d$) pomeron.  Instead, two of them have the
same pair of arguments.  (This can also be seen in the graphical representation
in Figure~\ref{fig:p3palpha} when one realises that the upper two lines have
the same coordinates as the outer lines below.)  This implies that $V^{(3)}$
can be obtained from $V^{(2,2)}$ by exchanging $\rho_d$ with the external
coordinate of one of the two lower pomerons which have the same arguments
($\rho_b$ or $\rho_c$ in~(\ref{eq:p3p3})), and simultaneously exchanging
{\hbox{$\nu_d\leftrightarrow-\nu_b$}} or {\hbox{$\nu_d\leftrightarrow-\nu_c$}},
respectively.  Then the pomeron $b$ resp.\ $c$ has the arguments which
previously none of the lower pomerons had, and pomeron $d$ has the same
arguments as one of the lower pomerons.  This is the situation in~$V^{(2,2)}$
(modulo permutation of the lower pomerons and/or renaming of the integration
variables).
\dmath1{
V^{(3)}_{\pom\to3\pom}&
\putunder{$\longleftarrow\!\!\!-\!\!\!-\!\!\!-\!\!\!-\!\!\!
-\!\!\!-\!\!\!-\!\!\!-\!\!\!-\!\!\!-\!\!\!-\!\!\!-\!\!\!\longrightarrow$}
{$\rho_d\leftrightarrow\rho_b, \nu_d\leftrightarrow-\nu_b$}
&V^{(2,2)}_{\pom\to3\pom}
&eq:V3toV22\cr\noalign{\vskip2mm}}
Therefore it can already be stated without any calculation that
$V^{(2,2)}$ and $V^{(3)}$ lead to the same type of integral.

In the following, we will calculate~$\Psi_{(3)}$.  $\Psi_{(2,2)}$ can then be
obtained by exchanging the coordinates and conformal dimension according
to~(\ref{eq:V3toV22}).  Inserting (\ref{eq:p3p3}) into the formula for $\Psi$
(\ref{eq:getpsi}), we get:
\dmath2{
\Psi_{(3)}&=&
-\frac{40}3 \frac{g^6}{16} 4 \frac2{(2\pi)^3} (4\,\nu_d^2+1)^2\;
|x|^{-\frac13+\frac23i(-2\nu_a+\nu_b+\nu_c-\nu_d)}\;
|1-x|^{\frac23+\frac23i(\nu_a-2\nu_b+\nu_c+2\nu_d)}
\cdot{}\crd&&{}\cdot
\psithree acadcd-a \psithree abbdad-b \psithree bccdbd-c \psithree bdcdbc+d
\cdot{}\crd&&\hskip -10mm {}\cdot
\sumabc\int \frac{d^2\rho_1\,d^2\rho_2\,d^2\rho_3\,}
				{|\rho_{12}|^2|\rho_{13}|^2|\rho_{23}|^2}
\cEz12a\cEz13b\cEz13c\eEz23d 
\cr&=&
-\frac{40}3 \frac{g^6}{16} 4 \frac2{(2\pi)^3} (4\,\nu_d^2+1)^2\;
\sumabc|x|^{-\frac13+\frac23i(-2\nu_a+\nu_b+\nu_c-\nu_d)}\;
|1-x|^{\frac23+\frac23i(\nu_a-2\nu_b+\nu_c+2\nu_d)}
\cdot{}\crd&&{}\cdot
\int \frac{d^2\rho_1\,d^2\rho_2\,d^2\rho_3\,}
				{|\rho_{12}|^2|\rho_{13}|^2|\rho_{23}|^2}
\onetworat acadcd-a \onethreerat abbdad-b
\cdot{}
\crd
&&\qquad\qquad\qquad{}\cdot
\onethreerat bccdbd-c \twothreerat bdcdbc+d
&eq:psip3p3\cr
&=:&
-\frac{40}3 \frac{g^6}{16} 4 \frac2{(2\pi)^3} (4\,\nu_d^2+1)^2\;
\sumabc \Upsilon(-\nu_b,-\nu_c,-\nu_a,\nu_d;x,x^*)\,.
}
The step from the first to the second expression is not trivial.  It is legal
only because the vertex without the $\Psi$ function is invariant under
permutations of the pomerons $a$, $b$ and $c$, as can be seen
from~(\ref{eq:our4point}).  (In fact it is symmetric under all permutations of
conformal fields if one takes care of the different sign of~$\nu_d$.)  It is
this part of the vertex which we have written into the sum over the
permutations, even though the symmetry is not obvious any more.  The sum over
the permutations also extends to~$x$ in the sense that it is transformed into a
different anharmonic ratio by a permutation of the coordinates.  In fact, all
six anharmonic ratios occur in the sum, some with a minus sign.

$\Upsilon$ is defined as follows:
\dmath2{
\Upsilon(-\nu_b,-\nu_c&&,-\nu_a,\nu_d;x,x^*)=
|x|^{-\frac13+\frac23i(-2\nu_a+\nu_b+\nu_c-\nu_d)}\;
|1-x|^{\frac23+\frac23i(\nu_a-2\nu_b+\nu_c+2\nu_d)}
\cdot{}\crd&&{}\cdot
\int \frac{d^2\rho_1\,d^2\rho_2\,d^2\rho_3\,}
				{|\rho_{12}|^2|\rho_{13}|^2|\rho_{23}|^2}
\onetworat acadcd-a \onethreerat abbdad-b
\cdot{}
\crd
&&\qquad\qquad\qquad{}\cdot
\onethreerat bccdbd-c \twothreerat bdcdbc+d\,.
&eq:Upsilon}
We have defined it so that the conformal dimensions of the two pomerons with
the same coordinate arguments come first.  It is symmetric in these first two
arguments provided the corresponding coordinates are exchanged as well, which
leads to a different anharmonic ratio, $1/x$ instead of~$x$.
\dmath1{
\Upsilon(-\nu_b,-\nu_c,-\nu_a,\nu_d;x,x^*)&=&
\Upsilon(-\nu_c,-\nu_b,-\nu_a,\nu_d;\frac1x,\frac1{x^*})
&eq:Upssymm}

The conformally invariant function $\Psi_{(2,2)}$ corresponding to~$V^{(2,2)}$
can be obtained from $\Psi_{(3)}$ by way of~(\ref{eq:V3toV22}).  The only
slight difficulty is the anharmonic ratio~$x$.  Exchanging the coordinates
$\rho_d\leftrightarrow\rho_b$ also changes the anharmonic ratio on which
$\Upsilon$ still depends:
\dmath2{
x=\crossrat abcdacbd
&\longrightarrow&
\crossrat adcbacdb=\crossrat adbcacbd=1-x\,,
\quad\hbox{according to~(\ref{eq:cross4}).}
}
So we get for $\Psi_{(2,2)}$:
\dmath2{
\Psi_{(2,2)}&=&
\frac{40}3 \frac{g^6}{16} 4 \frac2{(2\pi)^3} (4\,\nu_d^2+1)^2\;
\sumabc \Upsilon(-\nu_d,-\nu_c,-\nu_a,\nu_b;1-x,1-x^*)\,.
&eq:psip3p22\cr
}

\section{Summary of the $1\to3$ pomeron vertex from the irreducible $2\to6$ 
			gluon vertex}

This section will summarise the results of the calculations of the previous
section.  We have seen that the $1\to3$ pomeron vertex has the form of a
conformal four-point function:
\dmath2{
V_{\pom\to3\pom}(\{\rho_i\};\{\nu_i\})&=&\Psi(x,x^*;\{\nu_i\})\prod_{i<j}
|\rho_{ij}|^{2i\left(-\tilde\nu_i-\tilde\nu_j+\frac13\sum_k \tilde\nu_k\right)}
\,,
&eq:Vp3p4point}
where $x=\icrossrat abcdacbd$, $\tilde\nu_i=-\nu_i$ for $i=a,b,c$ and
$\tilde\nu_d=\nu_d$.

The freedom which remains in the otherwise fixed form of the vertex, the
function $\Psi$, can in this case be written as a sum of five terms:
\dmath2{
\Psi(x,x^*;\{\nu_i\})&=&
\Psi_{BFKL}(x,x^*;\{\nu_i\})+\Psi_{(1)}(x,x^*;\{\nu_i\})
\crl{eq:psip3psum}
&&{}+\Psi_{(2,1)}(x,x^*;\{\nu_i\})+\Psi_{(2,2)}(x,x^*;\{\nu_i\})
+\Psi_{(3)}(x,x^*;\{\nu_i\})\,.
}
The five terms of the function~$\Psi$ have the following form:
\dmath{2.5}
{
\Psi_{BFKL}&&(x,x^*;\{\nu_i\})=
\frac{40}3 \frac{g^6}{16} 4 \frac2{(2\pi)^3} \xi(\nu_d)\,(4\nu_d^2+1)^2\;
\Xi(-\nu_a,-\nu_b,-\nu_c,\nu_d;x,x^*)
\cr
\Psi_{(1)}&&(x,x^*;\{\nu_i\})=
-\frac{40}3 \frac{g^6}{16} 4 \frac2{(2\pi)^3}
(\xi(\nu_a)+\xi(\nu_b)+\xi(\nu_c)) \,(4\nu_d^2+1)^2\;
\Xi(-\nu_a,-\nu_b,-\nu_c,\nu_d;x,x^*)
\cr
\Psi_{(2,1)}&&(x,x^*;\{\nu_i\})=
\frac{40}3 \frac{g^6}{16} 4 \frac2{(2\pi)^3}(4\,\nu_d^2+1)^2\;
|x|^{-\frac13+\frac23i(-2\nu_a+\nu_b+\nu_c-\nu_d)}\;
|1-x|^{\frac23+\frac23i(\nu_a-2\nu_b+\nu_c+2\nu_d)}
\cdot{}\cr&&{}\cdot
\psithree acadcd-a \psithree abbdad-b \psithree bccdbd-c \psithree bdcdbc+d
\cdot{}\crl{eq:psip3pall}&&{}\cdot
\sumabc \int d^2\rho_1\,d^2\rho_2
\frac1{|\rho_{12}|^4}
\cEz12a\eEz12d\cdot{}
\cr
&&{}\cdot
\int d^2\rho_3\,\left[2\pi\,\delta^2(\rho_{23})\ln\epsilon+
        \frac{|\rho_{12}|^2}{|\rho_{13}|^2|\rho_{23}|^2}
        \theta\!\left(\frac{|\rho_{23}|}{|\rho_{13}|}-\epsilon\right)\right]
\cEz13b\cEz13c
\cr
\Psi_{(2,2)}&&(x,x^*;\{\nu_i\})=
\frac{40}3 \frac{g^6}{16} 4 \frac2{(2\pi)^3} (4\,\nu_d^2+1)^2\;
\sumabc \Upsilon(-\nu_d,-\nu_c,-\nu_a,\nu_b;1-x,1-x^*)
\cr
\Psi_{(3)}&&(x,x^*;\{\nu_i\})=
-\frac{40}3 \frac{g^6}{16} 4 \frac2{(2\pi)^3} (4\,\nu_d^2+1)^2\;
\sumabc \Upsilon(-\nu_b,-\nu_c,-\nu_a,\nu_d;x,x^*)
}
The sums over $\Upsilon$ in $\Psi_{(2,2)}$ and~$\Psi_{(3)}$ run over all six
simultaneous permutations of the conformal dimensions $\nu_{a/b/c}$ and the
corresponding coordinates $\rho_{a/b/c}$.  The latter entails that $x$ is
replaced by a different cross ratio according to the permutation of the
$\rho\,$s.  The two integral expressions $\Xi$ and~$\Upsilon$ are defined as
follows:
\dmath2{
\Xi&&(-\nu_a,-\nu_b,-\nu_c,\nu_d;x,x^*)=
|x|^{-\frac13+\frac23i(-2\nu_a+\nu_b+\nu_c-\nu_d)}\;
|1-x|^{\frac23+\frac23i(\nu_a-2\nu_b+\nu_c+2\nu_d)}
\cdot{}\cr&&\qquad\qquad{}\cdot
\int \frac{d^2\rho_1\,d^2\rho_2}{|\rho_{12}|^4}
\onetworat acadcd-a \onetworat abbdad-b
\cdot{}
\cr
&&\qquad\qquad\qquad\qquad\qquad{}\cdot
\onetworat bccdbd-c \onetworat bdcdbc+d
\crl{eq:xiups}
\Upsilon&&(-\nu_b,-\nu_c,-\nu_a,\nu_d;x,x^*)=
|x|^{-\frac13+\frac23i(-2\nu_a+\nu_b+\nu_c-\nu_d)}\;
|1-x|^{\frac23+\frac23i(\nu_a-2\nu_b+\nu_c+2\nu_d)}
\cdot{}\cr&&\qquad\qquad{}\cdot
\int \frac{d^2\rho_1\,d^2\rho_2\,d^2\rho_3\,}
				{|\rho_{12}|^2|\rho_{13}|^2|\rho_{23}|^2}
\onetworat acadcd-a \onethreerat abbdad-b
\cdot{}
\cr
&&\qquad\qquad\qquad\qquad\qquad\qquad\quad{}\cdot
\onethreerat bccdbd-c \twothreerat bdcdbc+d
}

\section{Conclusions from the existence of the vertex}

In the previous sections, we have derived the explicit form of a $1\to3$
pomeron vertex.  It was obtained from a $2\to6$ reggeised gluon vertex from the
integral equation for the six-gluon amplitude in Extended Generalised Leading
Logarithmic corrections to BFKL.  This gluon vertex, and hence the $1\to3$
pomeron vertex obtained by projection, is local in rapidity.

This has some implications.  Leading Logarithmic Order BFKL is known to be
equivalent to leading-order calculations in the dipole approach proposed by
Mueller~\cite{mueller,mueller2}.  In the dipole approach, scattering colour
dipoles are assumed to split up repeatedly into smaller ones, which then
interact by exchanging gluons.  This splitting up of the dipoles was found to
be equivalent mathematically to the gluon ladders the BFKL equation was
originally derived from.

How far this correspondence extends to higher orders is not yet known.  The
state of the art regarding this question is rather complex.  Peschanski has
derived a general $1\to n$ pomeron vertex from Mueller's dipole
approach~\cite{pesch}.  However, there are two qualifications to this
statement: For one thing, the higher vertices were obtained by an extrapolation
of the $1\to2$ pomeron vertex.  The other point of caution concerns the
expression for the $1\to2$ pomeron vertex which forms the basis for the
interpolation.  Though Peschanski starts out from the dipole formalism, the
extrapolated expression is very close to the form of the $1\to2$ pomeron vertex
obtained by Lotter~\cite{lotterphd} in the gluon ladder approach.  Hence
Peschanski's results constitute no proof that these vertices exist in the
dipole approach.  Braun and Vacca~\cite{brvac} have since argued, starting from
the generating functional of the $n$-fold dipole densities, that the vertices
of $1\to n$ pomerons for $n>2$ do not exist in the dipole approach.

In contrast to that, the results of the previous sections show that the $1\to3$
pomeron vertex does exist in the gluon ladder approach.  This constitutes a
discrepancy between the dipole and the gluon ladder approach and implies that
the correspondence of the two approaches does not extend beyond Leading
Logarithmic Order.

\section{The $1\to3$ pomeron vertices from reducible $2\to6$ gluon transitions}
\label{sec:reducible}

\subsection{Reducible $2\to6$ gluon transitions}
\label{sec:other2to6}

As was already mentioned at the beginning of Section~\ref{sec:V2to6}, the
$2\to6$ reggeised gluon vertex treated in the first part of this chapter comes
from the integral equation (\ref{eq:D6eq}) for a six-gluon amplitude which
occurs in unitarity corrections.  Besides $V_{2\to6}$, this equation contains
other components, some of which are also $2\to6$ gluon transitions local in
rapidity.  Unlike $V_{2\to6}$, however, these terms do not represent
irreducible $2\to6$ gluon vertices, and we will therefore not call them
``vertices'' but ``transitions''.  They can be expressed in terms of the
function $(VD_2)$ of which the $2\to4$ gluon vertex is composed.  They are
therefore related to the irreducible part of the four-gluon amplitude.

We will not consider the completely reggeising part of the six-gluon amplitude
here.  It is expected to yield a $1\to3$ pomeron vertex, just as the reggeising
part of the four-gluon amplitude gives rise to a $1\to2$ pomeron vertex.  The
latter vertex was discussed in~\cite{bart3p} and found to be subleading in the
large $N_c$ limit.

We will derive the $1\to3$ pomeron vertices contained in the partly reggeising
components in the second, third and fourth line of~(\ref{eq:D6eq}):
\dmath{1.5}{
\ldots&+&\sum f_{a_1a_2a_3}\,f_{a_4a_5a_6}\;L(1,2,3;4,5,6) \cr
&+&\sum d^{a_1a_2a_3a_4}\,\delta_{a_5a_6}\;I(1,2,3,4;5,6) &eq:other2to6\cr
&+&\sum d^{a_2a_1a_3a_4}\,\delta_{a_5a_6}\;J(1,2,3,4;5,6)+\ldots
}
The sums run over simultaneous permutations of colour indices and
momenta/coordinates.  However, the symmetry of both colour tensors and spatial
functions is limited.  These terms are not totally symmetric in the indices, as
$V_{2\to6}$ was.  This forces us to make a choice for the symmetries of the
three-pomeron state we use for the projection.  If, for instance, one of the
pomerons is distinguishable from the other two, the result of the projection
may depend on which pairs of arguments of the $2\to6$ gluon transition term it
couples to.  Which pomerons are distinguishable depends on where there are
$t$-channel cuts, which in turn depends on the physical process considered.
The existence of cuts also places restrictions on which arguments of the
transition terms each pomeron can couple to.  Since we have no particular
process in mind, we will for the larger part of this section choose a
superposition of three-pomeron states in which the pomerons' six arguments have
all possible permutations:
$$
\frac1{6!}\sum_{\pi\in\Pi(1,2,3,4,5,6)}
\Ez a^*(\pi_1,\pi_2)\,\Ez b^*(\pi_3,\pi_4)\,\Ez c^*(\pi_5,\pi_6)
$$
This is a completely symmetric state of six reggeised gluons, with the
additional requirement that they form three pomerons.  It implies that we
assume the pomerons to be indistinguishable.  This has the advantage that in
our projection we will encounter all the terms which can possibly result from a
projection onto an arbitrary three-pomeron state, provided we refrain from
making use of the symmetry to find cancellations.  Later, in
Section~\ref{sec:distinguishable}, we will also consider the case of
distinguishable pomerons.


Two of the terms in~(\ref{eq:other2to6}) contain the colour
tensor~$d^{a_1a_2a_3a_4}$.  It is defined as follows:
\dmath1{
d^{a_1a_2a_3a_4}&=&
tr(t^{a_1}t^{a_2}t^{a_3}t^{a_4})+tr(t^{a_4}t^{a_3}t^{a_2}t^{a_1})\,.
&eq:dabcd}
Here $t^a$ are the generators of $SU(N_c)$, half the Gell-Mann matrices for
$N_c=3$.  $d^{a_1a_2a_3a_4}$ obeys the following relation:
\dmath1{
d^{a_1a_2a_3a_4}&=&
\frac1{2N_c}\delta_{a_1a_2}\delta_{a_3a_4}
+\frac14(d_{ka_1a_2}d_{ka_3a_4} - f_{la_1a_2}f_{la_3a_4})\,.
&eq:dabcdrel}
$f_{abc}$ are the structure constants and $d_{abc}$ the symmetric structure
constants of the algebra.

The three spatial functions have the following form:
\dmath{2.5}{
L(1,2,3;4,5,6)&=&\frac{g^2}4 [(VD_2)(12,3;45,6)-(VD_2)(12,3;46,5)
+(VD_2)(12,3;4,56)
\crd&&\qquad{}-(VD_2)(13,2;45,6)+(VD_2)(13,2;46,5)-(VD_2)(13,2;4,56)
\crd&&\qquad{}+(VD_2)(1,23;45,6)-(VD_2)(1,23;46,5)+(VD_2)(1,23;4,56)]
\crl{eq:LIJ}
I(1,2,3,4;5,6)&=&g^2[-(VD_2)(1,234;5,6)-(VD_2)(123,4;5,6)+(VD_2)(14,23;5,6)]
\qquad\qquad\cr
J(1,2,3,4;5,6)&=&g^2[-(VD_2)(134,2;5,6)-(VD_2)(124,3;5,6)
\crd&&\qquad{}+(VD_2)(12,34;5,6)+(VD_2)(13,24;5,6)]
}
The function $(VD_2)$ can be expressed in terms of the function~$G$.
\dmath0{
(VD_2)(1,2;3,4)&=&\frac{g^2}2 \big[G(1,23,4)+G(2,13,4)+G(1,24,3)+G(2,14,3)
&eq:VD2\cr
&&\qquad{}-G(12,3,4)-G(12,4,3)-G(1,2,34)-G(2,1,34)+G(12,-,34)\big]\qquad
}
This formula is in fact analogous to (\ref{eq:WD2}), but since there are only
nine non-zero terms, the sum has been written out.  $(VD_2)$ is symmetric under
exchange of the arguments 1 and 2, and 3 and 4, respectively, and under
exchange of its two pairs of arguments.  This can be easily derived from the
symmetry of the $G$ function under exchange of its first and third argument.
\dmath2{
&(VD_2)(1,2;3,4)=(VD_2)(2,1;3,4)=(VD_2)(1,2;4,3)&
\crl{eq:VD2symm}
&(VD_2)(1,2;3,4)=(VD_2)(3,4;1,2)&
}

We will shortly derive the $1\to3$ pomeron vertices contained in the three
reducible $2\to6$ gluon transitions.  We will see that they are composed of the
same terms as the pomeron vertex obtained from~$V_{2\to6}$.  To determine which
vertex will dominate in the large $N_c$ limit, we will calculate the colour
constants for general~$N_c$.  Before that, however, we will rewrite the
irreducible vertex computed in Section~\ref{sec:Wp3p} for general $N_c$ and in
a concise graphical notation for the convolutions of the spatial part.

\subsection{The irreducible vertex for general $N_c$}

The colour constant of the irreducible vertex is $d_{abc}d_{abc}$.  For $N_c=3$
this is equal to $40/3$.  For general $N_c$, it can be derived from the
following relation:
\dmath2{
d_{abc}d_{abd}&=&\frac{N_c^2-4}{N_c}\,\delta_{cd}
\cr
\Rightarrow\qquad
d_{abc}d_{abc}&=&\frac1{N_c}(N_c^2-4)(N_c^2-1)\,.
}

Collecting all the prefactors, the vertex takes the following form:
\dmath1{
&\sumabc g^6\,\frac1{N_c}(N_c^2-4)(N_c^2-1)\,
\Bigg[-\frac14\Vallr+\frac14\Valrb+\frac14\Vallb-\frac14\Valbb
+\frac1{12}\frac1{N_c\,g^2}\Vbfkl\Bigg]\,.&
\cr&&&eq:Wresult}
The prefactors $\frac14$ of the $\alpha$ terms come from a factor 4 for the
four permutations~(\ref{eq:projind}), from the factor $\frac18$
in~(\ref{eq:WD2}) and the factor $\frac12$ in~(\ref{eq:G_symb}).  The prefactor
of the BFKL term differs because $K_{BFKL}$ already contains a factor~$N_cg^2$,
lacks a factor 2 relative to~$\alpha$, and because there is only one BFKL term
for six of the $\alpha$ terms.

The functions represented graphically correspond to the integrals we obtained
in Section~\ref{sec:Wp3p}.  They depend on the conformal dimensions of the four
pomerons and on the cross ratio of their coordinates.
\dmath{2.5}
{
\Vallr&&=
\frac2{(2\pi)^3} (4\,\nu_d^2+1)^2\;
\Upsilon(-\nu_b,-\nu_c,-\nu_a,\nu_d;x,x^*)
\cr
\Valrb&&=
\frac2{(2\pi)^3} (4\,\nu_d^2+1)^2\;
\Upsilon(-\nu_d,-\nu_c,-\nu_a,\nu_b;1-x,1-x^*)
\cr
\Vallb&&=
\frac2{(2\pi)^3}(4\,\nu_d^2+1)^2\;
|x|^{-\frac13+\frac23i(-2\nu_a+\nu_b+\nu_c-\nu_d)}\;
|1-x|^{\frac23+\frac23i(\nu_a-2\nu_b+\nu_c+2\nu_d)}
\cdot{}\cr&&{}\cdot
\psithree acadcd-a \psithree abbdad-b \psithree bccdbd-c \psithree bdcdbc+d
\cdot{}\crl{eq:graphdef}&&{}\cdot
\int d^2\rho_1\,d^2\rho_2
\frac1{|\rho_{12}|^4}
\cEz12a\eEz12d\cdot{}
\cr
&&{}\cdot
\int d^2\rho_3\,\left[2\pi\,\delta^2(\rho_{23})\ln\epsilon+
        \frac{|\rho_{12}|^2}{|\rho_{13}|^2|\rho_{23}|^2}
        \theta\!\left(\frac{|\rho_{23}|}{|\rho_{13}|}-\epsilon\right)\right]
\cEz13b\cEz13c
\cr
\Valbb&&=
\frac1{(2\pi)^3} \xi(\nu_a) \,(4\nu_d^2+1)^2\;
\Xi(-\nu_a,-\nu_b,-\nu_c,\nu_d;x,x^*)
\cr
\Vbfkl&&=
g^2N_c \frac2{(2\pi)^3} \xi(\nu_d)\,(4\nu_d^2+1)^2\;
\Xi(-\nu_a,-\nu_b,-\nu_c,\nu_d;x,x^*)
}
Note that the graphical symbols for the alpha terms, as defined here, are not
symmetric with respect to exchange of the pomeron indices.  This does not
constitute a problem here, since we have decided to use a three-pomeron state
which is a superposition of the six possible permutations.  In
Section~\ref{sec:distinguishable}, where we consider distinguishable pomerons,
we will write indices next to the pomeron lines.  In the immediately following
sections, however, the identities in the graphical notation are always to be
understood as modulo permutations of the indices of the three lower pomerons
$a$, $b$ and~$c$.  We will take care always to average over these permutations.

\subsection{The colour-antisymmetric $L$ terms}
\label{sec:Lterms}

The first set of reducible terms we will now deal with are the ones containing
the spatial function~$L$.  They have a colour structure given by the
antisymmetric structure constants.
\dmath2{
\sum f_{a_1a_2a_3}\,f_{a_4a_5a_6}\;L(1,2,3;4,5,6)&=&
f_{a_1a_2a_3}\,f_{a_4a_5a_6}\;L(1,2,3;4,5,6)
\crl{eq:Lterms}
&&{}+f_{a_1a_2a_4}\,f_{a_3a_5a_6}\;L(1,2,4;3,5,6)+\ldots
}
The sum runs over the same permutations as for the irreducible vertex, that is
the ten possibilities of dividing a set of six indices into two groups of
three.  The indices of the $f$ tensor are always in ascending order.

Unlike $V_{2\to6}$, this expression is not completely symmetric in the
reggeised gluon indices.  The $f$ tensors are symmetric only under cyclic
permutations and antisymmetric under anti-cyclic ones.  The spatial
function~$L$ is even less symmetric.  It can be worked out from its
form~(\ref{eq:LIJ}) in terms of the function~$G$ that it is invariant under the
exchange of its first group of three arguments with its second group of three
arguments and of the first and third argument in each group.

Because of the complete symmetry of the three-pomeron state we project onto, it
is sufficient to perform the projection of one of the terms in the
sum~(\ref{eq:Lterms}).  Therefore the sum over the ten permutations
in~(\ref{eq:Lterms}) can then be discarded and replaced by a factor ten.  To
find out which different terms result from the projection, we have a look at
the permutations of the six arguments of three pomerons.  There are $6!=720$
permutations of six arguments.  A factor $2^3=8$ can be divided out for the
exchange of each pomeron's two arguments.  Since they are symmetric, no
averaging over these possibilities is necessary.  The remaining 90
possibilities decompose into the 15 possibilities in which three identical
pomerons can couple to six positions and six permutations of the three
pomerons.  We will deal with them separately.

Because the colour tensor $f$ vanishes when contracted with a Kronecker delta,
each pomeron has to have one argument in each of the two groups of three.  This
is the case for 6 of the 15 possible ways of coupling them.\footnote{%
One pomeron has to be attached to the first index of the first $f$ tensor, or
equivalently the first argument of $L$.  Its other argument can be attached to
any argument in the second group, which gives three possibilities.  The pomeron
attached to the second argument in the first group has two possibilities, the
third just one.}
So only six of the fifteen possibilities of the pomerons' coupling contribute.
Since the function $L$ is not symmetric under exchange of the middle argument
in each group with one of the others (of the same group), a pomeron coupled to
this argument is distinguishable from the others.  Two different terms arise
depending on whether the same pomeron couples to the middle argument in each
group or not.  They can be represented graphically.
$$
\VLa\qquad\qquad\VLb
$$
Since the other two arguments within each group are exchangeable, these are the
only two distinguishable terms.  Two of the six terms are of the first type,
four of the second.

However, the antisymmetry of the $f$ tensors causes these terms to cancel out
in pairs.  A certain way of coupling three pomerons to two groups of three
arguments can be written as a permutation of the numbers one to three.  The
permutation maps 1 to 1, 2 or~3 depending on which argument of the second group
is attached to the same pomeron as the first argument of the first group, and
so on.  The colour constant corresponding to the permutation
$(\pi_1\pi_2\pi_3)$ is
$$
f_{a_1a_2a_3}f_{a_{\pi_1}a_{\pi_2}a_{\pi_3}}=
\chi(\pi)\;N_c(N_c^2-1),
$$
where $\chi(\pi)$ is the character of the permutation, $+1$ for even and $-1$
for odd permutations.  Exchanging the first and third argument in one argument
group of $L$ leaves the spatial expression unchanged (because of $L$'s
symmetry) but reverses the permutation and hence the sign of the colour factor.
Therefore, all terms come in pairs which cancel out.  The terms with the $L$
function do not give rise to a $1\to3$ pomeron vertex.

We could have obtained this result more quickly by invoking symmetry arguments.
However, our aim was to present all possible terms which might result from a
projection onto an arbitrary three-pomeron state.  In a projection onto a
three-pomeron state which is not completely symmetric, the above cancellations
do not occur and non-zero terms remain, see Section~\ref{sec:distinguishable}.

%

\subsection{The terms with the function $I$}
\label{sec:Iterms}

The sum of the terms with the function~$I$ runs over all fifteen ways of
dividing six arguments into one group of four and one group of two.  The
arguments are always in ascending order of the indices:
\dmath2{
\sum d^{a_1a_2a_3a_4}\,\delta_{a_5a_6}\;I(1,2,3,4;5,6) 
&=&
d^{a_1a_2a_3a_4}\,\delta_{a_5a_6}\;I(1,2,3,4;5,6) 
\crl{eq:Iterms}
&&{}+d^{a_1a_2a_3a_5}\,\delta_{a_4a_6}\;I(1,2,3,5;4,6) + \ldots
}

The terms containing the spatial function~$I$ have very little symmetry.  Both
the colour structure and the function~$I$ are invariant under exchange of the
last two arguments.  Besides, the colour tensor $d^{a_1a_2a_3a_4}$ is invariant
under cyclic permutations and under inversion of the order of the indices.  The
function~$I$ is symmetric under exchange of its first with its fourth and of
its second with its third argument, respectively.  This implies invariance
under reversal of the order of the first four arguments, a symmetry shared by
the colour tensor.  This and the exchange of the last arguments is their only
common symmetry.

Since the three-pomeron state we project onto is completely symmetric, we
average over the permutations of the three pomerons attached below the $2\to6$
gluon transition.  We can then again discard the sum over the fifteen $I$ terms
and replace them with a factor~15.  (This cancels a factor $\frac1{15}$ from
the average over the pomerons, so we have in effect replaced the sum over the
terms by a sum over the ways the pomerons couple.)

\subsubsection{The colour factors}

We will first calculate the possible colour factors which can occur in the
terms with the $I$ function.  We will see that all of them actually occur.  To
make it easier to match them to the spatial expressions, we will write them
in a graphical notation.  The colour structure of the terms is:
\dmath2{
d^{a_1a_2a_3a_4}\,\delta_{a_5a_6}&=&\ \Cddel
&eq:Icol}
Half-circles denote Kronecker deltas.  We will now calculate the possible
colour factors by attaching three of them from below in all possible ways.  For
instance, one pomeron (which has a singlet colour structure described by a
Kronecker delta) could couple to the first two legs, another one to the third
and fourth, and the third to the last two.  This expression factorises, and we
can treat each factor separately.
$$
\Cdpairs\ \Cdeldel=\Cdpairs\cdot\ \Cdeldel
$$
The second factor is the easier to evaluate.  It is the contraction of two
Kronecker deltas, ie the trace of the unit matrix in the adjoint representation:
\dmath1{
\Cdeldel&=&\delta_{ab}\delta_{ab}=N_c^2-1\,.
&eq:deldel}
The other factor can be computed with the help of relation~(\ref{eq:dabcdrel}).
\dmath2{
\Cdpairs&=&d^{abcd}\delta_{ab}\delta_{cd}
=\left[\frac1{2N_c}\delta_{ab}\delta_{cd}
+\frac14(d_{kab}d_{kcd} - f_{lab}f_{lcd})\right]
\,\delta_{ab}\delta_{cd}
\cr
&=&\frac1{2N_c}\delta_{ab}\delta_{cd}\,\delta_{ab}\delta_{cd}
=\frac1{2N_c}(N_c^2-1)^2
&eq:Cdpairs}
Both the three-index $d$ and $f$ tensors vanish when contracted with a Kronecker
delta in this way.  So we obtain the first colour factor:
\dmath1{
\Cdpairs\ \Cdeldel&=&\frac1{2N_c}(N_c^2-1)^3\,.
&eq:C3pairs}

The second possibility is for a pomeron to couple to the first two legs (again)
and for the other two to link the $d^{abcd}$ tensor to the Kronecker delta.
Due to the symmetry of the delta symbol, there is only one way of doing this:
$$
\Cddellink
$$
It is easy to see that this ``linking'' has the same result as attaching a
Kronecker delta to the two last indices of the $d^{abcd}$ tensor:
\dmath1{
\Clink&=&\delta_{cf}\delta_{de}\delta_{ef}=\delta_{cd}=\Clowdel
&eq:Clink}
So we get the same result as for a $d^{abcd}$ tensor contracted with two
Kronecker deltas (\ref{eq:Cdpairs}).
\dmath1{
\Cddellink&=&\Cdpairs=\frac1{2N_c}(N_c^2-1)^2
&eq:Cddellink}

Because the $d^{abcd}$ tensor is not completely symmetric in its indices, terms
in which the delta symbols attached to it cross are different from the ones we
had already.  We obtain for the $d^{abcd}$ tensor alone:
\dmath1{
\Cdcross&=&d^{abcd}\delta_{ac}\delta_{bd}
=\left[\frac1{2N_c}\delta_{ab}\delta_{cd}
+\frac14(d_{kab}d_{kcd} - f_{lab}f_{lcd})\right]
\delta_{ac}\delta_{bd}
\cr
&=&\frac1{2N_c}\delta_{aa}+\frac14(d_{kab}d_{kab}-f_{lab}f_{lab})
\cr
&=&\frac1{2N_c}(N_c^2-1)+\frac1{4N_c}(N_c^2-4)(N_c^2-1)-\frac14 N_c(N_c^2-1)
=-\frac1{2N_c}(N_c^2-1)\,.
\qquad&eq:Cdcross}
As for the uncrossed diagram, there are two different colour factors related to
this one.  In the first, the Kronecker delta from the transition term is
contracted with a pomeron separately from the $d^{abcd}$ tensor; in the second,
they are linked.  The factorising diagram gives the result:
\dmath1{
\Cdcross\ \Cdeldel&=&-\frac1{2N_c}(N_c^2-1)^2\,.
&eq:Cddelcross}
Because of (\ref{eq:Clink}), the non-factorising diagram gives the same result
as the contraction of the $d^{abcd}$ tensor alone.
\dmath1{
\Cddelcrlnk&=&\Cdcross=-\frac1{2N_c}(N_c^2-1)
&eq:Cddelcrlnk}

There are no further different colour factors.  All other diagrams are equal to
one of those already presented by way of the symmetry of the colour tensor.
For instance,
\dmath1{
\Cdbridge&=&\Cdpairs
&eq:Cdbridge}
because of the $d^{abcd}$ tensor's invariance under cyclic permutations.

\subsubsection{The spatial part}

The spatial part of this transition term is given by the function~$I$:
\dmath1{
I(1,2,3,4;5,6)&=&g^2
\big[-(VD_2)(1,234;5,6)-(VD_2)(123,4;5,6)+(VD_2)(14,23;5,6)\big]\,.
\qquad&eq:I}
We will now look at all possible ways of convoluting three pomeron wave
functions with it.

Let us first have a look at the terms in which one pomeron couples to the last
two arguments of~$I$.  There are three possibilities of coupling the remaining
two pomerons to four arguments.  They can be represented graphically.  A
half-circle denotes a pomeron amplitude which is convoluted with the respective
arguments of~$I$.  We get three diagrams:
$$
\VIa\qquad\VIb\ =\ \ \VIbb
$$
The legs of the $I$~function have been marked to indicate its symmetries:
Exchanging the pair of arguments marked in the same way, or the unmarked ones,
leaves it invariant.  However, the colour factor changes if one exchanges only
one of the two pairs of marked legs, since $d^{abcd}$ is invariant under
reversal of its indices but not under arbitrary permutations.  So the latter
two diagrams represent the same spatial convolutions but have different colour
factors.

Let us deal next with the terms in which the last two arguments of~$I$ are
connected by pomerons to two arguments which also form a pair (ie under the
exchange of which~$I$ also is invariant).  There are two possibilities here:
The other pair can either be the second and third argument, or the first and
fourth argument:
$$
\VIc\qquad\VId
$$
These terms have the same colour factor~(\ref{eq:Cdbridge}), taking into
account~(\ref{eq:Clink}).  From each of them a new diagram can be obtained by
exchanging the two rightmost legs.  These diagrams are equal in both colour
structure and spatial part since this exchange is a common symmetry of both the
colour tensor and~$I$.  So they can be accounted for by multiplying the
displayed terms by two.

There remain eight terms.  They can be obtained from each other by applying the
exchanges of arguments with respect to which the function~$I$ is invariant.
Therefore their spatial convolutions are all equal.  However, two different
colour factors occur, with four terms having the same factor.  The reason for
this is that $d^{abcd}$ is invariant only under reversal of its indices, which
amounts to exchanging the first with the fourth and the second with the third
index simultaneously.  Graphs which differ by only one of these exchanges have
different colour factors.  Here is one representative of each group, the
spatial part of which is equal:
$$
\VIe\ =\ \ \VIee
$$

We will now put the spatial diagrams and the colour factors together and sum up
all the terms.  In addition, we will write down the average over the
permutations of the pomerons, the only symmetry of the three-pomeron state
which we have not yet taken into account.  The result is the following sum of
the five different terms with the given prefactors:
\dmath2{
\frac16\sumabc\Bigg[&&\frac1{2N_c}(N_c^2-1)^3\;\VIa
+\left(\frac1{2N_c}(N_c^2-1)^3-\frac1{2N_c}(N_c^2-1)^2\right)\VIb
\cr&&
{}+2\frac1{2N_c}(N_c^2-1)^2\;\VIc
+2\frac1{2N_c}(N_c^2-1)^2\;\VId
\cr&&
{}+4\left(\frac1{2N_c}(N_c^2-1)^2-\frac1{2N_c}(N_c^2-1)\right)\VIe\Bigg]
\cr
{}=\frac16\sumabc&&\frac{N_c^2-1}{2N_c}\Bigg[
(N_c^2-1)^2\;\VIa + (N_c^2-2)(N_c^2-1)\;\VIb + 2\,(N_c^2-1)\;\VIc
\cr
&&\phantom{N_c^2-1}\;\;{}+ 2\,(N_c^2-1)\;\VId + 4\,(N_c^2-2)\;\VIe\Bigg]\,.
&eq:Igraphsum}

What now remains to be done is to express $I$ in terms of~$(VD_2)$, $(VD_2)$ in
terms of~$G$, and $G$ in terms of BFKL kernels and alpha functions.  We will
see that a $G$ function convoluted with three pomerons always leads to one or
the other.  Using~(\ref{eq:I}) and bearing in mind the
symmetries~(\ref{eq:VD2symm}) of $V$ and the fact that a pomeron wave function
with two equal coordinate arguments vanishes, we obtain:
\dmath2{
\VIa&=&0 \cr
\VIb&=&g^2\;\VVllr \cr
\VIc&=&-2\,g^2\;\VVlmR &eq:ItoV\cr
\VId&=&0 \cr
\VIe&=&g^2\Bigg(-\VVlmR+\VVlmA\Bigg)
}
The sum of all terms including the correct prefactors now reads:
\dmath1{
\frac16\sumabc g^2\frac{N_c^2-1}{2N_c}\Bigg[
(N_c^2-2)(N_c^2-1)\;\VVllr
-4\,(2N_c^2-3)\;\VVlmR
+4\,(N_c^2-2)\;\VVlmA\Bigg]\!.\qquad\qquad
&&&eq:Iterms-mid}

Now the three different terms with $V$ have to be expressed in terms of $G$ and
then as $\alpha$ or BFKL terms.  Before we do that, we will show that a $G$
function contracted with pomeron wave functions always is a sum of the two, ie
that the functions $b$ and $t$ cannot lead to anything other than the
well-known BFKL term.  Leaving aside the functions $a$ and $s$ which lead to
$\alpha$, $G$ is composed of three functions:
$$
G(1,2,3)=\frac{g^2}2
\big[2\,c(123)-2\,b(12,3)-2\,b(23,1)+t(12,3)+t(23,1)+\ldots\big]\,.
$$
The function $c$ does not play a role when one projects onto pomerons, since it
always identifies both coordinates of the pomerons.  The functions $b$ and $t$
may also identify two coordinates of a pomeron and thus vanish under
projection.  But if, for instance, no pomeron couples to the pair of arguments
2 and~3, one of the $b$ and $t$ functions remains.  In the simpler case with
only two pomerons, this looks like this:
\dmath1{
&&G(1,2,3)\otimes\Ez a^*(1,2)\,\Ez b^*(1,3) = {}\cr
&&
{}=\frac{g^2}2 \big({-}2\,b(23,1)+t(23,1)\big)\otimes\Ez a^*(1,2)\,\Ez b^*(1,3)
+\frac{g^2}2 \alpha(2,1,3)\otimes\Ez a^*(1,2)\,\Ez b^*(1,3)\,.
}
Since the pomeron wave functions are symmetric, only the symmetric part of $b$
and~$t$ survives the projection.  Therefore we are allowed to exchange their
arguments.  We symmetrise the expression in these arguments and add a $c$
function (whose projection is zero anyway) to obtain a complete BFKL kernel:
\dmath1{
\frac{g^2}2 \big({-}2\,b(23,1)&+&t(23,1)\big)\otimes\Ez a^*(1,2)\,\Ez b^*(1,3)
\cr
&=&\frac{g^2}2 \big({-}2\,b(2,1)+t(2,1)\big)\otimes\Ez a^*(1,2)\,\Ez b^*(1,2)
\cr
&=&\frac{g^2}2\frac12 \big(c(12)-2\,b(1,2)-2\,b(2,1)+t(1,2)+t(2,1)\big)
\otimes\Ez a^*(1,2)\,\Ez b^*(1,2)
\cr
&=&\frac12\frac1{N_c}
\big[\Delta_1\,\Delta_2\big(K_{BFKL}\otimes\Phi_2\big)(1,2)\big]
\otimes\Ez a^*(1,2)\,\Ez b^*(1,2)\,.
&eq:bfkl/2}
Note that the prefactor of the BFKL kernel is half that obtained when the
middle argument of $G$ is left empty, see~(\ref{eq:G_BFKL}).  The Laplacians
only serve to cancel propagators defined into the BFKL kernel.  The wave
function $\Phi_2$ is the amplitude attached to the kernel from above, which
will in our case be replaced by an~$\Ez{}$ as in Equation~\ref{eq:p3pbfkl}.

Now we can evaluate the three terms containing the $2\to4$ vertex function~$V$.
Taking into account the symmetry of $G$, one arrives at the result:
\dmath2{
\VVllr&=&4\frac{g^2}2\VGllr=g^4\;\Vallr \cr
\VVlmR&=&\frac{g^2}2\Bigg[2\;\VGllb-\VGllr-\VGlbb\Bigg] \crl{eq:IVtoG}
&=&\frac{g^4}4\Bigg[2\;\Vallb-\Vallr-\Valbb\Bigg] 
	+ \frac{g^2}4\frac1{N_c}\Vbfkl   \cr
\VVlmA&=&2\frac{g^2}2 \Bigg[\VGllr-\VGlrb\Bigg]
=\frac{g^4}2 \Bigg[\Vallr-\Valrb\Bigg]
}
In the second $V$ term, the first and the last $G$ term both contain a BFKL
term.  They partly cancel each other.

We now plug the formulas~(\ref{eq:IVtoG}) into (\ref{eq:Iterms-mid}) to compute
our final result.\footnote{In principle it would be sufficient now to average
over the cyclic permutations of the pomerons since due to the symmetry of
$\alpha$ only one pomeron can be distinguished from the others.  But we will
leave the sum over all permutation to make the comparison
with~(\ref{eq:Wresult}) easier.}
\dmath1{
\ldots&=&\sumabc\frac16\frac{g^6}8\frac{N_c^2-1}{N_c}
\Bigg[\big(4\,(N_c^2-2)(N_c^2-1)+4\,(2N_c^2-3)+2\cdot4\,(N_c^2-2)\big)\;\Vallr
\cr&&\qquad\qquad\qquad\qquad\qquad{}
-2\cdot4\,(N_c^2-2)\;\Valrb-2\cdot4\,(2N_c^2-3)\;\Vallb
\cr&&\qquad\qquad\qquad\qquad\qquad{}
-4\,(2N_c^2-3)\;\Valbb+4\,(2N_c^2-3)\frac1{N_cg^2}\Vbfkl\Bigg]
\cr
&=&\sumabc g^6\,\frac{N_c^2-1}{N_c}\,
\Bigg[\frac1{12}(N_c^4+N_c^2-5)\;\Vallr-\frac16(N_c^2-2)\;\Valrb
-\frac16(2N_c^2-3)\;\Vallb
\cr&&\qquad\qquad\qquad\qquad\qquad{}
+\frac1{12}(2N_c^2-3)\;\Valbb-\frac1{12}(2N_c^2-3)\frac1{N_cg^2}\Vbfkl\Bigg]
&eq:Iresult}
One can see that this vertex dominates over the one derived from the
irreducible $2\to6$ gluon vertex in the large $N_c$ limit: Here the first term
is of order~$N_c^5$, while (\ref{eq:Wresult}) is of order~$N_c^3$.  This
differs from the situation in the case of the $1\to2$ pomeron vertex, where the
irreducible vertex was $N_c$-leading (see~\cite{bart3p}).

\subsection{The terms with the function $J$}
\label{sec:Jterms}

The terms with the $J$ function are very similar to the $I$ terms treated in
the previous section.  The sum
\dmath2{
\sum d^{a_2a_1a_3a_4}\,\delta_{a_5a_6}\;J(1,2,3,4;5,6) 
&=&
d^{a_2a_1a_3a_4}\,\delta_{a_5a_6}\;J(1,2,3,4;5,6) 
\crl{eq:Jterms}
&&{}+d^{a_2a_1a_3a_5}\,\delta_{a_4a_6}\;J(1,2,3,5;4,6) + \ldots
}
runs over the same fifteen terms obtained by dividing six arguments into two
groups of four and two.  The only difference is in the colour structure: While
the indices of the $d^{abcd}$ tensor were in ascending order in the case of the
$I$ function, here the first two indices are interchanged.

The first steps of the derivation of the $J$ terms are exactly analogous to the
terms with~$I$.  The same colour factors occur.  The functions $I$ and $J$ have
the same symmetry, so the graphs which led to equal terms with the $I$ function
are also equal for the $J$ function.  The colour tensor is again symmetric
under simultaneous exchange of the arguments/colour indices 1 with 4, and 2
with 3.  Due to the difference in the colour tensor, this is brought about not
by its invariance under reversal of the order of the indices (as for the $I$
terms) but by invariance under cyclic permutations (by two positions, in this
case).

What is different, however, is the matching of colour factors to spatial
convolutions.  One can see this graphically by drawing a tensor $d^{bacd}$ as a
$d^{abcd}$ with crossed legs.  In the previous section we derived two different
contractions of the $d^{abcd}$ alone and found a third to be equal to one of
them.  One of these colour factors remains the same for the ``crossed''
$d^{abcd}$ tensor, but the other two are interchanged:
\dmath1{
\Cdxpairs&=&\Cdpairs=\frac1{2N_c}(N_c^2-1)^2 \cr
\Cdxcross&=&\Cdbridge=\Cdpairs=\frac1{2N_c}(N_c^2-1)^2 \cr
\Cdxbridge&=&\Cdcross=-\frac1{2N_c}(N_c^2-1)
}
The additional Kronecker delta affects the colour factors in the usual way: If
it is linked to the $d^{abcd}$ tensor (as in (\ref{eq:Cddellink})), the result
is the same as for the $d^{abcd}$ tensor alone, otherwise there is an
additional factor $(N_c^2-1)$.

We will give the colour factors for each term below, together with its
decomposition in terms of the $2\to4$ vertex function~$V$.  In contrast
to~(\ref{eq:ItoV}) we give all terms whose spatial part is equal but which
might differ in the colour factor separately.  As it happens, the two pairs of
equal terms also have equal colour structure.  The multiplicity of the seven
terms is given in the first column.
\dmath1{
1\cdot{}\qquad\VJa&=2g^2\;\VVllr\ \ \ \qquad\qquad\qquad\qquad
&\Cdxbridge\ \Cdeldel=\Cdcross\ \Cdeldel=-\frac1{2N_c}(N_c^2-1)^2
\cr
1\cdot{}\qquad\VJb&=g^2\;\VVllr\phantom{2}\ \ \ \qquad\qquad\qquad\qquad
&\Cdxpairs\ \Cdeldel=\Cdpairs\ \Cdeldel=\frac1{2N_c}(N_c^2-1)^3
\cr
1\cdot{}\qquad\VJbb&=g^2\;\VVllr\phantom{2}\ \ \ \qquad\qquad\qquad\qquad
&\Cdxcross\ \Cdeldel=\Cdpairs\ \Cdeldel=\frac1{2N_c}(N_c^2-1)^3
\cr
2\cdot{}\qquad\VJc&=2g^2\;\VVlmA\ \ \ \qquad\qquad\qquad\qquad
&\CJc=\Cdcross=-\frac1{2N_c}(N_c^2-1)
\cr
2\cdot{}\qquad\VJd&=-2g^2\;\VVlmR+2g^2\;\VVlmA\qquad
&\CJd=\Cdcross=-\frac1{2N_c}(N_c^2-1)
\cr
4\cdot{}\qquad\VJe&=-g^2\;\VVlmR+g^2\;\VVlmA\phantom{22}\qquad
&\CJe=\Cdpairs=\frac1{2N_c}(N_c^2-1)^2
\cr
4\cdot{}\qquad\VJee&=-g^2\;\VVlmR+g^2\;\VVlmA\phantom{22}\qquad
&\CJee=\Cdpairs=\frac1{2N_c}(N_c^2-1)^2
}
Adding them all up, we obtain:
\dmath2{
&&\frac16\sumabc g^2\frac{N_c^2-1}{2N_c}
\Bigg[2\,(N_c^2-2)(N_c^2-1)\;\VVllr-4\,(2N_c^2-3)\;\VVlmR+8\,(N_c^2-2)\;\VVlmA
\Bigg]\,.
}
Putting in (\ref{eq:IVtoG}) we get the result:
\dmath2{
\sumabc g^6\frac{N_c^2-1}{N_c}
\Bigg[&&\frac1{12}(2N_c^4-7)\;\Vallr-\frac13(N_c^2-2)\;\Valrb
-\frac16(2N_c^2-3)\Vallb
\cr&&
{}+\frac1{12}(2N_c^2-3)\;\Valbb-\frac1{12}(2N_c^2-3)\frac1{N_cg^2}\Vbfkl\Bigg]
\,.&eq:Jresult}
As was the case for the $I$ terms, the first term is of order $N_c^5$ and
therefore dominates in the large $N_c$ limit.  The other terms of the reducible
vertices and all of the irreducible one~(\ref{eq:Wresult}) were only of
order~$N_c^3$.

\subsection{Distinguishable pomerons}
\label{sec:distinguishable}

Having computed the projection of reducible $2\to6$ gluon transitions onto a
completely symmetric three-pomeron state, we will now have a look at the
opposite extreme: Three distinguishable pomerons which couple to specific
arguments of the transition terms.  We will choose as an example the
three-pomeron state in which pomeron $a$ couples to arguments 1 and~2, pomeron
$b$ to arguments 3 and~4, and pomeron $c$ to arguments 5 and~6.  This gives the
following three-pomeron state:
$$
\Ez a^*(1,2)\,\Ez b^*(3,4)\,\Ez c^*(5,6)
$$

The calculation of the projection is analogous to the projection onto a
symmetric three-pomeron state performed above.  The one additional difficulty
is that the terms are not only distinguished by whether a pomeron couples to a
given pair of arguments, but also which pomeron couples to them.  The pomerons
in the graphical notation have to be labelled $a$, $b$ or~$c$.  The identities
(\ref{eq:ItoV}) and (\ref{eq:IVtoG}) for the $I$ terms and analogous ones for
$L$ and $J$ have to be rewritten to take the distinctiveness of the pomerons
into account.  Other than that, the derivation is quite analogous to the
previous sections.  We will not present it in detail here, but give just the
results.  In all diagrams, two of the pomerons are interchangeable, either
because they couple to the same pair of arguments or because of the symmetry
of~$\alpha$.  Where necessary, the third pomeron is labelled in the diagrams.

The result for the $L$ terms is:
\dmath2{
\delta_{a_1a_2}\,&&\delta_{a_3a_4}\,\delta_{a_5a_6}\,
\left(\sum f_{a_1a_2a_3}\,f_{a_4a_5a_6}\;L(1,2,3;4,5,6)\right) \otimes
\Ez a^*(1,2)\,\Ez b^*(3,4)\,\Ez c^*(5,6)={}
\crd&&{}=
4\,N_c(N_c^2-1)
\frac{g^6}8 \Bigg[\frac12\sumabc\Vallr 
- \Valrbi a + \Valrbi b - \Valrbi c  
&eq:Ldistpom\cr\crd&&\quad{}
- \Vallbi a + \Vallbi b - \Vallbi c
- \Valbbi a + \Valbbi b - \Valbbi c \Bigg]
-(N_c^2-1)\,\frac{g^4}8\,\Vbfkl\,.
}
One can see that the coefficients of some terms in which the pomeron with
index~$b$ is special differ from those with indices $a$ and~$c$.  The reason
for this is that the pomeron $b$ couples to the arguments of~$L$ which are
distinguishable from the others, see Section~\ref{sec:Lterms}.

The projection of the terms with the functions $I$ and~$J$ is:
\dmath2{
\delta_{a_1a_2}\,&&\delta_{a_3a_4}\,\delta_{a_5a_6}\,
\left(\sum d^{a_1a_2a_3a_4}\,\delta_{a_5a_6}\;I(1,2,3,4;5,6)\right) \otimes
\Ez a^*(1,2)\,\Ez b^*(3,4)\,\Ez c^*(5,6)={}
\crd&&{}=
\frac1{2N_c}(N_c^2-1)^2\,(N_c^2+1)\,g^6\frac12\sumabc\Vallr
\crd&&\quad{}
+\frac1{2N_c}(N_c^2-1)^2\,g^6\Bigg[
-\Valrbi a-2\,\Valrbi b-\Valrbi c
-\Vallbi a-2\,\Vallbi b-\Vallbi c
\crd&&\quad{}
+\Valbbi a+\Valbbi c\Bigg]
-2\frac1{2N_c^2}(N_c^2-1)^2\,g^4\,\Vbfkl
&eq:Idistpom\cr
\delta_{a_1a_2}\,&&\delta_{a_3a_4}\,\delta_{a_5a_6}\,
\left(\sum d^{a_2a_1a_3a_4}\,\delta_{a_5a_6}\;J(1,2,3,4;5,6)\right) \otimes
\Ez a^*(1,2)\,\Ez b^*(3,4)\,\Ez c^*(5,6)={}
\crd&&{}=
\frac1{2N_c}(N_c^2-1)\,g^6\Bigg[
(N_c^4-3)\frac12\sumabc\Vallr
-(N_c^2-3)\Bigg(\;\Valrbi a + \;\Vallbi a\Bigg)
\crd&&\quad{}
- 2(N_c^2-1)\Bigg(\;\Valrbi b + \;\Vallbi b\Bigg)
- (N_c^2-3)\Bigg(\;\Valrbi c + \;\Vallbi c \Bigg) 
+ (N_c^2-1)\,\Valbbi a 
\crd&&\quad{}
- 2\,\Valbbi b + (N_c^2-1)\,\Valbbi c \Bigg]
-2\frac1{2N_c^2}(N_c^2-1)\,(N_c^2-2)\,g^4\,\Vbfkl
&eq:Jdistpom}
The respective $N_c$-leading term is symmetric in the pomerons.  The other
$\alpha$ terms are still symmetric in the pomerons $a$ and~$c$.  This was not
obvious a priori.  It is a consequence of the symmetries of $I$ and~$J$ and of
the fact that their two groups of arguments are ordered with ascending indices.

All expressions are of the same order in $N_c$ as in the result of the
projection onto the completely symmetric three-pomeron state.  This applies not
only to the sums of terms with $L$, $I$ and $J$, but also to each type of
diagram separately, even though the exact prefactors have changed.  In
particular, the same $\alpha$ diagram of the $I$ and $J$ terms dominates in the
large $N_c$ limit.

\cleardoublepage

\addcontentsline{toc}{part}{Appendix}

\thispagestyle{empty}

\vbox to \vsize
{
\vfill

\begin{center}
\Huge\bfseries Appendix
\end{center}

\vfill
}
\pagebreak

\appendix

\chapter{From correlation functions to Feynman graphs}
\label{app:tofeyn}

\section{Obtaining the graph classification from the correlator of path 
			integrals}
\label{sec:corrclass}

This section closes the gap left between Sections \ref{sec:pathapply}
and~\ref{sec:oddcolour}.  In Section \ref{sec:oddcolour} we switched to the
Feynman graph formulation of our problem and explained the graph classification
and the different prefactors on its basis.  It is interesting to see why this
is allowed and how the same factors follow from the continuation of our
derivation on the basis of the path integral formalism described in
Section~\ref{sec:posscatt}.

We shall now start out from the expression for the $S$-matrix and the expansion
of the gluon potential path integrals~${\bf V}_i^a$.  See
Section~\ref{sec:posscatt}, Equation~\ref{3A3} for a definition of the~${\bf
V}_i^a$.
\begin{eqnarray}
S\left(\mbf b,\left\{\mbf{x}\,_i^a\right\}\right) = 
				\frac1{36}\!
\left\langle\epsilon_{\alpha\beta\gamma}
\left({\bf V}^1_1\right)_{\alpha\alpha'}\left({\bf V}^2_1\right)_{\beta\beta'}
\left({\bf V}^3_1\right)_{\gamma\gamma'}\epsilon_{\alpha'\beta'\gamma'}
\epsilon_{\rho\mu\nu}\left({\bf V}^1_2\right)_{\rho\rho'}
\left({\bf V}^2_2\right)_{\mu\mu'}
\left({\bf V}^3_2\right)_{\nu\nu'}\epsilon_{\rho'\mu'\nu'}\right\rangle 
&& \quad
\\\noalign{\hskip2mm}
\label{eq:Vexpa_app}
\left({\bf V}^a_i\right)_{\alpha\beta} =
\delta_{\alpha\beta} - i g  \hat B_{a,i}^c \tau^c_{\alpha\beta}
-\frac12 g^2 
\hat B_{a,i}^c\hat B_{a,i}^{c'}(\tau^c\tau^{c'})_{\alpha\beta}
-\frac{i^3}{3!} g^3 
\hat B_{a,i}^c\hat B_{a,i}^{c'}\hat B_{a,i}^{c''}
(\tau^c\tau^{c'}\tau^{c''})_{\alpha\beta} + {\cal O}(g^4) 
&& \quad
\\\noalign{\hskip2mm}
\hat B_{a,i}^c \tau^c = \int_{\Gamma^a_i}dz^\mu~{\bf B}_\mu(z) 
\hskip 104mm
&& \quad
\end{eqnarray}
The expansion of the $S$-matrix in $g$ (or, equivalently, in the 
$\hat B_{a,i}^c)$ up to ${\cal O}(g^3)$ contains a huge number of 
terms. We have to extract those which represent perturbative odderon
exchange. Since they are of the order $g^6$, they all contain a 
correlator of six $\hat B_{a,i}^c$. 

In a perturbative calculation this correlator can be expanded in correlators of
two $\hat B_{a,i}^c$.  The reason for this is that even though the $\hat
B_{a,i}^c$ are expansion coefficients of path integrals over the gauge
potentials, they will ultimately lead to correlators of gluon potentials
themselves.  (This is shown in Section~\ref{sec:intlight} for the
two-correlator.)  These correlators in turn can be expanded in gluon
propagators in perturbative QCD.

We do not want terms in the expansion which contain correlators of two $\hat
B_{a,i}^c$ from the same baryon.  They represent a gluon exchange between two
quarks of the same baryon, ie a self-energy correction.  We are interested only
in the lowest-order odderon exchange in which three gluons are exchanged
between two baryons represented by three quarks.  Therefore we can restrict
ourselves to those terms with three $\hat B_{a,i}^c$ from each baryon: The
expansion of a six-correlator in which a different number of $\hat B_{a,i}^c$
comes from each hadron always contains self-energy terms.  We obtain six terms
of the expansion which are relevant for triple-gluon exchange:
\begin{eqnarray}
\label{eq:sixexpa}
\langle \hat B_{a,1}^c \,\hat B_{a',1}^{c'} \,\hat B_{a'',1}^{c''} \;
		\hat B_{b,2}^d \,\hat B_{b',2}^{d'} \,\hat B_{b'',2}^{d''} \rangle =
&\phantom{+\;}&
\langle \hat B_{a,1}^c       \,\hat B_{b,2}^d       \rangle \,
\langle \hat B_{a',1}^{c'}   \,\hat B_{b',2}^{d'}   \rangle \,
\langle \hat B_{a'',1}^{c''} \,\hat B_{b'',2}^{d''} \rangle
\nonumber\\ &+&
\langle \hat B_{a,1}^c       \,\hat B_{b,2}^d       \rangle \,
\langle \hat B_{a',1}^{c'}   \,\hat B_{b'',2}^{d''} \rangle \,
\langle \hat B_{a'',1}^{c''} \,\hat B_{b',2}^{d'}   \rangle
\nonumber\\ &+&
\langle \hat B_{a,1}^c       \,\hat B_{b',2}^{d'}   \rangle \,
\langle \hat B_{a',1}^{c'}   \,\hat B_{b,2}^{d}     \rangle \,
\langle \hat B_{a'',1}^{c''} \,\hat B_{b'',2}^{d''} \rangle
\nonumber\\ &+&
\langle \hat B_{a,1}^c       \,\hat B_{b',2}^{d'}   \rangle \,
\langle \hat B_{a',1}^{c'}   \,\hat B_{b'',2}^{d''} \rangle \,
\langle \hat B_{a'',1}^{c''} \,\hat B_{b,2}^{d}     \rangle
\nonumber\\ &+&
\langle \hat B_{a,1}^c       \,\hat B_{b'',2}^{d''} \rangle \,
\langle \hat B_{a',1}^{c'}   \,\hat B_{b,2}^{d}     \rangle \,
\langle \hat B_{a'',1}^{c''} \,\hat B_{b',2}^{d'}   \rangle
\nonumber\\ &+&
\langle \hat B_{a,1}^c       \,\hat B_{b'',2}^{d''} \rangle \,
\langle \hat B_{a',1}^{c'}   \,\hat B_{b',2}^{d'}   \rangle \,
\langle \hat B_{a'',1}^{c''} \,\hat B_{b,2}^{d}     \rangle
\nonumber\\ &+&\hbox{self-energy terms.}
\end{eqnarray}

The procedure now is to list every six-correlator containing three 
$\hat B_{a,i}^c$ from each baryon and expand them according 
to~(\ref{eq:sixexpa}). We will see that in many cases there appear
equivalent terms which can be summed up by calculating one of them
and multiplying them with a constant factor. The different terms
correspond one-to-one to the different Feynman graphs. 

The fact that some terms in the expansion of the $S$-matrix are equivalent is
one source of constant prefactors.  Other factors arise from the coefficients
in the expansion of ${\bf V}_i^a$~(\ref{eq:Vexpa_app}) and from the colour
structure.  The colour tensors and the corresponding factors have been treated
in Section~\ref{sec:oddcolour}.  They are listed again in
Table~\ref{tab:fact_app}.  The factors from the exponential series of ${\bf
V}_i^a$, which are straightforward, and the factors from equivalent terms,
which are quite tricky, will be discussed in the following.

For coping with the multitude of terms, we again employ the classification of
the odderon-proton coupling described in Section~\ref{sec:oddcolour}.
In the language of the above formulas, the three types of couplings
are distinguished by how many $\hat B_{a,i}^c$ come from each of the
three ${\bf V}_i^a$ of a baryon: \par\noindent
a) The third-order term of one of the ${\bf V}_i^a$ and the Kronecker delta
of the others contributes. \\
b) One second-order, one linear and one zeroth-order term come from the 
three expanded ${\bf V}_i^a$. \\
c) Each of the ${\bf V}_i^a$ contributes the linear term from its expansion.

One can see immediately from Equation~\ref{eq:Vexpa_app} that a type~(a) 
coupling entails a factor of $\frac1{3!}$, (b) a factor of $\frac12$ and 
(c) a factor of one. This gives the factors from the expansion of 
${\bf V}_i^a$ listed in Table~\ref{tab:fact_app}.

The derivation of the factors due to terms which represent the same Feynman 
graph and are numerically equal is more involved. Practically the only way to
obtain them is to write down at least a couple of terms for each graph type
and their expansion according to~(\ref{eq:sixexpa}). This will be done in 
the following for two examples, (aa) and (cc).

Two (a)-type couplings yield nine different six-correlators, depending on 
which quark the three gluons couple to (or, to put it differently, which
of the ${\bf V}_i^a$ contributes its third-order term). Here are some of them:
\begin{eqnarray}
\hbox{(aa):} &&
\langle \hat B_{1,1}^c \,\hat B_{1,1}^{c'} \,\hat B_{1,1}^{c''} \;
		\hat B_{1,2}^d \,\hat B_{1,2}^{d'} \,\hat B_{1,2}^{d''} \rangle
+ \langle \hat B_{2,1}^c \,\hat B_{2,1}^{c'} \,\hat B_{2,1}^{c''} \;
		  \hat B_{1,2}^d \,\hat B_{1,2}^{d'} \,\hat B_{1,2}^{d''} \rangle
\nonumber\\&&
{}+ \ldots + 
\langle \hat B_{3,1}^c \,\hat B_{3,1}^{c'} \,\hat B_{3,1}^{c''} \;
		\hat B_{3,2}^d \,\hat B_{3,2}^{d'} \,\hat B_{3,2}^{d''} \rangle\,.
\end{eqnarray}
They represent a three-gluon exchange between the first quark of the first
baryon and the first quark of the second baryon, between the second quark of 
the first baryon and the first quark of the second baryon, and so on to the
third quark of the first baryon and the third quark of the second baryon.
Let us have a closer look at the first of these six-correlators. It can be 
expanded in two-correlators according to~(\ref{eq:sixexpa}). 
\dmath{1.5}{
\hbox{(aa), 1$^{\rm st}$ term:}\quad
\langle \hat B_{1,1}^c \,\hat B_{1,1}^{c'} \,\hat B_{1,1}^{c''} \;
		\hat B_{1,2}^d \,\hat B_{1,2}^{d'} \,\hat B_{1,2}^{d''} \rangle =
&&\;
\langle \hat B_{1,1}^c       \,\hat B_{1,2}^d       \rangle \,
\langle \hat B_{1,1}^{c'}    \,\hat B_{1,2}^{d'}   \rangle \,
\langle \hat B_{1,1}^{c''}   \,\hat B_{1,2}^{d''} \rangle
\cr &+&\;
\langle \hat B_{1,1}^c       \,\hat B_{1,2}^d       \rangle \,
\langle \hat B_{1,1}^{c'}    \,\hat B_{1,2}^{d''} \rangle \,
\langle \hat B_{1,1}^{c''}   \,\hat B_{1,2}^{d'}   \rangle
\cr &+&\;
\langle \hat B_{1,1}^c       \,\hat B_{1,2}^{d'}   \rangle \,
\langle \hat B_{1,1}^{c'}    \,\hat B_{1,2}^{d}     \rangle \,
\langle \hat B_{1,1}^{c''}   \,\hat B_{1,2}^{d''} \rangle
\cr &+&\;
\langle \hat B_{1,1}^c       \,\hat B_{1,2}^{d'}   \rangle \,
\langle \hat B_{1,1}^{c'}    \,\hat B_{1,2}^{d''} \rangle \,
\langle \hat B_{1,1}^{c''}   \,\hat B_{1,2}^{d}     \rangle
\cr &+&\;
\langle \hat B_{1,1}^c       \,\hat B_{1,2}^{d''} \rangle \,
\langle \hat B_{1,1}^{c'}    \,\hat B_{1,2}^{d}     \rangle \,
\langle \hat B_{1,1}^{c''}   \,\hat B_{1,2}^{d'}   \rangle
\cr &+&\;
\langle \hat B_{1,1}^c       \,\hat B_{1,2}^{d''} \rangle \,
\langle \hat B_{1,1}^{c'}    \,\hat B_{1,2}^{d'}   \rangle \,
\langle \hat B_{1,1}^{c''}   \,\hat B_{1,2}^{d}     \rangle
\cr &+&\;\hbox{self-energy terms}  
\cr = 
&-&\Chi_{11}^3[\delta^{cd}\,\delta^{c'd'}\,\delta^{c''d''} + 
			   \delta^{cd}\,\delta^{c'd''}\,\delta^{c''d'} 
\cr&&\;\;
{}+ \delta^{cd'}\,\delta^{c'd}\,\delta^{c''d''}
+ \delta^{cd'}\,\delta^{c'd''}\,\delta^{c''d}
\cr&&\;\;
{}+ \delta^{cd''}\,\delta^{c'd}\,\delta^{c''d'}
+ \delta^{cd''}\,\delta^{c'd'}\,\delta^{c''d}]
\cr
&& \;\;{}+\hbox{self-energy terms}  
&eq:aa_1st}
Here we have used the fact that the colour part of the gluon propagator is a
Kronecker delta.  The configuration space part of the propagator has been
wrapped up in $\Chi_{ab}$ which will be calculated in the following section and
is defined as follows:
\begin{equation}
\label{eq:chidef_app}
\langle\hat B_{a,1}^c\,\hat B_{b,2}^d\rangle =
-\delta^{cd}\;\Chi_{ab}\,.
\end{equation}
It should now be obvious that all six terms in Equation~\ref{eq:aa_1st}
are equal: When the symmetric colour tensors of the odderon are contracted
with the delta symbols, it does not play a role which index of~$d^{cc'c''}$
is contracted with which index of~$d^{dd'd''}$. Therefore the first term
for graph type (aa) is equal to $-6\;\Chi_{11}^3$.

Let us now have a look at the graph type (cc). It contains only one term
from the expansion of the $S$-matrix since there is only one way to couple
three gluons to three different quarks: one to each. Its expansion in 
two-correlators is the following:
\begin{eqnarray}
\label{eq:cc_expa}
\hbox{(cc):}\quad
\langle \hat B_{1,1}^c \,\hat B_{2,1}^{c'} \,\hat B_{3,1}^{c''} \;
		\hat B_{1,2}^d \,\hat B_{2,2}^{d'} \,\hat B_{3,2}^{d''} \rangle =
&\phantom{+\;}&
\langle \hat B_{1,1}^c       \,\hat B_{1,2}^d       \rangle \,
\langle \hat B_{2,1}^{c'}    \,\hat B_{2,2}^{d'}   \rangle \,
\langle \hat B_{3,1}^{c''}   \,\hat B_{3,2}^{d''} \rangle
\nonumber\\ &+&
\langle \hat B_{1,1}^c       \,\hat B_{1,2}^d       \rangle \,
\langle \hat B_{2,1}^{c'}    \,\hat B_{3,2}^{d''} \rangle \,
\langle \hat B_{3,1}^{c''}   \,\hat B_{2,2}^{d'}   \rangle
\nonumber\\ &+&
\langle \hat B_{1,1}^c       \,\hat B_{2,2}^{d'}   \rangle \,
\langle \hat B_{2,1}^{c'}    \,\hat B_{1,2}^{d}     \rangle \,
\langle \hat B_{3,1}^{c''}   \,\hat B_{3,2}^{d''} \rangle
\nonumber\\ &+&
\langle \hat B_{1,1}^c       \,\hat B_{2,2}^{d'}   \rangle \,
\langle \hat B_{2,1}^{c'}    \,\hat B_{3,2}^{d''} \rangle \,
\langle \hat B_{3,1}^{c''}   \,\hat B_{1,2}^{d}     \rangle
\nonumber\\ &+&
\langle \hat B_{1,1}^c       \,\hat B_{3,2}^{d''} \rangle \,
\langle \hat B_{2,1}^{c'}    \,\hat B_{1,2}^{d}     \rangle \,
\langle \hat B_{3,1}^{c''}   \,\hat B_{2,2}^{d'}   \rangle
\nonumber\\ &+&
\langle \hat B_{1,1}^c       \,\hat B_{3,2}^{d''} \rangle \,
\langle \hat B_{2,1}^{c'}    \,\hat B_{2,2}^{d'}   \rangle \,
\langle \hat B_{3,1}^{c''}   \,\hat B_{1,2}^{d}     \rangle
\nonumber\\ &+&\hbox{self-energy terms.}
\end{eqnarray}
This time the six terms are all different: They differ not only in the
colour indices but also in which pairs of quarks exchange a gluon. They 
correspond one-to-one to the six Feynman graphs one obtains when assuming
that the three quarks of a baryon are distinguishable. This preposition 
is quite necessary since the quarks differ in their position according to 
our geometric model for the proton. 

The graph types (aa) and (cc) illustrate two extreme cases: All terms may be
equal because of the symmetric colour structure of the odderon, which leads to
a factor~6 for the expression obtained in the case (aa), or all may be
different, which leads to a factor~1 for (cc).  Graph types (ab) and (ac) also
fit into this simple scheme, resulting in six equal terms and therefore a
factor of~6 like~(aa).  For the type (bc), the expansion~(\ref{eq:sixexpa})
contains three pairs of equal terms, giving a factor~2 for each different term.

Two (b)-type couplings are the most interesting case. The six terms of the
expansion in two-correlators split into a set of two and a set of four
equivalent terms.  They correspond to the graph types (bb1) and~(bb2) whose
prefactors differ by a factor of~2.  The two equivalent terms are the ones with
two $\Chi$ functions with the same set of indices. They get a prefactor~2.
They correspond to Feynman graphs in which the pair of gluons coupling to the
same quark is the same at both ends, ie type (bb1).  The four equivalent terms
(bb2) contain $\Chi$ functions whose pairs of indices are all different.  They
get a factor~4.

The relative prefactors of the different graph types have now been
adequately explained on the basis of our derivation of high-energy 
scattering in position space. As Table~\ref{tab:fact_app} shows, the 
prefactors are the same as in the derivation from the Feynman graph 
formulation but differ in their composition~(confer Table~\ref{tab:prefact}). 
What remains to be done is to actually calculate the 
two-correlators~(\ref{eq:chidef_app}). This will be done in the 
following section.

\begin{table}[h]
\moveright 11mm \vbox{
\def\eol{\cr\noalign{\nointerlineskip}}
\def\m{\llap{$-$}}
\tabskip=0pt
\halign{
\vrule height2.75ex depth1.25ex # \tabskip=1em & #\hfil &#\vrule& 
	\hfil$#$\hfil &#\vrule& \hfil$#$\hfil &#\vrule& \hfil$#$\hfil &#\vrule& 
	\hfil $#$ & \hskip 60pt plus 1fil #\vrule \tabskip=0pt\cr\noalign{\hrule}
& Type && \hbox{Expansion of ${\bf V}_i^a$} && \hbox{Colour} &
		& \hbox{Equivalence} &&\omit\hfil $C$\rlap{${}_{\rm type}$} &\cr
\noalign{\hrule}
& (aa) && \frac1{36} && \frac14    && 6 && 1 \rlap{ (by definition)} &\eol
& (ab) && \frac1{12} && \m\frac18  && 6 && -\frac32 &\eol
& (ac) && \frac16    && \frac14    && 6 && 6        &\eol
& (bb1)&& \frac14    && \frac1{16} && 2 && \frac34  &\eol
& (bb2)&& \frac14    && \frac1{16} && 4 && \frac32  &\eol
& (bc) && \frac12    && \m\frac18  && 2 && -3       &\eol
& (cc) && 1          && \frac14    && 1 && 6        &\cr
\noalign{\hrule}
}}
\caption{Prefactors of different graph types originating from the expansion
		of ${\bf V}_i^a$, the colour structure and the summing up of 
		equivalent terms.}
\label{tab:fact_app}
\end{table}

\section{Integrating out the light cone coordinates}
\label{sec:intlight}

In the previous section, we obtained terms containing correlators of two
$\hat B_{a,i}^c$, the expansion coefficients of path integrals over the
gauge potentials. We now want to relate them to an expression in perturbative 
QCD which we know, such as the gluon propagator. To that end, we first
express the correlator in the expansion coefficients of the potentials 
themselves.
\begin{equation}
\langle\hat B_{a,1}^c\,\hat B_{b,2}^d\rangle=
\left\langle\int_{\Gamma_1^a} dx^\mu\,B_\mu^c(x)\;
			\int_{\Gamma_2^b} dy^\nu\,B_\nu^d(y)\right\rangle\,,
\end{equation}
where
\begin{equation}
\hat B_{a,i}^c \tau^c = \int_{\Gamma^a_i}dz^\mu~{\bf B}_\mu(z)
\quad\hbox{and}\quad
B_\mu^c(z) \tau^c = {\bf B}_\mu(z)\,.
\end{equation}
The paths $\Gamma_i^a$ run along the light cone, in opposite directions for
each scattering baryon.  We obtain
\begin{eqnarray}
\langle\hat B_{a,1}^c\,\hat B_{b,2}^d\rangle&=&
\left\langle\int dx^+\,B_+^c(\mbf x_1^a,x^+,0)\;
			\int dy^-\,B_-^d(\mbf x_2^b,0,y^-)\right\rangle\, \nonumber\\
&=&\int dx^+\int dy^-\langle B_+^c(\mbf x_1^a,x^+,0)\,
							B_-^d(\mbf x_2^b,0,y^-)\rangle
\nonumber\\
&=&\int dx^+\int dx^-\int dy^+\int dy^-
\langle B_+^c(\mbf x_1^a,x^+,x^-)\,B_-^d(\mbf x_2^b,y^+,y^-)\rangle
\,\delta(x^-)\,\delta(y^+)\,. \qquad\quad
\end{eqnarray}
$\mbf x_1^a$ and $\mbf x_2^b$ are the transverse coordinates of the paths
$\Gamma_1^a$ and $\Gamma_2^b$, respectively. Those paths are the trajectories
of a quark in each baryon. See Figure~\ref{fig:baryonpaths} for a graphical
representation of them.  The formulation in the last line allows us to
introduce into our calculations the gluon Green's function
$$
D_{\mu\nu}^{cd}(\mbf x-\mbf y, x^+-y^+, x^--y^-) =
(-i)^2\,\langle B_\mu^c(\mbf x,x^+,x^-)\,B_\nu^d(\mbf y,y^+,y^-)\rangle\,.
$$
It already contains the factors $-i$ which in our derivation come from the 
expansion of 
${\bf V}_i^a$. $D$ is translationally invariant, ie it depends only on the 
relative coordinates of the gluon potentials. We will use the Fourier 
transform $\tilde D_{\mu\nu}^{cd}(\mbf k, k^+, k^-)$ of $D$:
\begin{eqnarray}
\ldots&=& - \int dx^+\int dx^-\int dy^+\int dy^-
\int\frac{d^2\mbf k\,dk^+\,dk^-}{2\cdot(2\pi)^4}
{}\cdot \nonumber\\ &&\qquad\qquad{}\cdot
\tilde D_{+-}^{cd}(\mbf k, k^+, k^-)\;
e^{i(-\mbf k(\mbf x_1^a-\mbf x_2^b)+
	\frac12k^+(x^--y^-)+\frac12k^-(x^+-y^+) )}
\,\delta(x^-)\,\delta(y^+)
\nonumber\\\noalign{\vskip 1mm}
&=& - \int dx^+\int dy^-
\int\frac{d^2\mbf k\,dk^+\,dk^-}{2\cdot(2\pi)^4}
\tilde D_{+-}^{cd}(\mbf k, k^+, k^-)\;
e^{i(-\mbf k(\mbf x_1^a-\mbf x_2^b) - \frac12k^+y^- + \frac12k^-x^+ )}
\nonumber\\\noalign{\vskip 2mm}
&=& - \int\frac{d^2\mbf k\,dk^+\,dk^-}{2\cdot(2\pi)^4}
\tilde D_{+-}^{cd}(\mbf k, k^+, k^-)\;
e^{-i\mbf k(\mbf x_1^a-\mbf x_2^b)}\;
2\pi\cdot2\,\delta(k^+)\;\;2\pi\cdot2\,\delta(k^-)
\nonumber\\\noalign{\vskip 2mm}
&=& -2 \int\frac{d^2\mbf k}{(2\pi)^2}
\tilde D_{+-}^{cd}(\mbf k,0,0)\;
e^{-i\mbf k(\mbf x_1^a-\mbf x_2^b)}\,.
\end{eqnarray}
The additional factor 2 in the denominator of the Fourier integral reflects 
the integral measure of the light-cone coordinates, $dx^+\,dx^-=2\,dx^0\,dx^3$.
So we have arrived at the gluon propagator in transverse space:
\begin{eqnarray}
&& -2 \int\frac{d^2\mbf k}{(2\pi)^2}
\tilde D_{+-}^{cd}(\mbf k,0,0)\;
e^{-i\mbf k(\mbf x_1^a-\mbf x_2^b)}
 = 
-2\;\delta^{cd} \int\frac{d^2\mbf k}{(2\pi)^2} \frac{-i\,g_{+-}}{-\mbf k^2-m^2}
\,e^{-i\mbf k(\mbf x_1^a-\mbf x_2^b)}
\nonumber\\\noalign{\vskip 2mm}
&& = -\delta^{cd} \int\frac{d^2\mbf k}{(2\pi)^2} 
\frac i{\mbf k^2+m^2}  \,e^{-i\mbf k(\mbf x_1^a-\mbf x_2^b)}
 =: -\delta^{cd}\;\Chi_{ab}\,.
\end{eqnarray}
$\Chi$ is thereby defined as minus the spatial part of the gluon propagator in
transverse space.  The minus sign is conventional and represents a factor
$(-i)^2$ from the expansion~(\ref{eq:Vexpa_app}).  A non-zero gluon mass was
introduced to make the two-dimensional Fourier integral convergent.  In
gauge-invariant expressions which we will obtain later it can be set to 0
without causing a divergence.  All that remains to be done now is to perform
the Fourier transformation:
\begin{eqnarray}
\Chi_{ab}&=&  
	 \int \frac{d^2\mbf k}{(2\pi)^2}
		\frac i{\mbf k^2+m^2} \,e^{i\mbf k(\mbf x_1^a-\mbf x_2^b)}
	=\frac{i}{(2\pi)^2}\int\limits_0^\infty d\kappa\;\frac\kappa{\kappa^2+m^2}
	  \int\limits_0^{2\pi}d\varphi\; e^{i\kappa r \cos \varphi} \nonumber \\
&&=\frac{i}{(2\pi)^2}\int\limits_0^\infty d\kappa\;\frac\kappa{\kappa^2+m^2}\;
	2\pi\,J_0(\kappa r) 
	= \frac{i}{2\pi}\,K_0(mr)
	= \frac{i}{2\pi}\,K_0(m\,|\mbf x_1^a-\mbf x_2^b|)\,. \phantom{aaaa}
\label{eq:chiderive}
\end{eqnarray}
Here $\kappa=|\mbf k|$ and  $r=|\mbf x_1^a-\mbf x_2^b|$, the distance 
bridged by the gluon. To perform the second integration, the following 
integral relation between Bessel functions has been used:
\begin{eqnarray}
\int\limits_0^\infty\kappa^{\frac n2} (\kappa^2+c^2)^\lambda\; 
	J_{\frac n2-1}(\kappa r)\,d\kappa &=& \left(\frac2r\right)^{\lambda+1} 
	\frac{c^{\frac n2+\lambda}}{\Gamma(-\lambda)}\, K_{\frac n2+\lambda}(cr)
\\
n=2,\;\lambda=-1 &\Rightarrow& \int\limits_0^\infty\kappa(\kappa^2+c^2)^{-1}
	J_0(\kappa r)\,d\kappa = K_0(cr)\,.  \nonumber 
\label{eq:bessJtoK}
\end{eqnarray}

\chapter{The parameter values of the Donnachie-Landshoff fit}
\label{app:dlparams}

This appendix will give all the parameters of the Donnachie-Landshoff fit.
This will enable the interested reader to reproduce all our fits with the help
of the publication~\cite{dlfit}.

Most of the parameters were taken straight from~\cite{dlfit} and never
modified.  All the same, they are listed for the sake of completeness in
Table~\ref{tab:dlparams}.

\begin{table}[h]

\moveright .7in \vbox{
\def\m{\kern 1em}
\def\lowword#1{\vbox to 0pt{\hbox{\lower 2ex \hbox{#1}}\vss}}
\def\eol{
  \cr
  \noalign{\nointerlineskip}
  \omit\vrule & \omit && \omit\hrulefill 
		&& \omit\hrulefill & \omit\hrulefill &\cr
  \noalign{\nointerlineskip}}
\tabskip=0pt
\halign{\vrule width0pt height2.75ex depth1.25ex 
  \vrule # & \m #\hfil\m &#\vrule& \m$#$ \hfil\m &#\vrule &
  \qquad\qquad\hfil$#$\  & #\hfil\m & #\vrule\cr
\noalign{\hrule}
& Object exchanged && $ Parameter $ 
	&& \omit\span\hfil Numerical value &\cr\noalign{\hrule}
&  && \alpha_0-1=\epsilon 
			  && 0.08   &                         &\eol
& \lowword{Pomeron}
   && \alpha' && 0.25   & GeV$^{-2}$              &\eol
&  && \beta   && \sqrt2 & mb$^{1/4}$ GeV$^{-1/2}$ &\eol
&  && m_{\hbox{\scriptsize dipole}}^2 
			  && 0.71   & GeV$^2$                 &\cr\noalign{\hrule}
& 2 Pomerons && D\cdot\beta^2 
			  && 1.608  &                         &\cr\noalign{\hrule}
&  && \alpha_0 && 0.44  &           &\eol
& \lowword{Reggeon}
   && \alpha' && 0.93   & GeV$^{-2}$ &\eol
&  && A       && 7.8    &            &\eol
&  && B       && 2.1    &            &\cr\noalign{\hrule}
&  && t_0     && 0.3    & GeV$^2$    &\eol
& \lowword{Gluon}
   && t_1     && 0.3    & GeV$^2$    &\eol
&  && \lambda && \omit\span\hfil$ 2/t_1^3$            &\eol
&  && \tau    && \omit\span\hfil$ -\frac32\,t_1$ &\cr\noalign{\hrule}
& 3 Particles && A && 109.961 & fm$^{1/4}$ GeV$^{7/4}$ &\cr\noalign{\hrule}
}}
\caption{All parameters of the Donnachie-Landshoff fit.}
\label{tab:dlparams}
\end{table}

$m_{\hbox{\scriptsize dipole}}^2$ is the square of a dipole mass which appears
in the dipole form factor in Section~\ref{sec:dlpom}.  For double pomeron
exchange, the constant $D\cdot\beta^2$ is given instead of $D$ since this
combination is independent of $\beta$.  $D$ is the constant that ensures that
the pomeron and double pomeron contributions cancel in the dip region, and the
double pomeron amplitude is proportional to~$\beta^4$ while the pomeron is
proportional to~$\beta^2$.

$t_1$ is the gluon propagator cutoff, ie the value for $-t$ below which the
hyperbola $1/t$ is replaced by a parabola (see Figure~\ref{fig:dlprop}) or some
other cutoff function (see Section~\ref{sec:othercut}).  Its value depends on
the cutoff function used and the coupling constant.  The values for $t_1$ are
displayed in Table~\ref{tab:t1A}.  The cutoff parabola used by Donnachie and
Landshoff is parametrised by $\lambda$ and $\tau$.  Its form is $\lambda t
(t+\tau)$.

The last parameter $A$ represents the strength of the
coupling of the triple-particle exchanges. It is determined by fitting
the triple gluon exchange differential cross section to an asymptotic 
power law. It depends on $t_1$ and the coupling constant. It is also listed
in Table~\ref{tab:t1A}. 

The values given for $t_0$, $t_1$ and $A$ in Table~\ref{tab:dlparams} and in
the first line of Table~\ref{tab:t1A} apply to the unmodified
Donnachie-Landshoff fit which was used in nearly all calculations.  The others
were used when different cutoff functions for the gluon propagator were chosen
and for a running coupling.  Finally, some of them apply to calculations in
which we changed the coupling constant of the DL fit.  They were done before we
decided to use the DL fit with the unchanged coupling constant as a fixed
framework for the other odderon contributions.  All the same the relevant
parameters are given in Table~\ref{tab:t1A}.

\begin{table}[h]

\moveright .5in \vbox{
\def\eol{\cr\noalign{\nointerlineskip}}
\tabskip=0pt
\halign{\vrule width0pt height2.75ex depth1.25ex 
  \vrule #\tabskip=1em & #\hfil &#\vrule& \hfil#\hfil &#\vrule& \hfil#\hfil
  &#\vrule& \hfil#\hfil &#\vrule& \qquad\hfil#\tabskip=0pt &# \hfil\tabskip=1em
  &#\vrule\tabskip=0pt\cr
\noalign{\hrule}
& Cutoff && $\alpha_s$ && $t_0$ [GeV$^2$] && $t_1$ [GeV$^2$] && 
	\omit\hfil$A$ [fm$^{1/4}$ GeV$^{7/4}$]\span  &\cr\noalign{\hrule}
& Parabola && 0.3 && 0.3 && 0.3 && 109.&961  &\cr\noalign{\hrule}
& Parabola && 0.3 && 0.0003 && 0.31 && 109.&589  &\eol
& Parabola && 0.32&& 0.0003 && 0.31 && 104.&411  &\eol
& Parabola && 0.4 && 0.0003 && 0.31 &&  88.&321  &\eol
& Parabola && 0.5 && 0.0003 && 0.31 &&  74.&7105 &\eol
& Flat     && 0.4 && 0.0003 && 0.37 &&  88.&5942 &\eol
& Linear   && 0.4 && 0.0003 && 0.22 &&  88.&2452 &\eol
& Parabola && running && 0.0003 && 0.44 &&  91.&048 &\cr\noalign{\hrule}
}}
\caption{The values of $t_1$ and $A$ for different cutoff functions and 
		coupling constants.}
\label{tab:t1A}
\end{table}

\chapter{Derivatives of the conformal eigenfunctions \boldmath$E^{(\nu,n)}$}
\label{app:dE}

When proving the conformal transformation properties of the $G$~function, we
used the explicit form of derivatives of the conformal eigenfunction
$E^{(\nu,n)}$. These derivatives will be calculated in this appendix.

As a starting point, here is the explicit form of the conformal eigenfunction 
$E^{(\nu,n)}$ in  position space:
\begin{equation}
\label{eq:Enun_app}
E^{(\nu,n)}(\rho_{1},\rho_{2})=
\left(\frac{\rho_{12}}{\rho_{1a}\rho_{2a}}\right)^{\frac{1+n}2+i\nu}
\left(\frac{\rho_{12}^*}{\rho_{1a}^*\rho_{2a}^*}\right)^{\frac{1-n}2+i\nu}\,.
\end{equation}
$a$ is the external coordinate of the pomeron state which does not enter into
our calculations and which we will not write as an argument of $E^{(\nu,n)}$.

The differential operators which occur in the $G$~function are:
$\Delta_1\Delta_2$, $\nabla_1\Delta_2$ and a single~$\Delta_1$.  Expressed as
derivatives with respect to the complex coordinates, they become:
$16\,\del_1\del^*_1\del_2\del^*_2$, $8\,\del^*_1\del_2\del^*_2$ and
$4\,\del_1\del^*_1$.  Since none of the exponents of coordinates
in~(\ref{eq:Enun_app}) is~$-1$, we can completely separate the conjugated from
the unconjugated coordinates.%
\footnote{Otherwise, we might have to take relation~(\ref{eq:del1/rho}) into
account.}

We will first calculate the gradient with respect to one coordinate of the 
eigenfunction~$E^{(\nu,n)}$. 
\dmath2{
\nabla_1 E^{(\nu,n)}(\rho_{1},\rho_{2})
&=&2\,\del^*_1 E^{(\nu,n)}(\rho_{1},\rho_{2})
=2 \left(\frac{\rho_{12}}{\rho_{1a}\rho_{2a}}\right)^{\frac{1+n}2+i\nu}
\del^*_1
\left(\frac{\rho_{12}^*}{\rho_{1a}^*\rho_{2a}^*}\right)^{\frac{1-n}2+i\nu}
\cr
&=&2 \left(\frac{\rho_{12}}{\rho_{1a}\rho_{2a}}\right)^{\frac{1+n}2+i\nu}
\left(\frac{1-n}2+i\nu\right)
\left(\frac{\rho_{12}^*}{\rho_{1a}^*\rho_{2a}^*}\right)^{-\frac{1-n}2+i\nu}
\left(\frac1{\rho^*_{1a}\rho^*_{2a}}
		+\rho^*_{12}\frac{-\rho^*_{2a}}{(\rho^*_{1a}\rho^*_{2a})^2}\right)
\cr
&=&2 \left(\frac{\rho_{12}}{\rho_{1a}\rho_{2a}}\right)^{\frac{1+n}2+i\nu}
\left(\frac{1-n}2+i\nu\right)
\left(\frac{\rho_{12}^*}{\rho_{1a}^*\rho_{2a}^*}\right)^{-\frac{1-n}2+i\nu}
\frac1{{\rho^*_{1a}}^2}
\cr
&=&(1-n+2\,i\nu)\frac{\rho^*_{2a}}{\rho^*_{1a}\rho^*_{12}}
E^{(\nu,n)}(\rho_{1},\rho_{2})
&eq:nablE_app}
Performing the differentiation with respect to the non-conjugated coordinate 
yields an analogous result. Here it is together with the gradients with
respect to the other coordinate:%
\footnote{The symbol $\nabla^*$ is meaningful in the complex, though not in 
the vectorial notation.  See Section~\ref{sec:vect_compl} for its definition.}
\dmath2{
\nabla^*_1 E^{(\nu,n)}(\rho_{1},\rho_{2})&=&
(1+n+2\,i\nu)\frac{\rho_{2a}}{\rho_{1a}\rho_{12}}
E^{(\nu,n)}(\rho_{1},\rho_{2})
\cr
\nabla_2 E^{(\nu,n)}(\rho_{1},\rho_{2})&=&
-\,(1-n+2\,i\nu)\frac{\rho^*_{1a}}{\rho^*_{2a}\rho^*_{12}}
E^{(\nu,n)}(\rho_{1},\rho_{2})
&eq:nablE2_app\cr
\nabla^*_2 E^{(\nu,n)}(\rho_{1},\rho_{2})&=&
-\,(1+n+2\,i\nu)\frac{\rho_{1a}}{\rho_{2a}\rho_{12}}
E^{(\nu,n)}(\rho_{1},\rho_{2})
}
The derivatives with respect to~$\rho_2$ get a minus sign because the result is
expressed in~$\rho_{12}$ instead of~$\rho_{21}$.

Having successfully derived the gradient, we can immediately give the
Laplacians of a pomeron wave function~$E^{(\nu,n)}$. Since the differentiations
with respect to (un)conjugated variables add only factors of variables of the
same type, the two differentiations contained in a Laplacian are quite
independent. Despite the $\rho_{1a}$ and $\rho_{12}$ in the denominator of the
prefactor, the relation concerning the derivative of $1/\rho$
(\ref{eq:del1/rho}) must not be used.  Since $E^{(\nu,n)}$ contains additional
powers of both these variables, the total power is not $-1$, so their
derivative with respect to the complex conjugated variable vanishes. The
results are:
\dmath2{
\Delta_1 E^{(\nu,n)}(\rho_{1},\rho_{2})&=&
\big((1+2\,i\nu)^2-n^2\big)\frac{|\rho_{2a}|^2}{|\rho_{1a}|^2|\rho_{12}|^2}
E^{(\nu,n)}(\rho_{1},\rho_{2})
\crl{eq:laplE_app}
\Delta_2 E^{(\nu,n)}(\rho_{1},\rho_{2})&=&
\big((1+2\,i\nu)^2-n^2\big)\frac{|\rho_{1a}|^2}{|\rho_{2a}|^2|\rho_{12}|^2}
E^{(\nu,n)}(\rho_{1},\rho_{2})
}

The combination of a gradient and a Laplacian requires some more work.  Here
the gradient affects also the additional coordinate moduli
in~(\ref{eq:laplE_app}).  We use the product rule to obtain:
\dmath2{
\nabla_1\Delta_2 E^{(\nu,n)}(\rho_{1},\rho_{2})
&=&\nabla_1 \big((1+2\,i\nu)^2-n^2\big)
	\frac{|\rho_{1a}|^2}{|\rho_{2a}|^2|\rho_{12}|^2}
	E^{(\nu,n)}(\rho_{1},\rho_{2})
\cr
=\big((1+2\,i\nu)^2-n^2\big)&&\hskip -4.5mm\left(
2\left(\del^*_1\frac{|\rho_{1a}|^2}{|\rho_{2a}|^2|\rho_{12}|^2}\right)
+\frac{|\rho_{1a}|^2}{|\rho_{2a}|^2|\rho_{12}|^2}
(1-n+2\,i\nu)\frac{\rho^*_{2a}}{\rho^*_{1a}\rho^*_{12}}\right)
	E^{(\nu,n)}(\rho_{1},\rho_{2})\,.
}
The derivative of the first factor becomes:
$$
\del^*_1\frac{|\rho_{1a}|^2}{|\rho_{2a}|^2|\rho_{12}|^2}
=\frac{\rho_{1a}}{|\rho_{2a}|^2\rho_{12}}
\del^*_1\frac{\rho^*_{1a}}{\rho^*_{12}}
=\frac{\rho_{1a}}{|\rho_{2a}|^2\rho_{12}}
\left(\frac1{\rho^*_{12}}+\frac{-\rho^*_{1a}}{{\rho^*_{12}}^2}\right)
=\frac{-\rho_{1a}\rho^*_{2a}}{|\rho_{2a}|^2|\rho_{12}|^2\rho^*_{12}}
=\frac{|\rho_{1a}|^2}{|\rho_{2a}|^2|\rho_{12}|^2}
	\frac{-\rho^*_{2a}}{\rho^*_{1a}\rho^*_{12}}\,.
$$
This term can easily be added to the derivative of the second factor, giving
the constant prefactor $(-1-n+2\,i\nu)$.  After factorising the prefactor of
(\ref{eq:laplE_app}) and combining it with the new factor, we obtain the
result:
\begin{equation}
\nabla_1\Delta_2 E^{(\nu,n)}(\rho_{1},\rho_{2})
=-\big(4\,\nu^2+(n+1)^2\big)(1-n+2\,i\nu)
\frac{\rho^*_{2a}}{\rho^*_{1a}\rho^*_{12}}
\frac{|\rho_{1a}|^2}{|\rho_{2a}|^2|\rho_{12}|^2}
E^{(\nu,n)}(\rho_{1},\rho_{2}).
\end{equation}

The additional differentiation necessary to apply a double Laplacian to%
~$E^{(\nu,n)}$ is again independent of the one just performed and can be 
done analogously. The moduli of $\rho_{1a}$ and~$\rho_{2a}$ then cancel out. 
The result is:
\begin{equation}
\label{eq:LapLapE}
\Delta_1\Delta_2 E^{(\nu,n)}(\rho_{1},\rho_{2})
=\big(4\,\nu^2+(n+1)^2\big)\big(4\,\nu^2+(n-1)^2\big)
 \frac1{|\rho_{12}|^4} E^{(\nu,n)}(\rho_{1},\rho_{2})\,.
\end{equation}

\cleardoublepage

\cleardoublepage
\pagestyle{empty}

\chapter*{Thanks to$\,\ldots$}

First I would like to thank Hans G\"unter Dosch for accepting me as a PhD
student even though he was going to retire during the course of my work, for
supervising the first part of my PhD work and for always being available even
in retirement.

Second, but no less, thanks go to Carlo Ewerz for coming up with a very
interesting topic for the second part of my PhD work, for supervising and
supporting me during that part, for proof-reading this thesis, and for all the
jokes.

Furthermore I would like to thank Hans J\"urgen Pirner for agreeing to being my
co-supervisor.

Then a lot of thanks go to many more people which made working at our institute
a pleasure and/or helped me with my work through discussions and hints, in no
particular order and asking for leniency from anyone I might have forgotten:
Michael Doran, Frank Daniel Steffen, Otto Nachtmann, Timo Paulus, Arif Shoshi,
J\"org J\"ackel, Dietrich F\"othke, Felix Nagel, Michael Schmidt, Gregor
Sch\"afer, Claus Zahlten, Dennis Kostka, Dieter Gromes, Jan Schwindt, Lala
Adueva, Michael Thesen, Tobias Baier, Eduard Thommes, Florian Conrady, Tania
Robens, Matt Lilley, Kai Schwenzer, Xaver Schlagberger, Reimer K\"uhn, Eike
Bick,$\,\ldots$

Last but not least I want to thank Jochen Bartels for a brief but fruitful
discussion about the $1\to3$ pomeron vertex.

\def\thepage{} 

\cleardoublepage

\end{document}